%% file: thesis.tex
\pdfoutput=1
\documentclass[11pt,cm,twoside,dissertation,actual]{uhthesis2e}

\usepackage{amsmath} 
\usepackage{eurosym} 
\usepackage{amssymb} 
\usepackage{bm} 
\usepackage{graphicx} 
\usepackage{multicol} 
\usepackage{multirow} 
\usepackage{perpage} 
\usepackage{mcite} 
\usepackage{bbm} 
\usepackage{subfig} 
\usepackage{fancyvrb} 
\usepackage[number=none,toc=true]{glossary}

\MakePerPage[2]{footnote} 

\newcommand{\fsfI}{\widetilde{f}_1}
\newcommand{\fsfII}{\widetilde{f}_2}

\newcommand{\fsfl}{\widetilde{f}_L}
\newcommand{\fsfr}{\widetilde{f}_R}

\newcommand{\fstI}{\widetilde{t}_1}

\newcommand{\fscI}{\widetilde{c}_1}

\newcommand{\mgut}{M_{\mathrm{GUT}}}
\newcommand{\msusy}{M_{\mathrm{SUSY}}}
\newcommand{\boldy}{\mathbf{Y}^{}}
\newcommand{\boldyd}{{\mathbf{Y}}^{\dag}}
\newcommand{\tr}{\mathrm{Tr}}
\newcommand{\bbeta}{\bm{\beta}}
\newcommand{\bident}{\bm{\mathbbm{1}}}

\newcommand{\progrge}{\texttt{RGE\-FLAV}}
\newcommand{\progstd}{\texttt{STDEC}}
\newcommand{\progisa}{\texttt{ISA\-JET}}
\newcommand{\progisasug}{\texttt{ISA\-SUGRA}}
\newcommand{\inrge}{\texttt{INRGE.DAT}}
\newcommand{\fortran}{\texttt{FORTRAN}}

\def\dblone{\hbox{$\bm{1}\hskip -1.2pt\vrule depth 0pt height 1.6ex width 0.7pt
                  \vrule depth 0pt height 0.3pt width 0.12em$}}

\def\sn{s} 
\def\cs{c}

\newcommand{\by}{\mathbf{Y}}
\newcommand{\bu}{\mathbf{U}}
\newcommand{\bv}{\mathbf{V}}
\newcommand{\bw}{\mathbf{W}}
\newcommand{\bx}{\mathbf{X}}

\newcommand{\ba}{\mathbf{a}}

\newcommand{\bdf}{\mathbf{f}}
\newcommand{\bdl}{\bm{\lambda}}

\newcommand{\bft}{\tilde{\mathbf{f}}}
\newcommand{\bgt}{\tilde{\mathbf{g}}}
\newcommand{\bbet}{\bm{\beta}}
\newcommand{\bdm}{\mathbf{m}}
\newcommand{\bdLm}{\bm{\Lambda}}
\newcommand{\bmfx}{\left(\mathbf{m}_{X}-i\mathbf{m}'_{X}\right)}
\newcommand{\bmfxd}{\left(\mathbf{m}_{X}+i\mathbf{m}'_{X}\right)}
\newcommand{\bftuq}{(\tilde{\mathbf{f}}^Q_u)}
\newcommand{\bftuqnb}{\tilde{\mathbf{f}}^Q_u}
\newcommand{\bftur}{(\tilde{\mathbf{f}}^{u_R}_u)}
\newcommand{\bfturnb}{\tilde{\mathbf{f}}^{u_R}_u}
\newcommand{\bftdq}{(\tilde{\mathbf{f}}^Q_d)}
\newcommand{\bftdr}{(\tilde{\mathbf{f}}^{d_R}_d)}
\newcommand{\bftel}{(\tilde{\mathbf{f}}^L_e)}
\newcommand{\bfter}{(\tilde{\mathbf{f}}^{e_R}_e)}
\newcommand{\bgtpq}{(\tilde{\mathbf{g}}'^Q)}
\newcommand{\bgtpqnb}{\tilde{\mathbf{g}}'^Q}
\newcommand{\bgtpl}{(\tilde{\mathbf{g}}'^L)}
\newcommand{\bgtpur}{(\tilde{\mathbf{g}}'^{u_R})}
\newcommand{\bgtpurnb}{\tilde{\mathbf{g}}'^{u_R}}
\newcommand{\bgtpdr}{(\tilde{\mathbf{g}}'^{d_R})}
\newcommand{\bgtper}{(\tilde{\mathbf{g}}'^{e_R})}
\newcommand{\bgtq}{(\tilde{\mathbf{g}}^Q)}
\newcommand{\bgtqnb}{\tilde{\mathbf{g}}^Q}
\newcommand{\bgtl}{(\tilde{\mathbf{g}}^L)}
\newcommand{\bgtsq}{(\tilde{\mathbf{g}}_s^{Q})}

\newcommand{\bgtsur}{(\tilde{\mathbf{g}}_s^{u_R})}

\newcommand{\bgtsdr}{(\tilde{\mathbf{g}}_s^{d_R})}

\newcommand{\gtphu}{\tilde{g}'^{h_u}}
\newcommand{\gtphus}{(\tilde{g}'^{h_u})^{*}}
\newcommand{\gtphd}{\tilde{g}'^{h_d}}
\newcommand{\gtphds}{(\tilde{g}'^{h_d})^{*}}

\newcommand{\gthu}{\tilde{g}^{h_u}}
\newcommand{\gthd}{\tilde{g}^{h_d}}
\newcommand{\gthus}{(\tilde{g}^{h_u})^{*}}
\newcommand{\gthds}{(\tilde{g}^{h_d})^{*}}

\newcommand{\fuhu}{(\mathbf{f}_{u})}
\newcommand{\fuhut}{(\mathbf{f}_{u})^{T}}
\newcommand{\fuhus}{(\mathbf{f}_{u})^{*}}
\newcommand{\fdhd}{(\mathbf{f}_{d})}
\newcommand{\fdhdt}{(\mathbf{f}_{d})^{T}}
\newcommand{\fdhds}{(\mathbf{f}_{d})^{*}}
\newcommand{\fehd}{(\mathbf{f}_{e})}
\newcommand{\fehdt}{(\mathbf{f}_{e})^{T}}
\newcommand{\fehds}{(\mathbf{f}_{e})^{*}}
\newcommand{\tfuhut}{(\mathbf{f}^{h_{u}}_{u})^{T}}
\newcommand{\tfuhus}{(\mathbf{f}^{h_{u}}_{u})^{*}}
\newcommand{\tfdhdt}{(\mathbf{f}^{h_{d}}_{d})^{T}}
\newcommand{\tfdhds}{(\mathbf{f}^{h_{d}}_{d})^{*}}
\newcommand{\tfehdt}{(\mathbf{f}^{h_{d}}_{e})^{T}}
\newcommand{\tfehds}{(\mathbf{f}^{h_{d}}_{e})^{*}}
\newcommand{\fuul}{(\mathbf{f}_{u})}
\newcommand{\fuult}{(\mathbf{f}_{u})^{T}}
\newcommand{\fuuls}{(\mathbf{f}_{u})^{*}}
\newcommand{\fuur}{(\mathbf{f}_{u})}
\newcommand{\fuurt}{(\mathbf{f}_{u})^{T}}
\newcommand{\fuurs}{(\mathbf{f}_{u})^{*}}
\newcommand{\fudl}{(\mathbf{f}_{u})}

\newcommand{\fddl}{(\mathbf{f}_{d})}
\newcommand{\fddlt}{(\mathbf{f}_{d})^{T}}
\newcommand{\fddls}{(\mathbf{f}_{d})^{*}}
\newcommand{\fddr}{(\mathbf{f}_{d})}
\newcommand{\fddrs}{(\mathbf{f}_{d})^{*}}
\newcommand{\fddrt}{(\mathbf{f}_{d})^{T}}
\newcommand{\fdul}{(\mathbf{f}_{d})}

\newcommand{\feel}{(\mathbf{f}_{e})}
\newcommand{\feelt}{(\mathbf{f}_{e})^{T}}
\newcommand{\feels}{(\mathbf{f}_{e})^{*}}
\newcommand{\feer}{(\mathbf{f}_{e})}
\newcommand{\feers}{(\mathbf{f}_{e})^{*}}
\newcommand{\feert}{(\mathbf{f}_{e})^{T}}

\newcommand{\glp}{g'}
\newcommand{\gtwl}{g_{2}}
\newcommand{\gthl}{g_{3}}
\newcommand{\au}{(\mathbf{a}_{u})}
\newcommand{\aut}{(\mathbf{a}_{u})^{T}}
\newcommand{\aus}{(\mathbf{a}_{u})^{*}}
\newcommand{\ad}{(\mathbf{a}_{d})}
\newcommand{\adt}{(\mathbf{a}_{d})^{T}}
\newcommand{\ads}{(\mathbf{a}_{d})^{*}}
\newcommand{\bae}{(\mathbf{a}_{e})}
\newcommand{\baet}{(\mathbf{a}_{e})^{T}}
\newcommand{\baes}{(\mathbf{a}_{e})^{*}}

\newcommand{\mtsq}{\left|\tilde{\mu}\right|^{2}}

\newcommand{\mhusq}{m^{2}_{H_{u}}}
\newcommand{\mhdsq}{m^{2}_{H_{d}}}
\newcommand{\musq}{\left(\bdm^{2}_{U}\right)}
\newcommand{\mqsq}{\left(\bdm^{2}_{Q}\right)}
\newcommand{\mdsq}{\left(\bdm^{2}_{D}\right)}
\newcommand{\mlsq}{\left(\bdm^{2}_{L}\right)}
\newcommand{\mesq}{\left(\bdm^{2}_{E}\right)}
\newcommand{\h}{\theta_h}
\newcommand{\Hh}{\theta_H}
\newcommand{\Go}{\theta_{G^0}}

\newcommand{\Gp}{\theta_{G^+}}
\newcommand{\A}{\theta_{A}}

\newcommand{\Hp}{\theta_{H^+}}
\newcommand{\sh}{\theta_{\tilde{h}}}
\newcommand{\shram}{\theta_{\tilde{h}^{0}_{1}}}
\newcommand{\sH}{\theta_{\tilde{h}^{0}_{2}}}
\newcommand{\shc}{\theta_{\tilde{h}^{\pm}_1}}
\newcommand{\sHc}{\theta_{\tilde{h}^{\pm}_2}}
\newcommand{\sq}{\theta_{{\tilde{Q}_j}}}
\newcommand{\sqi}{\theta_{{\tilde{Q}_i}}}
\newcommand{\sqj}{\theta_{{\tilde{Q}_j}}}
\newcommand{\sqk}{\theta_{{\tilde{Q}_k}}}
\newcommand{\sql}{\theta_{{\tilde{Q}_l}}}
\newcommand{\su}{\theta_{{\tilde{u}_i}}}
\newcommand{\sui}{\theta_{{\tilde{u}_i}}}
\newcommand{\suj}{\theta_{{\tilde{u}_j}}}
\newcommand{\suk}{\theta_{{\tilde{u}_k}}}
\newcommand{\sul}{\theta_{{\tilde{u}_l}}}
\newcommand{\sdi}{\theta_{{\tilde{d}_i}}}
\newcommand{\sdj}{\theta_{{\tilde{d}_j}}}
\newcommand{\sdk}{\theta_{{\tilde{d}_k}}}
\newcommand{\sdl}{\theta_{{\tilde{d}_l}}}
\newcommand{\sLi}{\theta_{{\tilde{L}_i}}}
\newcommand{\sLj}{\theta_{{\tilde{L}_j}}}
\newcommand{\sll}{\theta_{{\tilde{L}_l}}}
\newcommand{\slk}{\theta_{{\tilde{L}_k}}}
\newcommand{\sei}{\theta_{{\tilde{e}_i}}}
\newcommand{\sej}{\theta_{{\tilde{e}_j}}}
\newcommand{\sek}{\theta_{{\tilde{e}_k}}}
\newcommand{\sel}{\theta_{{\tilde{e}_l}}}
\newcommand{\sbi}{\theta_{\tilde{B}}}
\newcommand{\swi}{\theta_{\tilde{W}}}
\newcommand{\sgl}{\theta_{\tilde{g}}}
\newcommand{\mgtphusq}{\left|\gtphu\right|^{2}}
\newcommand{\mgtphdsq}{\left|\gtphd\right|^{2}}
\newcommand{\mgthusq}{\left|\gthu\right|^{2}}
\newcommand{\mgthdsq}{\left|\gthd\right|^{2}}
\newcommand{\gev}{\mathrm{GeV}}
\newcommand{\tev}{\mathrm{TeV}}
\newcommand{\mtsfuhu}{\left(\tilde{\mu}^{*}\mathbf{f}^{h_{u}}_{u}\right)}
\newcommand{\mtsfdhd}{\left(\tilde{\mu}^{*}\mathbf{f}^{h_{d}}_{d}\right)}
\newcommand{\mtsfehd}{\left(\tilde{\mu}^{*}\mathbf{f}^{h_{d}}_{e}\right)}

\newcommand{\mtsfuhut}{\left(\tilde{\mu}^{*}\mathbf{f}^{h_{u}}_{u}\right)^{T}}
\newcommand{\mtsfdhdt}{\left(\tilde{\mu}^{*}\mathbf{f}^{h_{d}}_{d}\right)^{T}}
\newcommand{\mtsfehdt}{\left(\tilde{\mu}^{*}\mathbf{f}^{h_{d}}_{e}\right)^{T}}
\newcommand{\mtfuhus}{\left(\tilde{\mu}^{*}\mathbf{f}^{h_{u}}_{u}\right)^{*}}
\newcommand{\mtfdhds}{\left(\tilde{\mu}^{*}\mathbf{f}^{h_{d}}_{d}\right)^{*}}
\newcommand{\mtfehds}{\left(\tilde{\mu}^{*}\mathbf{f}^{h_{d}}_{e}\right)^{*}}
\newcommand{\msb}{$\overline{\mathrm{MS}}$}
\newcommand{\drb}{$\overline{\mathrm{DR}}$}
\newcommand{\mmsb}{\overline{\mathrm{MS}}}
\newcommand{\mdrb}{\overline{\mathrm{DR}}}
\newenvironment{itemise}{\begin{itemize}}{\end{itemize}}

\begin{document}

\title{Renormalisation Group Analysis of\\ Supersymmetric Particle Interactions}
\author{Andrew D. Box}
\degreemonth{December}
\degreeyear{2008}
\degree{Doctor of Philosophy}
\chair{Xerxes Tata}
\othermembers{
Sandip Pakvasa\\
Thomas Browder\\
John M. J. Madey\\
George Wilkens}
\numberofmembers{5}
\field{Physics}

\maketitle

\begin{frontmatter}

\signaturepage



\include{acknowledgements}

\include{abstract}

\tableofcontents

\listoftables

\listoffigures

\end{frontmatter}

\include{intro}

\include{susy}

\include{flavour}
\include{renorm}

\include{derivation}

\include{rgeflav}

\include{solution}

\include{stop}

\include{summary}

\appendix
\include{fullrges}

\include{drsmrge}

\include{smrgecheck}

\include{infile}
\include{orthfix}

\printglossary

\bibliographystyle{xtra/mythesis}
\bibliography{xtra/2bod,xtra/3bod,xtra/rge,xtra/susy,xtra/renorm,xtra/rgeflav,xtra/results,xtra/deriv,xtra/intro}
\newpage\pagestyle{empty}
\tiny$\dagger$
\end{document}

%% file: acknowledgements.tex
\begin{acknowledgements}

\noindent First of all I am most appreciative of the assistance I have received from my advisor, Xerxes Tata, whose tireless efforts, especially over the past six months, have been a key part of the completion of this work. In addition, the endless patience of my family, particularly my ex-fianc\'ee, have not gone without notice, and my brother's timely radio silence at a particularly difficult stage in his life has ensured that I will soon return to provide whatever help I can. It was necessary to change my committee at the last minute, and I am grateful to both David Bleecker and George Wilkens for their help in making the necessary changes. During the course of preparing our two papers, Xerxes and I had many fruitful conversations with H. Baer, D. Casta\~no, V. Cirigliano, A. Dedes, J. Ferrandis, S. Martin, K. Melnikov, A. Mustafayev and M. Vaughn. I am particularly thankful for Howie Baer's assistance and understanding, not only in connection with the workings of \progisa~but also on other matters. I hope this work will continue to be of some use after I have left Hawai`i.\\

\noindent Many of my close friends deserve a particular mention, although I only list here those who I believe have had a direct impact on this work: Assaf Azouri and Santa Moreno Gonzalez, Sandip Biswas, Jan Bruce, (Math) Matt Chasse, Eric Dodson, Peter, Sandy, Kaleb, Ailani and Nathan Grach, Mike and Sarah Hadmack, Diane Ibaraki, Stuart (Dolphin) Ibsen, Roger Kadala, Nick Kent, Margie Kidani, Jeremy and Emi (Big Bang Theory) Kowalczyk, Josie Nanao, Kurtis Nishimura, Jeff (yes, I'm done now) Perkins, Jamal Rorie, Stefanie Smith and John Montgomery, Kirika Uchida. Also, to all those from whom I copied my homework \mbox{---} thank you. Finally, a special mention should go to Kalpana and Kashmira Tata, who have been deprived of Xerxes during the weekends many times. I hope that any further disappearances will not be on my account.\\

\noindent It would be impossible to find a way to properly express my gratitude to all those who have helped me during the preparation of this work. You all have my prayers and I truly believe your kindness will be repaid in some way.

\end{acknowledgements}

%% file: abstract.tex
\begin{abstract}

\noindent In the Minimal Supersymmetric Standard Model (MSSM), there are numerous sources of flavour-violation in addition to the usual Kobayashi-Maskawa mixing matrix of the Standard Model. We reexamine the renormalisation group equations (RGEs) with a view to investigating flavour effects in a supersymmetric theory with an arbitrary flavour structure at some high scale. To incorporate (two-loop sized) threshold effects in the one-loop RGEs, we calculate the $\beta$-functions using a sequence of (non-supersymmetric) effective theories with heavy particles decoupled at the scale of their mass, keeping track of the fact that many couplings (such as gauge and gaugino couplings) which are equal in an exact supersymmetric theory may no longer be equal once the supersymmetry (SUSY) is broken. We find that this splitting, which is ignored in the literature, may be larger than two-loop terms that are included. In addition, gaugino couplings develop flavour structure that is absent without including decoupling effects. A program (to be incorporated into \progisa) has been developed, which includes flavour-violating couplings of superparticles and solves the two-loop threshold RGEs subject to specified high scale inputs. The weak scale flavour structure derived in this way can be applied to the study of flavour-changing decays of SUSY particles. As an illustration, we revisit the branching ratio of the flavour-violating decay of the top squark. We find that, in the minimal supergravity (mSUGRA) class of models, previous estimates for the width of this decay have been too large by a factor $10-25$. However, this decay rate is sensitive to the flavour structure of the high scale boundary conditions. We analyse the consequences of introducing non-universality in the high scale soft SUSY-breaking mass matrices and find that under these conditions the partial width can be altered by a large amount.

\end{abstract}

%% file: intro.tex
\chapter{Introduction}\label{ch:intro}

On the 10th September, 2008, the Conseil Europ\'{e}en pour la Recherche Nucl\'{e}aire (CERN)\glossary{name={CERN},description={European Organization for Nuclear Research}} began circulating a beam around the $27$~km diameter Large Hadron Collider (LHC), $175$~m below the border between Switzerland and France on the outskirts of Geneva  \cite{LHC}\glossary{name={LHC},description={Large Hadron Collider}}. Once fully commissioned, the LHC will collide beams of protons travelling at close to the speed of light (with $\gamma\simeq7400$) and analyse the products of these collisions.  At a total cost of around \officialeuro5~billion the experiments at the LHC will address some of the biggest questions of contemporary particle physics. Although a highly successful theory for almost 35 years, the Standard Model of particle physics (SM)\glossary{name={SM},description={Standard Model of particle physics}} (see, for example, \cite{HM,perk,ps}), which is the current state of the art, is generally recognised to be incomplete.

One potential discovery at the LHC is the Higgs boson, the elementary particle which has been postulated to give all other elementary particles their mass. Close study of the properties of the Higgs boson will give us some indication of the solutions to the inconsistencies of the SM, and it is the only particle in the SM that is yet to be observed experimentally. Additionally, the LHC will be looking for hints about the nature of dark matter \mbox{---} the unknown material that makes up over 80\% of the mass of the universe. Contrary to suggestions by the writers of \textit{Star Trek}, dark matter is not as obvious as a big purple cloud and has until now not been seen using standard optical techniques. We know, however, from studies of the dynamics of galactic rotation, that typical galaxies contain a substantial amount of dark matter out to large distances from the centre \cite{knapp}. The detectors at the LHC may be able to infer the presence of dark matter produced as a result of the particle collisions.

Returning to the problems suffered by the SM, one extremely popular solution involves introducing a novel new symmetry, or supersymmetry (SUSY)\glossary{name={SUSY},description={Supersymmetry}} \cite{wss,drees,*bin,martinprimer}, which links bosons and fermions, the fundamental groups to which all particles can be assigned. Since SUSY is a connection between these two groups, it involves introducing a bosonic partner for every SM fermion and vice-versa, thereby approximately doubling the total number of elementary particles. In addition, if we insist that SUSY is respected by all particle interactions, we find that the additional particles must have precisely the same mass, a fact not supported by experiment. Supersymmetry must therefore be broken in a way that we can introduce additional mass contributions for the SUSY partners without ruining the mechanisms which fix our SM blues (described later). When we break the symmetry in this manner, we say we have broken it \textit{softly}.

In the SM, the particle content of which is listed in Table~\ref{tab:SM}, the interactions available to the various matter particles are dictated by choosing certain symmetries that these interactions must obey. 
\begin{table} \centering
\includegraphics[viewport=120 100 690 500, clip, scale=0.5]{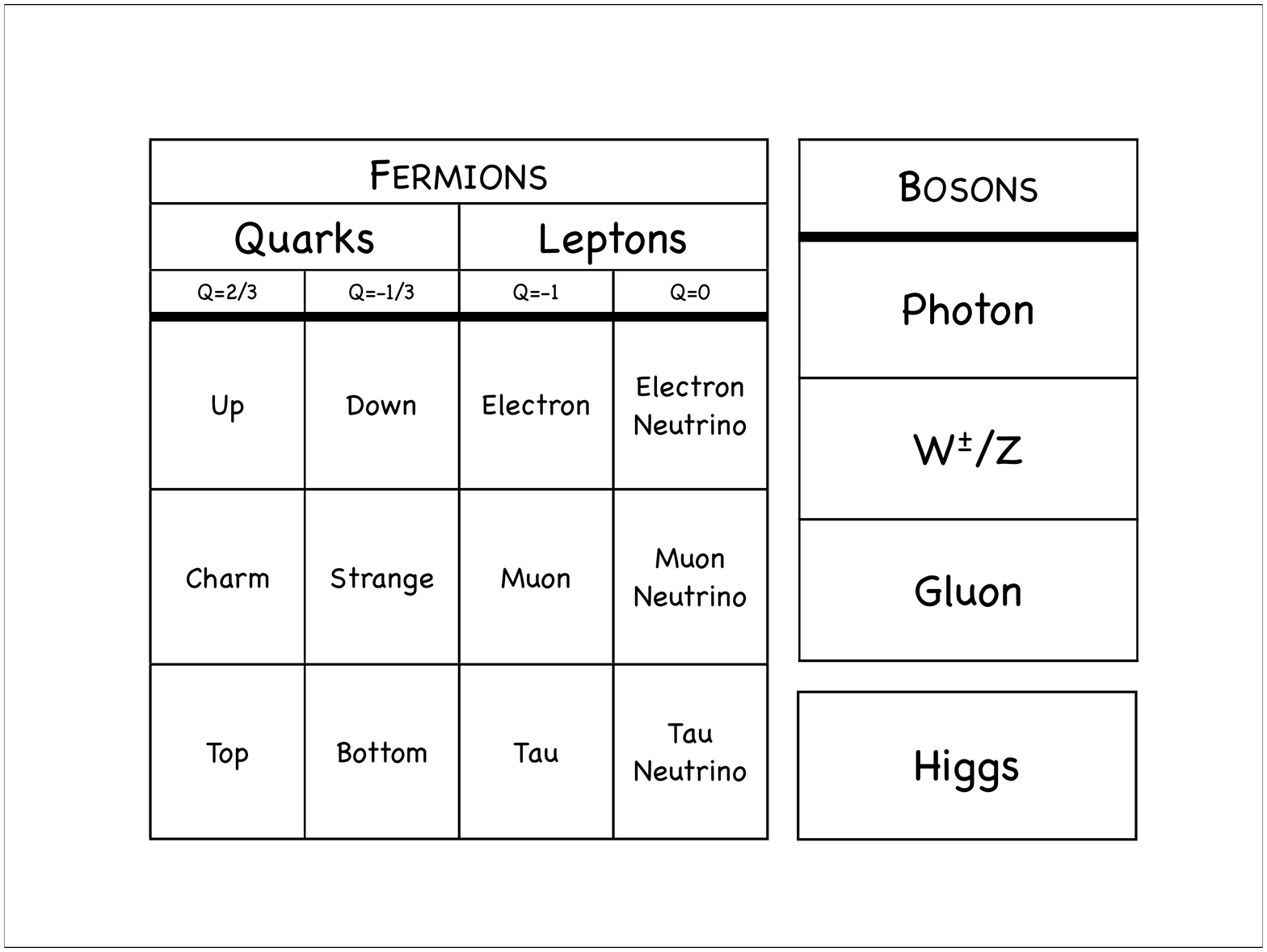} 
\caption[The elementary particles of the Standard Model.]{\small The elementary particles of the Standard Model, including matter fermions and gauge bosons. Also indicated is the charge of the matter fermions, $Q$.}\label{tab:SM}
\end{table}
There are three fundamental symmetries (or gauge groups) observed by the SM: colour (or $SU(3)$), weak isospin ($SU(2)$) and hypercharge ($U(1)$) \cite{ps}. The elementary particles can ``feel'' a force associated with each symmetry by interacting with the ``gauge bosons'': gluons (which carry colour), charged and neutral weak bosons and the photon. Each matter particle is introduced into the theory by specifying how it transforms under the various symmetries together with its couplings to the Higgs sector. The matter fermions are: leptons, which do not carry colour, and quarks, which interact with all three force carriers.

The various matter fermions (\textit{i.e.}, quarks and leptons) of the SM are collected into three ``generations'' corresponding to the three rows in the left-hand section of Table~\ref{tab:SM}. Within each generation of quarks there are two different ``flavours'' with differing charge. We find that fundamental (or tree-level) gauge interactions between quarks of different flavour are extremely restricted in the SM, with so-called ``flavour-changing'' interactions only taking place between quarks of different charge. Although this charged current flavour-changing interaction mainly takes place within generations, there is also a small coupling between quarks of different charge and generation. When we introduce SUSY, we find that there is a potential proliferation of flavour-changing interactions, and unless the coupling strengths are strongly suppressed such interactions would breach bounds for SM flavour-violation obtained by experiment \cite{gabb}. There must therefore be some mechanism by which SUSY contributions to SM flavour-violating processes are suppressed. Our aim is to investigate the flavour characteristics of SUSY theories in a general manner.

Interest in SUSY phenomenology was sparked once it was understood that its properties would allow meaningful predictions from theories at scales far in excess of those currently available (or likely to ever be available) in terrestrial experiments \cite{dim,*sak,*kaul}. SUSY can therefore provide us with a framework within which it is possible to obtain the low energy consequences of high scale inputs, or ans\"{a}tze. An early pointer to the possibility for a simplified high scale description of the rich structure of weak scale SUSY is the apparent unification of the three gauge couplings in the SM. Although they have very different values at low scales ($\sim100$~GeV), they are seen to come together \cite{amaldi} at approximately $10^{16}$~GeV if we introduce a SUSY theory\footnote{Since the masses of the SUSY particles are essentially unknown, the point at which SUSY influences the scale dependence of the couplings is not fixed. However, we have good reason to expect that SUSY particles reside in the region of $10^{3}$~GeV, as we shall see in the next chapter.} at around $10^{3}$~GeV.

Scale dependence appears in the couplings as a result of our renormalisation procedure. The tree-level interaction is augmented by higher-order corrections, coming from particles circulating in loops, and the parameters are ``renormalised'' by defining new parameters which are scale dependent. If we try to carry out perturbative calculations at the weak scale using parameters renormalised at a hierarchically larger scale, large logarithms will render the perturbative expansion invalid. Instead, we sum these large contributions using renormalisation group methods \cite{ps}. The couplings are said to `run' with scale and it is in this sense that we say the gauge couplings come to a point.

We can derive a system of coupled differential equations called the renormalisation group equations (RGEs)\glossary{name={RGE},description={Renormalisation Group Equation}} which predict how each Lagrangian parameter will depend on the scale. RGEs therefore play an important role in the study of the phenomenology of theories valid at high energy scales \cite{earlyrges}. Although softly broken weak scale SUSY contains a large number of parameters in addition to those of the SM, this is a result of our ignorance of the mechanism of SUSY breaking. Until this is understood, we resort to models of unified theories, where specific physics assumptions lead to models with more manageable numbers of independent parameters. The predictions these unified theories make for weak scale phenomenology are found by solving the RGEs to obtain the low energy interactions.

We are therefore in a position to consider various means of introducing flavour-violation at some specified high scale, using the RGEs to derive the weak scale parameters associated with our favourite theory. Once we have found the weak scale predictions of our high scale structure we will be able to investigate the implications for flavour-violation in the various decays and interactions of the theory. We must, however, remember that our boundary conditions will include not only the values of various unified SUSY couplings, but also the size of the weak scale SM interactions which reproduce (to a first approximation) the currently observed levels of flavour-violation.

In many models, the masses of SUSY particles are spread over a wide range. To account for the effect that this has on the form of the RGEs, we re-derive the RGEs for both the SM \cite{MVrgeI,*MVrgeII,*MVrgeIII,ACMPRWrge} and the Minimal Supersymmetric Standard Model (MSSM) \cite{MVrge,BBOrge,*Yam1,*Yam2,*Yam3,*JJ}, the SUSY extension which introduces the fewest additional particles and interactions. We construct a collection of effective theories with varying particle content depending on the scale at which they are valid. The high energy theory contains the same particle content as the MSSM, and we remove, or ``decouple'', particles from the effective theory at a series of thresholds until we obtain the SM as the effective theory valid when all SUSY particles are decoupled \cite{casno,sakis}.

We find that once we decouple at least one SUSY particle, the effective theory `knows' about the SUSY breaking so that couplings that were related in the supersymmetric limit are now distinct and have their own RGEs. As a result, the full system of RGEs is much larger than usually written in the literature. Specifically, the couplings of the gauginos (superpartners to the gauge bosons) to the fermion-sfermion system evolve differently from the gauge couplings, while the couplings of the higgsinos (superpartners to the Higgs bosons) evolve differently from the usual fermion ``Yukawa'' coupling. Indeed, gaugino couplings that were flavour-diagonal in the SUSY limit can now develop flavour-violating terms, which must be accounted for in the implementation of our decoupling procedure.

On completion of the program for a complete two-loop calculation of weak scale flavour structure, we will apply our results to a sample process \mbox{---} the decay of the lightest stop ($\tilde{t}_{1}$), a superpartner to the heaviest quark. Although the $\tilde{t}_{1}$ has interactions that would permit standard tree-level two-body decays, \textit{i.e.}, $\tilde{t}_{1}\rightarrow t\tilde{Z}_{i}$, where $\tilde{Z}_{i}$ is a neutralino (one of the additional SUSY fermions with zero charge), and $\tilde{t}_{1}\rightarrow b\tilde{W}_{i}$, where $\tilde{W}_{i}$ is a chargino (charged cousin of the neutralino), the stop may be lighter than the sum of the products, in which case such decays would be forbidden. However, if we allow for mixing between squarks of different flavours, the $\tilde{t}_{1}$ may be able to decay via its flavour-changing coupling to a charm quark and a neutralino. In this case the decay could compete with, or even exceed, the rate of other three-body \cite{wohr,*djouadi} (or four-body \cite{boehm}) decays, which are suppressed by phase space factors. 

\section*{Outline}

The next chapter outlines the various arguments for physics beyond the Standard Model. We construct the basic SUSY model, the MSSM\glossary{name={MSSM},description={Minimal Supersymmetric Standard Model}}, introducing its particle content and interactions. After discussing the breaking of both supersymmetry and the electroweak symmetry, we briefly mention efforts to detect SUSY particles in colliders and elsewhere.

In Chapter~\ref{ch:flavth} we write out our notation for the mixing of the quarks and their superpartners the squarks. This mixing is responsible for flavour-violating interactions and (for SM particles) experimentally well constrained. We also show the interactions responsible for squark mixing in the up-type sector. These relations will be necessary for our analysis of stop decay.

Chapter~\ref{ch:renorm} considers the general equations we will use to derive the RGEs of the MSSM with full thresholds. In this chapter we discuss our choice of renormalisation scheme, our way of writing the various mass eigenstates, and the consequences of allowing for broken SUSY in the RGEs.

The general relations from Chapter~\ref{ch:renorm} are applied to the MSSM in Chapter~\ref{ch:application}. We first write the interaction terms available in softly broken SUSY, and use them to derive samples of the full threshold RGEs, contained in full in Appendix~\ref{app:dlessRGEs} and Appendix~\ref{app:dfulRGEs}.

Chapter~\ref{ch:rgeflav} outlines our method for iteratively solving the RGEs, which was carried out by a \texttt{FORTRAN} code, \progrge. Along with a sample input file in Appendix~\ref{app:inrgeeg}, we write the general layout of the code, the specifics of the boundary conditions and the iterative procedure. Finally, we detail the expected output.

Once we have described the code used, Chapter~\ref{ch:results} presents some sample solutions for the various SM and SUSY couplings. We consider examples of all important SUSY couplings, and present some surprising consequences of a consistent treatment of flavour structure in the RGEs.

In Chapter~\ref{ch:stopdec}, the flavour-violating decay alluded to above is considered. We compare our result with previous literature and show some scenarios in which there are significant changes to the partial widths and branching ratios. This will serve as an example of the application of our work.

Finally, we summarise in Chapter~\ref{ch:summ}.

%% file: susy.tex
\chapter{Supersymmetry at the TeV Scale}\label{ch:SUSY}

The Standard Model of particle physics (see, for example, Ref.~\cite{HM}) is one of the most successful physical models to date in that its predictions have withstood experimental scrutiny over a wide range of energy scales. Right up to the present day \mbox{---} over 30 years since it appeared in its current form \mbox{---} it continues to withstand the assault of sustained investigation, being used as the yard-stick for experimental data.

Having said this, there are a number of cracks in its armour, which perhaps point to models with wider scope that include the SM as a low energy approximation. If we are to search for a model to replace the SM, we must ensure that the successes of the last 30 years are not ignored. A natural path to follow would therefore be to search for a new theory which contains the SM in the limit of validity of current experimental results. If we take this approach, there are a number of issues with the SM which may point us in the right direction:

\begin{enumerate}
\item Perhaps the most obvious limitation is the lack of an inclusion of gravitational effects. Since these effects become important when gravitational corrections begin to compete with other forces, conventional wisdom states that gravity need only be included at scales in the region of the reduced Planck mass, approximately $10^{18}~\gev$. Gravitational effects therefore continue to be of subordinate importance.\footnote{Many models have been developed \cite{hewett} which suggest this scale may be a gross overestimate, and that gravitational effects could become important at scales soon to be probed. In this respect, however, we will trust conventional wisdom.}
\item The Standard Model does not contain a candidate for the source of dark matter \mbox{---} the non-luminous matter responsible for a variety of astronomical effects including the anomalous rotation curves of most galaxies \cite{knapp}.
\item We have no guiding principle to inform our choice of gauge group and particle content, particle masses and mixings. All these `ingredients' of the SM must be chosen on an empirical basis.
\item The theory is not stable when we include scalar fields (which are part of the SM), and it is desirable to find some mechanism to prevent scalar masses from suffering corrections of the order of the Planck scale. This is the problem of fine-tuning within the SM.
\end{enumerate}
It is due to these issues that we can be certain of the existence of a new theory at energy scales we are yet to probe with experiment. This chapter puts forward some arguments for weak scale supersymmetry (SUSY) \cite{wss,drees,*bin,martinprimer} as an interesting candidate for a new theoretical framework that contains the SM at low energies. Sec.~\ref{sec:scale} contains a more detailed description of the fine-tuning problem and considers the scale at which SUSY may exist. This is followed by a discussion of why SUSY is considered by some to be a good candidate for the new physics.

\section{What is the Scale of the New Physics?}\label{sec:scale}

The key question that needs to be addressed when considering the presence of physics beyond the SM is at what scale the new physics becomes important. We know that no clear signature of new physics has appeared at currently probed scales (up to around a few hundred GeV), but the only other clear scale available is the Planck scale. Is there any indication of an intermediate scale? To answer this, we must take a closer look at the Higgs boson and the fine-tuning problem.

The Higgs boson, the scalar particle which is introduced to give mass to the other fundamental particles, is the final piece in the puzzle for the SM. Current experimental data \cite{lepwg,*kumar} suggests that the SM Higgs boson mass must lie somewhere between $100-150~\gev$, and it is widely believed to be within easy reach of the LHC \cite{LHC}, which is scheduled to resume operation at CERN early next year.

When we consider the radiative corrections to the Higgs boson propagator which arise from higher order contributions in perturbation theory, we find a number of quadratically divergent pieces coming from both boson and fermion loops. The computation of such contributions involves integrating the momentum in the loop up to a cutoff value, $\Lambda$, which can be interpreted as the scale at which the SM is no longer valid and where new physics is expected to intervene. The physical Higgs boson mass can therefore be expressed as the sum of the mass parameter in the Lagrangian, $m_{H_{0}}$, and the loop contributions, which enter with a coefficient, $c\sim1$,
\begin{equation}
m^{2}_{H}=m^{2}_{H_{0}}+\frac{c}{16\pi^{2}}\Lambda^{2}\;.
\end{equation}
Since the loop contributions are quadratically divergent, the corrections to the squared mass of the Higgs boson depend on the square of the cutoff, and therefore, without any new physics below the Planck scale, these corrections would be roughly 30 orders of magnitude larger than the squared mass itself.

Many theorists are concerned about the extremely fine cancellation required between the separate terms in the sum over contributing processes. In order to allay these concerns and make the mass stable, we would like the corrections to the mass of the Higgs boson to be of the same order as the mass itself, in other words $\Lambda\sim$~TeV. This is an extremely exciting suggestion, since we would therefore expect the LHC to not only find the Higgs boson, but also begin probing the scale at which we hope to find deviations from the predictions of the SM.

\section{Why SUSY is a Good Candidate}

We can choose to continue in a number of directions, explaining away the fine-tuning problem with various arguments. Some possibilities are:
\begin{enumerate}
\item \textbf{The Planck scale is not the scale at which gravity becomes important}. If there exist extra spatially compact dimensions in which gravitational effects can propagate \cite{hewett}, we would not observe this propagation at large distances, and gravity would appear weaker than it really is. Assuming the effects of the extra dimensions start becoming apparent at the TeV scale, this would provide the cut-off we are looking for, and we would need to incorporate gravitational effects into our theory. A scary prospect.
\item \textbf{There is no Higgs particle}. The Higgs particle is still to be discovered experimentally, so perhaps there is another mechanism by which the fundamental particles acquire mass. An example of a theory where this occurs is technicolour \cite{fahri,*lane}, in which the job of the Higgs particle is carried out by composites of a new set of fermions. Unfortunately, technicolour runs into problems since it appears to require more and more complicated structures to avoid issues with unreasonably large flavour-changing processes. The models become increasingly cumbersome and often conflict with precision electroweak data. At any rate, since the scale of the technicolour composites is $500-1000$~GeV, these new technicolour particles  should be within reach of the LHC.
\item \textbf{The world is finely tuned}. The cancellations in the contributions to the Higgs boson mass are exact to an extremely high precision to all orders in perturbation theory.
\end{enumerate}
If we decide we do not like any of these suggestions, we could try instead to find a way to make the quadratic divergences cancel for a symmetry reason, and this is the approach which leads to a discussion of supersymmetry.

When we separate out the divergent one-loop contributions to the Higgs boson mass and look at each term individually, we find that the bosonic loop contributions always appear with the opposite sign to the fermionic contributions. This is a general feature and the reason why SUSY is so interesting. If we arrange for our Lagrangian to obey a new symmetry, where the contributions from fermions and bosons are related, we can take advantage of the opposite signs to ensure that the quadratic divergences cancel \textit{to all orders in perturbation theory} \cite{dim,*sak,*kaul}. This cancellation of divergences is similar to the mechanism observed in the massless limit of Quantum Electrodynamics (QED), where large radiative corrections to the fermion mass are forbidden by chiral symmetry.

The essence of supersymmetry, then, is a connection between bosons and fermions. We introduce a transformation which connects the two states, and a new operator, $Q$, which generates the transformations, such that
\begin{equation}
Q\left|\mathrm{boson}\right>=\left|\mathrm{fermion}\right>\qquad\mathrm{and}\qquad Q\left|\mathrm{fermion}\right>=\left|\mathrm{boson}\right>\;.
\end{equation}
The operator Q is fermionic and therefore carries half-integer spin. We continue by building our theory out of supermultiplets, irreducible representations of the supersymmetry algebra. These supermultiplets contain both bosons and fermions, which are referred to as superpartners. When constructing a supersymmetric theory with the SM as a low energy effective theory, we immediately encounter the problem that there are no observed superpartners to ordinary matter. This leads us to roughly double the number of particles so that each SM particle gains a superpartner and as we build our theory we must be mindful of the fact that we need to arrange for these superpartners to be unobserved to date.

In addition to providing an elegant solution to the fine-tuning problem, there are a number of other issues which may be eased in a supersymmetric universe. An almost accidental consequence of the introduction of SUSY at approximately $1~\tev$ is to point towards some kind of high scale unification of the SM gauge groups. In some SUSY models various SUSY parameters are also unified at this scale, which may be an indication of the type of physics beyond our simple description of $\tev$ scale SUSY.

\section{The Minimal Supersymmetric Model}\label{sec:MSSMtheory}

Once we decide to pursue supersymmetry as the solution to some of our Standard Model problems, we may legitimately ask what a supersymmetrized SM would look like. To construct such a model we must promote our fields to superfields (indicated by a caret, as in $\hat{\mathcal{S}}$), promote our particle multiplets to particle supermultiplets, and introduce soft SUSY-breaking terms allowed by the symmetry. The resulting theory, known as the Minimal Supersymmetric Model (MSSM) \cite{wss}, represents the TeV scale limit of our supersymmetric universe, and is minimal in the sense that it contains the smallest possible number of additional particle states and interactions while keeping the SM as the low energy limit.

\subsection{Particle Content}

In much the same way as we choose the gauge groups of the SM, to construct a supersymmetric model we need to choose a group for our supermultiplets. So that we can step directly from the SM to a supersymmetric equivalent, this is chosen to be the same as the SM, namely $SU(3)_{C}\times SU(2)_{L}\times U(1)_{Y}$. This will fix the gauge supermultiplet to contain both the expected gauge bosons from the Standard Model in addition to their superpartners, the gauginos.

We then promote the SM fermion fields to superfields \mbox{---} one superfield for each chirality of SM fermion \mbox{---} and use the charge conjugates of the right-handed fermion fields so that all superfields are left-chiral. We now have spin 0 `sfermion' partners for each chirality state of each fermion, so we give them a subscript `L' or `R' to indicate which fermion they are partnered with. It is important to remember that since the sfermions are scalar particles they do not have chirality, and these subscripts are just labels to specify the fermion with which they are connected.

The final superfields we need to introduce are for the Higgs sector of the theory. We include a doublet Higgs superfield, $\hat{H}_{u}$, which contains the usual Higgs doublet and a spin $1/2$ doublet of higgsinos. We assign it weak hypercharge $Y=1$ and stipulate that it must transform as a $\mathbf{2}$ under $SU(2)_{L}$. In the SM, the Higgs doublet would be used to give mass via Yukawa interactions to the up-type quarks and its complex conjugate would have Yukawa interactions which would give mass to the down-type quarks and the charged leptons. In supersymmetry, such Yukawa interactions are specified by the so-called superpotential, $\hat{f}$, which contains the interactions of various chiral superfields. When we choose a form for the superpotential, supersymmetry prevents us from using both a field and its complex conjugate. For this reason, the SM Higgs mechanism cannot be precisely copied. Instead, we must introduce an additional Higgs doublet, called $\hat{H}_{d}$, which transforms as a $\mathbf{2}^{*}$ under $SU(2)_{L}$ and is assigned weak hypercharge $Y=-1$. We can now see that the labels on each Higgs superfield have been chosen to remind us of their role in the superpotential.

The final choices for the matter and Higgs content of the MSSM are shown in Table~\ref{tab:MSSMcont}.
\begin{table}
\begin{center}
\begin{tabular}{|c|cc|ccc|}\hline
\multicolumn{1}{|c}{} & \multicolumn{2}{|c|}{Content} & \multicolumn{3}{c|}{} \\
Superfield & Spin $1/2$ & Spin $0$ & $SU(3)_{C}$ & $SU(2)_{L}$ & $U(1)_{Y}$ \\\hline\hline
&&&&&\\
$\hat{Q}$ & $\left(\begin{array}{c}u_{L}\\[-5pt]d_{L}\end{array}\right)$ & $\left(\begin{array}{c}\tilde{u}_{L}\\[-5pt]\tilde{d}_{L}\end{array}\right)$ & $\mathbf{3}$ & $\mathbf{2}$ & $\frac{1}{3}$ \\
&&&&&\\
$\hat{U}^{c}$ & $\bar{u}_{R}$ & $\tilde{u}^{c}_{R}$ & $\mathbf{3}^{*}$ & $\mathbf{1}$ & $-\frac{4}{3}$ \\
&&&&&\\
$\hat{D}^{c}$ & $\bar{d}_{R}$ & $\tilde{d}^{c}_{R}$ & $\mathbf{3}^{*}$ & $\mathbf{1}$ & $\frac{2}{3}$ \\
&&&&&\\
$\hat{L}$ & $\left(\begin{array}{c}\nu_{eL}\\[-5pt]e_{L}\end{array}\right)$ & $\left(\begin{array}{c}\tilde{\nu}_{eL}\\[-5pt]\tilde{e}_{L}\end{array}\right)$ & $\mathbf{1}$ & $\mathbf{2}$ & $-1$ \\
&&&&&\\
$\hat{E}^{c}$ & $\bar{e}_{R}$ & $\tilde{e}^{c}_{R}$ & $\mathbf{1}$ & $\mathbf{1}$ & $2$ \\
&&&&&\\
$\hat{H}_{u}$ & $\left(\begin{array}{c}\Psi_{h^{+}_{u}}\\[-5pt]\Psi_{h^{0}_{u}}\end{array}\right)$ & $\left(\begin{array}{c}h^{+}_{u}\\[-5pt]h^{0}_{u}\end{array}\right)$ & $\mathbf{1}$ & $\mathbf{2}$ & $1$ \\
&&&&&\\
$\hat{H}_{d}$ & $\left(\begin{array}{c}\Psi_{h^{-}_{d}}\\[-5pt]\Psi_{h^{0}_{d}}\end{array}\right)$ & $\left(\begin{array}{c}h^{-}_{d}\\[-5pt]h^{0}_{d}\end{array}\right)$ & $\mathbf{1}$ & $\mathbf{2}^{*}$ & $-1$ \\\hline
\end{tabular}
\end{center}
\caption[The chiral superfields of the MSSM, with particle labels and gauge transformation properties.]{\small The chiral superfields of the MSSM, with particle labels and gauge transformation properties. Of course, for each quark and lepton there are three generations, only one of which is displayed.}\label{tab:MSSMcont}
\end{table}

\subsection{Interactions}

Taking the interaction of the particles with gauge bosons to be given by the minimal coupling prescription, our general SUSY Lagrangian is \cite{wss}:
\begin{equation}\label{eq:genlag}
\begin{split}
\mathcal{L}=&\sum_{i}(D_{\mu}\mathcal{S}_{i})^{\dagger}(D^{\mu}\mathcal{S}_{i})+\frac{i}{2}\sum_{i}\bar{\psi}_{i}\gamma_{\mu}D^{\mu}\psi_{i}+\sum_{\alpha,A}\left[\frac{i}{2}\bar{\lambda}_{\alpha A}\left(\gamma_{\mu}D^{\mu}\lambda\right)_{\alpha A}-\frac{1}{4}F_{\mu\nu\alpha A}F^{\mu\nu}_{\alpha A}\right]\\
&-\sqrt{2}\sum_{i,\alpha,A}\left(\mathcal{S}^{\dagger}_{i}g_{\alpha}t_{\alpha A}\bar{\lambda}_{\alpha A}\frac{1-\gamma_{5}}{2}\psi_{i}+\mathrm{h.c.}\right)\\
&-\frac{1}{2}\sum_{\alpha,A}\left[\sum_{i}\mathcal{S}^{\dagger}_{i}g_{\alpha}t_{\alpha A}\mathcal{S}_{i}+\xi_{\alpha A}\right]^{2}-\sum_{i}\left|\frac{\partial\hat{f}}{\partial\hat{\mathcal{S}}_{i}}\right|^{2}_{\hat{\mathcal{S}}=\mathcal{S}}\\
&-\frac{1}{2}\sum_{i,j}\bar{\psi}_{i}\left[\left(\frac{\partial^{2}\hat{f}}{\partial\hat{\mathcal{S}}_{i}\partial\hat{\mathcal{S}}_{j}}\right)_{\hat{\mathcal{S}}=\mathcal{S}}\frac{1-\gamma_{5}}{2}+\left(\frac{\partial^{2}\hat{f}}{\partial\hat{\mathcal{S}}_{i}\partial\hat{\mathcal{S}}_{j}}\right)^{\dagger}_{\hat{\mathcal{S}}=\mathcal{S}}\frac{1+\gamma_{5}}{2}\right]\psi_{j}\;,
\end{split}
\end{equation}
where $D^{\mu}$ is the appropriate gauge covariant derivative, $\mathcal{S}_{i}$ and $\psi_{i}$ are, respectively, the scalar and fermion components of the left-chiral superfield $\hat{\mathcal{S}_{i}}$. In addition, for each gauge group $\alpha$ with corresponding gauge index $A$, $F_{\mu\nu\alpha A}$ is the gauge field, $\lambda_{\alpha A}$ are the superpartners of the gauge bosons, $g_{\alpha}$ is the coupling strength, $t_{\alpha A}$ are the group generators, and $\xi_{\alpha A}$ are constants which are non-zero only for $U(1)$ gauge groups. The coupling strengths are: $g'$ for $U(1)$; $g$ for $SU(2)$; $g_{s}$ for $SU(3)$. In essence, \eqref{eq:genlag} is a master formula, representing the complete Lagrangian for renormalisable, supersymmetric gauge theories.

The first line of \eqref{eq:genlag} lists the expected kinetic energies with appropriate covariant derivatives. The next line contains the interaction of gauginos with the scalar and fermion components of the superfields. It describes how gauginos couple matter fermions to their superpartners, and Higgs scalars to higgsinos. The terms on the third line describe scalar quartic interactions, of which there are a large number due to the square occurring outside the sum on the first term. They will be important for the Higgs potential as well as describing the interactions between squarks and sleptons. The first set of terms are called D-terms, and the second set, which arise from the superpotential, are called F-terms. The final line of \eqref{eq:genlag} contains the interactions of a scalar with two fermions. These  give rise to the Yukawa terms of the SM quarks, and additional higgsino-sfermion-fermion ``Yukawa'' terms.

The next step is to choose a superpotential for the MSSM, which describes the interactions between the chiral superfields and, via the last line of \eqref{eq:genlag}, will contain the Yukawa terms for quarks. We must ensure that the operators we choose result in renormalisable interactions and obey all the gauge symmetries of the SM in addition to supersymmetry. Before writing down the most general superpotential possible, we will add an additional constraint, not motivated by supersymmetry, which is that all operators must conserve baryon number ($B$) and lepton number ($L$). Although this is achieved for free in the SM, there are valid renormalisable operators which conserve supersymmetry and would be included in the most general superpotential possible, but could give rise to, for example, a proton decay rate significantly higher than current experimental bounds.

One method for justifying the absence of these operators from the superpotential would be to introduce a new parity that forbids them. This parity can be defined (using $s$ for the spin of the field) as
\begin{equation}
R=(-1)^{3(B-L)+2s}\;,
\end{equation}
and is commonly referred to as $R$-parity \cite{farrar}. Under this symmetry, all SM particles are $R$-parity even, and all SUSY partners are $R$-parity odd. Note that although requiring $R$-parity conservation means that all allowed renormalisable operators will conserve $B$ and $L$, it is possible to construct $R$-parity conserving \textit{non-renormalisable} operators which do not conserve $B$ and $L$.

The introduction of $R$-parity conservation has profound implications for SUSY phenomenology. Namely, that SUSY particles must always be produced from SM collisions in pairs, that the decay state of a SUSY particle always contains at least one SUSY particle, and that the lightest SUSY particle is stable. This final implication, that there is an $R$-odd particle which is stable with a mass at the weak scale, may turn out particularly useful, as we shall see in Sec.~\ref{sec:experiment}.

The superpotential for the MSSM with R-parity conservation is:
\begin{equation}\label{eq:super}
\hat{f}=\mu\hat{H}^a_u\hat{H}_{da}+(\bdf_u)_{ij}\epsilon_{ab}\hat{Q}^a_i\hat{H}^b_u\hat{U}^c_j+(\bdf_d)_{ij}\hat{Q}^a_i\hat{H}_{da}\hat{D}^c_j+(\bdf_e)_{ij}\hat{L}^a_i\hat{H}_{da}\hat{E}^c_j\;,
\end{equation}
which follows the notation of \cite{wss}, using the superfields listed in Table~\ref{tab:MSSMcont}. The couplings, $(\bdf_{u,d,e})_{ij}$, are $(3\times3)$ Yukawa coupling matrices (where $i,j$ label the matter fermion or sfermion flavour) and $\mu$ is the Higgs self coupling. The indices $a,b$ are $SU(2)$ doublet labels and the matrix $\epsilon_{ab}$ is the completely antisymmetric tensor with $\epsilon_{12}=1$. This tensor is required in order that we combine the doublets $\hat{Q}$ and $\hat{H}_{u}$, which each transform as a $\mathbf{2}$ under $SU(2)$, in an $SU(2)$ invariant manner, and will introduce certain minus signs into the resulting Lagrangian.

Now that we have specified our particle content, supersymmetric interaction Lagrangian and superpotential, we can turn to the question of why SUSY particles have not yet been detected.
 
\subsection{Supersymmetry Breaking}

In a world where supersymmetry is an exact symmetry of nature, the superpotential given in \eqref{eq:super}, combined with the general Lagrangian for SUSY, \eqref{eq:genlag}, would be all we need to find the masses and interactions of the various particles in the theory. We would be able to calculate all tree-level processes and decays. Unfortunately, the Lagrangian we have constructed, being completely supersymmetric, describes SUSY particles which have precisely the same mass as their SM counterparts.\footnote{In this instance the matter fermions and their superpartners would be massless, but it is possible to construct extensions to the MSSM with exact SUSY in which they have non-zero masses. For more information, see the exercise on p.160 of Ref.~\cite{wss}.} If we lived in this exact SUSY universe, we would know about it already from experiment.

We are thus led to the conclusion that our exact SUSY Lagrangian must be augmented by terms which are not invariant under the SUSY transformations. In adding these terms, we must be careful not to lose the cancellation of divergences which was our initial motivation. Luckily, when the full analysis in carried out, we find that there are certain terms which result in mass differences between SM particles and their partners without introducing new quadratic divergences. These terms only introduce logarithmic divergences that are therefore not sufficiently large to resurrect the fine-tuning problem. If we add terms to the master formula which break the supersymmetry without destroying the cancellation of quadratic divergences, we are said to have broken SUSY \textit{softly}. In the MSSM, we construct a list of all such soft terms (soft masses as well as additional soft interactions) consistent with the symmetries of the underlying theory and add these to the rest of the Lagrangian \cite{girar}. Using, again, the fields listed in Table~\ref{tab:MSSMcont}, the soft supersymmetry-breaking (SSB)\glossary{name={SSB},description={Soft SUSY-Breaking}} terms are:
\begin{equation} \label{eq:soft}\begin{split}
\mathcal{L}_{\rm SSB}=\ &-\left\{\tilde{u}^\dagger_{Lk}(\mathbf{m}^2_{Q})_{kl}\tilde{u}_{Ll}
+\tilde{d}^\dagger_{Lk}(\mathbf{m}^2_{Q})_{kl}\tilde{d}_{Ll}
+\tilde{u}^\dagger_{Rk}(\mathbf{m}^2_{U})_{kl}\tilde{u}_{Rl}
+\tilde{d}^\dagger_{Rk}(\mathbf{m}^2_{D})_{kl}\tilde{d}_{Rl}\right.\\
&\quad\ +\tilde{\nu}^\dagger_{Lk}(\mathbf{m}^2_{L})_{kl}\tilde{\nu}_{Ll}
+\tilde{e}^\dagger_{Lk}(\mathbf{m}^2_{L})_{kl}\tilde{e}_{Ll}
+\tilde{e}^\dagger_{Rk}(\mathbf{m}^2_{E})_{kl}\tilde{e}_{Rl}\\[5pt]
&\left.\quad\ +m^2_{H_u}h^{0\dagger}_uh^0_u+m^2_{H_u}h^{+\dagger}_uh^+_u
+m^2_{H_d}h^{0\dagger}_dh^0_d+m^2_{H_d}h^{-\dagger}_dh^-_d\right\}\\[5pt]
&+\left\{\tilde{u}^\dagger_{Rk}(\mathbf{a}^T_u)_{kl}\tilde{u}_{Ll}h^0_u
-\tilde{u}^\dagger_{Rk}(\mathbf{a}^T_u)_{kl}\tilde{d}_{Ll}h^+_u
+\tilde{d}^\dagger_{Rk}(\mathbf{a}^T_d)_{kl}\tilde{d}_{Ll}h^0_d\right.\\
&\left.\quad\ +\tilde{d}^\dagger_{Rk}(\mathbf{a}^T_d)_{kl}\tilde{u}_{Ll}h^-_d
+\tilde{e}^\dagger_{Rk}(\mathbf{a}^T_e)_{kl}\tilde{\nu}_{Ll}h^-_d
+\tilde{e}^\dagger_{Rk}(\mathbf{a}^T_e)_{kl}\tilde{e}_{Ll}h^0_d+\mathrm{h.c.}\right\}\\[5pt]
&-\frac{1}{2}\left[M_1\bar{\lambda}_0\lambda_0+M_2\bar{\lambda}_p\lambda_p+M_3\bar{\tilde{g}}_A\tilde{g}_A\right]\\[5pt]
&-\frac{i}{2}\left[M'_1\bar{\lambda}_0\gamma_{5}\lambda_0+M'_2\bar{\lambda}_p\gamma_{5}\lambda_p+M'_3\bar{\tilde{g}}_A\gamma_{5}\tilde{g}_A\right]\\[5pt]
&+\left\{b(h^+_uh^-_d+h^0_uh^0_d)+\mathrm{h.c.}\right\},
\end{split}\end{equation}
where, in this instance, the roman letters $k,l$ are the flavour indices, and $p$ and $A$ are gauge group indices.
The Hermitian matrices $\mathbf{m}^{2}_{Q}$, $\mathbf{m}^{2}_{U}$, $\mathbf{m}^{2}_{D}$, $\mathbf{m}^{2}_{L}$ and $\mathbf{m}^{2}_{E}$  are labelled according to the scalar in the interaction and are called soft sfermion mass terms. The soft Higgs boson mass terms, $m^{2}_{H_u}$ and $m^{2}_{H_d}$, and the six gaugino mass parameters $M_{a}$, $M'_{a}$, are all real while the bilinear term, $b$, is complex. The trilinear terms, $\mathbf{a}^T_u$, $\mathbf{a}^T_d$, $\mathbf{a}^T_e$, are also matrices in flavour space. Their elements are frequently written as $\mathbf{A}^{T}_{kl}\mathbf{f}^{T}_{kl}$ which is especially useful in the context of models of the SUSY breaking mechanism. We have ignored additional trilinears that couple via the conjugate Higgs field and are usually given a coupling $\mathbf{c}_{kl}$ (\textit{e.g.},~$\tilde{d}^\dagger_{Rk}(\mathbf{c}^T_d)_{kl}\tilde{d}_{Ll}h^{*0}_u$). Although there is no principle forbidding the inclusion of these terms, they are negligible in many models.

The addition of a collection of SSB terms to the Lagrangian is a useful way of parametrizing the means by which the SUSY breaking is communicated to the superpartners, which is as yet unknown. There is a range of models that attempt to shed light on this mechanism \cite{mSUGRA,GMSB,AMSB,gaugino}, that typically involve the transmission of the breaking from a so-called ``hidden sector''. Such a model leads to a characteristic prediction for the pattern of SSB parameters in the Lagrangian renormalised at some high scale, $Q_{high}$.

One such model, which we will refer to regularly throughout this dissertation, is the minimal supergravity model (mSUGRA)\glossary{name={mSUGRA},description={Minimal Supergravity}} \cite{mSUGRA}. Within the framework of mSUGRA it is possible to obtain a simple set of `boundary conditions' at a so-called grand unification (GUT)\glossary{name={GUT},description={Grand Unified Theory}} scale, $\mgut\sim10^{16}~\gev$, defined to be the scale at which the running gauge couplings unify. We make the assumption that there is no new physics other than the MSSM between the weak scale and $\mgut$, and furthermore that the GUT scale boundary conditions consist of a small set of unified parameters. These include: a single value for the three gauge couplings, a unified gaugino mass $m_{1/2}$, and unified soft SUSY breaking scalar masses $m_{0}$ and trilinear parameters $A_{0}$. In addition, we require two parameters from the Higgs sector, and we shall see shortly that when we break the electroweak symmetry using the MSSM version of the Higgs mechanism we obtain relations between the various terms. We will be able to write $b$ in terms of $\tan{\beta}$, the ratio of the two Higgs field vacuum expectation values, $v_{u}/v_{d}$, and will also be able to fix $\mu^{2}$ so as to obtain the observed value of $M_{Z}$. Since fixing $\mu^{2}$ doesn't fix the overall sign of $\mu$, mSUGRA is therefore completely specified by the parameters
\begin{equation}\label{eq:mSUGRAin}
m_{0},\;m_{1/2},\;A_{0},\;\tan{\beta},\;\mathrm{sign}(\mu)\;.
\end{equation}

Notice that we have said nothing about the flavour structure of many of the couplings. Since we have a number of scalars with the same gauge quantum numbers, SUSY potentially contains many new sources of flavour-violation \cite{dim,*sak,*kaul,duncan} which can exceed experimental bounds. All the sfermion mass matrices and $\mathbf{a}$-parameters are matrices in flavour space and therefore it is expedient to find a procedure which will enable an estimation of their off-diagonal entries given our general high scale inputs from the various models. In mSUGRA, all off-diagonal entries in the soft masses are set to zero\footnote{Explicitly writing the flavour structure of the mSUGRA conditions we therefore have $\bdm^{2}_{Q,U,D,L,E}=m_{0}\dblone$ and $\ba_{u,d,e}=A_{0}\bdf_{u,d,e}$.} at $\mgut$, but this does not imply that they are also zero at the weak scale. The procedure we use to obtain phenomenological predictions at the weak scale, is discussed in more detail in Chapter~\ref{ch:renorm}.

Once we have used this procedure to obtain the weak scale Lagrangian, we will be in a position to find the states with definite mass. As in the SM, these physical states are not necessarily the same as the current eigenstates. We will describe the fermionic states formed by mixing between the charged winos and higgsinos as ``charginos'', $\tilde{W}_{i}$, where $i=1$ is the lighter and $i=2$ the heavier state. Similarly, we will describe the states formed by mixing between the bino, the neutral wino and neutral higgsinos as ``neutralinos'', $\tilde{Z}_{i}$, where the states are labelled $i=1,\dots4$. Squarks and sleptons that share the same quantum numbers will also mix, forming two eigenstates of each type and flavour. If we allow for an arbitrary flavour structure in all MSSM parameters, these flavour eigenstates will be a mixture of both $\tilde{f}_{L}$ and $\tilde{f}_{R}$ `chirality' states, and all three current flavour eigenstates, leading to the possibility for mixing between a total of six matter sfermions. For example, the lighter of the top squark mass eigenstates (labelled with a $1$ in analogy with the charginos and neutralinos) will be formed from the current eigenstates in the following way:
\begin{equation}\label{eq:t1state}
\tilde{t}_{1}=a_{1}\tilde{u}_{L}+a_{2}\tilde{c}_{L}+a_{3}\tilde{t}_{L}+a_{4}\tilde{u}_{R}+a_{5}\tilde{c}_{R}+a_{6}\tilde{t}_{R}\;.
\end{equation}
The method for constructing this eigenstate from the Lagrangian will be discussed in more detail later.

\subsection{Electroweak Symmetry Breaking}\label{sec:MSSMtheoryewsb}

In the SM, the gauge group is $SU(3)_{C}\times SU(2)_{L}\times U(1)_{Y}$ with $SU(2)_{L}\times U(1)_{Y}$ spontaneously broken to $U(1)_{em}$. We break the symmetry via a single complex spin zero $SU(2)$ doublet, the Higgs field, which acquires a vacuum expectation value (VEV)\glossary{name={VEV},description={Vacuum Expectation Value}}. One combination of gauge fields remains massless, and is identified with the photon, and the other combinations of gauge fields develop masses through the Higgs mechanism. The interaction Lagrangian is then rewritten in terms of the mass eigenstate gauge fields, and Yukawa interactions between the Higgs bosons and the SM fermions are used to include fermion mass terms.

In the MSSM, we use the same mechanism to develop gauge boson and SM fermion mass terms. Assuming the matter scalars do not develop non-zero VEVs, we minimise the scalar potential with respect to the neutral Higgs fields and define $\left<h^{0}_{u}\right>\equiv v_{u}$ and $\left<h^{0}_{d}\right>\equiv v_{d}$, where we take $v_{u}$ and $v_{d}$ to be real. The requirement that there is a stable minimum with at least one non-vanishing Higgs field VEV leads us to the following minimisation conditions:
\begin{subequations}
\begin{gather}
\label{eq:EWSBb}b=\frac{\left(m^{2}_{H_{u}}+m^{2}_{h_{d}}+2\mu^{2}\right)\sin{2\beta}}{2}\quad\mathrm{and}\\
\label{eq:EWSBmu}\mu^{2}=\frac{m^{2}_{H_{d}}-m^{2}_{H_{u}}\tan^{2}{\beta}}{\tan^{2}{\beta}-1}-\frac{M^{2}_{Z}}{2}\;,
\end{gather}
\end{subequations}
in which $\tan{\beta}=v_{u}/v_{d}$. $M_{Z}$ is the mass of the neutral combination of gauge fields left-over after construction of the photon, and is easily found to be given by $M^{2}_{Z}=\left(g'^{2}+g^{2}\right)\left(v^{2}_{u}+v^{2}_{d}\right)/2$.

We now have all the information we need (aside from certain subtleties) to construct the full Lagrangian for the MSSM complete with interactions and masses for all SUSY particles. Although this is certainly possible, it would clearly result in a lengthly formula. We will return to certain terms in this Lagrangian in later chapters, but before closing this chapter we take a brief diversion to consider how SUSY may make its first appearance on the experimental stage.

\section{Diversion: Experimental Signatures of SUSY}\label{sec:experiment}

It would be foolish to close this chapter, having set forth a complex theory full of many additional parameters, without at least some comment on the potential for discovery. 
Here we consider two methods of attack.

\subsection{Dark Matter Detection}\label{sec:DM}

As previously mentioned, the assumption of $R$-parity conservation in the MSSM leads us to the conclusion that there is an $R$-odd particle which is stable, the lightest supersymmetric particle, or LSP\glossary{name={LSP},description={Lightest Supersymmetric Particle}}. The LSP is either one of the matter sfermions or mixed higgsinos/gauginos at a mass smaller than $\sim1$~TeV, and its presence is another promising aspect to SUSY. SUSY particles produced in the Big Bang, which would eventually decay to SM particles and the LSP, may leave us with a relic abundance of LSPs that can account for some, if not all, of the dark matter. We are led to ask whether there is any way to narrow down the suspects on the list of possible LSPs by making the requirement that they are good dark matter candidates, \textit{i.e.}, in agreement with current astrophysical measurements.

Since all particles produced in the Big Bang were in thermal equilibrium, we can calculate their density at the time they ceased to interact with each other. This ``decoupling'' took place at different times for different particles, and when it happened the total number of these particles was fixed. We can reduce their decoupling density in line with the expansion of the Universe, providing us with an estimate of their relic density \mbox{---} the level observed today.

If the LSP is a charged particle it will be present in exotic atoms. These atoms would look like a very heavy isotope, and should be detected in experiments searching for anomalous isotopes. Searches for such isotopes \cite{hem} do not find abundances in line with the expectations for a TeV scale LSP \cite{wolf,*dover}. We can carry out the same argument for a coloured LSP, so that the only viable candidates remaining are the sneutrino and the lightest neutralino. The neutralino is currently favoured since sneutrinos are bounded by two complementary experiments. Doublet sneutrinos heavier than $25\ $GeV would already have been observed in direct DM detection experiments, and a sneutrino lighter than this would have been observed in detailed observations of the properties of the $Z$ boson at LEP.

There is still much excitement about the possibilities for a neutralino dark matter solution. In one way, we can think of the evidence for dark matter as being a strong signal pointing towards a SUSY model with a neutralino LSP, and it is fortunate that the neutralino is indeed the LSP in a wide range of models. The WMAP\glossary{name={WMAP},description={Wilkinson Microwave Anisotropy Probe}} experiment has made a very good estimate of the dark matter relic density which strongly constrains the MSSM parameter space. Perhaps this will be the area of experimental enquiry that will be the first to provide compelling evidence for physics beyond the SM.

\subsection{Colliders}

Colliders have so far been unable to offer evidence for the presence of TeV scale SUSY. They continue instead to place bounds on  sparticle masses and reduce the available parameter space for the various models of SUSY breaking. Of course, the extended reach of the LHC and any possible linear collider will either push these bounds further or offer our first glimpse of supersymmetric particles.

The cosmological arguments from the previous section, in which we used the assumption of $R$-parity conservation in the MSSM, strongly point toward the presence of a neutral LSP. In addition, $R$-parity conservation also implies that sparticles can only be produced in pairs, and only then if there is sufficient energy in the collision. The current bounds on SUSY particle masses would suggest, therefore, that the energies available at current or future colliders will only be sufficient to produce two sparticles per event. These two particles will undergo cascade processes and eventually decay into two LSPs, which, only interacting with normal matter through the exchange of other heavy (SUSY) particles, will escape detection \mbox{---} the LSP will behave like a heavy neutrino.

Since (even in hadron colliders) we know that the total transverse energy for the initial state particles is zero, a clear indication of a SUSY event would be an imbalance in the transverse energy of the final state. Although a missing transverse energy signal can also be due to SM processes such as neutrino production, in this case the amount of missing energy is usually significantly smaller. Detection of missing transverse energy events is considered an important ingredient in SUSY searches at colliders.

It should be noted that if the LHC is unsuccessful in finding SUSY, some of the elegant arguments in favour of supersymmetry at the weak scale will no longer apply. This would not rule out SUSY as a candidate for new physics at some high scale, but its rich phenomenology would be out of reach for the foreseeable future. For an overview of the current state of collider searches and the discovery potential at future colliders, the reader is encouraged to see Chapter 15 of Ref.~\cite{wss}.

If the LHC does indeed produce SUSY particles, the next step would be to begin to measure the particle spectrum and coupling strengths, and eventually try to uncover the origin of the SSB parameters. This would be a complex, but rewarding challenge.

%% file: flavour.tex
\chapter{Flavour-violating Interactions}\label{ch:flavth}

If the soft SUSY-breaking matrices of the MSSM (written in the basis in which the corresponding SM fermions have definite mass) contain large off-diagonal entries, they lead to flavour-changing effects that are not present in the SM. Up to now we have avoided any mention of the structure of these terms, but their inclusion in the MSSM is potentially a major problem for SUSY model builders. An attractive feature of mSUGRA is that the GUT scale inputs in \eqref{eq:mSUGRAin} contain no flavour off-diagonal couplings, effectively reducing any additional flavour-violating effects at the weak scale to acceptable bounds. We shall see that we can adapt the mSUGRA framework to carefully control the sources of flavour-violation in the weak scale effective theory.

This chapter briefly reviews flavour-changing effects in the quark sector before describing in more detail the situation in the MSSM and construction of the squark mass eigenstates, which dictate the size of flavour-violation in the squark sector.

\input{quarks}

\section{Squark Flavour Mixing}

In the MSSM the squarks have couplings to the Higgs sector equal to those of the quarks. They therefore likewise have flavour off-diagonal Yukawa terms in the Lagrangian. In addition, since we know SUSY is broken, the squarks also have flavour-violating couplings coming from SSB squark mass terms and $\mathbf{a}$-parameters. If there is no new source of flavour-violation in the MSSM, the KM matrix still governs the size of the mixing, and the theory is still independent of the specific choices for $\mathbf{V}_{L}$ and $\mathbf{V}_{R}$. However, the squarks will not necessarily be simultaneously diagonalisable with the quarks due to the SSB terms in the Lagrangian.

Large flavour dependence in the sfermion mass matrices could lead to large contributions to flavour-changing neutral current interactions, for example $b\rightarrow s\gamma$ and $\mu\rightarrow e\gamma$. The most stringent constraints on the flavour structure of supersymmetry come from studies of $K^{0}-\overline{K^{0}}$ and $D^{0}-\overline{D^{0}}$ \cite{pakvasa} mixing. If we naively estimate the off-diagonal entries of the SSB squark mass matrices to be of the same order as the diagonal entries, we would achieve SUSY contributions to the mixing that violate current bounds from experiment by many orders of magnitude unless sparticles are very heavy \mbox{---} excluding the model as a solution to the hierarchy problem. This is known as the SUSY flavour problem.

In addition to contributing to SM processes, the off-diagonal elements of the SSB masses and $\mathbf{a}$-parameters will result in squark eigenstates which can couple to all three quark flavours. In particular, the squarks may experience flavour-violating decays including our sample decay in which the squark eigenstate that is predominantly stop can decay to a charm quark (or an up quark) and a neutralino.

In order to calculate the size of flavour-violating decays of squarks, we must work with mass eigenstates of both quarks and squarks. We construct a matrix of all mass terms for squarks in the Lagrangian in the basis where the up-type quarks are diagonal, and then diagonalize this matrix to find the squark eigenstates. Note that in the next section, we drop the basis labels on the couplings and fields on the understanding that we write everything in the ``standard'' current basis. This choice allows us to easily identify the coupling of the $c$ quark mass eigenstate to the up-type squarks.

\subsection{The Squark Mass Matrix}\label{sec:sqmass}

The current basis Lagrangian for the MSSM contains several mass terms for the up-type squarks that can be collected into a $(6\times6)$ matrix, ${\left( \bm{\mathcal{M}}^2_{\tilde{u}} \right)}_{(ia)(jb)}$, such that
\begin{equation}\label{eq:currsqmass}
\mathcal{L} \ni -\left( {\tilde{u}}^{\dag}_{Li}, {\tilde{u}}^{\dag}_{Ra} \right) {\left( \bm{\mathcal{M}}^2_{\tilde{u}} \right)}_{(ia)(jb)} \left( \begin{array}{c} \tilde{u}_{Lj} \\ \tilde{u}_{Rb} \end{array} \right)\;,
\end{equation}
where the indices $i,j$ label left-handed squarks and $a,b$ label right-handed squarks. These indices can be expanded out to show the current basis squarks to be $\tilde{u}_{Li}=\left(\tilde{u}_{L},\tilde{c}_{L},\tilde{t}_{L}\right)$ and $\tilde{u}_{Ra}=\left(\tilde{u}_{R},\tilde{c}_{R},\tilde{t}_{R}\right)$. Considering the individual contributions to this matrix, we can write ${\left( \bm{\mathcal{M}}^2_{\tilde{u}} \right)}_{(ia)(jb)}$ in terms of $(3\times3)$ sub-matrices
\begin{equation}\label{eq:umixmat}
\left(\bm{\mathcal{M}}^{2}_{\tilde{u}}\right)_{(ia)(jb)}\equiv\left(\begin{array}{cc}\left(\bm{\mathcal{M}}^{2}_{LL}\right)_{ij}&\left(\bm{\mathcal{M}}^{2}_{LR}\right)_{ib}\\\left(\bm{\mathcal{M}}^{2}_{LR}\right)^{\dagger}_{aj}&\left(\bm{\mathcal{M}}^{2}_{RR}\right)_{ab}\end{array}\right)\;,
\end{equation}
and find that
\begin{subequations}
\begin{align}
\left(\bm{\mathcal{M}}^{2}_{LL}\right)_{ij} = & {(\bdm^2_{Q})}_{ij}+ v_u^{2} {\left({\bdf^{*}_u}\bdf^{T}_u\right)}_{ij}+\left(\frac{g^{\prime2}}{12}-\frac{g^{2}_{2}}{4}\right)\left(v_{u}^{2}-v_{d}^{2}\right)\bm{\delta}_{ij}\\
{(\bm{\mathcal{M}}^{2}_{RR})}_{ab} = & {(\bdm^2_{U})}_{ab}+ v_u^{2}{\left(\bdf^{T}_u{\bdf^{*}_u}\right)}_{ab}-\frac{g^{\prime2}}{3}\left(v_{u}^{2}-v_{d}^{2}\right)\bm{\delta}_{ab}\\
\left(\bm{\mathcal{M}}^{2}_{LR}\right)_{ib}=&-v_{u}\left(\ba_{u}\right)^{*}_{ib}+v_{d}\left(\mu^{*}\mathbf{f}_{u}\right)^{*}_{ib}\;.
\end{align}
\end{subequations}

The Yukawa coupling terms will be (almost) diagonal in the quark mass basis\footnote{We will discuss the extent to which the Yukawa matrices deviate from diagonal in later chapters.} and therefore not contribute significantly to the flavour-violating terms. If ${\bdm}^2_{Q}$, ${\bdm}^2_{U}$ and ${\ba}_{u}$ are all diagonal in the same basis as the Yukawa couplings there is no mixing between generations, but we still have mixing between sfermions with the same charge in the same generation. In this scenario, we can write the mixed sfermions in terms of the unmixed sfermions via a ($2\times2$) mixing matrix
\begin{equation} \label{eq:sfermix}
\left( \begin{array}{c} \fsfl \\ \fsfr \end{array} \right)
=
\left( \begin{array}{rr}
\cos{\theta_f} & \sin{\theta_f} \\
-\sin{\theta_f} & \cos{\theta_f}
\end{array} \right)
\left( \begin{array}{c} \fsfI \\ \fsfII \end{array} \right)\;,
\end{equation}
which depends on a single rotation angle, $\theta_{f}$.

As mentioned in Ch.~\ref{ch:SUSY}, there is a further simplification in many models where the terms $A_{u,c,t}$ are related to the matrix $\mathbf{a}_{u}$ by the following relation:
\begin{equation} \label{eq:asnfs}
{\mathbf{a}}_u=
\left(
\begin{array}{ccc}
A_u f_u & 0 & 0 \\
0 & A_c f_c & 0 \\
0 & 0 & A_t f_t \\
\end{array}
\right)\;.
\end{equation}
Using this substitution, the mixing angle for \textit{up-type} squarks is given by
\begin{equation}
\tan{\theta_f} = \frac{ m^2_{\fsfl} + m^2_f + M^2_Z \cos{2\beta} \left( \frac{1}{2} - \frac{2}{3} \sin^2{\theta_W} \right) - m^2_{\fsfI} }{ m_f \left( -A_f + \mu \cot{\beta} \right) }\;,
\end{equation}
and the masses of the eigenstates $\fsfI $ and $\fsfII$ are
\begin{equation}
\begin{split}
&m^2_{{\widetilde{f}_{1,2}}}=\frac{1}{2}\left(m^2_{\fsfl}+m^2_{\fsfr}\right)+\frac{1}{4}M^2_Z\cos{2\beta}+m^2_f \\
& \mp{\left\{{\left[\frac{1}{2}\left(m^2_{\fsfl}-m^2_{\fsfr}\right)+M^{2}_{Z}\cos{2\beta}\left(\frac{1}{4}-\frac{2}{3}\sin^{2}{\theta_{w}}\right)\right]}^2+m^2_f{\left(-A_f+\mu\cot{\beta}\right)}^2\right\}}^{1/2}\;.
\end{split}
\end{equation}
It is straightforward to derive similar equations for the masses and mixings of the down-type squarks and sleptons.

\subsection{Flavour-violating Decays of Up-type Squarks}

It should be clear that in general, if ${\bdm}^2_{Q}$, ${\bdm}^2_{U}$ and ${\ba}_{U}$ are not all diagonal in the same basis as the Yukawa coupling matrices, all three flavours will mix with each other. In this case we label the mass eigenstates with the name of the current basis squark to which they are most closely aligned, and with a number indicating their relative mass. For the up-squarks, $\fstI$ is defined as the lighter of the two squarks with the most amount of stop, $\fscI$ is the lighter of the two squarks with the most amount of scharm and so on. 

To find the mass eigenstates, we diagonalise the $(6\times6)$ matrix given in (\ref{eq:umixmat}) by rotating the squarks by a unitary matrix, $\bm{\mathcal{U}}$. This matrix can be split into left and right sections
\begin{equation}
\bm{\mathcal{U}}=\left(\begin{array}{c}\bm{\mathcal{U}}_{L}\\\bm{\mathcal{U}}_{R}\end{array}\right)\;,
\end{equation}
where both $\bm{\mathcal{U}}_{L}$ and $\bm{\mathcal{U}}_{R}$ are \textit{non-unitary} $(3\times6)$ matrices. Denoting the squark mass eigenstates with the superscript `$\tilde{M}$' and the label $\alpha$, we can write the generalisation of \eqref{eq:sfermix} as
\begin{equation}\label{eq:squarkrot}
\left(\begin{array}{c}\tilde{u}_{Li}\\\tilde{u}_{Ra}\end{array}\right)=\left(\begin{array}{c}\left.\bm{\mathcal{U}}_{L}\right._{i\alpha}\\\left.\bm{\mathcal{U}}_{R}\right._{a\alpha}\end{array}\right)\tilde{u}^{\tilde{M}}_{\alpha}\;,
\end{equation}
so that $\tilde{u}^{\tilde{M}}_{\alpha}=\left.\bm{\mathcal{U}}^{\dagger}_{L}\right._{\alpha i}\tilde{u}_{Li}+\left.\bm{\mathcal{U}}^{\dagger}_{R}\right._{\alpha a}\tilde{u}_{Ra}$. When we transform the Lagrangian into the squark mass basis, we will use the relation:
\begin{equation}
\left(\left.\tilde{u}^{\dagger}_{L}\right._{i},\left.\tilde{u}^{\dagger}_{R}\right._{a}\right)=\tilde{u}^{\tilde{M}\dagger}_{\alpha}\left(\left.\bm{\mathcal{U}}^{\dagger}_{L}\right._{\alpha i},\left.\bm{\mathcal{U}}^{\dagger}_{R}\right._{\alpha a}\right)\;.
\end{equation}
and define the squark mass basis to be in the order $\left(\tilde{t}_{1},\tilde{t}_{2},\tilde{c}_{1},\tilde{c}_{2},\tilde{u}_{1},\tilde{u}_{2}\right)$ so that the $\tilde{t}_{1}$ is the state $\tilde{u}^{\tilde{M}}_{1}$ (which is not necessarily the lightest of the up-type squark mass eigenstates). Eq.~\eqref{eq:squarkrot} is essentially the inversion of \eqref{eq:t1state}, which symbolically lists the components of one of the mass eigenstates, the $\tilde{t}_{1}$. We can confirm that if the up-type squarks are not simultaneously diagonalisable with the up-type quarks, \textit{i.e.}, the squark mass matrix has non-zero off-diagonal entries in the ``standard'' current basis, the $\tilde{t}_{1}$ will contain a certain amount of $\tilde{c}$ (in addition to $\tilde{t}$ and $\tilde{u}$) and therefore be able to couple to, for example, a $c$ quark, corresponding to non-zero values of either one or both of $a_{2}$ and $a_{5}$ in \eqref{eq:t1state}.
%

In the mSUGRA framework, the SSB mass matrices are proportional to the unit matrix at some high energy scale ($\mgut$) and the $\mathbf{a}$-parameters are proportional to the corresponding Yukawa coupling matrices, such that these Yukawa couplings are the only source of flavour-violation in the theory. However, the form of the SSB parameters will be altered by radiative corrections, as embodied by their renormalisation group evolution, and the resulting weak scale terms will not generally be simultaneously diagonalisable with the corresponding Yukawa matrices.

In order to estimate the effect of radiative corrections on the squark mass terms, and calculate the squark mass eigenstates, we must consider the dependence on scale of the various couplings in our Lagrangian. Our discussion therefore turns to the renormalisation group equations, which will tell us how the relevant weak scale couplings are related to our GUT scale inputs.

%% file: quarks.tex
\section{Quark Flavour Mixing}\label{sec:quarks}

It is well known that in the Standard Model, the current eigenstate quarks (the states that take part in weak interactions), with fixed quantum numbers, are not simultaneously mass eigenstates. The $K^{+}$, for example, is a combination of a $u$ quark and an $\overline{s}$ anti-quark, and is observed to decay $63\%$ of the time to $\mu^{+}\nu_{\mu}$. Since the interaction Lagrangian of the SM only contains transitions between quarks of differing flavour from the same generation (\textit{i.e.}, $u\leftrightarrow d$, $c\leftrightarrow s$ and $t\leftrightarrow b$), we know from the observation of both $\pi^{+}(u\bar{d})$ and $K^{+}(u\bar{s})$ decays that the quark mass eigenstates must be a superposition of the current eigenstates.

When we write down our interaction Lagrangian, there is no principle to prevent the Yukawa matrices, which couple left- and right-handed quarks with Higgs bosons, from being off-diagonal in the current basis. Therefore, our Lagrangian may contain transitions between the various current eigenstate quarks of the same type, which is exactly what we need to reproduce the above behaviour. In order to find the mass eigenstates we write down all the mass terms in our theory and diagonalise them using a bi-unitary transformation which describes the connection between the current basis and mass basis as follows.

\subsection{Quark Yukawa Matrix Diagonalisation}

From the master equation, (\ref{eq:genlag}), and the superpotential, (\ref{eq:super}), the mass term for the up quarks is
\begin{equation}\label{eq:upyuk}
\mathcal{L}\ni-\bar{u}_{j}(\bdf_{u})^{T}_{ji}h^{0}_{u}P_{L}u_{i}\;,
\end{equation}
where the Higgs field acquires a VEV, $v_{u}$. The expression is altered slightly in the SM since the $h^{0}_{u}$ field is not the Higgs field of the Standard Model, and the subsequent relation between the MSSM Yukawa couplings ($\mathbf{f}_{u,d,e}$) and SM Yukawa couplings ($\bm{\lambda}_{u,d,e}$) will be discussed in more detail in Ch.~\ref{ch:renorm}.

Eq.~\eqref{eq:upyuk} is written in the current (or interaction) basis, where the Yukawa matrix may have large off-diagonal entries. We now diagonalise the Yukawa coupling by writing
\begin{equation}\label{eq:yukrot}
\bdf^{\mathrm{diag}}_{u}=\mathbf{V}^{T}_{L}(u)\bdf_{u}\mathbf{V}^{*}_{R}(u)\;,
\end{equation}
where the rotation matrices $\mathbf{V}_{L,R}(u)$ are unitary, and requiring that $\bdf^{\mathrm{diag}}_{u}$ has vanishing off-diagonal entries. Denoting the mass eigenstates with a superscript $M$, it is clear from (\ref{eq:upyuk}) that the relation between current and mass basis quarks is
\begin{equation}\label{eq:qrot}
u_{Li}=[\mathbf{V}_{L}(u)]_{ij}u^{M}_{Lj}\qquad;\qquad u_{Ri}=[\mathbf{V}_{R}(u)]_{ij}u^{M}_{Rj}\;.
\end{equation}
We can write similar rotations for the down-type quarks using rotation matrices $\mathbf{V}_{L,R}(d)$ to obtain an equivalent relation to \eqref{eq:yukrot} for $\mathbf{f}_{d}$.

All we have done is to change the basis. To see whether the matrices $\mathbf{V}_{L,R}(u,d)$ are physically meaningful, we must consider each term in the Lagrangian. For the SM, the Higgs interactions are diagonalised simultaneously with the quark masses. In addition, the neutral current interaction, which is by definition flavour diagonal in the current basis, remains flavour diagonal in the mass basis due to the unitarity of the rotation matrices. It is only the charged current interaction which causes us trouble. We write
\begin{equation}
\mathcal{L}\ni -\frac{g}{\sqrt{2}}\bar{u}_{j}\gamma^{\mu}W^{+}_{\mu}\delta_{ji}P_{L}d_{i}+\mathrm{h.c.}
\end{equation}
which only allows transitions in the current basis between flavours of the \textit{same generation}. We transform to the mass basis using (\ref{eq:qrot}), and the equivalent transformation for the down-type quarks, to obtain
\begin{equation}
\mathcal{L}\ni -\frac{g}{\sqrt{2}}\bar{u}^{M}_{j}\gamma^{\mu}W^{+}_{\mu}[\mathbf{V}_{L}(u)]^{\dagger}_{jk}[\mathbf{V}_{L}(d)]_{ki}P_{L}d^{M}_{i}+\mathrm{h.c.}
\end{equation}
Since the right-handed quarks do not take part in the charged current interaction, the right-handed rotations do not enter. In addition, only one combination of the left-handed matrices, namely ${{\mathbf{V}}^{\dagger}_L}(u){\mathbf{V}}_L(d)$, is physical. This combination is usually referred to as the Kobayashi-Maskawa (KM)\glossary{name={KM},description={Kobayashi-Maskawa (mixing matrix)}} matrix \cite{KM}, \textit{i.e.},
\begin{equation}\label{eq:vudkm}
\mathbf{K}\equiv{{\bf{V}}^\dagger_L}(u) {\bf{V}}_L(d).
\end{equation}

The SM Lagrangian in the mass basis only depends upon the ${\mathbf{V}}$-matrices via the KM matrix. We can therefore see that all the flavour structure in the quark sector of the Standard Model is dictated by this combination alone. Since it is formed of two unitary matrices, it is itself unitary and would therefore be completely described by three angles and three phases. Using our freedom to make field rotations, we can reduce this to three angles and a single phase, and the KM matrix is usually parametrised as \cite{pdg}
\begin{equation}\label{eq:kmparam}
\mathbf{K}=\left(\begin{array}{ccc}c_{12}c_{13}&s_{12}c_{13}&s_{13}e^{-i\delta_{13}}\\-s_{12}c_{23}-c_{12}s_{23}s_{13}e^{i\delta_{13}}&c_{12}c_{23}-s_{12}s_{23}s_{13}e^{i\delta_{13}}&s_{23}c_{13}\\s_{12}s_{23}-c_{12}c_{23}s_{13}e^{i\delta_{13}}&-c_{12}s_{23}-s_{12}c_{23}s_{13}e^{i\delta_{13}}&c_{23}c_{13}\end{array}\right)\;.
\end{equation}
For the numerical results that we present, we take $s_{12}=0.2243$, $s_{13}=0.0037$ and $s_{23}=0.0413$, and $\delta_{13}=60^\circ$.

We note for later reference that since flavour-violation in the SM enters only via the KM matrix, physical observables can only depend on ${\bf{V}}_L(u)$ and ${\bf{V}}_L(d)$ through $\mathbf{K}$ as specified by \eqref{eq:vudkm} and must be independent of ${\bf{V}}_R(u)$ and ${\bf{V}}_R(d)$.

\subsection{Choice of Basis}\label{sec:basischoice}

Although it is tempting to choose to work from now on in a basis in which all the Yukawa matrices are diagonal\footnote{We shall see that if the Yukawa coupling matrices are taken to be diagonal at some specific scale, for example $m_{t}$, they will not in general be diagonal at other scales.}, this approach is not a practical option. Since the quark mass basis is defined with different transformations for the up-type and down-type quarks this choice would not preserve the $SU(2)_{L}$ symmetry. Instead, we choose a current basis in which only one of the Yukawa matrices, $\mathbf{f}_{u}$ or $\mathbf{f}_{d}$, is diagonal.

In order that the underlying supersymmetry is preserved, we must rotate the squark multiplets by the same transformations as the quarks, although this does not mean that the squarks will also be in their mass eigenstate basis. As a consequence, the SSB masses and $\mathbf{a}$-parameters transform from the general current basis to the basis in which one of the Yukawa matrices are diagonal as
\begin{gather}
\label{eq:amix}{({\ba}_{u,d})}^T={{\mathbf{V}}_R}(u,d){{\left({\ba}^{M}_{u,d}\right)}}^T{{\mathbf{V}}^{\dagger}_L}(u,d) \\
\label{eq:mqmix}{\bdm}^{2}_Q={\mathbf{V}}_L(q){\left({\bdm}^{2}_Q\right)^{M}}{{\mathbf{V}}^{\dagger}_L}(q) \\
\label{eq:mudmix}{\bdm}^{2}_{U,D}={\mathbf{V}}_R(u,d){\left({\bdm}^{2}_{U,D}\right)^{M}}{{\mathbf{V}}^{\dagger}_R}(u,d)\;,
\end{gather}
where we use a $(q)$ to denote the rotation of the $SU(2)_{L}$ doublet squarks, and set $q=u(d)$ in the case that the up (down) Yukawa couplings are diagonal.

With a view to our study of the flavour-violating decay of the $\tilde{t}_{1}$ in Ch.~\ref{ch:stopdec}, we choose here a ``standard'' current basis in which it is the up-type Yukawa coupling matrix that is diagonal. In so doing, we have fixed $\mathbf{V}_{L,R}(u)=\mathbf{V}_{R}(d)=\dblone$, and $\mathbf{V}_{L}(d)=\mathbf{K}$ as required by \eqref{eq:vudkm}. The relation between the various parameters in the ``standard'' current basis and the basis in which the down-type Yukawa matrices are diagonal is relatively straightforward to derive. We drop the superscript `M', instead using $(u)$ and $(d)$ to signify whether $\mathbf{f}_{u}$ or $\mathbf{f}_{d}$ is diagonal, and find that
\begin{gather}
\label{eq:yukbasis}(\mathbf{f}_{u,d})^{T}(d)=(\mathbf{f}_{u,d})^{T}(u)\mathbf{K}\\
\label{eq:aubasis}(\mathbf{a}_{u,d})^{T}(d)=(\mathbf{a}_{u,d})^{T}(u)\mathbf{K}\\
\label{eq:mqbasis}\mathbf{m}^{2}_{Q}(d)=\mathbf{K}^{\dagger}\mathbf{m}^{2}_{Q}(u)\mathbf{K}\\
\mathbf{m}^{2}_{U,D}(d)=\mathbf{m}^{2}_{U,D}(u)\;.
\end{gather}
Remember that the $\mathbf{V}_{L,R}(u,d)$ diagonalise the quark Yukawa matrices and are fixed by the relation between the current basis and the quark mass basis. In this regard, the final relation above follows from the fact that the right-handed squarks are singlets under $SU(2)$.

%% file: renorm.tex
\chapter{Renormalisation Group Equations}\label{ch:renorm}

With the MSSM, we have put in place a model with mass scales that are stable under radiative corrections \cite{dim,*sak,*kaul}, which, however, contains a large number of additional parameters. We can hope that this proliferation of parameters simply reflects our ignorance of the mechanism of SUSY breaking, and that these arise from a simpler underlying theory valid at some high scale. In the absence of knowledge of such a theory, our approach, therefore, is to introduce more manageable frameworks, valid at the high scale, which predict the MSSM masses and couplings.

It is tempting to imagine that we could use the GUT scale parameters of a model such as mSUGRA to directly calculate the interactions and decays that take place at the weak scale. However, we are unable to carry out such a calculation due to the dependence on scale of the fields, couplings and masses of the theory. This scale dependence means that any perturbative calculation in terms of some coupling, $g$, renormalised at the high scale, would contain large logarithms, $\ln{(\mgut/M_{Z})}$, which would invalidate perturbation theory through the appearance of powers of $\tfrac{g^{2}}{4\pi}\ln{(\mgut/M_{Z})}\sim1$ for a fixed value of $g^{2}$. We must therefore obtain the weak scale parameters \textit{before} computing the various transition amplitudes, since in this manner we will sum all the large logs that would have invalidated the perturbative expansion.

Scale dependence in the parameters arises as a result of our ``renormalisation'' procedure. Although the so-called `bare' parameters are perturbatively divergent quantities, we shift the divergence into \textit{counterterms}. Once we have reparametrised the theory, the dependence of the renormalised quantities on the scale, $Q$, is described by the Callan-Symanzik $\beta$-function, defined as
\begin{equation}
\beta_{g}=Q\frac{\partial g}{\partial Q}\;,
\end{equation}
and we calculate the $\beta$-functions for the various parameters by considering all diagrams that contribute to the renormalisation. We then say that the coupling constants and masses in the theory `run' with scale according to the renormalisation group equations, which provides us with a convenient method to obtain the weak scale parameters from our high scale inputs.

Given the RGEs for all dimensionless (gauge and Yukawa) and dimensionful (soft mass and trilinear) parameters we will be able to study the flavour-violation in models with arbitrary values of high scale SSB parameters. The RGEs for the SM \cite{MVrgeI,*MVrgeII,*MVrgeIII,ACMPRWrge} and the MSSM \cite{MVrge,BBOrge,*Yam1,*Yam2,*Yam3,*JJ} are already known at the two-loop level. If we insist on achieving two-loop accuracy, however, we must also be prepared to model the transition from a fully supersymmetric theory, with RGEs as in the MSSM, to a SM theory with SM RGEs. The effect of this transition in the one-loop RGEs can be as large as two-loop effects and perhaps even larger if there is a sizable mass difference between the various SUSY particles. The task of estimating the effect of including this transition has been attempted in existing literature \cite{casno,sakis}. We detail our extension to this work by introducing the procedure for deriving the RGEs in a general field theory and then moving on to consider the problems encountered when applying these RGEs to the MSSM-SM transition.

\section{Gauge coupling running and unification}\label{sec:gaugerun}

The derivation of the RGEs for the gauge couplings in a general theory is laid out in many places \cite{ps,MVrgeI,*MVrgeII,*MVrgeIII}. For a theory with a number of scalars and fermions in different representations, the one-loop general result is \cite{gross,*politzer}
\begin{equation} \label{eq:gaugerge}
\left(4\pi\right)^{2}\left.\beta_{g}\right|_{1-\mathrm{loop}}=
-g^{3}\left[\frac{11}{3}C(G)-\frac{2}{3}\sum_{fermions}S(R_{F})
-\frac{1}{3}\sum_{scalars}S(R_{S})\right],
\end{equation}
where $C(G)$ is the quadratic Casimir for the adjoint representation of the associated Lie algebra, and $S(R_F)$ and $S(R_S)$, respectively, are the Dynkin indices for the representations $R_F$, $R_S$ under which the fermions and (complex) scalars transform. For the Lie algebra of $SU(N)$, $C(G)=N$, while $S(R) =1/2$ for the fundamental $N$-dimensional representation, and $S(R)=N$ for the adjoint representation. For the $U(1)_Y$ gauge coupling $g'$, $C(G)=0$ while $S(R)=(Y/2)^{2}$.

Of course, we obtain different RGEs for the gauge couplings with SM field content compared to MSSM field content and at some scale our theory must change from pure SM running to pure MSSM running. With $i$ running from $1,\dots3$ to indicate the different gauge groups, the RGEs are
\begin{equation}\label{eq:gbeta}
\beta_{i}=\frac{g^{3}_{i}}{16\pi^{2}}b_{i}\;.
\end{equation}
For the particle content of the SM
\begin{equation}
\left(b_{1},b_{2},b_{3}\right)=\left(\frac{41}{10},-\frac{19}{6},-7\right),
\end{equation}
and for the MSSM
\begin{equation}
\left(b_{1},b_{2},b_{3}\right)=\left(\frac{33}{5},1,-3\right).
\end{equation}
In (\ref{eq:gbeta}) we have used a scaled $U(1)$ gauge coupling\footnote{This change is motivated by grand unification considerations which fix the normalization of the $U(1)_{Y}$ generator.}, $g_{1}$, which is related to $g'$ by $g_{1}=\sqrt{5/3}g'$.

\begin{figure} \centering
\includegraphics[viewport=45 50 710 525, clip, scale=0.45]{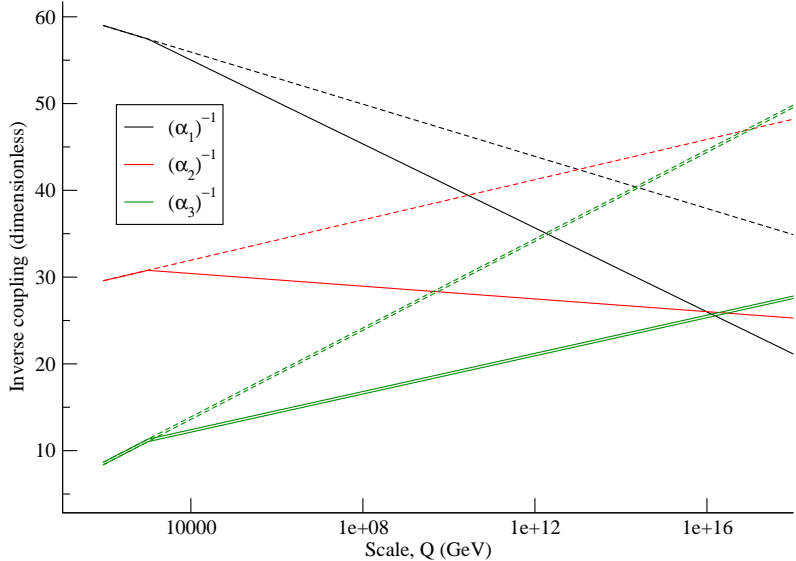} 
\caption[The running of the inverse gauge coupling strength, $\alpha^{-1}$, for all three gauge groups.]{\small The running of the inverse gauge coupling strength, $\alpha^{-1}_{i}$, for all three gauge groups, $i$. The solid lines contain a switch to MSSM running at $1\ $TeV, the dashed lines show SM running over all scales. Two lines are plotted for $\alpha^{-1}_{3}$ to indicate the limits of experimental uncertainty.} \label{fig:gaugeunif} 
\end{figure}

In Fig.~\ref{fig:gaugeunif} we give an example of the one-loop running of the gauge couplings in two simple scenarios.
Beginning with the experimental values of the gauge couplings at $M_{Z}$, we evolve their values up to the GUT scale using the RGEs. We plot the running of $\alpha^{-1}_{i}$ where $\alpha_{i}=g^{2}_{i}/4\pi$, and have drawn two lines for $g_{3}$ to indicate the relatively large experimental uncertainty in this value.

The dashed lines show the evolution in the SM all the way to the high scale. In contrast, the solid lines use a transition at an approximate SUSY scale, $M_{\mathrm{SUSY}}=1$~TeV where SM running gives way to the MSSM. Remarkably, the couplings appear to unify (or at least come extremely close to unification) at a scale somewhere in the range $Q=10^{16}$~GeV \cite{lang}. This outcome is clearly dependent on where we choose to implement the change to SUSY running, but the general result strongly suggests some sort of unification of the couplings and hence new physics at this high scale.

Note that even with the uncertainty in $\alpha_{s}(M_{Z})$ taken into account, there is no hope of unification within the SM. The appearance of (near-)unification only when we introduce MSSM effects at $M_{\mathrm{SUSY}}$ is regarded by some as indirect evidence of SUSY at the weak scale.

We mention briefly here that the one-loop RGEs (in the full MSSM) for the gaugino couplings, $M_{1}$, $M_{2}$ and $M_{3}$, are proportional to the $\beta$-functions for their respective gauge couplings. Since the constant of proportionality is the same in each case, this has the consequence that, in models where both the gauge couplings and the gaugino masses unify at the GUT scale, we obtain the RGE invariant relation
\begin{equation}\label{eq:gauginounif}
\frac{\alpha_{1}}{M_{1}}=\frac{\alpha_{2}}{M_{2}}=\frac{\alpha_{3}}{M_{3}}\;,
\end{equation}
which is valid at one-loop order. In Ch.~\ref{ch:results} we shall consider deviations from this relation in the two-loop MSSM with thresholds.

\section{RGEs for a General Field Theory}

Our starting point for including threshold effects in the RGEs for the MSSM will be the general results in the seminal papers by Machacek and Vaughn \cite{MVrgeI,*MVrgeII,*MVrgeIII} for the dimensionless couplings, and the paper by Luo, Wang, and Xiao \cite{luo} for the dimensionful couplings. These RGEs are written in a general gauge field theory, which is not necessarily supersymmetric, and uses two-component notation for fermions and real scalar fields. Since most phenomenologists are more familiar with four-component spinor fields, we convert their work to apply to four-component fermions (Majorana or Dirac) and complex scalars.

We denote our four-component spinors by $\Psi_{Di}$ for Dirac fields and $\Psi_{Mi}$ for Majorana fields. The indices, $i,j,k,...$ are fermion field indices which carry information about field type (quark, lepton, gaugino, higgsino) and also carry information about flavour and colour. Similarly, complex scalar fields are denoted by $\Phi_{a}$ where $a,b,...$ are scalar indices running over squark, slepton, and Higgs fields. In the context of the MSSM with conserved $R$-parity we will write our general Lagrangian density without including baryon or lepton number violating Yukawa terms, or terms which only appear in the MSSM after the spontaneous breakdown of electroweak symmetry. With this in mind, the four-component Lagrangian takes the form
\begin{equation}\begin{split}\label{eq:fourlag}
\mathcal{L}_{(4)} = &\ \frac{i}{2}\bar{\Psi}_{j} \gamma^\mu D_\mu \Psi_j + \left(D_{\mu}\Phi_{a}\right)^{\dagger}\left(D^{\mu}\Phi_{a}\right) - \frac{1}{4}F_{\mu\nu A}F^{\mu\nu}_A\\
&-\frac{1}{2}\left[\left(\mathbf{m}_{X}\right)_{jk}\bar{\Psi}_{Mj}\Psi_{Mk}+i\left(\mathbf{m}'_{X}\right)_{jk}\bar{\Psi}_{Mj}\gamma_{5}\Psi_{Mk}\right]\\
&+\left[\frac{1}{2!}\bm{\mathcal{B}}_{ab}\Phi_a\Phi_b+\mathrm{h.c.}\right]-\mathbf{m}^{2}_{ab}\Phi^\dagger_a\Phi_b\\
&-\left[\left(\bu^{1}_{a}\right)_{jk}\bar{\Psi}_{Dj}P_L\Psi_{Dk}\Phi_a+\left(\bu^{2}_{a}\right)_{jk}\bar{\Psi}_{Dj}P_L\Psi_{Dk}\Phi^\dagger_a\right.\\
&\quad\ +\left(\bv_{a}\right)_{jk}\bar{\Psi}_{Dj}P_L\Psi_{Mk}\Phi_a+\left(\bw_{a}\right)_{jk}\bar{\Psi}_{Mj}P_L\Psi_{Dk}\Phi^\dagger_a\\
&\left.\quad\ +\frac{1}{2}\left(\bx^{1}_{a}\right)_{jk}\bar{\Psi}_{Mj}P_L\Psi_{Mk}\Phi_a+\frac{1}{2}\left(\bx^{2}_{a}\right)_{jk}\bar{\Psi}_{Mj}P_L\Psi_{Mk}\Phi^\dagger_a+\mathrm{h.c.}\right]\\
&+\left[\frac{1}{2!}\Phi^\dagger_a\mathbf{H}_{abc}\Phi_b\Phi_c+\mathrm{h.c.}\right]-\frac{1}{2!}\frac{1}{2!}\bm{\Lambda}_{abcd}\Phi^\dagger_a\Phi^\dagger_b\Phi_c\Phi_d-\left[\frac{1}{3!}\bm{\Lambda}'_{abcd}\Phi^{\dagger}_{a}\Phi_{b}\Phi_{c}\Phi_{d}+\mathrm{h.c.}\right]\;.
\\
\end{split}\end{equation}
In \eqref{eq:fourlag}, there are a number of matrices with a variety of symmetry properties under interchange of indices. The matrices $\bu^{1}_{a}$, $\bu^{2}_{a}$, $\bv_{a}$ and $\bw_{a}$ have no particular symmetry or Hermiticity properties, but $\bx^{1}_{a}$ and $\bx^{2}_{a}$ are symmetric under $j\leftrightarrow k$ because of the symmetry properties of the Majorana spinor bilinears \cite{wss}. Finally among the collection of dimensionless parameters, the $\bdLm_{abcd}$ and $\bdLm'_{abcd}$ are symmetric under $a\leftrightarrow b$ and/or $c\leftrightarrow d$ for $\bdLm$ and under interchanges of $b$, $c$, $d$ for $\bdLm'$.

The dimensionful parameters in \eqref{eq:fourlag} also respect certain symmetries. The scalar mass squared matrix, $\bdm^{2}$, is Hermitian, while the scalar bilinear matrix, $\bm{\mathcal{B}}$, is symmetric. The CP-conserving and CP-violating fermion bilinears, $\bdm_{X}$ and $\bdm'_{X}$ respectively, are both real and symmetric, and finally the trilinear coupling $\mathbf{H}_{abc}$ is symmetric under $b\leftrightarrow c$. We continue by writing the four-component general RGEs for the dimensionless parameters from \eqref{eq:fourlag}.

\subsection{Dimensionless Parameters}\label{sec:dless}

After converting the general RGEs in \cite{MVrgeI,*MVrgeII,*MVrgeIII} to four-component notation, the $\beta$-functions for the Yukawa coupling matrices in \eqref{eq:fourlag} are found to be \cite{RGE1},
\begin{equation}\label{eq:U1}\begin{split}
\left(4\pi\right)^2\left.\bbet_{U^1_a}\right|_{1-\mathrm{loop}}=&\frac{1}{2}\left[\left(\bu^1_b\bu^{1\dagger}_b+\bu^2_b\bu^{2\dagger}_b+\bv_b\bv^\dagger_b\right)\bu^1_a\right.\\
&\left.\quad+\bu^1_a\left(\bu^{1\dagger}_b\bu^1_b+\bu^{2\dagger}_b\bu^2_b+\bw^\dagger_b\bw_b\right)\right]\\
&+2\left[\bu^1_b\bu^{2\dagger}_a\bu^2_b+\bu^2_b\bu^{2\dagger}_a\bu^1_b+\bv_b\bx^{2\dagger}_a\bw_b\right]\\
&+\bu^1_b\mathrm{Tr}\left\{\left(\bu^{1\dagger}_b\bu^1_a+\bu^{2\dagger}_a\bu^2_b\right)+\frac{1}{2}\left(\bx^{1\dagger}_b\bx^1_a+\bx^{2\dagger}_a\bx^2_b\right)
\right\}\\
&+\bu^2_b\mathrm{Tr}\left\{\left(\bu^{2\dagger}_b\bu^1_a+\bu^{2\dagger}_a\bu^1_b\right)+\frac{1}{2}\left(\bx^{2\dagger}_b\bx^1_a+\bx^{2\dagger}_a\bx^1_b\right)\right\}\\
&-3g^2\left[\bu^1_a\mathbf{C}^L_2(F)+\mathbf{C}^R_2(F)\bu^1_a\right]\\[10pt]
&+\frac{1}{2}\bv_a\left(\bv^\dagger_b\bu^1_b+\bx^{2\dagger}_b\bw_b\right)+2\bu^1_b\bw^\dagger_a\bw_b\\
&+\bu^{1}_{b}\mathrm{Tr}\left\{\bw^\dagger_a\bw_b+\bv^\dagger_b\bv_a\right\}\;; 
\end{split}\end{equation}
\begin{equation}\label{eq:U2}\begin{split}
\left(4\pi\right)^2\left.\bbet_{U^2_a}\right|_{1-\mathrm{loop}}=&\frac{1}{2}\left[\left(\bu^1_b\bu^{1\dagger}_b+\bu^2_b\bu^{2\dagger}_b+\bv_b\bv^\dagger_b\right)\bu^2_a\right.\\
&\left.\quad+\bu^2_a\left(\bu^{1\dagger}_b\bu^1_b+\bu^{2\dagger}_b\bu^2_b+\bw^\dagger_b\bw_b\right)\right]\\
&+2\left[\bu^1_b\bu^{1\dagger}_a\bu^2_b+\bu^2_b\bu^{1\dagger}_a\bu^1_b+\bv_b\bx^{1\dagger}_a\bw_b\right]\\
&+\bu^2_b\mathrm{Tr}\left\{\left(\bu^{2\dagger}_b\bu^2_a+\bu^{1\dagger}_a\bu^1_b\right)+\frac{1}{2}\left(\bx^{2\dagger}_b\bx^2_a+\bx^{1\dagger}_a\bx^1_b\right)\right\}\\
&+\bu^1_b\mathrm{Tr}\left\{\left(\bu^{1\dagger}_b\bu^2_a+\bu^{1\dagger}_a\bu^2_b\right)+\frac{1}{2}\left(\bx^{1\dagger}_b\bx^2_a+\bx^{1\dagger}_a\bx^2_b\right)\right\}\\
&-3g^2\left[\bu^2_a\mathbf{C}^L_2(F)+\mathbf{C}^R_2(F)\bu^2_a\right]\\[10pt]
&+\frac{1}{2}\left(\bu^{2}_{b}\bw^\dagger_b+\bv_b\bx^{1\dagger}_b\right)\bw_a+2\bv_{b}\bv^\dagger_a\bu^2_b\\
&+\bu^{2}_{b}\mathrm{Tr}\left\{\bw^\dagger_b\bw_a+\bv^\dagger_a\bv_b\right\}\;.
\end{split}\end{equation}
Since the Yukawa matrices represent the coupling of a scalar to two Dirac fermions, we will use these equations to derive the familiar $\bdf_{u,d,e}$ Yukawa coupling matrices. In addition
\begin{equation}\label{eq:X1}\begin{split}
\left(4\pi\right)^2\left.\bbet_{X^1_a}\right|_{1-\mathrm{loop}}=&\frac{1}{2}\left[\left(\bw_{b}\bw^{\dagger}_b+\bv^{T}_{b}\bv^{*}_{b}+\bx^{1}_{b}\bx^{1\dagger}_{b}+\bx^{2}_{b}\bx^{2\dagger}_{b}\right)\bx^{1}_{a}\right.\\
&\left.\quad+\bx^{1}_{a}\left(\bw^{*}_{b}\bw^{T}_{b}+\bv^{\dagger}_{b}\bv_{b}+\bx^{1\dagger}_{b}\bx^{1}_{b}+\bx^{2\dagger}_{b}\bx^{2}_{b}\right)\right]\\
&+2\left[\bw_{b}\bu^{2\dagger}_{b}\bv_{b}+\bv^{T}_{b}\bu^{2*}_{a}\bw^{T}_{b}+\bx^{1}_{b}\bx^{2\dagger}_{a}\bx^{2}_{b}+\bx^{2}_{b}\bx^{2\dagger}_{a}\bx^{1}_{b}\right]\\
&+\bx^1_b\mathrm{Tr}\left\{\left(\bu^{1\dagger}_b\bu^1_a+\bu^{2\dagger}_a\bu^2_b\right)+\frac{1}{2}\left(\bx^{1\dagger}_b\bx^1_a+\bx^{2\dagger}_a\bx^2_b\right)
\right\}\\
&+\bx^2_b\mathrm{Tr}\left\{\left(\bu^{2\dagger}_b\bu^1_a+\bu^{2\dagger}_a\bu^1_b\right)+\frac{1}{2}\left(\bx^{2\dagger}_b\bx^1_a+\bx^{2\dagger}_a\bx^1_b\right)\right\}\\
&-3g^2\left[\bx^1_a\mathbf{C}^L_2(F)+\mathbf{C}^R_2(F)\bx^1_a\right]\\[10pt]
&+\frac{1}{2}\left[\left(\bw_{b}\bu^{2\dagger}_{b}+\bx^{1}_{b}\bv^{\dagger}_{b}\right)\bv_{a}+\bv^{T}_a\left(\bu^{2*}_{b}\bw^{T}_{b}+\bv^{*}_{b}\bx^{1}_{b}\right)\right]\\
&+2\left[\bw_{b}\bw^{\dagger}_{a}\bx^{1}_{b}+\bx^{1}_{b}\bw^{*}_{a}\bw^{T}_{b}\right]+\bx^{1}_{b}\mathrm{Tr}\left\{\bw^\dagger_a\bw_b+\bv^\dagger_b\bv_a\right\}\;;
\end{split}\end{equation}
\begin{equation}\label{eq:X2}\begin{split}
\left(4\pi\right)^2\left.\bbet_{X^2_a}\right|_{1-\mathrm{loop}}=&\frac{1}{2}\left[\left(\bw_{b}\bw^{\dagger}_{b}+\bv^{T}_{b}\bv^{*}_{b}+\bx^{1}_{b}\bx^{1\dagger}_{b}+\bx^{2}_{b}\bx^{2\dagger}_{b}\right)\bx^{2}_{a}\right.\\
&\left.\quad+\bx^{2}_{a}\left(\bw^{*}_{b}\bw^{T}_{b}+\bv^{\dagger}_{b}\bv_{b}+\bx^{1\dagger}_{b}\bx^{1}_{b}+\bx^{2\dagger}_{b}\bx^{2}_{b}\right)\right]\\
&+2\left[\bw_{b}\bu^{1\dagger}_{a}\bv_{b}+\bv^{T}_{b}\bu^{1*}_{a}\bw^{T}_{b}+\bx^{1}_{b}\bx^{1\dagger}_{a}\bx^{2}_{b}+\bx^{2}_{b}\bx^{1\dagger}_{a}\bx^{1}_{b}\right]\\
&+\bx^2_b\mathrm{Tr}\left\{\left(\bu^{2\dagger}_b\bu^2_a+\bu^{1\dagger}_a\bu^1_b\right)+\frac{1}{2}\left(\bx^{2\dagger}_b\bx^2_a+\bx^{1\dagger}_a\bx^1_b\right)\right\}\\
&+\bx^1_b\mathrm{Tr}\left\{\left(\bu^{1\dagger}_b\bu^2_a+\bu^{1\dagger}_a\bu^2_b\right)+\frac{1}{2}\left(\bx^{1\dagger}_b\bx^2_a+\bx^{1\dagger}_a\bx^2_b\right)\right\}\\
&-3g^2\left[\bx^2_a\mathbf{C}^L_2(F)+\mathbf{C}^R_2(F)\bx^2_a\right]\\[10pt]
&+\frac{1}{2}\left[\left(\bv^{T}_{b}\bu^{1*}_{b}+\bx^{2}_{b}\bw^{*}_{b}\right)\bw^{T}_{a}+\bw_{a}\left(\bu^{1\dagger}_{b}\bv_{b}+\bw^{\dagger}_{b}\bx^{2}_{b}\right)\right]\\
&+2\left[\bv^{T}_{b}\bv^{*}_{a}\bx^{2}_{b}+\bx^{2}_{b}\bv^{\dagger}_{a}\bv_{b}\right]+\bx^{2}_{b}\mathrm{Tr}\left\{\bw^\dagger_b\bw_a+\bv^\dagger_a\bv_b\right\}\;,
\end{split}\end{equation}
which are the couplings of scalars to two Majorana fermions and will therefore describe the RGEs of Higgs-higgsino-gaugino interactions. Note that the property $\bx^{1,2}_a=(\bx^{1,2}_a)^{T}$, exhibited by the Lagrangian \eqref{eq:fourlag}, is observed by $\bbet_{X^{1,2}_a}$. Also, in \eqref{eq:U1}-\eqref{eq:X2} we have separated out terms which vanish in the context of the $R$-parity conserving MSSM and written them on the final two lines of each equation. These terms vanish because $\bw_{a}$ and $\bv_{a}$ are zero for $a$ values for which $\bu^{1,2}_{a}$ and $\bx^{1,2}_{a}$ are non-zero. Finally,
\begin{equation}\label{eq:Va}\begin{split}
\left(4\pi\right)^2\left.\bbet_{V_a}\right|_{1-\mathrm{loop}}=&\frac{1}{2}\left[\left(\bu^{1}_{b}\bu^{1\dagger}_{b}+\bu^{2}_{b}\bu^{2\dagger}_{b}+\bv_{b}\bv^{\dagger}_{b}\right)\bv_{a}\right.\\
&\left.\quad+\bv_{a}\left(\bw^{*}_{b}\bw^{T}_{b}+\bv^{\dagger}_{b}\bv_{b}+\bx^{1\dagger}_{b}\bx^{1}_{b}+\bx^{2\dagger}_{b}\bx^{2}_{b}\right)\right]\\
&+2\left[\bu^{1}_{b}\bw^{\dagger}_{a}\bx^{2}_{b}+\bu^{2}_{b}\bw^{\dagger}_{a}\bx^{1}_{b}+\bv_{b}\bw^{*}_{a}\bw^{T}_{b}\right]\\
&+\bv_b\mathrm{Tr}\left\{\bw^{\dagger}_{a}\bw_{b}+\bv^{\dagger}_{b}\bv_{a}\right\}-3g^2\left[\bv_a\mathbf{C}^L_2(F)+\mathbf{C}^R_2(F)\bv_a\right]\\[10pt]
&+\frac{1}{2}\left[\left(\bu^{2}_{b}\bw^{\dagger}_{b}+\bv_{b}\bx^{1\dagger}_{b}\right)\left(\bx^{1}_{a}+\bx^{2}_{a}\right)+\left(\bu^{1}_{a}+\bu^{2}_{a}\right)\left(\bu^{1\dagger}_{b}\bv_{b}+\bw^{\dagger}_{b}\bx^{2}_{b}\right)\right]\\
&+2\left[\bu^{2}_{b}\left(\bu^{1\dagger}_{a}+\bu^{2\dagger}_{a}\right)\bv_{b}+\bv_{b}\left(\bx^{1\dagger}_{a}+\bx^{2\dagger}_{a}\right)\bx^{2}_{b}\right]\\
&+\bv_{b}\mathrm{Tr}\left\{\left[\bu^{1\dagger}_{b}\left(\bu^{1}_{a}+\bu^{2}_{a}\right)+\left(\bu^{1\dagger}_{a}+\bu^{2\dagger}_{a}\right)\bu^{2}_{b}\right]\right.\\
&\left.\qquad\qquad+\frac{1}{2}\left[\bx^{1\dagger}_{b}\left(\bx^{1}_{a}+\bx^{2}_{a}\right)+\left(\bx^{1\dagger}_{a}+\bx^{2\dagger}_{a}\right)\bx^{2}_{b}\right]\right\}\;;
\end{split}\end{equation}
\begin{equation}\label{eq:Wa}\begin{split}
\left(4\pi\right)^2\left.\bbet_{W_a}\right|_{1-\mathrm{loop}}=&\frac{1}{2}\left[\left(\bw_{b}\bw^{\dagger}_{b}+\bv^{T}_{b}\bv^{*}_{b}+\bx^{1}_{b}\bx^{1\dagger}_{b}+\bx^{2}_{b}\bx^{2\dagger}_{b}\right)\bw_{a}\right.\\
&\left.\quad+\bw_{a}\left(\bu^{1\dagger}_{b}\bu^{1}_{b}+\bu^{2\dagger}_{b}\bu^{2}_{b}+\bw^{\dagger}_{b}\bw_{b}\right)\right]\\
&+2\left[\bv^{T}_{b}\bv^{*}_{a}\bw_{b}+\bx^{1}_{b}\bv^{\dagger}_{a}\bu^{2}_{b}+\bx^{2}_{b}\bv^{\dagger}_{a}\bu^{1}_{b}\right]\\
&+\bw_b\mathrm{Tr}\left\{\bw^{\dagger}_{b}\bw_{a}+\bv^{\dagger}_{a}\bv_{b}\right\}-3g^2\left[\bw_a\mathbf{C}^L_2(F)+\mathbf{C}^R_2(F)\bw_a\right]\\[10pt]
&+\frac{1}{2}\left[\left(\bw_{b}\bu^{2\dagger}_{b}+\bx^{1}_{b}\bv^{\dagger}_{b}\right)\left(\bu^{1}_{a}+\bu^{2}_{a}\right)+\left(\bx^{1}_{a}+\bx^{2}_{a}\right)\left(\bv^{\dagger}_{b}\bu^{1}_{b}+\bx^{1\dagger}_{b}\bw_{b}\right)\right]\\
&+2\left[\bw_{b}\left(\bu^{1\dagger}_{a}+\bu^{2\dagger}_{a}\right)\bu^{1}_{b}+\bx^{1}_{b}\left(\bx^{1\dagger}_{a}+\bx^{2\dagger}_{a}\right)\bw_{b}\right]\\
&+\bw_{b}\mathrm{Tr}\left\{\left[\bu^{2\dagger}_{b}\left(\bu^{1}_{a}+\bu^{2}_{a}\right)+\left(\bu^{1\dagger}_{a}+\bu^{2\dagger}_{a}\right)\bu^{1}_{b}\right]\right.\\
&\left.\qquad\qquad+\frac{1}{2}\left[\bx^{2\dagger}_{b}\left(\bx^{1}_{a}+\bx^{2}_{a}\right)+\left(\bx^{1\dagger}_{a}+\bx^{2\dagger}_{a}\right)\bx^{1}_{b}\right]\right\}\;,
\end{split}\end{equation}
are the RGEs for squark-quark-gaugino and squark-quark-higgsino interactions. In this instance, the last three lines of \eqref{eq:Va} and \eqref{eq:Wa} contain terms which will be zero when applied to the $R$-parity conserving MSSM because each of
$\mathbf{U}^{1,2}_a$ and $\mathbf{X}^{1,2}_a$ vanish for the values of $a$ for which $\mathbf{V}_a$ and $\mathbf{W}_a$ do not.

Eqs. \eqref{eq:U1}-\eqref{eq:Wa} contain the quadratic Casimirs $\mathbf{C}^{L}_{2}(F)$ and $\mathbf{C}^{R}_{2}(F)$. Along with $\mathbf{C}^{M}_{2}(F)$, to be used in the next section, they are given by $\mathbf{C}_{2}(F)=\mathbf{t}^{A}\mathbf{t}^{A}$, with $\mathbf{t}^{A}$ the group generators for the reducible representation that includes all fermion fields. The superscripts $L$, $R$, $M$ indicate the contributions from left-handed, right-handed and Majorana fermions respectively. Flavour is carried not only by some of the indices of the various matrices, but can also be carried by $a$, which may also carry a colour index. As a result, it is not necessarily true that a trace in \eqref{eq:U1}-\eqref{eq:Wa} (which is a trace over fermion types) will translate into a trace over flavours in the MSSM.

The process of converting the general RGEs to four-component notation involves significantly increasing the number of terms. Writing the Yukawa part of the general two-component Lagrangian as
\begin{equation}\begin{split}\label{eq:twolag}
\mathcal{L}_{(2)}\ni&\ i\psi^\dagger_p\sigma^\mu D_\mu\psi_p + \frac{1}{2}D_\mu\phi_a D^\mu\phi_a -\frac{1}{4}F_{\mu\nu A}F^{\mu\nu}_A-\left(\frac{1}{2}\by^a_{pq}\psi^T_p\zeta\psi_q\phi_a+\mathrm{h.c.}\right)\;,
\end{split}\end{equation}
the two-component RGE for the Yukawa couplings \cite{MVrgeI,*MVrgeII,*MVrgeIII,luo} is\footnote{Eq.~(\ref{eq:mv}) is slightly modified from that in Ref.~\cite{MVrgeI,*MVrgeII,*MVrgeIII}. We have written the $\mathbf{Y}_2^T(F)$ instead of $\mathbf{Y}_2^\dagger(F)$ in the first term and symmetrized the trace with respect to $a$ and $b$. The second modification also appears in Ref.~\cite{luo}, while the first one preserves the symmetry of the Yukawa coupling matrix.}
\begin{equation}\begin{split}\label{eq:mv}
\left(4\pi\right)^2\left.\bbet_Y^a\right|_{1-\mathrm{loop}}=&\tfrac{1}{2}\left[\by^T_2(F)\by^a+\by^a\by_2(F)\right]+2\by^b\by^{a\dagger}\by^b\\[2pt]
&+\by^b\mathrm{Tr}\left\{\tfrac{1}{2}\left(\by^{b\dagger}\by^a+\by^{a\dagger}\by^b\right)\right\}-3g^2\left\{\mathbf{C}_2(F),\by^a\right\}\;. 
\end{split}\end{equation}
where, ${\mathbf Y}_2(F)={\mathbf Y}^{b\dagger}{\mathbf Y}^b$. Clearly, the matrices $\mathbf{Y}^a$, $\mathbf{Y}_2(F)$ and $\mathbf{C}_2(F)$ all have the same dimensionality, determined by the total number of two-component fermion fields in the system. Although the four-component RGEs of \eqref{eq:U1}-\eqref{eq:Wa} look longer and more cumbersome, much work needed to obtain the MSSM RGEs has been done and our general equations therefore represent an intermediate step in this process. The reader is strongly encouraged to consult Ref.~\cite{RGE1}, not least because it contains a detailed discussion of the method used to convert between notations.

Before concluding this section, we note that we have not listed the RGEs for the dimensionless $\bdLm_{abcd}$ and $\bdLm'_{abcd}$, all of which (in the MSSM) correspond to SUSY counterparts of standard SM terms. We anticipate that threshold corrections to these terms will be unimportant for our analysis since they will not enter two-body (flavour-violating) squark decays except at the loop-level. In what follows we have set all these quartic couplings equal to the ``square'' of either the usual Yukawa couplings or the corresponding gauge couplings, their value in the SUSY limit.

\subsection{Dimensionful Parameters}

As with the dimensionless parameters, we use the general equations in Ref.~\cite{luo} to derive the full threshold RGEs for the dimensionful parameters. In the general Lagrangian of \eqref{eq:fourlag} we write the fermion mass terms as $\bdm^{(\prime)}_{X}$, with the subscript `$X$' indicating that they are mass terms for Majorana fermions. We do not write terms like $\bdm^{(\prime)}_{U}$, $\bdm^{(\prime)}_{V}$ or $\bdm^{(\prime)}_{W}$ since gauge invariance precludes the corresponding fermion bilinears in the $R$-parity conserving MSSM.

Applying a similar conversion from two- to four-component fermion spinors and from real to complex scalars, we find \cite{RGE2} (writing $t=\ln{Q}$)
\begin{equation}\label{eq:RGEmx}\begin{split}
\left(4\pi\right)^2\frac{d\mathbf{m}_X}{dt}=&\frac{1}{4}\left[\left(\bw_b\bw^\dagger_b+\bv^T_b\bv^*_b+\bx^{1}_{b}\bx^{1\dagger}_{b}+\bx^{2}_{b}\bx^{2\dagger}_{b}\right)\bmfx\right.\\
&\quad\ +\bmfxd\left(\bw_b\bw^\dagger_b+\bv^T_b\bv^*_b+\bx^{1}_{b}\bx^{1\dagger}_{b}+\bx^{2}_{b}\bx^{2\dagger}_{b}\right)\\
&\quad\ +\bmfx\left(\bw^*_b\bw^T_b+\bv^\dagger_b\bv_b+\bx^{1\dagger}_{b}\bx^{1}_{b}+\bx^{2\dagger}_{b}\bx^{2}_{b}\right)\\
&\left.\quad\ +\left(\bw^*_b\bw^T_b+\bv^\dagger_b\bv_b+\bx^{1\dagger}_{b}\bx^{1}_{b}+\bx^{2\dagger}_{b}\bx^{2}_{b}\right)\bmfxd\right]\\
&+\left[\bx^{1}_{b}\bmfxd\bx^{2}_{b}+\bx^{2}_{b}\bmfxd\bx^{1}_{b}\right.\\
&\left.\quad\ +\bx^{2\dagger}_{b}\bmfx\bx^{1\dagger}_{b}+\bx^{1\dagger}_{b}\bmfx\bx^{2\dagger}_{b}\right]\\
&+\frac{1}{4}\left[\bx^{1}_{b}\mathrm{Tr}\left\{\bx^{1\dagger}_{b}\bmfx+\bmfxd\bx^{2}_{b}\right\}\right.\\
&\qquad\ +\bx^{2}_{b}\mathrm{Tr}\left\{\bx^{2\dagger}_{b}\bmfx+\bmfxd\bx^{1}_{b}\right\}\\
&\qquad\ +\bx^{1\dagger}_{b}\mathrm{Tr}\left\{\bmfxd\bx^{1}_{b}+\bx^{2\dagger}_{b}\bmfx\right\}\\
&\left.\qquad\ +\bx^{2\dagger}_{b}\mathrm{Tr}\left\{\bmfxd\bx^{2}_{b}+\bx^{1\dagger}_{b}\bmfx\right\}\right]\\
&-6g^2\mathbf{C}^{M}_{2}(F)\mathbf{m}_X\;, 
\end{split}\end{equation}
and 
\begin{equation}\label{eq:RGEmxp}\begin{split}
\left(4\pi\right)^2\frac{d\mathbf{m}'_X}{dt}=&\frac{i}{4}\left[\left(\bw_b\bw^\dagger_b+\bv^T_b\bv^*_b+\bx^{1}_{b}\bx^{1\dagger}_{b}+\bx^{2}_{b}\bx^{2\dagger}_{b}\right)\bmfx\right.\\
&\quad\ -\bmfxd\left(\bw_b\bw^\dagger_b+\bv^T_b\bv^*_b+\bx^{1}_{b}\bx^{1\dagger}_{b}+\bx^{2}_{b}\bx^{2\dagger}_{b}\right)\\
&\quad\ +\bmfx\left(\bw^*_b\bw^T_b+\bv^\dagger_b\bv_b+\bx^{1\dagger}_{b}\bx^{1}_{b}+\bx^{2\dagger}_{b}\bx^{2}_{b}\right)\\
&\left.\quad\ -\left(\bw^*_b\bw^T_b+\bv^\dagger_b\bv_b+\bx^{1\dagger}_{b}\bx^{1}_{b}+\bx^{2\dagger}_{b}\bx^{2}_{b}\right)\bmfxd\right]\\
&+i\left[\bx^{1}_{b}\bmfxd\bx^{2}_{b}+\bx^{2}_{b}\bmfxd\bx^{1}_{b}\right.\\
&\left.\qquad-\bx^{2\dagger}_{b}\bmfx\bx^{1\dagger}_{b}-\bx^{1\dagger}_{b}\bmfx\bx^{2\dagger}_{b}\right]\\
&+\frac{i}{4}\left[\bx^{1}_{b}\mathrm{Tr}\left\{\bx^{1\dagger}_{b}\bmfx+\bmfxd\bx^{2}_{b}\right\}\right.\\
&\qquad\ +\bx^{2}_{b}\mathrm{Tr}\left\{\bx^{2\dagger}_{b}\bmfx+\bmfxd\bx^{1}_{b}\right\}\\
&\qquad\ -\bx^{1\dagger}_{b}\mathrm{Tr}\left\{\bmfxd\bx^{1}_{b}+\bx^{2\dagger}_{b}\bmfx\right\}\\
&\left.\qquad\ -\bx^{2\dagger}_{b}\mathrm{Tr}\left\{\bmfxd\bx^{2}_{b}+\bx^{1\dagger}_{b}\bmfx\right\}\right]\\
&-6g^2\mathbf{C}_{2}^M(F)\mathbf{m}'_X\;,
\end{split}\end{equation}
where, as mentioned after the dimensionless RGEs, $\mathbf{C}_{2}^M(F)$ is the quadratic Casimir for the Majorana fermions. Note that, as with the RGEs for $X^{1,2}_{a}$, \eqref{eq:RGEmx} and \eqref{eq:RGEmxp} are symmetric in line with the property of $\bdm^{(\prime)}_{X}$ in \eqref{eq:fourlag}. Also, in the MSSM with $R$-parity conservation, the trace terms in (\ref{eq:RGEmx}) and (\ref{eq:RGEmxp}) vanish. This is because, as seen in Eq.~(\ref{eq:fourlag}), $\bx$ only connects Higgsinos to gauginos, while $\mathbf{m}^{(\prime)}_{X}$ never connects Higgsinos to gauginos. Therefore, a trace of the multiple of these two matrices is always zero.

Writing the scalar quadratic Casimir as $C_{2}(S)=\mathbf{t}^{A}\mathbf{t}^{A}$, the RGE for the trilinear couplings is
\begin{equation}\begin{split}
\left(4\pi\right)^2\frac{d\mathbf{H}_{abc}}{dt}=&
\left[2\left(\bdLm_{afbe}+\bdLm'_{fabe}\right)\mathbf{H}_{ecf}+
\bdLm'_{abef}\mathbf{H}^{*}_{cef}+\left(b\leftrightarrow c\right)\right]\\
&+\bdLm_{efbc}\mathbf{H}_{aef}+2\bdLm'_{ebcf}\left(\mathbf{H}^{*}_{eaf}
+\mathbf{H}_{fae}\right)\\
&+2\mathrm{Tr}\left\{\left(\bdm_{X}-i\bdm'_{X}\right)\left[\left(\bv^
\dagger_a\left\{\bu^1_b\bw^\dagger_c+\bv_b\bx^{2\dagger}_c\right\}+
\left(\bx^{1\dagger}_a+\bx^{2\dagger}_a\right)\bx^1_b\bx^{2\dagger}_c
\right.\right.\right.\\
&\qquad\qquad\qquad\qquad\qquad+\bw^*_b\bw^T_a\bx^{2\dagger}_c+
\bx^{2\dagger}_b\left\{\bw_a\bw^\dagger_c+\left(\bx^1_a+\bx^2_a\right)
\bx^{2\dagger}_c\right\}\\
&\qquad\qquad\qquad\qquad\qquad\left.+\bw^*_b\bu^{1T}_c\bv^*_a+
\bx^{2\dagger}_b\left\{\bv^T_c\bv^*_a+\bx^1_c\left(\bx^{1\dagger}_a+
\bx^{2\dagger}_a\right)\right\}\right)\\
&\qquad\qquad\qquad\qquad\quad\ \left.\left.+\left(b\leftrightarrow c
\right)\right]\right\}\\
&+2\mathrm{Tr}\left\{\left(\bw^T_a\bmfxd\left\{\bv^T_b\bu^{2*}_c+
\bx^1_b\bw^*_c\right\}\right.\right.\\
&\qquad\qquad+\left(\bx^1_a+\bx^2_a\right)\bmfxd\bx^1_b\bx^{2\dagger}_c
+\bv_b\bmfxd\bw_a\bu^{2\dagger}_c\\
&\qquad\qquad+\bx^1_b\bmfxd\left\{\bw_a\bw^\dagger_c+\left(\bx^1_a+
\bx^2_a\right)\bx^{2\dagger}_c\right\}\\
&\qquad\qquad+\bv_b\bmfxd\bx^1_c\bv^\dagger_a\\
&\qquad\qquad\left.+\bx^1_b\bmfxd\left\{\bv^T_c\bv^*_a+\bx^1_c
\left(\bx^{1\dagger}_a+\bx^{2\dagger}_a\right)\right\}\right)\\
&\qquad\quad\ \left.+\left(b\leftrightarrow c\right)\right\}\\
&+\mathrm{Tr}\left\{\bu^{2\dagger}_{a'}\left(\bu^1_a+\bu^2_a\right)+
\left(\bu^{1\dagger}_a+\bu^{2\dagger}_a\right)\bu^1_{a'}+\bv^\dagger_a
\bv_{a'}+\bw^\dagger_{a'}\bw_a\right.\\
&\qquad\quad\left.+\frac{1}{2}\left\{\bx^{2\dagger}_{a'}\left(\bx^1_a+
\bx^2_a\right)+\left(\bx^{1\dagger}_a+\bx^{2\dagger}_a\right)\bx^1_{a'}
\right\}\right\}\mathbf{H}_{a'bc}\\
&+\mathrm{Tr}\left\{\bu^{1\dagger}_{b'}\bu^1_b+\bu^{2\dagger}_b
\bu^2_{b'}+\bv^\dagger_{b'}\bv_b+\bw^\dagger_b\bw_{b'}\right.\\
&\left.\qquad\qquad\qquad\qquad\qquad\qquad+\frac{1}{2}\left
\{\bx^{1\dagger}_{b'}\bx^1_b+\bx^{2\dagger}_b\bx^2_{b'}\right\}\right\}
\mathbf{H}_{a{b'}c}\\
&+\mathrm{Tr}\left\{\bu^{2\dagger}_{b'}\bu^1_b+\bu^{2\dagger}_b
\bu^1_{b'}+\frac{1}{2}\left\{\bx^{2\dagger}_{b'}\bx^1_b+
\bx^{2\dagger}_b\bx^1_{b'}\right\}\right\}\left(\mathbf{H}_{b'ac}+
\mathbf{H}^*_{cab'}\right)\\
&+\mathrm{Tr}\left\{\bu^{1\dagger}_{c'}\bu^1_c+\bu^{2\dagger}_c
\bu^2_{c'}+\bv^\dagger_{c'}\bv_c+\bw^\dagger_c\bw_{c'}\right.\\
&\left.\qquad\qquad\qquad\qquad\qquad\qquad+\frac{1}{2}\left
\{\bx^{1\dagger}_{c'}\bx^1_c+\bx^{2\dagger}_c\bx^2_{c'}\right\}\right\}
\mathbf{H}_{ac'b}\\
&+\mathrm{Tr}\left\{\bu^{2\dagger}_{c'}\bu^1_c+\bu^{2\dagger}_c
\bu^1_{c'}+\frac{1}{2}\left\{\bx^{2\dagger}_{c'}\bx^1_c+
\bx^{2\dagger}_c\bx^1_{c'}\right\}\right\}\left(\mathbf{H}_{c'ab}+
\mathbf{H}^*_{bac'}\right)\\
&-3g^2\left[C_2(a)+C_2(b)+C_2(c)\right]\mathbf{H}_{abc}\;, \label{eq:4compH}
\end{split}\end{equation}
where, again, it can be seen that the $b\leftrightarrow c$ symmetry is preserved by the RGE as it must be. Note that in the $R$-parity conserving MSSM we can choose (without loss of generality) for the first index of $H_{abc}$ to apply to a sfermion. This means that $\bu^{1,2}_{a}$ and $\bx^{1,2}_{a}$ are always zero, and the corresponding terms drop out from the equation.

Finally, since in the $R$-parity conserving MSSM we never have non-zero entries in the Lagrangian corresponding to both $\bdm^{2}_{ab}$ and $\bm{\mathcal{B}}_{ab}$ for the same $a,b$, we write the RGE for a combination of these two terms.
\begin{equation}\label{eq:4compmb}\begin{split}
\left(4\pi\right)^2\frac{d\left[\bdm^{2}_{ab}-\bm{\mathcal{B}}_{ab}\right]}{dt}=&\left\{2\left(\bdLm_{afbe}+\bdLm'_{fabe}\right)\bdm^{2}_{ef}-\bdLm'_{abef}\bm{\mathcal{B}}_{ef}^*-\left(\bdLm_{efab}+\bdLm'^{*}_{baef}\right)\bm{\mathcal{B}}_{ef}\right\}\\
&+\left\{\mathbf{H}_{aef}\mathbf{H}^*_{bef}+2\left(\mathbf{H}^*_{eaf}+\mathbf{H}_{fae}\right)\mathbf{H}_{ebf}\right\}\\
&-2\left[2\ \mathrm{Tr}\left\{\left[\bv^T_b\bv^*_a+\bw_a\bw^\dagger_b+\left(\bx^1_a+\bx^2_a\right)\bx^{2\dagger}_b\right.\right.\right.\\
&\qquad\qquad\quad\ \left.\left.+\bx^1_b\left(\bx^{1\dagger}_a+\bx^{2\dagger}_a\right)\right]\bmfx\bmfxd\right\}\\
&\qquad\ +\mathrm{Tr}\left\{\left(\bx^1_a+\bx^2_a\right)\bmfxd\bx^1_b\bmfxd\right\}\\
&\qquad\ \left.+\mathrm{Tr}\left\{\left(\bx^{1\dagger}_a+\bx^{2\dagger}_a\right)\bmfx\bx^{2\dagger}_b\bmfx\right\}\right]\\
&-3g^2\left[C_2(a)+C_2(b)\right]\left(\bdm^{2}_{ab}-\bm{\mathcal{B}}_{ab}\right)\\
&+\left[\mathrm{Tr}\left\{\bu^{2\dagger}_{a'}\left(\bu^1_a+\bu^2_a\right)+\left(\bu^{1\dagger}_a+\bu^{2\dagger}_a\right)\bu^1_{a'}+\bv^\dagger_a\bv_{a'}+\bw^\dagger_{a'}\bw_a\right.\right.\\
&\qquad\quad\ \left.+\frac{1}{2}\bx^{2\dagger}_{a'}\left(\bx^1_a+\bx^2_a\right)+\frac{1}{2}\left(\bx^{1\dagger}_a+\bx^{2\dagger}_a\right)\bx^1_{a'}\right\}\bdm^{2}_{a'b}\\
&\quad\ \ -\mathrm{Tr}\left\{\bu^{1\dagger}_{a'}\left(\bu^1_a+\bu^2_a\right)+\left(\bu^{1\dagger}_a+\bu^{2\dagger}_a\right)\bu^2_{a'}\right.\\
&\qquad\qquad\ +\bv^\dagger_{a'}\bv_a+\bw^\dagger_a\bw_{a'}\\
&\qquad\qquad\ \left.+\frac{1}{2}\bx^{1\dagger}_{a'}\left(\bx^1_a+\bx^2_a\right)+\frac{1}{2}\left(\bx^{1\dagger}_a+\bx^{2\dagger}_a\right)\bx^2_{a'}\right\}\bm{\mathcal{B}}_{a'b}\\
&\quad\ \ +\mathrm{Tr}\left\{\bu^{1\dagger}_{b'}\bu^1_b+\bu^{2\dagger}_b\bu^2_{b'}+\bv^\dagger_{b'}\bv_b\right.\\
&\qquad\qquad\ \left.+\bw^\dagger_b\bw_{b'}+\frac{1}{2}\bx^{1\dagger}_{b'}\bx^1_b+\frac{1}{2}\bx^{2\dagger}_b\bx^2_{b'}\right\}\left(\bdm^{2}_{ab'}-\bm{\mathcal{B}}_{ab'}\right)\\
&\quad\ \ +\mathrm{Tr}\left\{\bu^{2\dagger}_{b'}\bu^1_b+\bu^{2\dagger}_b\bu^1_{b'}+\frac{1}{2}\bx^{2\dagger}_{b'}\bx^1_b+\frac{1}{2}\bx^{2\dagger}_b\bx^1_{b'}\right\}\\
&\left.\qquad\qquad\qquad\qquad\qquad\qquad\qquad\qquad\qquad\times\left(\bdm^{2}_{ab'}-\bm{\mathcal{B}}_{ab'}\right)^*\right]\;.
\end{split}\end{equation}
In a similar manner to the RGE for $H_{abc}$ above, when $a$ and $b$ are sfermion indices, $\bu^{1,2}_{a,b}$ and $\bx^{1,2}_{a,b}$ are zero.  Likewise, when $a$ and $b$ label Higgs fields, $\bv_{a,b}$ and $\bw_{a,b}$ are zero.  The derivation of the MSSM RGEs is, therefore, considerably less cumbersome than it appears at first sight.

It is worth repeating before ending this section that we have written the RGEs for both the dimensionless and dimensionful couplings with regard only to the $R$-parity conserving MSSM. As such, we have removed terms which are zero in this context. Complete RGEs derived from the Lagrangian in \eqref{eq:fourlag} can be found in Refs.~\cite{RGE1}~and~\cite{RGE2} which also contain a more detailed description of the procedure for obtaining our general RGEs from the two-component ones.

\section{Choice of Renormalisation Scheme}\label{sec:choiceren}

The removal of divergences as part of the renormalisation process requires us to choose a regularisation scheme that is consistent with all the symmetries in our theory. Although the modified minimal subtraction (\msb) scheme is most commonly used in the Standard Model, the modified dimensional reduction (\drb) scheme is more appropriate for supersymmetric theories.

As we will describe in Ch.~\ref{ch:rgeflav}, we use the \msb~gauge couplings at the weak scale as our boundary conditions, and these will therefore need to be converted to \drb, using the prescription in, for example, Ref.~\cite{mvregdep}:
\begin{align}
\label{eq:a1msdr}\frac{1}{\alpha_{1}^{\mmsb}}=&\frac{1}{\alpha_{1}^{\mdrb}}\\
\frac{1}{\alpha_{2}^{\mmsb}}=&\frac{1}{\alpha_{2}^{\mdrb}}+\frac{1}{6\pi}\\
\label{eq:a3msdr}\frac{1}{\alpha_{3}^{\mmsb}}=&\frac{1}{\alpha_{3}^{\mdrb}}+\frac{1}{4\pi}\;.
\end{align}
In addition to converting our weak scale inputs, we must also consider the effects of the relations \eqref{eq:a1msdr}-\eqref{eq:a3msdr} on the RGEs. We find that the additional contributions that appear when we convert the one-loop RGEs are of two-loop order and have the effect of changing the various coefficients in the two-loop RGEs. Therefore, since we wish to keep all two-loop sized effects, we must ensure that all RGEs we use are converted to the \drb~scheme.

The invariance of the one-loop RGEs themselves under \msb-\drb~conversion means that we are free to use our adaptation of the general RGEs from Refs.~\cite{MVrgeI,*MVrgeII,*MVrgeIII,luo} (which use the \msb~scheme) to derive new one-loop RGEs with threshold effects included. However, our two-loop RGEs are taken from existing literature, and the SM RGEs must therefore be converted. In addition to the gauge couplings, the Yukawa couplings also differ between the two scehemes. We use the symbol $\bm{\lambda}_{u,d,e}$ (as opposed to $\bdf_{u,d,e}$) to differentiate the SM Yukawa matrix from the MSSM one\footnote{The two Yukawa matrices are of course related, something we will discuss in Sec.~\ref{sec:masseig} when we consider the Higgs boson mass eigenstates.} and convert the two-loop RGEs \cite{casno} using the following relations \cite{mvregdep}
\begin{align}
\bm{\lambda}^{\mmsb}_{u}&=\bm{\lambda}^{\mdrb}_{u}\left\{1-\frac{1}{120}\frac{g^{2}_{1}}{16\pi^{2}}-\frac{3}{8}\frac{g^{2}_{2}}{16\pi^{2}}+\frac{4}{3}\frac{g^{2}_{3}}{16\pi^{2}}\right\}\\
\bm{\lambda}^{\mmsb}_{d}&=\bm{\lambda}^{\mdrb}_{d}\left\{1-\frac{13}{120}\frac{g^{2}_{1}}{16\pi^{2}}-\frac{3}{8}\frac{g^{2}_{2}}{16\pi^{2}}+\frac{4}{3}\frac{g^{2}_{3}}{16\pi^{2}}\right\}\\
\bm{\lambda}^{\mmsb}_{e}&=\bm{\lambda}^{\mdrb}_{e}\left\{1+\frac{9}{40}\frac{g^{2}_{1}}{16\pi^{2}}-\frac{3}{8}\frac{g^{2}_{2}}{16\pi^{2}}\right\}\;,
\end{align}
where we omit scheme labels from the $g^{2}_{i}$ terms since we are only keeping two-loop sized differences. The RGEs for two-loop running of the gauge and Yukawa couplings are listed in Appendix~\ref{app:drbar}, and since the two-loop terms of the Yukawa couplings depend on the SM Higgs quartic coupling, $\lambda$, its RGE is also given. We do not convert the $\lambda$ RGE to \drb~since its appearance at the two-loop level will mean corrections would only be of three-loop order. A simple check of the $g_{3}$ coefficient in the \drb~RGEs for Yukawa couplings is outlined in  Appendix~\ref{app:smrgecheck}, where we compare our result with that given in Ref.~\cite{BFMTdr2ms}.

Having developed our one-loop RGEs with threshold corrections, we augment these with two-loop terms without threshold corrections from the RGEs for the SM \cite{casno} and the MSSM \cite{MVrge}. We change the two-loop RGEs from MSSM to SM at a single threshold, $m_{H}$, so that our running consists of one-loop RGEs with full thresholds and two-loop \drb~RGEs with a single threshold. Threshold effects in the two-loop RGEs are expected to be of three-loop order and therefore small enough to be neglected in our analysis.

\section{The MSSM to SM Transition}\label{sec:smmssmtrans}

We now turn to consider the transition from SM to MSSM evolution. In Fig.~\ref{fig:gaugeunif} we plotted the SM running below $1\ $TeV and MSSM running above. This is a single `threshold' where our effective theory transitions from being the SM, with SM field content, to being softly broken SUSY. The coincidence of unification coupled with the fact that SUSY is stable to radiative corrections strongly suggests that softly broken SUSY may be the correct effective theory all the way to $\sim10^{16}\ $GeV.

Considering that we wish to integrate the RGEs to full two-loop accuracy, we are led to consider the errors introduced by our approximation of a single threshold. We first write the change in some quantity, $q$, due to the one-loop running over the whole range $M_{Z}<Q<\mgut$ as
\begin{equation*}
\delta^{(1)}_{q}\sim\beta_{q}\ln{\left(\frac{\mgut}{M_{Z}}\right)}\;,
\end{equation*}
where $\beta_{q}$ represents the RGE for $q$, and then compare order of magnitude estimates of the corrections to this change from two-loop terms versus threshold corrections.
\begin{description}
\item[Two-loop:] The two-loop contribution to the weak scale value of $q$, $\delta^{(2)}_{q}$, is smaller than the one-loop contribution to the running, $\delta^{(1)}_{q}$, by approximately a factor of $1/16\pi^{2}$, so that the size of the two-loop contribution can be symbolically written as
\begin{equation*}
\delta^{(2)}_{q}\sim\frac{1}{16\pi^{2}}\delta^{(1)}_{q}
\end{equation*}
\item[Thresholds:] Smeared out thresholds will only affect the running at most over the range from the heaviest SUSY particle ($m_{\mathrm{HSP}}$) to the lightest SUSY particle ($m_{\mathrm{LSP}}$). In the intermediate range, the $\beta$-function for the one-loop running will be neither equal to the MSSM value nor the SM value but will be the same order of magnitude. We can therefore estimate the correction to the one-loop weak scale value of $q$, $\delta^{(th)}_{q}$, to be
\begin{equation*}
\delta^{(th)}_{q}\sim\beta_{q}\ln{\left(\frac{m_{\mathrm{HSP}}}{m_{\mathrm{LSP}}}\right)}\sim\delta^{(1)}_{q}\ln{\left(\frac{m_{\mathrm{HSP}}}{m_{\mathrm{LSP}}}\right)}/\ln{\left(\frac{\mgut}{M_{Z}}\right)}\;.
\end{equation*}
\end{description}
We can see that if $\ln{\left(m_{\mathrm{HSP}}/m_{\mathrm{LSP}}\right)}\sim1$, which is a very likely scenario, these two contributions will be comparable. We therefore find that we must include more detailed threshold effects to one-loop order, since these can easily be of the size of two-loop corrections, or even more sizable if the mass spectrum of SUSY is spread over a large range.

One method to improve on the single threshold approximation is to include Higgs boson and SUSY thresholds as step functions in the RGEs, an approach which has been considered before \cite{casno,sakis} and is also used here. We follow Ref.~\cite{casno} and make the assumption that each particle with mass $M_{i}$ is included in the effective theory if $Q>M_{i}$ and excluded if $Q<M_{i}$. Using this approach, the $\beta$-functions are as in the MSSM at a scale above the masses of all SUSY particles. As we move down in scale, the particles are decoupled individually until we have only SM particles in the theory and the $\beta$-functions reduce to those of the SM. The reduction to the RGEs of the MSSM and SM in the appropriate limits will be a check on our work.

Our RGEs must be able to describe the slow transition from the MSSM to the SM via a sequence of effective theories with changing numbers of particles. To this end, we separate out the contribution to the RGEs from each particle and identify the terms with a $\theta_{\mathcal{P}}$ for each particle $\mathcal{P}$ \cite{casno,sakis}, such that
\begin{equation*}\begin{split}
\theta_\mathcal{P} =&\ 1 \ \ {\rm if} \ Q > M_{\mathcal{P}} \\
  &\ 0 \ \ {\rm if} \ Q < M_{\mathcal{P}}\;.
\end{split}\end{equation*}
In this manner, we obtain the MSSM limit of our equations by setting all $\theta_{\mathcal{P}}=1$ and the SM limit when all $\theta_{\mathcal{P}}$ corresponding to decoupled SUSY particles are set equal to zero.

If the separation between mass scales is not large, the errors introduced by decoupling the various particles at the same point, say $M_{\mathrm{SUSY}}$, are small. Since the threshold effects enter only logarithmically in the RGE solutions, we have seen that any error would be of $\mathcal{O}\left(\frac{1}{16\pi^{2}}\ln{\frac{M_{\mathrm{SUSY}}}{M_{i}}}\right)$ and therefore less important as $M_{i}\rightarrow M_{\mathrm{SUSY}}$. As a result, we see that it is permissible to find only approximate mass eigenstates, as long as we ensure that the size of the splitting between masses is small enough to introduce only a negligible amount of error compared to the two-loop corrections.

Attempting to follow this route uncovers two issues:
\begin{enumerate}
\item How do we define the `mass eigenstates' that are decoupled?
\item What happens once SUSY is explicitly broken by the decoupling of the heaviest SUSY particle?
\end{enumerate}
The following two sections consider each of these questions in turn.

\subsection{Mass Eigenstates}\label{sec:masseig}

Although it is possible to develop a scheme in which particles are all decoupled from the effective theory at a given point, this procedure involves its own complications. If, for example, we had run the gauge couplings in Fig.~\ref{fig:gaugeunif} all the way to $M_{Z}$ according to the MSSM RGEs, we would have to apply a correction to the weak scale couplings proportional to the logarithm of $M_{\mathrm{SUSY}}/M_{Z}$. In the more realistic scenario where the SUSY particles have a variety of masses in the range of $1$~TeV, it would be necessary to apply a number of different corrections to the weak scale values. Also, the larger the difference between $M_{\mathrm{SUSY}}$ and $M_{Z}$ the less accurate this procedure becomes.

In many cases it is therefore simplest to decouple the various SUSY particles at the scale of their mass. If we are to do this we must rewrite the Lagrangian in terms of fields which are in their (approximate) mass basis, so we continue by considering the mass eigenstates of each sector of the MSSM in turn. The one exception to our general approach is the top quark, whose mass is $\sim172$~GeV. Since we wish to maintain $SU(2)$ invariance, and decoupling the top would break this, we decouple this quark at the scale $M_{Z}$ and apply a fixed correction to the gauge couplings and Yukawa matrices to compensate \cite{weing,*hallg,*ovrutg,*chetg,*noteg}. This allows us to keep our RGEs $SU(2)$ invariant over the whole range we consider, \textit{i.e.}, $M_{Z}<Q<M_{\mathrm{GUT}}$.

\subsubsection{Higgs Sector}

In the Higgs sector the rotation to the mass eigenstate basis is particularly important. Since the MSSM contains two Higgs doublets, $h_{u}$ and $h_{d}$, we have a number of mass eigenstates, labelled $h$, $H$, $A$, and $H^{\pm}$, which are combinations of the $h_{u}$ and $h_{d}$ fields. It is not clear, therefore, where we would place the transition from a theory including the $h_{d}$ field, to one where this field is absent. In contrast, it is clear where to decouple, for example, the $A$ field, since it has a definite mass. Moreover, we can write our fields in such a way that the lightest field, $h$, does the job of the SM Higgs boson and remains in the theory all the way down to the weak scale.

In view of the discussion just before the beginning of this section on the definition of the approximate mass eigenstates, we can simplify our treatment of the Higgs sector considerably. We find that threshold effects are only important if the mass scale of the non-SM Higgs bosons is much larger than $m_{h}$. In this scenario, diagonalization of the various mass matrices in the Higgs sector leads us to conclude that $m_{A}\simeq m_{H}\simeq m_{H^{\pm}}>>m_{h}$ with $h$ approximately the SM Higgs particle. We rearrange the doublets $h_{u}$ and $h_{d}$ into two new doublets which transform as $\mathbf{2}$s under $SU(2)$ and have positive weak hypercharge:
\begin{eqnarray}
\label{eq:hrot}\left(\begin{array}{c}G^{+}\\\mathsf{h}\end{array}\right)&=\sn\left(\begin{array}{c}h^{+}_{u}\\[5pt]h^{0}_{u}\end{array}\right)+\cs\left(\begin{array}{c}h^{-*}_{d}\\[5pt]h^{0*}_{d}\end{array}\right)\\[5pt] 
\label{eq:Hrot}\left(\begin{array}{c}H^{+}\\\mathcal{H}\end{array}\right)&=\cs\left(\begin{array}{c}h^{+}_{u}\\[5pt]h^{0}_{u}\end{array}\right)-\sn\left(\begin{array}{c}h^{-*}_{d}\\[5pt]h^{0*}_{d}\end{array}\right),
\end{eqnarray}
where the electrically neutral, {\it complex} fields $\mathsf{h}$ and
$\mathcal{H}$ are given by,
\begin{equation}\label{eq:hHbasis}
\mathsf{h}=\frac{h+iG^{0}}{\sqrt{2}}\qquad\qquad\mathcal{H}=\frac{-H+iA}{\sqrt{2}}\;,
\end{equation}
and $\sn=\sin{\beta}$ and $\cs=\cos{\beta}$. Here, $G^{0}$ and $G^{+}$ are the would-be Goldstone bosons which are incorporated into the longitudinal components of the heavy gauge bosons as a result of the Higgs mechanism.

The heavy doublet (containing $A$, $H$ and $H^{\pm}$) will be decoupled at a common scale, $m_{H}$. At this stage some terms in the Lagrangian drop out, and to continue running we must switch from considering the Yukawa matrices, $\mathbf{f}_{u,d,e}$, which couple to both $h_{u}$ and $h_{d}$, to running only that component of the Yukawa couplings which couples to the remaining field, $\mathsf{h}$. A sample term in the MSSM Lagrangian which couples the up-type quarks to the Higgs doublets via the Yukawa matrices is
\begin{equation}
\mathcal{L}\ni-\bar{u}_{j}(\mathbf{f}_{u})^{T}_{ji}h^{0}_{u}P_{L}u_{i}\equiv-\bar{u}_{j}(\mathbf{f}_{u})^{T}_{ji}(\sn\mathsf{h}+\cs\mathcal{H})P_{L}u_{i}\;.
\end{equation}
We can see that at scales below $m_{H}$, the coefficient of the operator that remains in the theory is $\sn\mathbf{f}_{u}$, which we will identify as the Standard Model Yukawa coupling, denoted previously by $\bm{\lambda}_{u}$. Similarly, the other SM Yukawa matrices are $\bm{\lambda}_{d}=\cs\mathbf{f}_{d}$ and $\bm{\lambda}_{e}=\cs\mathbf{f}_{e}$. We will write our RGEs in such a way so that as we pass the threshold for the heavy Higgs particles we switch from evolving the MSSM Yukawa couplings ($\mathbf{f}_{u,d,e}$) to evolving the SM Yukawa couplings ($\bm{\lambda}_{u,d,e}$).

Note that the rotation defined in \eqref{eq:hrot} and \eqref{eq:Hrot} depends on a \textit{fixed} angle, $\tan{\beta}$ \mbox{---} the same angle as in the standard mSUGRA parameters. We therefore define the relation between $\bm{\lambda}_{u,d,e}$ and $\mathbf{f}_{u,d,e}$ to be valid only at the scale $Q=m_{H}$, along with the mSUGRA input for $\tan{\beta}$. At this scale, $\tan{\beta}\equiv v_{u}/v_{d}$.

Our\label{`Our'} rotation to the approximate Higgs mass basis is unimportant when $Q>m_{H}$ since all we have done is make a field redefinition. The point is that by making the rotation we are able to identify the Higgs field that remains in the theory below $m_{H}$. As a consequence, we are also able to determine the specific combinations of original Lagrangian parameters which remain in the effective theory. We will return to this point when we discuss the derivation of the MSSM RGEs in more detail.

\subsubsection{Neutralinos and Charginos}

Since decoupling effects are unimportant if the scale of the decoupling particle is of the order of $M_{Z}$, threshold corrections due to charginos and neutralinos will only be important if either $\left|\mu\right|$ or the SSB gaugino masses are significantly larger than the weak scale. In this scenario, higgsinos and gauginos will have a small mixing and the mass eigenstates are well approximated to be the bino, three winos and two higgsino doublets given by
\begin{equation}\label{eq:inorot}
\tilde{h}_{1,2}=\frac{\psi_{h_{d}}\mp\psi_{h_{u}}}{\sqrt{2}}\;.
\end{equation}
Both higgsinos have the same mass, decoupling at the scale $Q=\left|\mu\right|$, and therefore the rotation of the fields is largely irrelevant for this study. Having said this, it is important to note that the rotation that we perform on the higgsinos is not the same as that for their spin 0 partners. This appears to be different from the approach used elsewhere in the literature \cite{casno,sakis}.

Finally, note that there is the potential for some of the fermion fields to have negative mass eigenvalues. In this case, we must redefine the fields \cite{wss} $\Psi_{k}\rightarrow\left(i\gamma_{5}\right)^{\theta_{k}}\Psi_{k}$, where $\theta_{k}=0(1)$ for positive (negative) eigenvalues, so that the mass term in the Lagrangian has the standard form. We have checked that the RGEs are independent of these additional $\gamma_{5}$ factors.

\subsubsection{Squark (and Slepton) Sector}\label{sec:sqdec}

We have already described in some detail the potential for mixing among squarks (and sleptons) of equal charge. Ignoring the small effect on the eigenstate from flavour-violating contributions, the left- and right-handed squarks will mix with each other so that the mass eigenstates of the up-type squarks, for example, are formed from $\tilde{u}_{L}-\tilde{u}_{R}$, $\tilde{c}_{L}-\tilde{c}_{R}$ or $\tilde{t}_{L}-\tilde{t}_{R}$. This mixing is intrinsically an $SU(2)$ breaking effect, since the $SU(2)$ doublet superfield $\hat{Q}$ contains both $\tilde{u}_{L}$ and $\tilde{d}_{L}$ which will form parts of mass eigenstates with \textit{different} mass. If we are to truly describe the effects from such mixing, we would have to introduce new $SU(2)\times U(1)$ violating operators that would appear when the squarks acquire mass, and derive corresponding RGEs for these terms.

We argue, however, that to the extent that our calculation only depends logarithmically on the location of the thresholds, we can ignore left-right mixing in the squark sector and decouple them at the scale of the eigenvalues of the respective SSB mass matrices. This approximation only breaks down when the off-diagonal entries in the squark mass matrices result in large cancellations, causing the physical mass of the lighter squark eigenvalue to become significantly smaller than either of the diagonal entries. We therefore do not claim to be accurately describing such a situation, which would require full consideration of the $SU(2)$ breaking terms referred to above.

As with the other particles, we must identify the approximate mass scale at which to decouple each squark. Since we are neglecting $SU(2)\times U(1)$ breaking effects when pinpointing these thresholds, they will be located solely by the eigenvalues of the various SSB mass matrices. Approaching the highest squark threshold from above, we will reach a point where the largest eigenvalue becomes less than the scale, $Q_{0}$. At this point, we decouple the corresponding squark, keeping all others in our effective theory below $Q_{0}$. To implement the decoupling we rotate to the basis in which the corresponding SSB matrix is diagonal at $Q=Q_{0}$ using the appropriate one of
\begin{subequations}\begin{eqnarray}
\left(\begin{array}{c}\tilde{u}_{L}\\\tilde{d}_{L}\end{array}\right)&=&\mathbf{R}_{Q}\left(\begin{array}{c}\tilde{u}^{M}_{L}\\\tilde{d}^{M}_{L}\end{array}
\right)\;,\\ \tilde{u}_{R}&=&\mathbf{R}_{u}\tilde{u}^{M}_{R}\;,\\
\tilde{d}_{R}&=&\mathbf{R}_{d}\tilde{d}^{M}_{R}\;, \ \ \ {\rm for\ 
squarks, or}\\
\left(\begin{array}{c}\tilde{e}_{L}\\\tilde{\nu}_{L}\end{array}\right)&=&\mathbf{R}_{L}\left(\begin{array}{c}\tilde{e}^{M}_{L}\\\tilde{\nu}^{M}_{L}\end{array}\right)\;,\\
\tilde{e}_{R}&=&\mathbf{R}_{e}\tilde{e}^{M}_{R} \ \ \ {\rm for\ 
sleptons.}
\end{eqnarray}\end{subequations}
The unitary rotation matrices, $\mathbf{R}_\bullet$, are chosen to diagonalize the
Hermitian SSB squark (or slepton) mass matrices, for example:
\begin{equation}
\left(\mathbf{R}^{\dagger}_{Q}\mathbf{m}^{2}_{Q}\mathbf{R}_{Q}\right)_{ij}=\left(\mathbf{m}^{2}_{Q}\right)^{\mathrm{diag}}_{ij}=(\mathbf{m}^{2}_{Q})^{\mathrm{diag}}_{ij}\bm{\delta}_{ij}\;. 
\end{equation}
With this in mind, the RGEs will be developed such that the contributions from each squark, $\tilde{q}_{k}$ (with $\tilde{q}=\tilde{Q}_{L}$, $\tilde{u}_{R}$ or $\tilde{d}_{R}$) with mass basis index $k$, will be identified with a $\theta_{\tilde{q}_{k}}$. This identification requires that we write our RGEs in the current basis where the various SSB mass matrices are diagonal. In any other current basis, the $\theta_{\tilde{q}_{k}}$ become matrices in flavour space, $\bm{\Theta}_{q}$. Taking $\sqk$, for example, we can rotate back to the original current basis using
\begin{equation}\label{eq:Thetadef}
(\bm{\Theta}_{Q})_{ij}=(\mathbf{R}_{Q}\bm{\Theta}^{\mathrm{diag}}_{Q}\mathbf{R}^{\dagger}_{Q})_{ij}=\theta_{{\tilde{Q}_{k}}}(\mathbf{R}_{Q})_{ik}\bm{\delta}_{kl}(\mathbf{R}^{\dagger}_{Q})_{lj}
\end{equation}
It should be clear that when $\sqk=1$ for all $k$, $\bm{\Theta}_{Q}=\dblone$ so that in this regime we can use our RGEs in the current basis of our choosing without first rotating to the squark mass basis.

Below the scale of decoupling of the heaviest squark, we continue to evolve the RGEs in our new current basis in which the heaviest squark decoupled. The eigenvector and mass of the decoupled squark are frozen and stored for use in connection with the physical squark states, but the $\theta_{\tilde{q}_{k}}$ for the set of three squarks in question will ensure that the decoupled squark no longer contributes to the running. Although we continue calculating the running based upon three active squarks, the decoupled squark does not contribute, and the next squark will decouple at the heavier of the two eigenvalues of the $(2\times2)$ sub-matrix in the space orthogonal to the decoupled eigenvector.

Finally, once two of the three squarks in the set have decoupled, the eigenvector of the remaining squark is fixed to be orthogonal to the other two. Thus only the eigenvalue evolves, and the squark is decoupled from the effective theory at the scale $Q=Q_{0}$ for this eigenvalue. Naturally, once all the squarks have decoupled, $\theta_{\tilde{q}_{k}}=0$ for all $\tilde{q}_{k}$  and equivalently all matrices $\bm{\Theta}_{q}=\bm{0}$ so that any rotation from the original current basis is no longer necessary.

\subsection{`Broken' SUSY}\label{sec:broken}

The second question on our list of issues was that of the effect of explicitly breaking SUSY with the decoupling of the heaviest SUSY particle. Aside from the issues with changing operators in the Lagrangian as mentioned above (see the paragraph on page~\pageref{`Our'} at the end of the Higgs sector discussion), there is an additional effect which must be accounted for in a full two-loop analysis of the RGEs. This is the fact that when the heaviest particle is decoupled from our effective theory there is no longer any requirement that the couplings of SUSY particles remain the same as those of their SM counterparts. In fact, over the scale that they remain in the theory before they themselves decouple, their couplings have the potential to become significantly different, and through the RGEs appreciably affect the running of their SM counterparts.


With this effect in mind, we must carry out our derivation of the RGEs allowing for differences in the RGEs for the SUSY couplings compared with those for their SM counterparts. In the SUSY limit, the scalar-fermion-fermion couplings of squarks and quarks to gauginos and higgsinos are equal to the gauge couplings and usual Yukawa matrices respectively. However, once the heaviest SUSY particle decouples, we find that the effective theory `knows' about the SUSY breaking and as we shall see these SUSY `Yukawa' couplings retain some contributions which are removed from their SM counterparts.

For the gaugino couplings we must allow for flavour off-diagonal terms to develop, and we therefore write the couplings, which were diagonal and equal to the corresponding gauge coupling in the SUSY limit, as the matrices $\tilde{\mathbf{g}}^{\Phi}_{i}$, where $\Phi$ denotes the squark that takes part in the interaction and $i$ denotes the gauge group. Similarly, the higgsino couplings become different to their SM counterparts, $\mathbf{f}_{u,d,e}$, and we label them $\tilde{\mathbf{f}}^{\Phi}_{u,d,e}$, again denoting the relevant scalar with a $\Phi$.

The situation is best illustrated with an example. The coupling of a squark to a quark via an $SU(2)$ gaugino is restricted in the MSSM to be equal to the quark-quark-W coupling. They are both proportional to the $SU(2)$ gauge coupling, $g_{2}$, with some fixed coefficient. In addition, the RGE (to one-loop) for $g_{2}$ depends only on $g_{2}$ and nothing else. However, there is nothing conceptually to prevent the squark-quark-gaugino coupling from being different to $g_{2}$ once the effective theory contains broken SUSY. In this regime we therefore have a new coupling which we will label (following the discussion above) with a tilde, and also a superscript $\Phi$ to signify the scalar field in the interaction. For the $SU(2)$ coupling of left-handed squarks and quarks, we will call this coupling $\bgt^{Q}$.

\begin{figure}
\centering
\vspace{0.5cm}
\subfloat[][]{\label{fig:gtqgraph1}\includegraphics[scale=0.5]{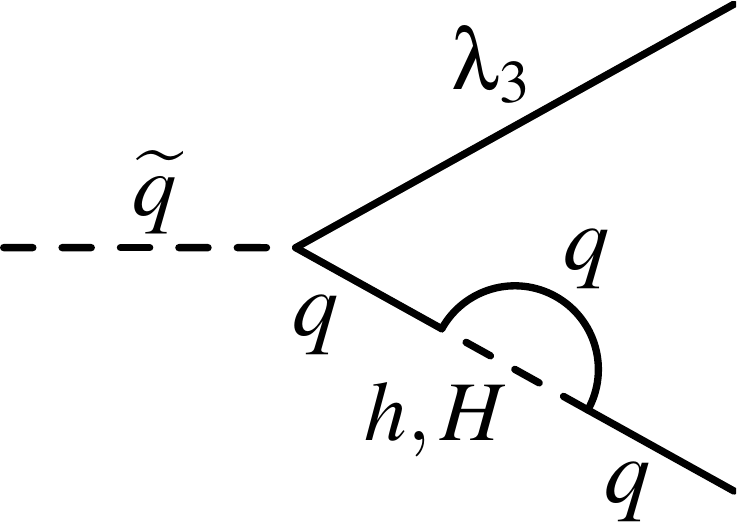}}
\hspace{20pt}
\subfloat[][]{\label{fig:gtqgraph2}\includegraphics[scale=0.5]{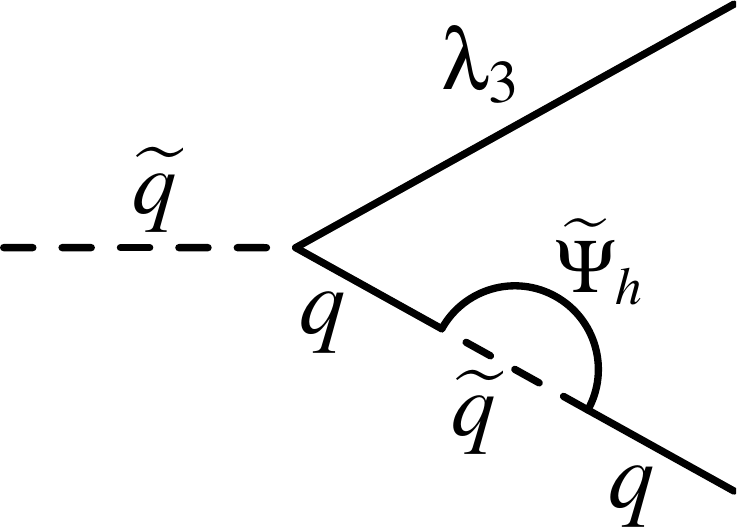}}\\
\subfloat[][]{\label{fig:gtqgraph3}\includegraphics[scale=0.5]{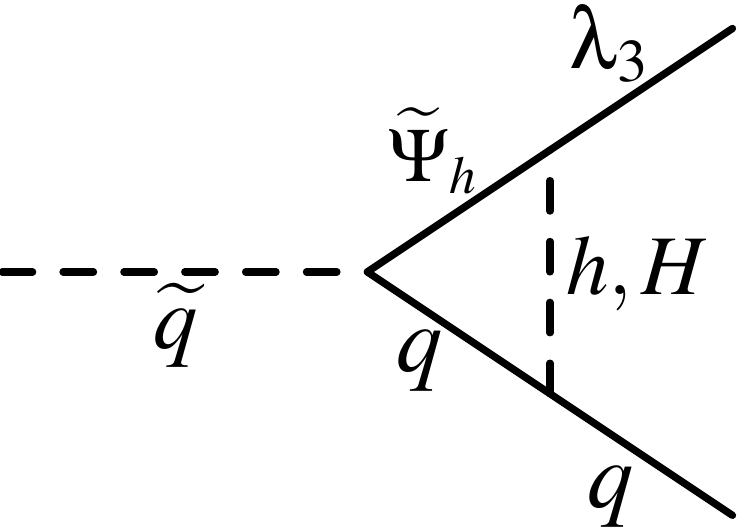}}
\hspace{20pt}
\subfloat[][]{\label{fig:gtqgraph4}\includegraphics[scale=0.5]{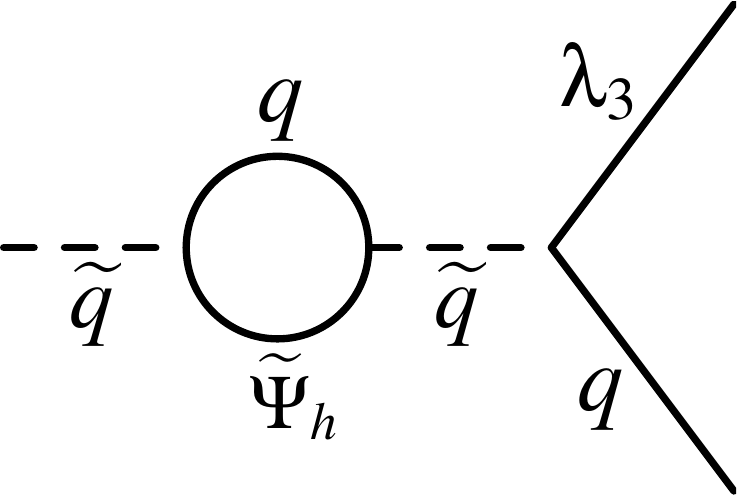}}
\caption[Feynman diagrams that contribute to the RGE for the squark-quark-gaugino coupling, $\bgt^{Q}$.]{\small Feynman diagrams which contribute to the RGE for the squark-quark-gaugino coupling, $\bgt^{Q}$.}
\label{fig:gtqgraph}
\end{figure}

In the notation for the additional coupling, we use a bold font since there is also the possibility of this coupling becoming a matrix in flavour space. The MSSM limit of this coupling is $g_{2}$, which couples quarks and squarks of the same flavour. However, once we are in a broken SUSY regime, the RGEs allow for the possibility of flavour off-diagonal terms to appear. To see why, consider the $\bgt^{Q}$ RGE (\ref{app:bgtq}) which will be derived in Ch.~\ref{ch:application}. The Feynman diagrams in Fig.~\ref{fig:gtqgraph} show all entries in the RGE that depend on the (flavour off-diagonal) Yukawa couplings.
We find that these four contributions to the one loop RGE for $\bgt^{Q}$ cancel in the MSSM limit since they appear as
\begin{equation*}
\mathrm{RGE}\ni(\ref{fig:gtqgraph1})+(\ref{fig:gtqgraph2})-4(\ref{fig:gtqgraph3})+2(\ref{fig:gtqgraph4})=0\;.
\end{equation*}
However, if we are in a regime where our effective theory contains all the SUSY particles except higgsinos, we see that only (\ref{fig:gtqgraph1}) remains in the RGE, which now contains flavour off-diagonal Yukawa coupling terms. We see similar features in the RGEs for the other sfermion-fermion-gaugino couplings.

Off-diagonal terms in the squark-quark-gluino couplings induce flavour-violating decays of squarks, $\tilde{q}_{i}\rightarrow q_{j}\tilde{Z}_{k}$, even if the only source of flavour-violation in our theory is from the KM matrix via Yukawa couplings. Although the partial widths for these decays will be small, they may be important for light squarks for which standard tree-level decays, \textit{i.e.}, $\tilde{q}_{i}\rightarrow q_{j}\tilde{Z}_{i}$ and $\tilde{q}_{i}\rightarrow q'_{j}\tilde{W}_{i}$, are kinematically forbidden. Therefore, the decay to a bino-like lightest neutralino, which is the LSP in many phenomenological models, is of particular interest. Although in this case it may seem that $\bgt'$ would be the most relevant coupling this will depend on the size of off-diagonal couplings from other sources.

To illustrate the potential for sizable off-diagonal terms in these tilde-couplings, we preempt our discussion in Ch.~\ref{ch:application}-\ref{ch:results} and examine the solution of the RGEs to two loop order using a simplified scenario for sparticle decoupling. In this scenario, we artificially place all thresholds at one of two points: the heavy Higgs bosons and gluinos at $2$~TeV, and the sfermions and electroweak gauginos at $600$~GeV. This helps to express the general features introduced by keeping the broken SUSY couplings separate from their SM counterparts. The theory we present is supersymmetric above $2$~TeV, contains SM particles, squarks, sleptons, charginos and neutralinos above $600\ $GeV, and reduces to the SM below this scale.

\begin{figure}[t] \centering
\includegraphics[viewport=20 50 725 525, clip, scale=0.45]{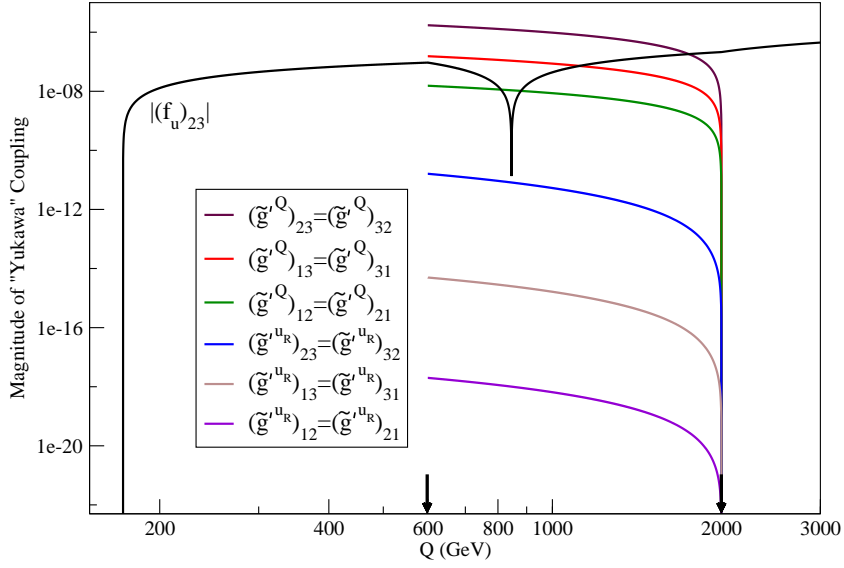} 
\caption[The evolution of the magnitudes of the off-diagonal elements of $\bgt^{\prime Q}$ and $\bgt^{\prime u_{R}}$ for a simplified threshold scenario.]{\small The evolution of the magnitudes of the off-diagonal elements of $\bgt^{\prime Q}$ and $\bgt^{\prime u_{R}}$ for our simplified threshold scenario where the thresholds (denoted by arrows) are clustered at $2\ $TeV and $600\ $GeV. Also shown for comparison is the running of the $(2,3)$ element of the up Yukawa coupling (see Sec.~\ref{sec:Yukrun}). Displayed in the ``standard'' current basis as discussed in Ch.~\ref{ch:flavth}, with the legend in the same order (top to bottom) as the curves.} \label{fig:gtpoffdiag}
\end{figure}

With these simplifications, the magnitudes of the couplings $\bgt^{\prime Q}$ and $\bgt^{\prime u_{R}}$ are plotted in Fig.~\ref{fig:gtpoffdiag}.
The arrow at $2\ $TeV indicates the first threshold where the heavy Higgs particles and gluinos decouple from the theory. At this point, the boundary condition on $\bgt^{\prime \Phi}$ is
\begin{equation*}
\bgt^{\prime \Phi}_{ij}(Q=2\ \mathrm{TeV})=g'\delta_{ij}\;.
\end{equation*}
The curves end at $600\ $GeV since this is the point where the squarks and gauginos decouple from the theory and the couplings no longer exist. As the vertical scale is logarithmic, and varies over a large range, the $(2,3)$ element of the up-quark Yukawa coupling matrix $\bdf_{u}$ is also plotted.\footnote{Note that the features of the running of $\bdf_{u}$ will be discussed in more detail in Ch.~\ref{ch:results}.} The $(2,3)$ element is the largest of the off-diagonal entries of $\bdf_{u}$ and we can see that there is potential for the off-diagonal entries of $\bgt$ to considerably exceed this coupling. It is therefore essential that the tilde-couplings are included in any discussion of flavour physics, especially when the KM matrix is the only source of flavour-violation via the Yukawa coupling matrices.

Also seen in Fig.~\ref{fig:gtpoffdiag} is that the off-diagonal elements of the gaugino coupling matrices are symmetric under interchange of the two indices. This symmetry is not exact, but deviations are not visible on the plot. In fact, the elements in question are equal to within a few parts per mille. This approximate Hermiticity is observed in the gaugino coupling RGEs when the diagonal elements of the gaugino coupling matrices are large with respect to the off-diagonal elements, and the difference between the Higgs and higgsino coupling matrices, $\bdf_{u,d,e}$ and $\tilde{\mathbf{f}}^{\Phi}_{u,d,e}$, can be ignored.

We move on to detail how the concepts in this chapter are applied to the MSSM and the way in which the subsequent RGEs are used to calculate the running for the couplings as in Fig.~\ref{fig:gtpoffdiag}. After considering these points, we will return (in Ch.~\ref{ch:results}) to further examine specific examples of the running of the various parameters of the MSSM in both our simplified spectrum and more realistic scenarios.

%% file: derivation.tex
\chapter{Application to the MSSM}\label{ch:application}

The previous chapter lists the general RGEs for both the dimensionless and dimensionful parameters in our general Lagrangian of Eq.~\eqref{eq:fourlag}. This general Lagrangian was constructed with a view to the $R$-parity conserving MSSM which will be our effective theory between the highest SUSY threshold and $\mgut$. The next step is to write a Lagrangian, using \eqref{eq:genlag} as a guide, for the specific particle content and gauge groups of the MSSM.

Due to the splitting, below the highest SUSY threshold, between couplings that are otherwise equal in the MSSM, we must place different couplings for these terms in the Lagrangian on the understanding that they will be equal to their usual SM counterparts above the highest threshold. This requirement will serve as a boundary condition for some of the dimensionless SUSY couplings, as described later. In addition, keeping in mind that we wish to derive RGEs for the MSSM with full thresholds for SUSY particles, we must be careful to retain information about the source of each contribution by introducing a $\theta_{\mathcal{P}}$ for each SUSY particle as described in Sec.~\ref{sec:smmssmtrans}.

\section{Interactions}\label{sec:int}

Before deriving the RGEs, we first write out all interaction terms in the $R$-parity conserving MSSM applicable to our analysis, using the general supersymmetric Lagrangian in \eqref{eq:genlag}. We begin by obtaining the scalar-fermion-fermion interactions, which make up the $\bu^{1,2}_{\Phi}$, $\bv_{\Phi}$, $\bw_{\Phi}$ and $\bx^{1,2}_{\Phi}$ matrices. The superpotential, \eqref{eq:super}, is used to evaluate the dimensionless couplings that govern the interactions of quarks and squarks with Higgs bosons and higgsinos, which are given by
\begin{equation}\label{eq:higgsinoint}\begin{split}
\mathcal{L}\ni\ &-\left[\bar{u}_j(\bdf_u)^T_{ji}h^0_uP_Lu_i-\bar{u}_j(\bdf_u)^T_{ji}h^+_uP_Ld_i+\bar{d}_j(\bdf_d)^T_{ji}h^-_dP_Lu_i+\bar{d}_j(\bdf_d)^T_{ji}h^0_dP_Ld_i\right.\\
&\left.\quad\ +\bar{e}_j(\bdf_e)^T_{ji}h^-_dP_L\nu_i+\bar{e}_j(\bdf_e)^T_{ji}h^0_dP_Le_i+\mathrm{h.c.}\right]\\
&-\left[\bar{\Psi}_{h^0_u}\tilde{u}^\dagger_{Rj}(\tilde{\bdf}^{u_R}_u)^T_{ji}P_Lu_i
+\bar{u}_{j}(\tilde{\bdf}^{Q}_u)^T_{ji}\tilde{u}_{Li}P_L\Psi_{h^0_u}
-\bar{\Psi}_{h^+_u}\tilde{u}^\dagger_{Rj}(\tilde{\bdf}^{u_R}_u)^T_{ji}P_Ld_i\right.\\
&\quad\ -\bar{u}_{j}(\tilde{\bdf}^{Q}_u)^T_{ji}\tilde{d}_{Li}P_L\Psi_{h^+_u}
+\bar{\Psi}_{h^-_d}\tilde{d}^\dagger_{Rj}(\tilde{\bdf}^{d_R}_d)^T_{ji}P_Lu_i
+\bar{d}_{j}(\tilde{\bdf}^{Q}_d)^T_{ji}\tilde{u}_{Li}P_L\Psi_{h^-_d}\\
&\quad\ +\bar{\Psi}_{h^0_d}\tilde{d}^\dagger_{Rj}(\tilde{\bdf}^{d_R}_d)^T_{ji}P_Ld_i
+\bar{d}_{j}(\tilde{\bdf}^{Q}_d)^T_{ji}\tilde{d}_{Li}P_L\Psi_{h^0_d}
+\bar{\Psi}_{h^-_d}\tilde{e}^\dagger_{Rj}(\tilde{\bdf}^{e_R}_e)^T_{ji}P_L\nu_i\\
&\left.\quad\ +\bar{e}_{j}(\tilde{\bdf}^{L}_e)^T_{ji}\tilde{\nu}_{Li}P_L\Psi_{h^-_d}
+\bar{\Psi}_{h^0_d}\tilde{e}^\dagger_{Rj}(\tilde{\bdf}^{e_R}_e)^T_{ji}P_Le_i
+\bar{e}_{j}(\tilde{\bdf}^{L}_e)^T_{ji}\tilde{e}_{Li}P_L\Psi_{h^0_d}+\mathrm{h.c.}\right]\;.
\end{split}\end{equation}
Notice that we write the interactions of quarks with the Higgs bosons as the usual superpotential Yukawa matrices. In contrast, the sfermion interactions have a tilde and are labelled with the sfermion that enters the interaction. Above all SUSY thresholds $\tilde{\bdf}^{\Phi}_{u,d,e}=\bdf_{u,d,e}$, but below this scale, all the `Yukawa' matrices may have different values.

In addition, \eqref{eq:genlag} describes scalar-fermion-fermion interactions involving gauginos, which are given by,
\begin{equation}\label{eq:gauginoint}\begin{split}
\mathcal{L}\ni\ &-\frac{1}{\sqrt{2}}\left\{\left(\tilde{u}^\dagger_{Lj},\tilde{d}^\dagger_{Lj}\right)\mathbf{G}_{Q}P_L\left(\begin{array}{c}u_i\\d_i\end{array}\right)+\left(\tilde{\nu}^\dagger_{Lj},\tilde{e}^\dagger_{Lj}\right)\mathbf{G}_{L}P_L\left(\begin{array}{c}\nu_i\\e_i\end{array}\right)\right.\\[5pt]
&\qquad\quad\ \ +\bar{u}_j\bgtpur_{ji}(-\tfrac{4}{3})\tilde{u}_{Ri}P_L\lambda_0+\bar{d}_j\bgtpdr_{ji}(\tfrac{2}{3})\tilde{d}_{Ri}P_L\lambda_0\\[5pt]
&\left.\qquad\quad\ \ +\bar{e}_j\bgtper_{ji}(2)\tilde{e}_{Ri}P_L\lambda_0+\mathrm{h.c.}\right\}\\[5pt]
&-\sqrt{2}\left\{(\tilde{\mathbf{g}}_{s}^{\tilde{q}_{L}})_{ji}(-i)^{\theta_{\tilde{g}}}\tilde{q}^\dagger_{Lj}\bar{\tilde{g}}_{A}\tfrac{\lambda_A}{2}P_Lq_i
-(\tilde{\mathbf{g}}_{s}^{\tilde{q}_{R}})_{ji}(-i)^{\theta_{\tilde{g}}}
\bar{q}_{j}\tfrac{\lambda_A}{2}P_L \tilde{g}_A \tilde{q}_{Ri} +\mathrm{h.c.}\right\}\\
&-\frac{1}{\sqrt{2}}\left\{\left(h^{+\dagger}_u,h^{0\dagger}_u\right)\mathbf{G}_{h_{u}}P_L\left(\begin{array}{c}\Psi_{h^+_u}\\\Psi_{h^0_u}\end{array}\right)+\left(h^{-\dagger}_d,h^{0\dagger}_d\right)\mathbf{G}_{h_{d}}P_L\left(\begin{array}{c}\Psi_{h^-_d}\\\Psi_{h^0_d}\end{array}\right)+\mathrm{h.c.}\right\}\;,\\
\end{split}\end{equation}
where we have defined the following matrices:
\begin{equation}
\mathbf{G}_{Q}=\left(\begin{array}{cc}\bgtq_{ji}\bar{\lambda}_3+\frac{1}{3}\bgtpq_{ji}\bar{\lambda}_0&\bgtq_{ji}\left(\bar{\lambda}_1-i\bar{\lambda}_2\right)\\[2pt]\bgtq_{ji}\left(\bar{\lambda}_1+i\bar{\lambda}_2\right)&-\bgtq_{ji}\bar{\lambda}_3+\frac{1}{3}\bgtpq_{ji}\bar{\lambda}_0\end{array}\right)
\end{equation}
\begin{equation}
\mathbf{G}_{L}=\left(\begin{array}{cc}\bgtl_{ji}\bar{\lambda}_3-\bgtpl_{ji}\bar{\lambda}_0&\bgtl_{ji}\left(\bar{\lambda}_1-i\bar{\lambda}_2\right)\\[2pt]\bgtl_{ji}\left(\bar{\lambda}_1+i\bar{\lambda}_2\right)&-\bgtl_{ji}\bar{\lambda}_3-\bgtpl_{ji}\bar{\lambda}_0\end{array}\right)
\end{equation}
\begin{equation}\label{eq:Ghu}
\mathbf{G}_{h_{u}}=\left(\begin{array}{cc}\gthu\bar{\lambda}_3+\gtphu\bar{\lambda}_0&\gthu\left(\bar{\lambda}_1-i\bar{\lambda}_2\right)\\[2pt]\gthu\left(\bar{\lambda}_1+i\bar{\lambda}_2\right)&-\gthu\bar{\lambda}_3+\gtphu\bar{\lambda}_0\end{array}\right)
\end{equation}
\begin{equation}\label{eq:Ghd}
\mathbf{G}_{h_{d}}=\left(\begin{array}{cc}-\gthd\bar{\lambda}_3-\gtphd\bar{\lambda}_0&\gthd\left(-\bar{\lambda}_1-i\bar{\lambda}_2\right)\\[2pt]\gthd\left(-\bar{\lambda}_1+i\bar{\lambda}_2\right)&\gthd\bar{\lambda}_3-\gtphd\bar{\lambda}_0\end{array}\right)\;.
\end{equation}
and hypercharge and $SU(2)$ gauginos are denoted $\lambda_{0}$ and $\lambda_{1,2,3}$ respectively. Below the highest SUSY threshold, not only can the $\tilde{\mathbf{g}}^{\Phi}$ develop different values to their corresponding gauge boson couplings, but they are also able to obtain a non-trivial flavour structure, as pointed out in Sec.~\ref{sec:broken}. In a similar manner to the higgsino couplings, above all SUSY thresholds the gaugino couplings must equal their gauge boson counterparts, so that the off-diagonal elements are zero, \textit{i.e.}, $\mathbf{\tilde{g}}^{\prime \Phi}=g^{\prime}\times\dblone$, \textit{etc}.

The interaction terms for the dimensionful parameters originate mainly in the SSB part of the Lagrangian, \eqref{eq:soft}. In addition to these terms there are mass terms for the higgsinos which arise from the superpotential via (\ref{eq:genlag})
\begin{equation}\begin{split}
\mathcal{L}\ni&-\frac{1}{2}\left\{\frac{1}{2}\left(\mu+\mu^{*}\right)\left[\bar{\Psi}_{h^{0}_{u}}\Psi_{h^{0}_{d}}+\bar{\Psi}_{h^{0}_{d}}\Psi_{h^{0}_{u}}+\bar{\Psi}_{h^{+}_{u}}\Psi_{h^{-}_{d}}+\bar{\Psi}_{h^{-}_{d}}\Psi_{h^{+}_{u}}\right]\right\}\\
&+\frac{i}{2}\left\{\frac{i}{2}\left(\mu^{*}-\mu\right)\left[\bar{\Psi}_{h^{0}_{u}}\gamma_{5}\Psi_{h^{0}_{d}}+\bar{\Psi}_{h^{0}_{d}}\gamma_{5}\Psi_{h^{0}_{u}}+\bar{\Psi}_{h^{+}_{u}}\gamma_{5}\Psi_{h^{-}_{d}}+\bar{\Psi}_{h^{-}_{d}}\gamma_{5}\Psi_{h^{+}_{u}}\right]\right\}\;.
\end{split}\end{equation}
As mentioned previously, both higgsino mass states are taken to have a mass, $\left|\mu\right|$, and are therefore written with a single threshold, $\sh$.

Finally, there are there are additional terms derived from the following parts of (\ref{eq:genlag}):
\begin{equation}\label{eq:quartlag}
\mathcal{L}\ni-\frac{1}{2}\sum_{A}\left|\sum_{i}\mathcal{S}^{\dagger}_{i}g_{\alpha}t_{\alpha A}\mathcal{S}_{i}\right|^{2}-\sum_{i}\left|\frac{\partial\hat{f}}{\partial\hat{\mathcal{S}}_{i}}\right|^{2}_{\hat{\mathcal{S}}=\mathcal{S}}\;.
\end{equation}
These result in scalar bilinear, trilinear and quartic terms. In particular, there are bilinear Higgs interactions with coupling strength $\left|\tilde{\mu}\right|^{2}$ arising from the superpotential, which is given a tilde to differentiate it from the $\mu$ in the higgsino mass terms. As an illustration, if we take the second term in (\ref{eq:quartlag}) and choose to differentiate with respect to the up-type Higgs superfield, \textit{i.e.},~$\hat{\mathcal{S}}=\hat{H}_{u}$, we see that
\begin{equation}\label{eq:quarteg}\begin{split}
\mathcal{L}\ni&-\left|\tilde{\mu}\right|^{2}h^{0\dagger}_{d}h^{0}_{d}-\tilde{u}^{\dagger}_{Rk}\tilde{u}^{\dagger}_{Ll}\left(\bdf_{u}\right)^{T}_{kn}\left(\bdf_{u}\right)^{*}_{lm}\tilde{u}_{Rm}\tilde{u}_{Ln}\\
&-\left(\tilde{u}^{\dagger}_{Lk}\left(\tilde{\mu}^{*}\bdf^{h_{u}}_{u}\right)^{*}_{kl}\tilde{u}_{Rl}h^{0}_{d}+\mathrm{h.c.}\right)\;.
\end{split}\end{equation}
In accord with our discussion at the end of Sec.~\ref{sec:dless}, we have not placed additional labels on the Yukawa matrices that appear in the quartic part of \eqref{eq:quarteg} as we are not keeping track of the difference between this term and the ``square'' of the corresponding Yukawa coupling. Conversely, $\left(\tilde{\mu}^{*}\bdf^{h_{u}}_{u}\right)_{kl}$ is a trilinear coupling, and as such we will derive an RGE for this term independently of $\mu$ or $\bdf_{u}$ (the usual up quark Yukawa coupling). This term enters the up-squark mass matrix as a left-right coupling term, and is therefore relevant for determining the squark mass eigenstates. Below the heaviest SUSY threshold, the coupling will evolve differently to the multiple $(\mu^{*}\times\bdf_{u})$ due to differences between the respective RGEs, and in this case, where the term had its origin in the derivative of the superpotential with respect to $\hat{H}_{u}$ we label the $\bdf_{u}$ with $h_{u}$.

\section{Gauge Coupling RGEs}\label{sec:gaugerges}

The one-loop RGEs for the gauge couplings including threshold effects are well known. We rewrite them here without quark thresholds since we will decouple the top at $Q=M_{Z}$, and always work in the range $Q>M_{Z}$. This means that we avoid the complications associated with $SU(2)$ breaking, which was also our motivation for ignoring left-right mixing when deciding upon the location of the squark thresholds.

Since there is no flavour structure in the one-loop gauge RGEs, the squark thresholds can be simplified using
\begin{equation*}
N_{\tilde{f}}=\sum_{i=1}^{3}\theta_{\tilde{f_{i}}}\;.
\end{equation*}
Applying \eqref{eq:gaugerge} to the MSSM particle content we obtain the familiar RGEs
\begin{equation}\begin{split}\label{eq:g1bet}
\left(4\pi\right)^{2}\left.\beta_{g_{1}}\right|_{1-\mathrm{loop}}=&g^{3}_{1}\left[4+\frac{1}{30}N_{\tilde{Q}}+\frac{4}{15}N_{\tilde{u}_{R}}+\frac{1}{15}N_{\tilde{d}_{R}}+\frac{1}{10}N_{\tilde{L}}+\frac{1}{5}N_{\tilde{e}_{R}}\right.\\
&\left.\quad\;+\frac{1}{10}\left(\h+\Hh\right)+\frac{2}{5}\sh\right]
\end{split}\end{equation}
\begin{equation}\begin{split}\label{eq:g2bet}
\left(4\pi\right)^{2}\left.\beta_{g}\right|_{1-\mathrm{loop}}=&g^{3}\left[-\frac{22}{3}+4+\frac{1}{2}N_{\tilde{Q}}+\frac{1}{6}N_{\tilde{L}}
+\frac{1}{6}\left(\h+\Hh\right)+\frac{2}{3}\sh+\frac{4}{3}\theta_{\tilde{W}}\right]
\end{split}\end{equation}
\begin{equation}\begin{split}\label{eq:g3bet}
\left(4\pi\right)^{2}\left.\beta_{g_{s}}\right|_{1-\mathrm{loop}}=&g^{3}_{s}\left[4+\frac{1}{3}N_{\tilde{Q}}+\frac{1}{6}N_{\tilde{u}_{R}}+\frac{1}{6}N_{\tilde{d}_{R}}+2\theta_{\tilde{g}}-11\right]\;.
\end{split}\end{equation}
Here, as in Sec.~\ref{sec:gaugerun}, $g_{1}$ is the scaled hypercharge gauge coupling that unifies with the $SU(2)$ and $SU(3)$ couplings when the MSSM is embedded in a GUT: $g'^{2}=\frac{3}{5}g^{2}_{1}$. It should be clear that the one-loop RGEs for the gauge couplings are a set of three differential equations, which form a closed set. As we introduce the RGEs for each group of couplings, we will increase the size of our closed set of differential equations until we have the whole system of RGEs for the MSSM.

\section{Yukawa and Yukawa-type Couplings}

There are many RGEs for scalar-fermion-fermion couplings in the MSSM. Glancing at our Lagrangian terms for the softly broken MSSM, \eqref{eq:higgsinoint} and \eqref{eq:gauginoint}, we see that we must find RGEs for: the usual Yukawa couplings to Higgs bosons, $\bdf_{u}$, $\bdf_{d}$, and $\bdf_{e}$; the couplings of higgsinos to the various fermion-sfermion pairs, $\bft^{Q}_{u}$, $\bft^{Q}_{d}$, $\bft^{L}_{e}$, $\bft^{u_{R}}_{u}$, $\bft^{d_{R}}_{d}$, $\bft^{e_{R}}_{e}$; hypercharge  gaugino couplings, $\bgt'^{Q}$, $\bgt'^{L}$, $\bgt'^{u_{R}}$, $\bgt'^{d_{R}}$, $\bgt'^{e_{R}}$, $\tilde{g}'^{h_{u}}$, $\tilde{g}'^{h_{d}}$; the $SU(2)$ gaugino couplings, $\bgt^{Q}$, $\bgt^{L}$, $\tilde{g}^{h_{u}}$, $\tilde{g}^{h_{d}}$; and finally, the gluino couplings,  $\bgt_{s}^{Q}$, $\bgt_{s}^{u_{R}}$, $\bgt_{s}^{d_{R}}$. The RGEs for \textit{all} these couplings must be taken together, along with those for the usual gauge couplings in Sec.~\ref{sec:gaugerges}, to form a closed set.

In order to apply the general RGEs to the particle content of the MSSM we must construct the various scalar-fermion-fermion coupling matrices from \eqref{eq:fourlag}. Not only do we need to keep track of particle type, we must also remember the presence of three particle flavours for the Dirac fermions and sfermions, and it should be stressed that in this context, counting of different flavours is included in our count of particle types.  We will write our final RGEs as matrices in flavour space, in accord with the notation we have used in our MSSM Lagrangian.

We use a basis for the fermion content such that the Dirac fermions are \linebreak$\{u_{i},d_{i},\nu_{i},e_{i}\}$ and the Majorana fermions are $\{{\tilde{h}}_1^0,{\tilde{h}}_2^0,{\tilde{h}}_1^\pm,{\tilde{h}}_2^\pm, \lambda_{0},\lambda_{1},\lambda_{2},\lambda_{3},\tilde{g}_{A}\}$, where $i=1,\dots3$ is the flavour index for the Dirac fermions, and ${\tilde{h}}_{1,2}^0$ and ${\tilde{h}}_{1,2}^\pm$ are the neutral and charged components of the Majorana spinor ${\tilde{h}}_{1,2}$ introduced in (\ref{eq:inorot}). In the MSSM, $\bu^{1}_{\Phi}$ and $\bu^{2}_{\Phi}$ will be $(4\times4)$ blocks of $(3\times 3)$ matrices when $\Phi$ is one of the Higgs bosons in (\ref{eq:hrot}) or (\ref{eq:Hrot}). Similarly, since flavour is carried by the sfermion scalar index, $\Phi=\tilde{f}_i$, $\bv_{\tilde{f}_{i}}$ will be a $(4\times9)$ matrix where  the number of rows  can be further expanded to  show each of the three matter fermion flavours. Likewise, $\bw_{\tilde{f}_{i}}$ a $(9\times4)$ matrix where now the number of columns can be similarly expanded to exhibit these flavours. Thus, when fully written out, $\bv_{\tilde{f}_{i}}$  is a $(12\times9)$ matrix, while $\bw_{\tilde{f}_{i}}$ is a $(9\times12)$ matrix. Finally, $\bx^{1}_{\Phi}$ and $\bx^{2}_{\Phi}$ will both be $(9\times9)$ matrices. In working out the size of these matrices, we have suppressed the colour index, $A$, which otherwise expands the size of the matrices due to additional quark entries.

Within the $R$-parity conserving MSSM, the matrices $\bu_{\Phi}^{1,2}$ and $\bx_{\Phi}^{1,2}$ are non-zero only for $\Phi=\{\mathsf{h},\mathcal{H},G^{+},H^{+}\}$ and the $\bv_{\Phi}$ and $\bw_{\Phi}$ matrices are non-zero only for $\Phi=\{\tilde{u}_{L},\tilde{d}_{L},\tilde{e}_{L},\tilde{\nu}_{L},\tilde{u}_{R},\tilde{d}_{R},\tilde{e}_{R}\}$. The matrices can be readily worked out from \eqref{eq:higgsinoint} and \eqref{eq:gauginoint}, and the derivation is tedious but straightforward. More detail can be found in Ref.~\cite{RGE1}.

\subsection{Comparison with Earlier Literature: Part I}\label{sec:litcomp}

Before moving on to calculate the RGEs for the usual Yukawa couplings, we pause here to compare our method to that used in the previous literature. As these earlier studies did not keep the distinction between our tilde-couplings and the standard SM couplings, we perform an initial derivation for the up-type Yukawa coupling in the same manner. Having rotated the Higgs fields in the Lagrangian, the coupling of the up quarks to the light Higgs field, $\mathsf{h}$, is $\sn\bdf_{u}$ and it is this RGE that we derive next.

Note that while we find it instructive to write the RGE including separate thresholds for $A$, $H$, $H^{\pm}$, and the four higgsinos, ${\tilde{h}}_{1,2}^0$ and ${\tilde{h}}_{1,2}^\pm$, we require that $SU(2)$ remains unbroken. This corresponds to placing all the Higgs boson (and separately, higgsino) thresholds at the same scale, since it is only when we enforce this restriction that the couplings related by $SU(2)$ have the same RGE, as we would expect.

Ignoring the separation of tilde terms in the Lagrangian, we use \eqref{eq:U1} to obtain
\begin{equation} \label{eq:fake}\begin{split}
{\left(4\pi\right)}^2\frac{d{\left(\sn \bdf_u\right)}_{ij}}{dt}=&\frac{3s}{2}\left[\frac{\sn^2}{3}\left(\h+\Go+\Gp\right)+\frac{\cs^2}{3}\left(\Hh+\A+\Hp\right)\right](\bdf_u\bdf^\dagger_u\bdf_u)_{ij}\\
&+\frac{\sn}{4}\left[\left\{\shram+\shc+\sH+\sHc\right\}\sqk{(\bdf_u\bdf^\dagger_u)}_{ik}{(\bdf_u)}_{kj}\right.\\
&\left.\qquad+\left\{\shram+\sH\right\}\suk{(\bdf_u)}_{ik}{(\bdf^\dagger_u\bdf_u)}_{kj}\right]\\
&+\sn\left[\sn^2\left(\h-\Go\right)+\cs^2\left(\Hh-\A\right)\right](\bdf_u\bdf^\dagger_u\bdf_u)_{ij}\\
&+\frac{\sn}{2}\left[\cs^2\Gp+\sn^2\Hp-4\cs^2\left(\Gp-\Hp\right)\right](\bdf_d\bdf^\dagger_d\bdf_u)_{ij}\\
&+\frac{\sn}{4}\left[\left\{\shc+\sHc\right\}\sdk{(\bdf_d)}_{ik}(\bdf^\dagger_d\bdf_u)_{kj}\right]\\
&+\sn{(\bdf_u)}_{ij}\left[3\left(\sn^2\h+\cs^2\Hh\right)\mathrm{Tr}\{\bdf^\dagger_u\bdf_u\}+\cs^2\left(\h-\Hh\right)\mathrm{Tr}\{3\bdf^\dagger_d\bdf_d+\bdf^\dagger_e\bdf_e\}\right]\\
&-{(\bdf_u)}_{ij}\left[\frac{3}{5}g^2_1\left\{\sn\frac{17}{12}-\sn\left(\frac{1}{36}\sq+\frac{4}{9}\su\right)\sbi\right.\right.\\
&\qquad\qquad\qquad\ -\left(\frac{\sn}{4}\left[{\left(1+2\sn\cs\right)}\shram+{\left(1-2\sn\cs\right)}\sH\right]\h\right.\\
&\qquad\qquad\qquad\ +\frac{\cs}{4}\left(\cs^{2}-\sn^{2}\right)\left(\shram-\sH\right)\Hh\\
&\left.\left.\qquad\qquad\qquad\ +\frac{1}{2}\left[\left(\sn+\cs\right)\shram-\left(\cs-\sn\right)\sH\right]\left\{\frac{1}{3}\sq-\frac{4}{3}\su\right\}\right)\sbi\right\}\\
&\qquad\quad\quad\;\;+g^2_2\left\{\sn\frac{9}{4}-\sn\frac{3}{4}\sq\swi-\left(\frac{\sn}{4}\left[{\left(1+2\sn\cs\right)}\shram+{\left(1-2\sn\cs\right)}\sH\right.\right.\right.\\
&\left.\qquad\qquad\qquad\quad+2\shc+2\sHc\right]\h+\frac{\cs}{4}\left(\cs^{2}-\sn^{2}\right)\left(\shram-\sH\right)\Hh\\
&\qquad\qquad\qquad\quad\left.\left.-\frac{1}{2}\left[\left(\sn+\cs\right)\shram-\left(\cs-\sn\right)\sH+2\sn\left(\shc+\sHc\right)\right]\sq\right)\swi\right\}\\
&\left.\qquad\quad\quad\;\;+g^2_3\left\{8\sn-\sn\frac{4}{3}\left(\sq+\su\right)\sgl\right\}\right]\;.
\end{split}\end{equation}
In all cases of a repeated index, $k$, there is an implied sum, including the case of three repeated indices where one is a $\theta_{\tilde{f}_{k}}$, as in, \textit{e.g.},~$\sqk{(\bdf_u\bdf^\dagger_u)}_{ik}{(\bdf_u)}_{kj}$. Since we have a collection of effective theories valid at various scales, with the SM and the MSSM as the limits at the weak scale and the high scale respectively, the RGE in \eqref{eq:fake} must reduce to the SM and the MSSM under the appropriate conditions. We will consider each case in turn:
\begin{description}
\item[MSSM] The MSSM limit is achieved when the effective theory contains all SUSY particles in addition to their SM counterparts. This corresponds to setting $\theta_{\mathcal{P}}=1$ for all $\mathcal{P}$. Since the rotation angle, $\tan{\beta}$, is fixed\footnote{In the case that the rotation is scale-dependent, we see from \eqref{eq:hrot} and \eqref{eq:Hrot} that the rotated fields themselves have an explicit scale dependence. Although our starting RGEs from Refs.~\cite{MVrgeI,*MVrgeII,*MVrgeIII,luo} do not take scale dependence of the fields into account, it should be clear that the rotation cannot change the RGEs above all thresholds since it is just a field redefinition. We can therefore safely conclude that the modification of the RGEs in the case of scale dependent fields would be precisely that necessary to cancel the contribution from removing the $\sin{\beta}$ from the differential on the left-hand side.}, we can safely remove it from the differential on the left-hand side. This means that, since the right-hand side is proportional to $\sin{\beta}$ (in the limit we are considering), the $\sin{\beta}$ cancels and we recover the equation for $\bdf_{u}$.
\item[SM] The SM limit is achieved when the effective theory only contains SM particles (including, of course, the SM-like Higgs particle, $\mathsf{h}$). To recover this limit, we therefore set all $\theta_{\mathcal{P}}=0$ except $\h$, $\Go$ and $\Gp$. When we do this, we see that all dependence on the MSSM Yukawa coupling matrices, $\bdf_{u,d,e}$, gives way to the SM Yukawa couplings, $\bm{\lambda}_{u,d,e}$, and we recover the SM RGE for $\bm{\lambda}_{u}$ \cite{ACMPRWrge,smrge}.
\end{description}

Before proceeding to our comparison with Refs.~\cite{casno,sakis}, we note that if the higgsino threshold is at a lower scale than the heavy Higgs boson threshold, so that $\sh=1$ at the same time as $\Hh=0$, the RGE cannot be written only in terms of SM Yukawa couplings. The higgsino couplings, which we will soon write as $\tilde{\bdf}^{\Phi}_{u,d,e}$, remain in the theory and enter the RGE without a dependence on the Higgs rotation angle.

Turning to our comparison with Ref.~\cite{casno}, we first note that our MSSM and SM limits agree, as expected. To compare the full RGE structure, we take a common threshold for the higgsinos and find agreement will all terms apart from the $SU(2)$ and $U(1)$ gauge couplings. We find, 
\begin{equation}\nonumber\begin{split}
{\left(4\pi\right)}^2\frac{d{\left(s \bdf_u\right)}_{ij}}{dt}&\ni\ -s{(\bdf_u)}_{ij}\left[\frac{3}{5}g^2_1\left\{\frac{17}{12}-\left(\frac{1}{36}\sq+\frac{4}{9}\su\right)\sbi\right.\right.\\
&\left.\qquad\qquad\quad\qquad\quad-\left(\frac{1}{2}\h+\frac{1}{3}\sq-\frac{4}{3}\su\right)\shram\sbi\right\}\\
&\left.\qquad\qquad\quad\ \ +g^2_2\left\{\frac{9}{4}-\frac{3}{4}\sq\swi-\left(\frac{3}{2}\h-3\sq\right)\shram\swi\right\}\right]\;, 
\end{split}\end{equation}
to be contrasted with~\cite{casno},
\begin{equation}\nonumber\begin{split}
{\left(4\pi\right)}^2\frac{d{\left(s \bdf_u\right)}_{ij}}{dt}&\ni\ -s{(\bdf_u)}_{ij}\left[\frac{3}{5}g^2_1\left\{\frac{17}{12}+\frac{3}{4}\shram-\left(\frac{1}{36}\sq+\frac{4}{9}\su+\frac{1}{4}\shram\right)\sbi\right\}\right.\\
&\left.\qquad\qquad\quad\ \ +g^2_2\left\{\frac{9}{4}+\frac{9}{4}\shram-\frac{3}{4}\left(\sq+\shram\right)\swi\right\}\right]\;,
\end{split}\end{equation}
with the down-type Yukawa coupling RGEs exhibiting a similar disagreement.

In contrast to our method and that of Ref.~\cite{casno}, Ref.~\cite{sakis} appears to write the RGEs in a basis in which the Higgs fields are unrotated. It is therefore difficult to construct the appropriate relationship between their Higgs thresholds and the fields $\mathsf{h}$, $\mathcal{H}$, $G^{+}$ and $H^{+}$. As a consequence it is hard to see how to reduce the RGEs in Ref.~\cite{sakis} to the SM, or to carry out a comparison with our result. The RGEs do, however, reduce to the MSSM when the appropriate limit is taken. We will return to compare our results for the dimensionful parameters to Ref.~\cite{sakis} in Sec.~\ref{sec:litcompII}.

\subsection{Full Yukawa Coupling RGEs with Thresholds} 

We must emphasise that \eqref{eq:fake} is not the correct RGE since it does not retain information about the splitting between the tilde-couplings and their usual SM counterparts. Using \eqref{eq:U1} to carry out the full derivation once more, retaining information about the tilde-couplings and setting single thresholds for both the heavy Higgs bosons, $\Hh$, and the higgsinos, $\sh$, the RGE for the coupling of the up-type quarks to the Higgs bosons becomes
\begin{equation}\label{eq:real}\begin{split}
{\left(4\pi\right)}^2\frac{d(\sn\bdf_u)_{ij}}{dt}=&\frac{\sn}{2}\left\{3\left[\sn^2\h+\cs^2\Hh\right](\bdf_u\bdf_u^\dagger)_{ik}+\left[\cs^2\h+\sn^2\Hh\right](\bdf_d\bdf_d^\dagger)_{ik}\right.\\
&\left.\quad+4\cs^2\left[-\h+\Hh\right](\bdf_d\bdf_d^\dagger)_{ik}\right\}(\bdf_u)_{kj}\\
&+\sn(\bdf_u)_{ik}\left[\sh\sql\bftuq^\dagger_{kl}\bftuq_{lj}+\frac{4}{9}\sbi\sul\bgtpur^*_{kl}\bgtpur^T_{lj}\right.\\
&\left.\qquad\qquad\quad+\frac{4}{3}\sgl\sul\bgtsur^*_{kl}\bgtsur^T_{lj}\right]\\
&+\frac{\sn}{4}\left[2\sh\suk\bftur_{ik}\bftur^\dagger_{kl}+2\sh\sdk\bftdr_{ik}\bftdr^\dagger_{kl}\right.\\
&\qquad+3\swi\sqk\bgtq^T_{ik}\bgtq^*_{kl}+\frac{1}{9}\sbi\sqk\bgtpq^T_{ik}\bgtpq^*_{kl}\\
&\left.\qquad+\frac{16}{3}\sgl\sqk\bgtsq^T_{ik}\bgtsq^*_{kl}\right](\bdf_u)_{lj}\\
&+\sn\sh\sqk\left[-3\swi\gthus\bgtq^T_{ik}+\frac{1}{3}\sbi\gtphus\bgtpq^T_{ik}\right]\bftuq_{kj}\\
&-\frac{4}{3}\sn\sbi\sh\suk\gtphus\bftur_{ik}\bgtpur^T_{kj}\\
&+\sn(\bdf_u)_{ij}\left[\left(\sn^2\h+\cs^2\Hh\right)\mathrm{Tr}\{3\bdf_u^\dagger\bdf_u\}+\cs^2\left(\h-\Hh\right)\mathrm{Tr}\{3\bdf_d^\dagger\bdf_d+\bdf_e^\dagger\bdf_e\}\right]\\
&+\frac{\sn}{2}\sh(\bdf_u)_{ij}\left\{3\swi\left[\mgthusq\left(\sn^2\h+\cs^2\Hh\right)+\mgthdsq\left(\cs^2\h-\cs^2\Hh\right)\right]\right.\\
&\left.\qquad\qquad\qquad+\sbi\left[\mgtphusq\left(\sn^2\h+\cs^2\Hh\right)+\mgtphdsq\left(\cs^2\h-\cs^2\Hh\right)\right]\right\}\\
&-\sn(\bdf_u)_{ij}\left[\frac{17}{12}g'^2+\frac{9}{4}g^2_2+8g^2_3\right]\;.
\end{split}\end{equation}
In a similar manner to \eqref{eq:fake}, three repeated indices indicate a sum, as in, for example, the term $\sn(\bdf_u)_{ik}\sh\sql\bftuq^\dagger_{kl}\bftuq_{lj}$. This is the same for the RGEs in Appendix~\ref{app:dlessRGEs} where we repeat the $\bdf_{u}$ RGE along with the RGEs for $\bdf_{d}$, $\bdf_{e}$, all their tilde counterparts, $\tilde{\mathbf{f}}$, and the gaugino `Yukawa' couplings, $\tilde{\mathbf{g}}$. In addition, we repeat that the decoupling of the matter sfermions takes place in the (approximate) sfermion mass basis. The RGE is written with this in mind, \textit{i.e.},~using $\theta_{\tilde{f}_{i}}$, so that, when the sfermions have partially decoupled, it should be applied in the current basis where the specific sfermion soft mass matrix is diagonal.\footnote{As stated in Sec.~\ref{sec:masseig}, when the three flavours of sfermions in a group are either all present or all absent from the theory, it is unnecessary to rotate to the applicable mass basis. This is because if all are present the rotation is just a field redefinition, and therefore cannot affect the RGEs, and if all are decoupled the whole term is absent from the RGE.}

The reduction of the RGE in \eqref{eq:real} to both the MSSM and SM limits is clear, with $\bm{\lambda}_{u}=\sn\bdf_{u}$ the corresponding coupling below $Q=m_{H}$. If some SUSY particles remain in the theory after the heavy Higgs bosons have decoupled, we find that the factors of $\sin{\beta}$ and $\cos{\beta}$ are exactly those needed to combine with the usual Yukawa matrices\footnote{The reader will also note that there are additional factors of $\sin{\beta}$ and $\cos{\beta}$ which combine with Higgs-gaugino-higgsino couplings, as we will discuss in Sec.~\ref{sec:higgaug}.}, $\bdf_{u,d,e}$, to form $\bdl_{u,d,e}$. It should be noted, however, that this is not true of the tilde-couplings, $\tilde{\mathbf{f}}$, which would remain unchanged in this case. Again, the term $\sn(\bdf_u)_{ik}\sh\sql\bftuq^\dagger_{kl}\bftuq_{lj}$ illustrates this effect, since below $m_{H}$ we can write it as $(\bdl_u)_{ik}\sh\sql\bftuq^\dagger_{kl}\bftuq_{lj}$.

Before moving on to consider the higgsino and gaugino couplings, we repeat that the reader should be vigilant when applying the general RGEs in \eqref{eq:U1}\mbox{--}\eqref{eq:Wa}. The matrix multiplication that takes place in these equations is over the full matrices which include all possible fermion types. In contrast, the matrix multiplication in $\eqref{eq:real}$ and in Appendices~\ref{app:dlessRGEs} and~\ref{app:dfulRGEs} takes place over the \textit{subspace} of flavours. This means that not all trace terms in \eqref{eq:U1}-\eqref{eq:Wa} lead to traces in the RGEs of the MSSM, and, conversely, not all traces in the RGEs are a result of traces in the general equations. For elaboration of this difference, see the comments below (36) of Ref.~\cite{RGE1}.

\subsection{RGEs for Higgsino and Gaugino `Yukawa' Couplings}\label{sec:higgaug}

The RGEs for the couplings of squarks to quarks via gauginos and higgsinos are contained in the general RGEs of \eqref{eq:Va} and \eqref{eq:Wa}. In eq.~\eqref{eq:real} we allowed for the possibility that this coupling (and the other gaugino couplings to matter fermions) can develop a non-trivial flavour structure, as discussed in Sec.~\ref{sec:broken}. This, of course, only happens at scales below the highest SUSY threshold, since above this scale we apply our boundary condition: $\tilde{\mathbf{g}}^{\Phi}_{i}=g_{i}\times\dblone$, where in this instance $i$ indicates the various gauge groups. It is essential to keep track of such non-trivial flavour structure at the weak scale since these couplings are exactly the ones that enter the decays of squarks and sleptons.

As an example, we write the RGE for the coupling of the right-handed up-type squark to up quarks and the $U(1)$ gaugino, $\tilde{\mathbf{g}}^{\prime u_{R}}$:
\begin{equation}\begin{split}\label{eq:gurtilde}
{\left(4\pi\right)}^2\frac{d\bgtpur_{ij}}{dt}=&\left[\sn^2\h+\cs^2\Hh\right](\bdf_u^T\bdf_u^*)_{ik}\bgtpur_{kj}\\
&+\left[\frac{4}{9}\sbi\suk\bgtpur_{ik}\bgtpur^\dagger_{kl}+\frac{4}{3}\sgl\suk\bgtsur_{ik}\bgtsur^\dagger_{kl}\right.\\
&\left.\qquad+\sh\sqk\bftuq^T_{ik}\bftuq^*_{kl}\right]\bgtpur_{lj}\\
&+\frac{1}{2}\sbi\bgtpur_{ij}\left[\frac{1}{3}\sql\bgtpq^\dagger_{kl}\bgtpq_{lk}+\sll\bgtpl^\dagger_{kl}\bgtpl_{lk}\right.\\
&\qquad\qquad\qquad+\frac{8}{3}\suk\bgtpur^\dagger_{kl}\bgtpur_{lk}+\frac{2}{3}\sdk\bgtpdr^\dagger_{kl}\bgtpdr_{lk}\\
&\left.\qquad\qquad\qquad+2\sek\bgtper^\dagger_{kl}\bgtper_{lk}\right]\\
&+\frac{1}{2}\sbi\sh\bgtpur_{ij}\left\{\left[\sn^2\h+\cs^2\Hh\right]\mgtphusq+\left[\cs^2\h+\sn^2\Hh\right]\mgtphdsq\right\}\\
&-3\sbi\sh\left[\sn^2\h+\cs^2\Hh\right]\gtphu(\bdf_u)^T_{ik}\bftur^*_{kj}\\
&-\sh\sqk\bftuq^T_{ik}\bgtpq_{kl}\bftur^*_{lj}+2\sh\suk\bgtpur_{ik}\bftur^T_{kl}\bftur^*_{lj}\\
&+\suk\bgtpur_{ik}\left[\frac{8}{9}\sbi\bgtpur^\dagger_{kl}\bgtpur_{lj}+\frac{8}{3}\sgl\bgtsur^\dagger_{kl}\bgtsur_{lj}\right]\\
&-\bgtpur_{ij}\left[\frac{4}{3}g'^2+4g^2_3\right]\;.
\end{split}\end{equation}
As with the RGE for $\bdf_{u}$, care must be taken with the trace terms, which may not be immediately obvious in \eqref{eq:gurtilde}. An example of such a term is
\begin{equation*}
\frac{1}{2}\sbi\bgtpur_{ij}\left[\frac{1}{3}\sql\bgtpq^\dagger_{kl}\bgtpq_{lk}\right]\;,
\end{equation*}
in which the trace is not explicitly written since we need to keep information about the position of the squark thresholds. Note that when we are checking reduction to the MSSM, the trace terms contribute an additional factor of three to account for the sum over flavours.

In all the RGEs we take the approach that we do not insert explicit thresholds for external particles. This means that for the $\tilde{\mathbf{g}}^{\prime u_{R}}$ RGE we do not include a $\theta_{\tilde{u}}$ or $\swi$ when the term involves external forms of these particles. For example, the term $-\sh\sqk\bftuq^T_{ik}\bgtpq_{kl}\bftur^*_{lj}$, which comes from $2\bv_{b}\bw^{*}_{a}\bw^{T}_{b}$ in (\ref{eq:Va}), contains no $\theta_{\tilde{u}_{j}}$, even though we have the coupling $\bftur^*_{lj}$. When solving the RGEs, we do, however, freeze each coupling at the scale where the heaviest particle decouples. Thus, for $\tilde{\mathbf{g}}^{\prime u_{R}}$, if the wino is heavier than the up-type squark in question, the value of the coupling stops running when the wino decouples, and vice-versa.

The RGEs for the couplings of higgsinos and gauginos to Higgs bosons, \textit{i.e.},~the interaction terms \eqref{eq:Ghu} and \eqref{eq:Ghd}, are derived from the general expressions in \eqref{eq:X1} and \eqref{eq:X2}. Since they are couplings with Higgs bosons, the rotation to the Higgs boson mass basis introduces a $\sin{\beta}$ or a $\cos{\beta}$ in the same way as the usual Yukawa couplings. As a result, we find that below $m_{H}$, when the heavy Higgs fields decouple from our theory, we must use the combination of the coupling with the appropriate $\sn$ or $\cs$, \textit{e.g.},~$\sn\tilde{g}^{h_{u}}$. Glancing back at \eqref{eq:real} and \eqref{eq:gurtilde} we see that the Higgs-gaugino-higgsino couplings always appear with the appropriate angle. This change to the subset of couplings that are associated with the Higgs field remaining in the effective theory is not only important for the Yukawa and $\tilde{g}^{h_{u},h_{d}}$ couplings, but will appear again when we turn to discuss the trilinear couplings and Higgs boson scalar mass terms.

\section{RGEs for Gaugino Mass Terms}\label{sec:GHP}

To derive the RGEs for gaugino mass parameters we begin by constructing the matrices $\bdm^{(\prime)}_{{X}}$ from the $M^{(\prime)}_{1,2,3}$ and $\mu$ terms appearing in our MSSM Lagrangian. The higgsino mass term, $\mu$, should not be confused with the Higgs parameter, $\tilde{\mu}$, which we will consider shortly.

For the rotated Lagrangian, with the higgsinos defined as in \eqref{eq:inorot}, the matrices $\bdm^{(\prime)}_{X}$ are diagonal and the derivation is remarkably simple. The RGE for $M_{2}$, for example, is
\begin{equation}\label{eq:M2rgeill}\begin{split}
{\left(4\pi\right)}^2\frac{dM_2}{dt}=&M_2\swi\left[3\sqk\bgtq_{kl}\bgtq^\dagger_{lk}+\slk\bgtl_{kl}\bgtl^\dagger_{lk}+\sh\mgthusq\left(\sn^2\h+\cs^2\Hh\right)\right.\\
&\left.\qquad\quad+\sh\mgthdsq\left(\cs^2\h+\sn^2\Hh\right)\right]\\[5pt]
&+2\sn\cs\left(-\h+\Hh\right)\sh\left[\gthd\mu^*\gthu+(\gthd)^{*}\mu(\gthu)^{*}\right]-12\swi M_2g^2_2\;,
\end{split}\end{equation}
where, once again, we write $\theta$'s for all internal particles. The $\swi$ in the first and last terms may seem superfluous, but they are included to indicate to the reader that these terms are the result of loops which contain an internal $\tilde{W}$ Majorana fermion. Note that the RGEs for the gaugino and higgsino mass terms, when added to those for the gauge couplings and the full set of Yukawa couplings, once again form a closed set.

The reader may be struck by the second term on the right-hand side, proportional to $\mu$, whose appearance seems odd at first sight.  It originates in the terms in (\ref{eq:RGEmx}) where $\bdm^{(\prime)}_{X}$ is sandwiched between two $\bx^{1,2}_{b}$ matrices. Since $\bx^{1,2}_{b}$ connects gauginos to higgsinos, and the external fields are themselves gauginos, the $\bdm^{(\prime)}_{X}$ is necessarily a higgsino mass term. Notice that this term is a threshold effect: it only enters the running of $M_{2}$ below $m_{H}$ and only if the higgsinos decouple below this scale. The light and heavy Higgs boson doublets in (\ref{eq:hrot}) and (\ref{eq:Hrot}), respectively, make equal but opposite contributions to the evolution of $M_2$ which indeed cancel above all thresholds.  The appearance of $\mu$ is a general feature of the electroweak gaugino mass RGEs \mbox{---} listed in full in Appendix~\ref{app:dfulRGEs}, (\ref{app:rgem1})-(\ref{app:rgem2p}). For much the same reasons, the RGE for the complex parameter $\mu$, (\ref{app:rgemu}), develops a dependence on the electroweak gaugino mass parameters, $M^{(\prime)}_{1}$ and $M^{(\prime)}_{2}$, if the masses are ordered appropriately. Finally, we point out that although the RGE in \eqref{eq:M2rgeill} develops a dependence on the complex parameter $\mu$, it nevertheless preserves the reality of $M_{2}$ as specified in the Lagrangian of Eq.~\eqref{eq:soft}.

\section{Trilinear Coupling RGEs}\label{sec:trideriv}

In the $R$-parity conserving MSSM, with interactions listed in \eqref{eq:higgsinoint}-\eqref{eq:quartlag}, our collection of complex scalar fields consists of $\left\{\mathsf{h},\mathcal{H},G^{+},H^{+},\tilde{u}_{Li},\tilde{d}_{Li},\tilde{e}_{Li},\tilde{\nu}_{Li},\tilde{u}_{Ri}\right\}$, where we have written the Higgs fields in their mass basis and the index $i$ runs over all three flavours (the colour index is suppressed). We will use all these fields when constructing the the matrices $\mathbf{H}_{abc}$, $\bdm^{2}_{ab}$, $\bm{\mathcal{B}}_{ab}$, $\bdLm_{abcd}$ and $\bdLm'_{abcd}$. The resulting $\mathbf{H}_{abc}$ is a $\left(25\times25\times25\right)$ array, whose entries are mostly zero. 

Following on from our discussion regarding the Higgs rotation in Sec.~\ref{sec:masseig}, we must take a few moments to consider how the rotation determines the combinations of parameters that remain in the theory below $m_{H}$, when the heavy Higgs fields have decoupled. In addition to requiring us to redefine the quark Yukawa couplings ($\bdf_{u,d,e}\rightarrow\bdl_{u,d,e}$) and the Higgs-gaugino-higgsino couplings, we also discover that we must restrict the trilinear couplings and Higgs boson mass terms. For example, when the two trilinear terms which couple $\tilde{u}_{L}$ and $\tilde{u}_{R}$ to the Higgs bosons are rotated into the $\left(\mathsf{h},\mathcal{H}\right)$ basis, we see that
\begin{equation}\begin{split}
\mathcal{L}&\ni\tilde{u}^\dagger_{Rk}(\mathbf{a}^T_u)_{kl}\tilde{u}_{Ll}h^0_u-\tilde{u}^{\dagger}_{Rk}\mtsfuhut_{kl}\tilde{u}_{Ll}h^{0*}_{d}\\
&=\tilde{u}^\dagger_{Rk}(\mathbf{a}^T_u)_{kl}\tilde{u}_{Ll}\left(\sn\mathsf{h}+\cs\mathcal{H}\right)-\tilde{u}^{\dagger}_{Rk}\mtsfuhut_{kl}\tilde{u}_{Ll}\left(\cs\mathsf{h}-
\sn\mathcal{H}\right)\;,
\end{split}\end{equation}
which form the entries of the trilinear matrix\footnote{It should be noted that we have chosen the index, $a$, of $\mathbf{H}_{abc}$ to be a sfermion as per the comments below \eqref{eq:4compH}.} $\mathbf{H}_{\tilde{u}_{Rk},\mathsf{h},\tilde{u}_{Ll}}$ and $\mathbf{H}_{\tilde{u}_{Rk},\mathcal{H},\tilde{u}_{Ll}}$. Upon decoupling the field $\mathcal{H}$, we are left with only $\mathbf{H}_{\tilde{u}_{Rk},\mathsf{h},\tilde{u}_{Ll}}$ and this part of the Lagrangian becomes
\begin{equation}\label{eq:trileg}
\mathcal{L}\ni\tilde{u}^\dagger_{Rk}\left[\sn(\mathbf{a}^T_u)_{kl}-\cs\mtsfuhut_{kl}\right]\tilde{u}_{Ll}\mathsf{h}.
\end{equation}
Although in the complete MSSM (indeed, at all scales above $m_{H}$) we can sensibly talk about the evolution of the constituent pieces $(\mathbf{a}^T_u)_{kl}$ and $\mtsfuhu_{kl}$ separately, for $Q<m_H$ it is no longer possible to write RGEs for both $\ba_{u}$ and $\tilde{\mu}^{*}\bdf^{h_{u}}_{u}$. We must instead talk only about the single combination $\mathbf{H}_{\tilde{u}_{Rk},\mathsf{h},\tilde{u}_{Ll}}$ that remains in the effective theory below $Q=m_H$.

Before using \eqref{eq:4compH} to construct the RGE, we must also compose the quartic coupling matrices, $\bdLm_{abcd}$ and $\bdLm'_{abcd}$, which are $\left(25\times25\times25\times25\right)$ arrays whose entries are given by \eqref{eq:quarteg}. Our task is greatly simplified by noting that the $\bdLm'_{abcd}$-type couplings are absent above all SUSY thresholds in the $R$-parity conserving MSSM since in this case the number of daggered and undaggered fields is always the same (remember that these originate from the absolute square of a quadratic operator). We find that $\bdLm'$-type couplings only arise because of the way we define the linear combinations of the Higgs rotations, \eqref{eq:hrot} and \eqref{eq:Hrot}, in order to take into account the decoupling of the heavy Higgs bosons.

It is now straightforward to use \eqref{eq:4compH} to derive the various trilinear RGEs, and for the coupling of up-type squarks to the light Higgs boson we obtain:\\

\noindent${\left(4\pi\right)}^2\frac{d\left[\sn\au_{ij}-\cs\mtsfuhu_{ij}\right]}{dt}$
\begin{equation*}\begin{split}
=&\suk\left\{\h\left[\sn\au_{ik}-\cs\mtsfuhu_{ik}\right]\left[\frac{2\glp^{2}}{3}\left(\cs^{2}-\sn^{2}\right)\delta_{kj}+2\sn^{2}\left[\fuul^{\dagger}\fuul\right]_{kj}\right]\right.\\
&\qquad\left.+\Hh\sn\cs\left[\cs\au_{ik}+\sn\mtsfuhu_{ik}\right]\left[-\frac{4\glp^{2}}{3}\delta_{kj}+2\left[\fuul^{\dagger}\fuul\right]_{kj}\right]\right\}\\
&+\sul\sqk\left[-2\left(\frac{\glp^{2}}{9}+\frac{4\gthl^{2}}{3}\right)\delta_{ik}\delta_{lj}+6\fuhu_{ij}\fuhu^{\dagger}_{lk}\right]\left[\sn\au_{kl}-\cs\mtsfuhu_{kl}\right]\\
&+2\sqk\left\{\h\left[\left(\frac{\glp^{2}}{12}-\frac{3\gtwl^{2}}{4}\right)\left(\sn^{2}-\cs^{2}\right)\delta_{ik}+2\sn^{2}\left[\fuur\fuur^{\dagger}\right]_{ik}-\cs^{2}\left[\fddr\fddr^{\dagger}\right]_{ik}\right]\right.\\
\end{split}\end{equation*}
\begin{equation}\begin{split}\label{eq:RGEtriu}
&\qquad\qquad\times\left[\sn\au_{kj}-\cs\mtsfuhu_{kj}\right]+\Hh\sn\cs\left[\left(\frac{\glp^{2}}{6}-\frac{3\gtwl^{2}}{2}\right)\delta_{ik}+2\left[\fuur\fuur^{\dagger}\right]_{ik}\right.\\
&\qquad\qquad\left.\left.+\left[\fddr\fddr^{\dagger}\right]_{ik}\right]\times\left[\cs\au_{kj}+\sn\mtsfuhu_{kj}\right]\right\}\\
&+2\Hh\sdk\left[\sn\ad_{ik}+\cs\mtsfdhd_{ik}\right]\left[\fddl^{\dagger}\fudl\right]_{kj}\\
&+\frac{2}{3}\sbi\sn\left(M_{1}-iM'_{1}\right)\left(\sh(\gtphu)^{*}\bgtpq^{*}_{ik}\bftur_{kj}-\frac{4}{3}\bgtpq^{*}_{ik}(\bdf_{u})_{kl}\bgtpur^{*}_{kj}\right.\\
&\left.\qquad\qquad\qquad\qquad\quad\ \ -4\sh\bftuq_{ik}\bgtpur^{*}_{kj}(\gtphu)^{*}\right)\\
&-\frac{32}{3}\sgl\sn\left(M_{3}-iM'_{3}\right)\bgtsq^{*}_{ik}(\bdf_{u})_{kl}\bgtsur^{*}_{lj}\\
&-6\swi\sh\sn\left(M_{2}-iM'_{2}\right)(\gthu)^{*}\bgtq^{*}_{ik}\bftur_{kj}\\
&+\frac{2}{3}\sbi\sh\cs\mu^{*}\gtphd\left(4\bftuq_{ik}\bgtpur^{*}_{kj}-\bgtpq^{*}_{ik}\bftur_{kj}\right)\\
&+6\sh\swi\cs\mu^{*}\gthd\bgtq^{*}_{ik}\bftur_{kj}-4\sh\cs\mu^{*}\bftdq_{ik}(\bdf^{\dagger}_{d})_{kl}\bftur_{lj}\\
&+\suk\left[\sn\au_{ik}-\cs\mtsfuhu_{ik}\right]\\
&\qquad\qquad\qquad\qquad\times\left[\frac{8}{9}\sbi\bgtpur^{T}_{kl}\bgtpur^{*}_{lj}+\frac{8}{3}\sgl\bgtsur^{T}_{kl}\bgtsur^{*}_{lj}+2\sh\bftur^{\dagger}_{kl}\bftur_{lj}\right]\\
&+\h\left[3\sn^{2}(\bdf^{\dagger}_{u})_{kl}(\bdf_{u})_{lk}+\cs^{2}\left\{3(\bdf^{\dagger}_{d})_{kl}(\bdf_{d})_{lk}+(\bdf^{\dagger}_{e})_{kl}(\bdf_{e})_{lk}\right\}\right]\left[\sn\au_{ij}-\cs\mtsfuhu_{ij}\right]\\
&+\Hh\sn\cs\left[3(\bdf^{\dagger}_{u})_{kl}(\bdf_{u})_{lk}-\left\{3(\bdf^{\dagger}_{d})_{kl}(\bdf_{d})_{lk}+(\bdf^{\dagger}_{e})_{kl}(\bdf_{e})_{lk}\right\}\right]\left[\cs\au_{ij}+\sn\mtsfuhu_{ij}\right]\\
&+\frac{1}{2}\sh\left\{\h\left[\cs^{2}\left(\sbi\mgtphdsq+3\swi\mgthdsq\right)+\sn^{2}\left(\sbi\mgtphusq+3\swi\mgthusq\right)\right]\right.\\
&\qquad\quad\;\;\times\left[\sn\au_{ij}-\cs\mtsfuhu_{ij}\right]+\Hh\sn\cs\left[-\left(\sbi\mgtphdsq+3\swi\mgthdsq\right)\right.\\
&\qquad\quad\;\;\left.\left.+\left(\sbi\mgtphusq+3\swi\mgthusq\right)\right]\times\left[\cs\au_{ij}+\sn\mtsfuhu_{ij}\right]\right\}\\
&+\sql\left[\sh\bftuq_{ik}\bftuq^{\dagger}_{kl}+\sh\bftdq_{ik}\bftdq^{\dagger}_{kl}+\frac{1}{18}\sbi\bgtpq^{*}_{ik}\bgtpq^{T}_{kl}+\frac{3}{2}\swi\bgtq^{*}_{ik}\bgtq^{T}_{kl}\right.\\
&\left.\qquad\quad+\frac{8}{3}\sgl\bgtsq^{*}_{ik}\bgtsq^{T}_{kl}\right]\left[\sn\au_{lj}-\cs\mtsfuhu_{lj}\right]\\
&-3\left\{\left(\frac{1}{36}\sqi+\frac{4}{9}\suj+\frac{1}{4}\h\right)g'^{2}+\frac{3}{4}\left(\sqi+\h\right)g^{2}_{2}+\frac{4}{3}\left(\sqi+\suj\right)g^{2}_{3}\right\}\\
&\qquad\qquad\qquad\qquad\qquad\qquad\qquad\qquad\qquad\qquad\qquad\qquad\times\left[\sn\au-\cs\mtsfuhu\right]_{ij}\;.
\end{split}\end{equation}
The RGE for the corresponding coupling, $\left[\cs\au_{ij}+\sn\mtsfuhu_{ij}\right]$, to the heavier Higgs doublet can be obtained in the same manner. By taking linear combinations of these RGEs, we can obtain the separate RGEs\footnote{The $s$ and $c$ can be taken out of the derivatives on the left-hand side as noted in Sec.~\ref{sec:litcomp} when discussing reduction to the MSSM.} for $\ba_u$ and $\tilde{\mu}^*\bdf_u^{h_u}$. Since the coupling $\tilde{\mu}^*\bdf_u^{h_u}$ always occurs as a product, it is not possible to obtain the RGEs for the individual factors. Of course, above all thresholds, we must have $\tilde{\mu}=\mu$ and
$\bdf_u^{h_u}=\bdf_u$.  We have checked that with these replacements, our RGEs reduce to the MSSM RGEs \cite{MVrge} if we put all $\theta_i=1$ and take care to sum over all internal flavours.

Before closing this section we point out that in the $R$-parity conserving MSSM the one-loop RGEs for the trilinear couplings do not contain dependence on the SSB scalar masses, a fact which is still true in the two-loop RGEs. This means that the complete set of coupled RGEs now includes the trilinear couplings, but the scalar mass RGEs are not needed except when setting the location of the various scalar thresholds.

\section{Scalar Mass RGEs}

Once we have derived the RGEs for the scalar mass parameters we will have completed the full set of RGEs for the $R$-parity conserving MSSM (with the approximation that the $\bdLm_{abcd}$ and $\bdLm'_{abcd}$ couplings are equal to their usual SM counterparts). The scalar mass RGEs depend on all the interaction terms in our general Lagrangian, \eqref{eq:fourlag}, including the matrices $\bdm_{ab}$ and $\bm{\mathcal{B}}_{ab}$.  Most of these entries can be written down directly from the SSB Lagrangian, \eqref{eq:soft}, but there are also $\left|\tilde{\mu}\right|^{2}$ bilinears coming from the Lagrangian in Eq.~\eqref{eq:quartlag} including the term written out in Eq.~\eqref{eq:quarteg}. Note that when the Lagrangian is written in terms of the rotated Higgs fields, there are no $\bm{\mathcal{B}}_{ab}$ terms, so that the RGE in \eqref{eq:4compmb} contains fewer terms.

\subsection{Mass Terms for Higgs Bosons}\label{sec:higgmassderiv}

As mentioned in connection with the trilinear RGE derivation, the mass terms for Higgs bosons in the Lagrangian become restricted when we are in the regime $Q<m_{H}$ and the heavy Higgs fields have therefore decoupled. The complete set of mass terms in the rotated Higgs basis are 
\begin{equation}\begin{split}\label{eq:higgsterms}
\mathcal{L}\ni&-\left[\sn^{2}\left(m^{2}_{H_{u}}+\left|\tilde{\mu}\right|^{2}\right)+\cs^{2}\left(m^{2}_{H_{d}}+\left|\tilde{\mu}\right|^{2}\right)-\sn\cs\left(b+b^{*}\right)\right]\mathsf{h}^\dagger\mathsf{h}\\
&-\left[\cs^{2}\left(m^{2}_{H_{u}}+\left|\tilde{\mu}\right|^{2}\right)+\sn^{2}\left(m^{2}_{H_{d}}+\left|\tilde{\mu}\right|^{2}\right)+\sn\cs\left(b+b^{*}\right)\right]\mathcal{H}^\dagger\mathcal{H}\\
&-\left[\sn\cs \left(m^{2}_{H_{u}}+\left|\tilde{\mu}\right|^{2}\right)-\sn\cs
\left(m^{2}_{H_{d}}+\left|\tilde{\mu}\right|^{2}\right)-\cs^{2}b+\sn^{2}b^{*}\right]\mathsf{h}^\dagger\mathcal{H}\\
&-\left[\sn\cs \left(m^{2}_{H_{u}}+\left|\tilde{\mu}\right|^{2}\right)-\sn\cs
\left(m^{2}_{H_{d}}+\left|\tilde{\mu}\right|^{2}\right)+\sn^{2}b-\cs^{2}b^{*}\right]\mathcal{H}^\dagger\mathsf{h}\;.
\end{split}\end{equation}
out of which only the first will remain in the effective theory below $m_{H}$. For $Q>m_{H}$, we can obtain RGEs for all four couplings and thereby derive separate RGEs for the real parameters $\left(m^{2}_{H_{u}}+\left|\tilde{\mu}\right|^{2}\right)$ and $\left(m^{2}_{H_{d}}+\left|\tilde{\mu}\right|^{2}\right)$, and the complex parameter, $b$, by taking the appropriate linear combinations.

We see from \eqref{eq:higgsterms} that the term $\left|\tilde{\mu}\right|^{2}$ only appears in the Lagrangian in combination with either $m^{2}_{H_{u}}$ or $m^{2}_{H_{d}}$, so that we cannot derive an RGE for them separately. Naturally, above all thresholds supersymmetry requires that $\mu=\tilde{\mu}$, so that we can use the RGE for $\mu$, \eqref{app:rgemu}, to derive the RGEs for $\mhusq$ and $\mhdsq$, but these will no longer be valid below the scale of the heaviest SUSY particle. We can, however, use this to check the reduction of our RGEs to the MSSM limit when all $\theta_{i}$ are set equal to unity.

Applying \eqref{eq:4compmb} to the light Higgs boson soft mass term, we obtain the RGE:\\

\noindent${\left(4\pi\right)}^{2}\frac{d\left[\sn^{2}M^{2}_{H_{u}}+\cs^{2}M^{2}_{H_{d}}-\sn\cs\left(b+b^{*}\right)\right]}{dt}$
\begin{equation*}\begin{split}
=&\frac{3}{2}\h\left[\glp^{2}+\gtwl^{2}\right]\left(\cs^{2}-\sn^{2}\right)^{2}\left[\sn^{2}M^{2}_{H_{u}}+\cs^{2}M^{2}_{H_{d}}-\sn\cs\left(b+b^{*}\right)\right]\\
&+\Hh\left[\glp^{2}\left(-\cs^{4}+4\sn^{2}\cs^{2}-\sn^{4}\right)+6\sn^{2}\cs^{2}\gtwl^{2}\right]\left[\cs^{2}M^{2}_{H_{u}}+\sn^{2}M^{2}_{H_{d}}+\sn\cs\left(b+b^{*}\right)\right]\\
&-6\h\Hh\left[\glp^{2}+\gtwl^{2}\right]\sn\cs\left(\cs^{2}-\sn^{2}\right)\left[\sn\cs\left\{M^{2}_{H_{u}}-M^{2}_{H_{d}}\right\}-\frac{1}{2}\left(\cs^{2}-\sn^{2}\right)\left(b+b^{*}\right)\right]\\
&+\suk\sul\left[-2\glp^{2}\left(\sn^{2}-\cs^{2}\right)\delta_{lk}+6\sn^{2}\left[\fuult\fuuls\right]_{lk}\right]\musq_{kl}\\
&+\sqk\sql\left[\glp^{2}\left(\sn^{2}-\cs^{2}\right)\delta_{lk}+6\sn^{2}\left[\fuurs\fuurt\right]_{lk}+6\cs^{2}\left[\fddrs\fddrt\right]_{lk}\right]\mqsq_{kl}\\
&+\sdk\sdl\left[\glp^{2}\left(\sn^{2}-\cs^{2}\right)\delta_{lk}+6\cs^{2}\left[\fddlt\fddls\right]_{lk}\right]\mdsq_{kl}\\
&+\slk\sll\left[-\glp^{2}\left(\sn^{2}-\cs^{2}\right)\delta_{lk}+2\cs^{2}\left[\feers\feert\right]_{lk}\right]\mlsq_{kl}\\
&+\sek\sel\left[\glp^{2}\left(\sn^{2}-\cs^{2}\right)\delta_{lk}+2\cs^{2}\left[\feelt\feels\right]_{lk}\right]\mesq_{kl}\\
&+6\suk\sql\left[\sn\au_{lk}-\cs\mtsfuhu_{lk}\right]\left[\sn\au^{\dagger}_{kl}-\cs\mtsfuhu^{\dagger}_{kl}\right]\\
&+\slk\sll\left[-\glp^{2}\left(\sn^{2}-\cs^{2}\right)\delta_{lk}+2\cs^{2}\left[\feers\feert\right]_{lk}\right]\mlsq_{kl}\\
&+\sek\sel\left[\glp^{2}\left(\sn^{2}-\cs^{2}\right)\delta_{lk}+2\cs^{2}\left[\feelt\feels\right]_{lk}\right]\mesq_{kl}\\
\end{split}\end{equation*}
\begin{equation}\begin{split}\label{eq:RGEmhud}
&+6\suk\sql\left[\sn\au_{lk}-\cs\mtsfuhu_{lk}\right]\left[\sn\au^{\dagger}_{kl}-\cs\mtsfuhu^{\dagger}_{kl}\right]\\
&+6\sqk\sdl\left[\cs\ad_{lk}-\sn\mtsfdhd_{lk}\right]\left[\cs\ad^{\dagger}_{kl}-\sn\mtsfdhd^{\dagger}_{kl}\right]\\
&+2\slk\sel\left[\cs\bae_{lk}-\sn\mtsfehd_{lk}\right]\left[\cs\bae^{\dagger}_{kl}-\sn\mtsfehd^{\dagger}_{kl}\right]\\
&-2\sh\left|\mu\right|^{2}\left\{\sbi\left[\sn^{2}\mgtphusq+\cs^{2}\mgtphdsq\right]+3\swi\left[\sn^{2}\mgthusq+\cs^{2}\mgthdsq\right]\right\}\\
&-2\sh\left\{\sbi\left(M^{2}_{1}+M'^{2}_{1}\right)\left[\sn^{2}\mgtphusq+\cs^{2}\mgtphdsq\right]\right.\\
&\left.\qquad\quad\ +3\swi\left(M^{2}_{2}+M'^{2}_{2}\right)\left[\sn^{2}\mgthusq+\cs^{2}\mgthdsq\right]\right\}\\
&-\frac{1}{2}\left\{-4\sh\sbi\sn\cs\mu^{*}\gtphu\gtphd\left(M_{1}+iM'_{1}\right)-12\sh\swi\sn\cs\mu^{*}\gthu\gthd\left(M_{2}+iM'_{2}\right)\right\}\\
&-\frac{1}{2}\left\{-4\sh\sbi\sn\cs\mu(\gtphu)^{*}(\gtphd)^{*}\left(M_{1}-iM'_{1}\right)-12\sh\swi\sn\cs\mu(\gthu)^{*}(\gthd)^{*}\left(M_{2}-iM'_{2}\right)\right\}\\
&-\h\left(\frac{3g'^{2}}{2}+\frac{9g^{2}_{2}}{2}\right)\left[\sn^{2}M^{2}_{H_{u}}+\cs^{2}M^{2}_{H_{d}}-\sn\cs\left(b+b^{*}\right)\right]\\
&+\h\left\{\sn^{2}\left[6\bdf^{*}_{u}\bdf^{T}_{u}\right]_{kk}+\cs^{2}\left[6\bdf^{*}_{d}\bdf^{T}_{d}+2\bdf^{*}_{e}\bdf^{T}_{e}\right]_{kk}+\sbi\sh\left[\sn^{2}\mgtphusq+\cs^{2}\mgtphdsq\right]\right.\\
&\left.\qquad\;\;+3\swi\sh\left[\sn^{2}\mgthusq+\cs^{2}\mgthdsq\right]\right\}\left[\sn^{2}M^{2}_{H_{u}}+\cs^{2}M^{2}_{H_{d}}-\sn\cs\left(b+b^{*}\right)\right]\\
&+\Hh\sn\cs\left\{\left[6\bdf^{*}_{u}\bdf^{T}_{u}\right]_{kk}-\left[6\bdf^{*}_{d}\bdf^{T}_{d}+2\bdf^{*}_{e}\bdf^{T}_{e}\right]_{kk}+\sbi\sh\left[\mgtphusq-\mgtphdsq\right]\right.\\
&\left.\qquad\qquad+3\swi\sh\left[\mgthusq-\mgthdsq\right]\right\}\left[\sn\cs\left\{M^{2}_{H_{u}}-M^{2}_{H_{d}}\right\}-\frac{1}{2}\left(\cs^{2}-\sn^{2}\right)\left(b+b^{*}\right)\right]\;,
\end{split}\end{equation}
where we have used $M^{2}_{H_{u}}\equiv\left(\mhusq+\mtsq\right)$ and $M^{2}_{H_{d}}\equiv\left(\mhdsq+\mtsq\right)$. A discussion of the origin of each of the terms in this RGE can be found in Ref.~\cite{RGE2}. It is perhaps informative to point out here, however, that all terms in \eqref{eq:RGEmhud} up to the trilinear terms derive from the single quartic, $\bdLm_{afbe}\bdm^{2}_{ef}$ in \eqref{eq:4compmb}. All other quartic terms are zero either because $\bdLm'_{abcd}$ vanishes when $a=b$ or due to $\bm{\mathcal{B}}_{ab}$ having all vanishing entries in the Higgs mass basis. Also of note is that the terms in this group that are proportional to $g^{\prime2}$ contain the so-called $S$-term from the RGEs in Ref.~\cite{wss,MVrge} and elsewhere. The term is noteworthy due to the fact that if it is zero at one scale (which is true at $\mgut$ in many models) it will remain zero at least in the one-loop renormalisation.

\subsection{Squark and Slepton SSB Mass Terms}

The RGEs for the sfermion soft mass terms are somewhat simpler on account of the fact that they have no Higgs fields in their operator in the Lagrangian. Otherwise, the derivation follows in a similar fashion to the mass terms for Higgs bosons above. We point out that all sfermion soft mass RGEs, \eqref{app:mqamh}-\eqref{app:meamh} and \eqref{app:mqbmh}-\eqref{app:mebmh}, contain a dependence on the usual Yukawa matrices without a corresponding $\sn$ or $\cs$ to use in the conversion to $\bdl_{u,d,e}$ below $m_{H}$. These terms have their origin in the quartic couplings, which we have set equal to the corresponding square of $\bdf_{u,d,e}$, and would therefore have their running described by the quartic RGEs, not considered here. In solving the RGEs, since we do not have a valid RGE for $\bdf_{u,d,e}$, we use the value of the Yukawa couplings frozen at $m_{H}$ when $Q<m_{H}$.

\section{Comparison with Earlier Literature: Part II}\label{sec:litcompII}

As a short conclusion to this chapter we note that Ref.~\cite{casno} only includes RGEs with thresholds for the Yukawa couplings, $\bdf_{u,d,e}$. It is possible, however, to carry out a final comparison with Ref.~\cite{sakis}, which writes out the RGEs for the Higgs boson, higgsino and gaugino mass terms, and the $(3,3)$ component of all the trilinear couplings and SSB scalar masses. 

We have checked that, except for terms involving couplings of Higgs boson fields, the RGEs that we obtain agree with those in Ref.~\cite{sakis}. We found it difficult to compare contributions involving Higgs boson fields since, as noted in Sec.~\ref{sec:litcomp}, the RGEs of Ref.~\cite{sakis} appear to have been written \textit{without any rotation} of these fields, making it impossible to abstract the relationship between the thresholds for the Higgs bosons $h$, $A$, $H$ and $H^{\pm}$ in Ref.~\cite{sakis} and the thresholds in the RGEs listed in Appendix~\ref{app:dlessRGEs} and Appendix~\ref{app:dfulRGEs}.

%% file: rgeflav.tex
\chapter{Deriving the Weak Scale Couplings from the RGEs: \progrge}\label{ch:rgeflav}

Now that we have the one-loop RGEs for the MSSM with full threshold effects from Chapter~\ref{ch:application}, we are in a position to combine these with the two-loop RGEs for the MSSM \cite{MVrge} and the SM \cite{ACMPRWrge} converted to the \drb-scheme as in Appendix~\ref{app:drbar}. As described at the end of Sec.~\ref{sec:choiceren}, our two-loop RGEs do not contain (numerically higher order) threshold effects, and, for the gauge and Yukawa couplings, we use the MSSM form above $Q=m_{H}$ and the SM form below $Q=m_{H}$. We solve this system of equations given our weak scale Standard Model inputs and GUT scale ansatz using \fortran~computer code to iteratively reach a numerical solution.

The code has been designed to be incorporated into the public release software, \progisa~\cite{isajet}\glossary{name={ISAJET},description={a Monte Carlo particle event generator}}, which is a Monte Carlo  program that simulates $pp$, $\bar{p}p$, and $e^+e^-$ interactions at high energies. \progrge~is the main program, which sets up the problem and calls the subroutines necessary to iteratively solve the RGEs. It will be employed as a subroutine of \progisasug, a section of \progisa~that can currently be used to solve the flavour-diagonal RGEs with a variety of different boundary conditions and then, through a call to an additional subroutine, \texttt{ISASUSY}, calculate the decay modes. 

This chapter aims to provide an overall description of the code and the various issues encountered in the procedure. After outlining the broad approach, we will consider each segment of the code individually, from the input file, through the choice and application of the various boundary conditions, to the final output.

\section*{General Outline}

Since \progrge~has been designed to be used in combination with \progisasug, this programme must be initialised first with equivalent mSUGRA inputs, and \progrge~is called at the end of the \progisasug~code. A general outline of the order in which \progrge~carries out the various steps is:
\begin{enumerate}
\item Read the input file \inrge~that contains all the choices available to the user.
\item Introduce gauge couplings, light quark Yukawa couplings, third generation Yukawa couplings from \progisa, and the SM Higgs field VEV all at $M_{Z}$.
\item Run the gauge couplings and diagonal Yukawa matrices to $m_{t}$. At this scale, insert \progisa's value for $f_{t}(m_{t})$ and rotate the Yukawa coupling matrices into the current basis using user defined rotations.
\item Run up to the high scale using appropriate thresholds, whose initial values are obtained from \progisa, to transform SM running into MSSM running.
\item Insert high scale boundary conditions. It is after this step that we begin the iterative loop.
\item Run back down to the weak scale, decoupling particles as their thresholds are passed.
\item At $m_{H}$, change over to the new set of RGEs, with restricted Lagrangian terms, as a result of the decoupling of the heavy Higgs boson terms. Save the values of the couplings at $m_{H}$ for use when running back up.
\item At $\msusy$, defined as $\sqrt{m_{\tilde{t}_{L}}m_{\tilde{t}_{R}}}$, apply the electroweak symmetry breaking conditions and radiative corrections to the third generation Yukawa couplings \cite{pierce}.
\item Once the running reaches $m_{t}$ rotate back to the basis in which the Yukawa matrices are diagonal, reset $f_{t}(m_{t})$ to the value obtained from \progisa~and continue running to $M_{Z}$. From now on $f_{t}$ remains in the theory all the way to $M_{Z}$ and will therefore change the running of the various couplings from that obtained during the first upwards run.
\item Reset the values of the gauge and all Yukawa couplings other than $f_{t}$.
\item Run back up to the high scale, applying the rotation (and boundary condition on $f_{t}$) at $m_{t}$ and the thresholds as before.
\item Reset the high scale values of the $\ba$-parameters and SSB mass matrices. Iterate until a solution with pre-selected accuracy is found.
\item On the final run, stop at $m_{H}$, the heavy Higgs threshold, and output the couplings at this scale.
\item At present, the code then calls \progstd, which is discussed in Sec.~\ref{sec:std}.
\end{enumerate}
The code must run as part of the \progisa~distribution because there are some common blocks which must be filled for \progrge~to execute properly. These common blocks are:
\begin{itemise}
\item \texttt{SUGPAS}~and \texttt{SUGMG}: Contain the initial values of the various thresholds.
\item \texttt{WKYUK}: Contains the weak scale third generation Yukawa couplings.
\item \texttt{BSG}: Contains the radiative corrections to the Yukawa couplings.
\item \texttt{RGEMS}: Contains the value of $\msusy$~from \progisa~and the value of $\mu$ at this scale.
\end{itemise}

The subtlety in initialising \progisa~is that equivalent mSUGRA inputs are needed since the interface has been developed to work with the mSUGRA-based executable \progisasug. The user may use non-universal GUT scale boundary conditions according to the choices given by \progisasug~and available in the input file (discussed next), but the user must also enter approximate mSUGRA conditions before \progrge~can begin the iterative process.

Once \progisasug~has called \progrge~the signs of the numerical values of $A_{0}$ and $\mu$ are flipped. This is because \progrge~has been developed for inputs that use the notation found in Ref.~\cite{wss}, and the sign changes are required to fix notational differences. Some signs are again flipped in the stop decay routine (\progstd) and this will be discussed later.

\section{The Input File}
Once \progrge~is called at the end of \progisa, it reads the input from \inrge, a sample of which is shown in Appendix~\ref{app:inrgeeg}. The choices, numbered in the same order as the input file, are:
\begin{enumerate}
\item Whether to use the complete two-loop equations. If the following line is `\verb+0+', only one-loop RGEs will be calculated, but threshold effects will still be included. Choosing to only calculate one-loop running speeds up the programme by around 15\%.
\item This item allows the user to choose between complex inputs and real inputs. If the next line is `\verb+1+', all inputs must be in complex form as in Appendix~\ref{app:inrgeeg} and the KM matrix will include a phase. If `\verb+0+'~is entered, any optional inputs which follow must be single numbers representing one real entry each. Otherwise the programme will print an error message and continue no further. Choosing to use a real KM matrix and real boundary conditions dramatically speeds up the running of the RGEs.
\item When the answer to Question 2.~is `\verb+1+' the user may wish to enter a phase for $\mu$ as opposed to a simple sign. If the  user answers yes, the optional setting for \texttt{THETA} on the next line is used, otherwise this value is ignored, and a sign for $\mu$ consistent with the \progisa~input is used.
\item If the response to Question 4.~is `\verb+1+' the programme will not attempt to change the thresholds from those obtained by \progisa. Otherwise, thresholds for: the left- and right-handed squarks and sleptons (\textit{i.e.}, the eigenvalues of the SSB sfermion mass matrices), the higgsino ($\mu$), the bino ($M_{1}$) and the winos ($M_{2}$), will be set during the running.
\item The mSUGRA GUT scale inputs used by \progisa~are automatically passed to \progrge, but they are only all used if the user sets this entry to `\verb+1+'. If so, reading of the input file moves to Question 17. Otherwise, reading continues to the non-universal inputs, $6.-16.$
\end{enumerate}

Inputs $6.-16.$ are the alternative GUT scale inputs. We will deal with these inputs in more detail in Sec.~\ref{sec:gutins}. After taking care of the GUT scale inputs, the file continues with the choices relating to the weak scale rotations $\mathbf{V}_{L,R}(u,d)$.
\begin{enumerate}\setcounter{enumi}{16}
\item The first choice in the final section is the basis in which to output the results. Since $SU(2)$ is broken in the quark mass basis, we only use this basis for the weak scale inputs. Instead, the output is in a current basis in which either the up- or down-type quarks are diagonal at $m_{t}$, thereby retaining $SU(2)$ invariance. Note that when interfacing with \progstd, described in Sec.~\ref{sec:std}, the answer to Question 17.~\textit{must} be `\verb+1+' since we wish to associate the $\tilde{c}_{L,R}$ squarks with the $c$ quark mass eigenstate.
\item This question asks whether the user should enter their own choice of general rotation matrices, as described in Sec.~\ref{sec:genu}. These matrices will be used to rotate the mass basis quark Yukawa matrices into the current basis and therefore this corresponds to a choice of current basis. Note, however, that the output is always in a current basis where either the up- or down-type quarks are diagonal at $m_{t}$. If the user chooses `\verb+1+', the inputs for Question 21.~will be read next. If not, the file proceeds to the next question, which allows for more basic choices for the rotation matrices.
\item Here the user chooses the form of $\mathbf{V}_{L}(u)$. Either they can choose to have $\mathbf{V}_{L}(u)=\mathbf{K}$, or $\mathbf{V}_{L}(u)=\dblone$. The programme will ensure that $\mathbf{V}_{L}(d)$ is fixed so that the appropriate combination of these two matrices is the correct KM matrix, according to \eqref{eq:vudkm}.
\item If the KM matrix is the only source of flavour-violation, the form of the $\mathbf{V}_{R}(u,d)$ matrices is unimportant. In this entry the user is given a restricted choice for the matrices $\mathbf{V}_{R}(u,d)$ which serve as default entries if the response to Question 18.~is `\verb+0+'. Either both $\mathbf{V}_{R}(u)$ and $\mathbf{V}_{R}(d)$ are the unit matrix or $\mathbf{V}_{R}(u)=\mathbf{K}^{\dagger}$ and $\mathbf{V}_{R}(d)=\mathbf{K}$.
\item The final entries are the choice of parameters which will be used to define the rotation matrices if the response to Question 18.~is `\verb+1+'. If the user has chosen to have real running in Question 2.~the phase in these inputs is ignored. For more detail on the way these inputs are converted into unitary matrices, see Sec.~\ref{sec:genu}.
\end{enumerate}
We now move on to consider some of the above inputs in more detail.

\section{Entering a General Unitary Matrix}\label{sec:genu}

As mentioned above, Question 21.~of \inrge~allows the user to define their own $\mathbf{V}_{L,R}(u,d)$ matrices. Since the KM matrix must still be correct, $\mathbf{V}_L(u)$, $\mathbf{V}_R(u)$, $\mathbf{V}_R(d)$ are taken in as inputs and $\mathbf{V}_L(d)$ is then fixed by requiring that (\ref{eq:vudkm}) is satisfied.

In order to guarantee the unitary nature of the rotation\footnote{The matrices ${\bf V}_{L,R}(u)$ should be numerically unitary to high accuracy. Otherwise, inverting the up Yukawa couplings from this new current basis to our ``standard'' basis will leave residual off-diagonal elements rather than zero even at $Q=m_t$. If the size of these elements is comparable to the values of the smallest off-diagonal elements at values of $Q$ substantially away from $m_t$, it is clear that our solutions will be dominated by the error from the non-unitarity of the $\mathbf{V}_{L,R}(u)$ matrices. A similar consideration applies to $\mathbf{V}_R(d)$.}, the input file reads three angles ($\alpha$, $\beta$ and $\gamma$) and a phase ($\delta$) using a similar parametrisation to that used for the KM matrix as given in \eqref{eq:kmparam}. Each unitary matrix, $\mathbf{U}$, is therefore given by
\begin{equation}\label{eq:rotin}
\mathbf{U}=\left(\begin{array}{ccc}
1 & 0 & 0 \\
0 & \cs_{\gamma} & \sn_{\gamma} \\
0 & -\sn_{\gamma} & \cs_{\gamma} \\
\end{array}\right)
\left(\begin{array}{ccc}
\cs_{\beta} & 0 & \sn_{\beta}e^{-i\delta} \\
0 & 1 & 0 \\
-\sn_{\beta}e^{i\delta} & 0 & \cs_{\beta} \\
\end{array}\right)
\left(\begin{array}{ccc}
\cs_{\alpha} & \sn_{\alpha} & 0 \\
-\sn_{\alpha} & \cs_{\alpha} & 0 \\
0 & 0 & 1 \\
\end{array}\right)\;,
\end{equation}
where $\sn_{\alpha}=\sin{\alpha}$, $\cs_{\alpha}=\cos{\alpha}$, etc. Note that this is not the most general unitary matrix possible, which would include an additional two phases. However, it is considered that for the time being this choice of inputs will suffice for most practical applications, and the addition of two phases would require only a minor change to the code.

\section{Weak Scale Boundary Conditions}

Once \progisa~calls the main program, and the input file has been read, \progrge~fixes the weak scale boundary conditions and runs them up to the GUT scale using the subroutine \texttt{UPMZMHIGH}. For the gauge sector, we take as our input the current PDG values \cite{pdg}
\begin{gather*}
\alpha^{-1}_{em}(M_{Z})=127.925\pm0.016\;;\;\alpha_{s}(M_{Z},\mmsb)=0.1176\pm0.002\;;\\
\sin^{2}{\theta_{W}}(M_{Z},\mmsb)=0.23119\pm0.00014\;. 
\end{gather*}
These are the couplings extracted using the effective theory with the electroweak gauge bosons and the top quark integrated out at $Q=M_{Z}$. In order to use the SM evolution for $Q>M_{Z}$, we must match these couplings to those of the full SM, which to two-loop accuracy implies that the SM gauge couplings in the \msb~scheme are given by \cite{weing,*hallg,*ovrutg,*chetg,*noteg},
\begin{subequations}\begin{align}
\frac{1}{\alpha_{1}(M_{Z})}=&\frac{3}{5}\left[\frac{1-\sin^{2}{\theta_{W}}(M_{Z})}{\alpha_{em}(M_{Z})}\right]+\frac{3}{5}\left[1-\sin^{2}{\theta_{W}}(M_{Z})\right]4\pi\Omega(M_{Z})\\
\frac{1}{\alpha_{2}(M_{Z})}=&\frac{\sin^{2}{\theta_{W}}(M_{Z})}{\alpha_{em}(M_{Z})}+\sin^{2}{\theta_{W}}(M_{Z})4\pi\Omega(M_{Z})\\
\frac{1}{\alpha_{3}(M_{Z})}=&\frac{1}{\alpha_{s}(M_{Z})}+4\pi\Omega_{3}(M_{Z})\;,
\end{align}\end{subequations}
where
\begin{subequations}\begin{align}
\Omega(\mu)=&\frac{1}{24\pi^{2}}\left[1-21\ln{\left(\frac{M_{W}}{\mu}\right)}\right]+\frac{2}{9\pi^{2}}\ln{\left(\frac{m_t}{\mu}\right)}\\
\Omega_{3}(\mu)=&\frac{2}{24\pi^{2}}\ln{\left(\frac{m_{t}}{\mu}\right)}\;.
\end{align}\end{subequations}
Notice that in order to preserve the $SU(2)$ symmetry of the effective theory down to $Q=M_Z$, we have, as mentioned in Sec.~\ref{sec:masseig}, integrated out the top quark at $Q=M_Z$ rather than at its mass as we do for all other particles.  This is the origin of the $\ln(m_t/\mu)$ terms in the matching conditions for the gauge couplings above. Since we decouple all SUSY particles as well as the additional Higgs bosons at the scale of their mass, we do not get corresponding jumps in the gauge couplings as these decouple. Our method \mbox{---} which is also used in \progisa~\mbox{---} has an important advantage that it ``sums the logs of the ratio of any large sparticle or Higgs boson mass to $M_Z$'', in contrast to the frequently used procedure that applies MSSM evolution all the way down to $M_Z$, and then corrects for this via a ``single-step evolution'' (between the heavy scale and $M_Z$) to take into account the difference between the running in the MSSM and in the SM.

Next, we convert the values of these gauge couplings in the \msb~scheme to their corresponding values in the \drb~scheme using the relations \eqref{eq:a1msdr}-\eqref{eq:a3msdr} and use the results as boundary conditions at $Q=M_Z$ when solving the RGEs.

For the Yukawa couplings, we begin with the quark masses at $Q=M_{Z}$ (the masses of the light quarks and leptons at $M_{Z}$ can be found in Ref.~\cite{fusaoka}), and convert to SM Yukawa coupling matrices using $v_{SM}=248.6/\sqrt{2}$~GeV as in Ref.~\cite{pierce}. The masses of the
first two generations of quarks have substantial error, which leads to a corresponding error in the Yukawa coupling. The third
generation quark masses are more precisely known \mbox{---} we take the top pole mass $m_{t}=172$~GeV \mbox{---} and in practice the values of the third generation Yukawa couplings are taken from \progisa, with bottom and tau Yukawa couplings at the scale $Q=M_{Z}$ and the top Yukawa coupling at $Q=m_{t}$. In extracting these Yukawa couplings we include SUSY radiative corrections \cite{pierce} at $M_{\mathrm{SUSY}}$ obtained by \progisa~during its execution, with inter-generation quark mixing neglected.

These diagonal Yukawa couplings, which are in the ``quark mass basis'' are run to $m_{t}$ with no flavour structure since it is a good approximation that the running is mainly due to the strong coupling. At $m_{t}$ all three Yukawa matrices are rotated to the user's choice of current basis using the SM version of \eqref{eq:yukrot} and the corresponding relation for $\bdl_{d}$.

Running then continues to the GUT scale with a basic RGE subroutine, \texttt{RGE215}, which only contains the RGEs necessary for running the gauge and Yukawa couplings, and implements only rudimentary thresholds. When we reach the scale $Q=m_{H}$ in the course of running up, we switch from SM Yukawa matrices ($\bdl_{u,d,e}$) to MSSM Yukawa matrices ($\bdf_{u,d,e}$) and from the SM VEV ($v_{SM}$) to the VEVs in the two Higgs doublet model:
\begin{equation*}
v_{SM}\equiv\sqrt{v^{2}_{u}+v^{2}_{d}}
\end{equation*}

The subroutine \texttt{HIGHIN} takes care of the boundary conditions at the high scale as discussed next.

\section{Boundary Conditions at the High Scale}\label{sec:gutins}

The running is deemed to have reached the GUT scale when $\alpha_{1}(Q)-\alpha_{2}(Q)$ becomes negative, unless the user responded to Question 6.~with a fixed $M_{\mathrm{HIGH}}$, in which case the running terminates at the value that was chosen in \inrge.

Since our purpose is to simulate flavour physics of sparticles in as general a way as possible, subject to experimental constraints that seem to suggest that that flavour physics is largely restricted by the structure of the Yukawa coupling matrices, we thought it would be useful to first seek a general parametrisation for SSB parameters that does not introduce a new source of flavour-violation, but allows for non-universality of model parameters. A different source of flavour-violation can easily be incorporated by allowing for additional, arbitrary contributions to the SSB mass and trilinear parameter matrices. We use the (s)quark sector to illustrate our arguments, but almost identical considerations will apply to (s)leptons (except that in this case we would also have to include lepton number and lepton flavour-violating matrices in the singlet (s)neutrino sector).

Within the framework of the $R$-parity conserving MSSM, the SSB matrices ${\bdm}^2_{U,D,Q}$ and ${\ba}_{u,d}$ potentially include new sources of flavour-violation, not included in the superpotential Yukawa couplings. In order not to introduce a new source of flavour-violation, these SSB matrices must be diagonal in the same superfield basis (where the SM fermions and their scalar superpartners are rotated by the same matrices) that the superpotential Yukawa interactions (renormalised at the same high scale as the SSB parameters) are diagonal.\footnote{This is {\it not} equivalent to the requirement that the Yukawa coupling matrix commute with the corresponding ${\ba}$-parameter matrix because these non-Hermitian matrices are diagonalised by bi-unitary and not by unitary transformations.} Of course, it is impossible to simultaneously diagonalize ${\bdf}_u$ and ${\bdf}_d$, but what we mean is that the SSB mass matrices that describe the mixing of both left- and right-handed up type squarks, and their trilinear couplings must be diagonal in the basis that the matrix ${\bdf}_u$ is diagonal, and likewise for the down sector. However, since $SU(2)$ symmetry dictates that the ${\bdm}^{2}_{\tilde{u}_L}$ and ${\bdm}^{2}_{\tilde{d}_L}$ SSB matrices must be identical, this matrix must be proportional to the unit matrix ($\dblone$) in order to remain diagonal, both when the up- or the down-type Yukawa coupling matrix is diagonal. In contrast, the matrices ${\bdm}_U^2$ and ${\ba}_u$ (${\bdm}_D^2$ and ${\ba}_d$) can be functions of the up (down) type Yukawa coupling matrices (and their Hermitian adjoints) chosen in such a way that these matrices are simultaneously diagonal when we transform these to the basis where the corresponding superpotential Yukawa coupling matrix is diagonal. 

To find the most general parametrisation of the ${\bdm}_{U,D}^2$ and ${\ba}_{u,d}$ matrices of the type that we are looking for, we first note from \eqref{eq:amix}-\eqref{eq:mudmix} that under field ``rotations'', these, respectively, transform in the same way as the matrices $\left({\bdf}^T_{u,d}{\bdf}^*_{u,d}\right)^n$ and ${\bdf}_{u,d}\left({\bdf}_{u,d}^\dagger {\bdf}_{u,d}\right)^n$, where $n$ is any integer. Thus any linear combination of these matrices (with $n=0,1,2,...$) is guaranteed to be diagonal in the basis that ${\bdf}_{u.d}$ is diagonal (at the high scale at which we enter the SSB parameters). The only question, then, is just how many terms we need to allow in the linear combination to guarantee the most general form for the SSB matrices, so that flavour-violation enters only through the superpotential Yukawa coupling matrices. This is easiest to see in the diagonal basis for the Yukawa couplings. The SSB matrices are also diagonal in this basis, and so are completely specified by $n_g$ diagonal elements, where $n_g$ is the number of generations. Transforming to a general basis does not alter the number of parameters that we need: we thus know that we must have $n_g$ terms in each of the linear combinations for ${\bdm}^2_{U,D}$ and for ${\ba}_{U,D}$ that we discussed above. For the MSSM with $n_g=3$ generations, we thus parametrize the SSB sfermion mass and ${\ba}$-parameter matrices at the high scale as,
\begin{subequations}\label{eq:GUTbound}
\begin{align}
\bdm^2_{Q,L}&=m^2_{\{Q,L\}0}\dblone+\mathbf{T}_{Q,L}\\
\bdm^2_{U,D,E}&=m^2_{\{U,D,E\}0}[c_{U,D,E}\dblone+R_{U,D,E}\bdf^T_{u,d,e}
\bdf^*_{u,d,e}+S_{U,D,E}(\bdf^T_{u,d,e}\bdf^*_{u,d,e})^2]+\mathbf{T}_{U,D,E}\label{eq:GUTboundsferm}\\
\ba_{u,d,e}&=\bdf_{u,d,e}[A_{\{u,d,e\}0}\dblone+W_{u,d,e}\bdf^\dagger_{u,d,e}\bdf_{u,d,e}+X_{u,d,e}(\bdf^\dagger_{u,d,e}\bdf_{u,d,e})^2]+\mathbf{Z}_{u,d,e}\;,
\label{eq:GUTboundtri}
\end{align}
\end{subequations}
where $c_{U,D,E}=0$ or $1$, and ${\bf f}_{u,d,e}$ are the superpotential Yukawa coupling matrices {\it in an arbitrary current basis}
at the same scale at which the SSB parameters of the model are specified. The matrices $\mathbf{T}_{Q,L,U,D,E}$ and $\mathbf{Z}_{u,d,e}$ have been introduced only to allow for additional sources of flavour-violation not contained in the Yukawa couplings. Setting $\mathbf{T}_{Q,L,U,D,E}=\mathbf{Z}_{u,d,e}=\bm{0}$ gives us the most general parametrisation of the three-generation $R$-parity conserving MSSM where the Yukawa coupling matrices are the sole source of flavour-violation.

Note that this collection of inputs (excluding the matrices $\mathbf{T}_{Q,L,U,D,E}$ and $\mathbf{Z}_{u,d,e}$) is a special case of the minimal flavour violation principle \cite{mfvorig1,*mfvorig2,*mfvsmall,*mfvfat} in that we assume the GUT scale SSB parameters do not introduce new sources of flavour-violation. Since our conditions are not invariant under renormalisation group evolution, the weak scale form of the SSB parameters will not be the same as in \eqref{eq:GUTbound}, but will, however, be described by the more general minimal flavour-violation ansatz.

Questions 9.~to~16. in \inrge~allow the user to choose arbitrary values of all the input coefficients above, subject to the constraint that $\mathbf{T}_{Q,L,U,D,E}$ are Hermitian. The familiar universal mSUGRA boundary conditions are reproduced by setting $c_{U,D,E}=1$; $m^{2}_{\{Q,L\}0}=m^{2}_{\{U,D,E\}0}=m^{2}_{0}$; $A_{\left\{u,d,e\right\}0}=A_{0}$; $R_{U,D,E}=S_{U,D,E}=W_{u,d,e}=X_{u,d,e}=0$; $\mathbf{T}_{Q,L}=\mathbf{T}_{U,D,E}=\mathbf{Z}_{u,d,e}=\bm{0}$ in (\ref{eq:GUTbound}). 

The remaining GUT scale inputs \mbox{---} namely, those for the gaugino and Higgs boson scalar masses \mbox{---} are simple numbers, which can be either given by the mSUGRA parameters passed from \progisa, or entered by the user in Questions. 7.~and~8. of \inrge.

\section{Electroweak Symmetry Breaking}

After fixing the high scale parameters, the programme proceeds to the subroutine \texttt{DOWNMHIGHMZ}, which runs the entire collection of RGEs contained in the subroutine \texttt{RGE646}. At each step, the code checks to see whether the scale is below $\msusy$ and, at the point $\msusy$ is passed, applies the electroweak breaking conditions using the subroutine \texttt{DOWNMSCOND}.

As discussed in Sec.~\ref{sec:MSSMtheoryewsb}, it is common practice to use the observed value of $M^{2}_{Z}$ to determine the value of $\mu^{2}$ via \eqref{eq:EWSBmu}, but our inclusion of threshold effects comes with additional complications in this case. Since $\msusy$ is at the scale $\sqrt{m_{\tilde{t}_{L}}m_{\tilde{t}_{R}}}$, the electroweak symmetry breaking conditions will always be applied at a scale smaller than the mass of the heaviest SUSY particle. As a result, $\mu$, the higgsino mass parameter, is no longer equal to $\tilde{\mu}$, the parameter that enters the Higgs potential. Moreover, the Higgs potential depends only on $M^{2}_{H_{u}}\equiv\left(\mhusq+\mtsq\right)$ and $M^{2}_{H_{d}}\equiv\left(\mhdsq+\mtsq\right)$, so that it is not possible to separate $\left|\tilde{\mu}\right|^2$ from the SSB parameters $m_{H_u}^2$ and $m_{H_d}^2$ that we specify at the high scale. Notice, however, that we can define the relations
\begin{gather*}
\left(M^{2}_{H_{u}}+M^{2}_{H_{d}}\right)\equiv m^{2}_{H_{u}}+m^{2}_{H_{d}}+2\left|\tilde{\mu}\right|^{2}\quad\mathrm{and}\\
\left(M^{2}_{H_{d}}-M^{2}_{H_{u}}\right)\equiv m^{2}_{H_{d}}-m^{2}_{H_{u}}\;.
\end{gather*}
so that the tree-level minimisation conditions of the Higgs potential can be written as,
\begin{gather}
b=\sn\cs\left(M^{2}_{H_{u}}+M^{2}_{H_{d}}\right)\label{eq:EWSB2}\\
\begin{split}
\left(M^{2}_{H_{u}}+M^{2}_{H_{d}}\right)=-\frac{1}{\cos{2\beta}}\left(M^{2}_{H_{d}}-M^{2}_{H_{u}}\right)-\frac{1}{2}\left(g'^{2}+g^{2}\right)\left(v^{2}_{u}+v^{2}_{d}\right)\label{eq:EWSB}\;.
\end{split}
\end{gather}
The second of these fixes the sum $(M^{2}_{H_{u}}+M^{2}_{H_{d}})$ in terms of the difference.  Since we know this difference at the high scale, we can evolve this down to $M_{\rm SUSY}$ (along with other SSB parameters) during  the iterative process that we use to solve the RGEs. At $Q=M_{\mathrm{SUSY}}$ we use (\ref{eq:EWSB}) to solve for $(M^{2}_{H_{u}}+M^{2}_{H_{d}})$, which can be evolved back to the high scale. We then use the sum to fix $\tilde{\mu}=\mu$ at the high scale, reset the difference, $(M^{2}_{H_{u}}+M^{2}_{H_{d}})$, to its input value, and iterate as per the discussion in Sec.~\ref{sec:iterate}. The value of the higgsino parameter $\mu$ can then be obtained at all scales using (\ref{app:rgemu}).

Finally, with a small adjustment to the usual relation in Eq.~\eqref{eq:EWSBb}, the $b$-parameter can be eliminated in favour of $\tan{\beta}$ using \eqref{eq:EWSB2}. Although the $b$-parameter is complex in general, our decision to make $v_{u}$ and $v_{d}$ real and positive\footnote{We can always make a gauge transformation such that just the lower component of $H_u$ has a VEV, and that this VEV is real and positive. Then the minimization of the scalar potential in the Higgs sector requires that the VEV of $H_d$ is aligned; \textit{i.e.}, is also in its lower component. This alignment is a result of dynamics. Finally, we can redefine the phase of the doublet superfield ${\hat{H}}_d$ so that $v_d$ is real and positive. This is not compulsory, but is the customary practice that allows us to define $\tan\beta$ to be real and positive.} requires that $b$ also be real and positive at the scale $\msusy$. Our parameter does, however, retain the ability to develop complex parts as a result of the running, and will not necessarily remain real at all scales.

Up to this point, we have ignored another potential complication that arises if $\msusy<m_{H}$. In this case, the heavy particles of the Higgs sector have decoupled by the time we apply the electroweak symmetry breaking conditions, and we only have the light doublet in the effective theory that we use to calculate the RGEs. In this case, the heavy Higgs doublet mass term $\left[\cs^{2}\left(m^{2}_{H_{u}}+\left|\tilde{\mu}\right|^{2}\right)+\sn^{2}\left(m^{2}_{H_{d}}+\left|\tilde{\mu}\right|^{2}\right)+\sn\cs\left(b+b^{*}\right)\right]$ and the mixing terms, $\left[\sn\cs\left(m^{2}_{H_{u}}+\left|\tilde{\mu}\right|^{2}\right)-\sn\cs\left(m^{2}_{H_{d}}+\left|\tilde{\mu}\right|^{2}\right)+\sn^{2}b-\cs^{2}b^{*}\right]$ (and its complex conjugate) from Eq.~\eqref{eq:higgsterms}, together with $\tan\beta$, are frozen at their values at $Q=m_H$, while the light doublet mass parameter, $\left[\sn^{2}\left(m^{2}_{H_{u}}+\left|\tilde{\mu}\right|^{2}\right)+\cs^{2}\left(m^{2}_{H_{d}}+\left|\tilde{\mu}\right|^{2}\right)-\sn\cs\left(b+b^{*}\right)\right]$,  along with $v_{\rm SM}$, continue to evolve to $M_{\rm SUSY}$. The three frozen coefficients together with the evolved mass term for the light doublet must therefore be used to solve for $\left(m^{2}_{H_{d}}+\left|\tilde{\mu}\right|^{2}\right)$, $\left(m^{2}_{H_{u}}+\left|\tilde{\mu}\right|^{2}\right)$ and the complex $b$-parameter. We can then find a solution in the same manner as for $\msusy>m_H$.

Before closing this section, we should add that although we have discussed EWSB\glossary{name={EWSB},description={Electroweak Symmetry Breaking}} conditions only at tree-level, in practice we minimize the one-loop effective potential including effects of third generation Yukawa couplings, but ignoring all flavour-mixing effects. These corrections, which effectively shift the Higgs boson SSB mass squared parameters by $\Sigma_u$ and $\Sigma_d$, respectively, are evaluated by replacing $f_{t,b,\tau}$ in the standard relations by the (3,3) element of the corresponding Yukawa matrices in the basis where they are diagonal at $m_{t}$, and with the dimensionful parameters also replaced by the (3,3) element of the corresponding matrix (or the appropriate frozen value).

\section{Iterative Stage}\label{sec:iterate}

Now that the boundary conditions are defined at each of the three relevant scales, the iteration can begin. The subroutines \texttt{DOWNMHIGHMZ} and \texttt{UPMZMHIGH2} implement the running in each direction. The iteration takes place a fixed number of times, chosen so that the RGEs reach a stable solution.

Unless the user has answered `\verb+1+' to using fixed thresholds from \progisa~in Question 4.~of \inrge, the programme will alter all the thresholds on each downwards run, except those for the gluinos and heavy Higgs fields, which are fixed at the locations obtained by \progisa.

The running is carried out as follows:

\subsubsection{Downwards running}

The subroutine first runs from the GUT scale to the highest threshold with the number of steps given by the variable \texttt{NSTEP}. During the $i$th iteration, this variable is has the value $100\times(1.6)^{i}$, until we reach iteration number $5$ at which point it becomes fixed at $100\times(1.6)^{5}$. If this is the first iteration, the thresholds are taken to be the \progisa~thresholds. Running continues between thresholds (inserting the boundary conditions at $m_{H}$ and $\msusy$ when necessary) with the number of steps given by
\begin{equation}
\mathrm{Number\ of\ Steps}=\frac{\left|\log{\left(Q_{1}/Q_{2}\right)}\right|}{\log{(M_{HIGH}/m_{t})}}\times(25\times\mathtt{NSTEP})\;,
\end{equation}
where $Q_{1}$ and $Q_{2}$ are the scales of the two thresholds between which the running is being carried out. The factor $25$ was chosen to ensure enough sampling between the thresholds without unnecessarily slowing down processing time.

At each step, the SSB sfermion mass matrices are diagonalised. If the user has not fixed the thresholds to be the same as those obtained from \progisa, the derived eigenvalues are checked to see if any matter sfermions have decoupled. When one of these sparticles does decouple, the subroutine \texttt{CHDEC} carries out the following procedure:
\begin{enumerate}
\item Set the corresponding $\theta_{\tilde{f}_{i}}$ to zero.
\item Store the eigenvectors of the mass matrix so that the rotation between the diagonal basis at the decoupling scale and the original current basis is saved.
\item Store all the entries of the mass matrix itself so that they can be used as a boundary condition when running up.
\item Call the subroutine \texttt{REMSF} so that, in the basis where the mass matrix is diagonal, the entry corresponding to the decoupled particle is set to zero. This ensures that the eigenvector for this particle is removed from the original current basis matrix and cannot influence further downward running.
\end{enumerate}
The subroutine continues to run down until it reaches $m_{t}$, where it rotates from the current basis back to the basis in which the Yukawa matrices are diagonal.\footnote{Rather than diagonalise the Yukawa matrices at this scale, \progrge~uses the rotation matrices, $\mathbf{V}_{L,R}(u,d)$ that were used on the first upwards running. Practically, this means that the rotation matrices are one of the boundary conditions on our iterative running.} Running resumes using the SM \drb~RGEs in \texttt{SMRGEDR}, \textit{without decoupling the top quark}, to the scale $M_{Z}$.

The integration of the RGEs is carried out in the same way as \progisa, \textit{i.e.}, by the \texttt{CERNLIB} routine \texttt{RKSTP}. The RGE subroutine, \texttt{RGE646}, contains RGEs for all couplings with and without tildes and with full thresholds for the one-loop running. The quartic couplings are entered separately, but they are set to be equal to their SM counterparts since the RGEs for the quartics are unavailable at this time. In addition, \texttt{RGE646} contains the two-loop terms from the RGEs, which depend only on the MSSM values of the couplings. In order to obtain an estimate of the two-loop contributions, the pure MSSM RGEs are solved even below all thresholds and these MSSM couplings are used for the two-loop level running of the SUSY couplings. This is acceptable since we are only trying to achieve two-loop level accuracy, and threshold effects in the two-loop terms are numerically much smaller.

Once we have decoupled at least one of the matter sfermions, the right-hand side of the RGEs are calculated in the basis where the squark/slepton matrices are diagonal at the decoupling scale. This is still a current basis, since the quarks and leptons are rotated by the same amount as the squarks and sleptons. \texttt{RGE646} rotates all couplings into this basis when calculating the right-hand side of the RGEs and, at the end of the subroutine, rotates the result back to the original current basis.

Note that if the location of the thresholds is altered every iteration, the Yukawa couplings are unable to reach a convergent solution. This is because moving the thresholds can disrupt the fine cancellation that is required to obtain vanishing values at $m_{t}$ for the off-diagonal elements of the Yukawa matrices in their mass basis. We therefore only allow \progrge~to change the locations of the thresholds for the first five iterations. This ensures that the Yukawa couplings converge as closely as allowed by the numerical accuracy of the machine.

\subsubsection{Upwards running}

Before commencing the run back up to the GUT scale, the boundary conditions at $M_{Z}$ \mbox{---} the gauge couplings and Yukawa coupling matrices in the quark mass basis \mbox{---} are reset. Running then continues to $m_{t}$, where the top quark Yukawa coupling is reset and the Yukawa matrices are rotated into the current basis, and then continues again until the first threshold above $m_{t}$ is reached.

The upwards running makes no changes to the thresholds. At each step, the subroutine checks whether a threshold has been passed. If so, the $\theta$ for the threshold in question is set to $1$, and if this is a matter sfermion threshold the soft mass matrix is set to the value which was saved at this point during the downward run.

The RGE subroutine \texttt{RGE646} is used just as with downwards running. When we have some but not all matter sfermions present in the theory, we rotate to the basis where the soft mass matrix for the sfermions in question is diagonal at the scale of decoupling. We know what this rotation is since it too was saved in the previous run down.

Once all the thresholds have been passed, the RGEs are equivalent to the standard MSSM RGEs and running continues in a straightforward manner until the high scale is reached. Residual inaccuracies in the running mean that the tilde-couplings are not precisely equal to their non-tilde counterparts once we have passed the highest SUSY threshold. Since these differences can feed back into the other couplings via the RGEs, we set the tilde-couplings (except $\tilde{\mu}$) equal to the usual SUSY couplings once the highest SUSY threshold has been passed. Also, if the answer to Question 6. is `\verb+0+', since the scale at which the gauge couplings unify may be altered by the location of the thresholds, we allow the running to continue past the unification scale from the previous iteration, and increase the number of steps for this iteration so that the step-size remains constant.

We have checked that if we do not reset the weak scale boundary conditions, and instead use the final values from the previous call to \texttt{DOWNMHIGHMZ}, the upwards running is precisely the same as the downwards running.

\section{Technical Aside: Numerical Accuracy of the Rotation}

Before concluding this chapter with a discussion of the output, we digress momentarily to draw attention to a technical issue which is important if the squark thresholds are not entirely degenerate. As discussed earlier, for values of $Q$ below the highest sfermion threshold we need to rotate at each step to the basis where the SSB squark mass squared matrices are diagonal, which, in turn, requires us to obtain the unitary matrix $\mathbf{R}_{\bullet}$ that relates our ``standard'' basis to this ``squark mass basis''. After evaluating the right-hand side of the RGEs, we always rotate back to our ``standard'' basis using $\mathbf{R}^{\dagger}_{\bullet}$. While this is straightforward in principle, the practical problem is that when two squark eigenvalues become very close \mbox{---} this is always the case in mSUGRA \mbox{---} $\mathbf{R}_{\bullet}\mathbf{R}^{\dagger}_{\bullet}$ has non-zero off-diagonal elements due to numerical noise at the $10^{-10}$ level. This ruins the delicate cancellations necessary to obtain the tiny magnitudes of some of the off-diagonal elements of the squark mass matrices that we will see in the next chapter. The subroutine \texttt{ORTH} is called in several places to fix this problem. A further discussion of this effect and the procedure we use to mitigate against the numerical instability can be found in Appendix~\ref{sec:orthfix}.

\section{\progrge~Output}

The iterative section exits at the high scale after resetting the GUT scale boundary conditions. In order to provide useful output, the code makes one final downward run to $m_{H}$. This scale was chosen due to its significance as a point where a number of the operators in the Lagrangian change, however, the output scale could have been chosen to be anywhere between the two extremes of the running. It is expected that $m_{H}$ will be fairly close to the scale at which the user will be using the couplings for their calculations.

Before passing on the output to an external routine, all couplings are rotated to the current basis chosen by the user in Question 17.~of \inrge, where either the up- or the down-type Yukawa couplings are diagonal at $m_{t}$. For any specific calculation the user can then simply evolve these couplings to higher or lower scales as desired, without the need to iterate. We will use the output, in the basis where the up-type Yukawa coupling is diagonal, in the subroutine \progstd~described in Sec.~\ref{sec:std}, to calculate the $\tilde{t}_{1}$ decay rate.

%% file: solution.tex
\chapter{Sample Solutions to the RGEs}\label{ch:results}

Having constructed the RGEs for all dimensionless and dimensionful parameters of the MSSM (except the scalar quartic couplings), and developed a program to solve these RGEs subject to SM and GUT scale boundary conditions, we are now in a position to consider their solutions.

We will focus on mSUGRA as a simple model for GUT scale inputs, and discuss the consequences of splitting between SUSY couplings and their SM counterparts. In addition, since our eventual goal is to study flavour-changing decays of squarks, we will consider the flavour structure of the couplings, and how certain aspects of our general GUT scale inputs in \eqref{eq:GUTbound} can affect the off-diagonal elements of the various couplings renormalised at the scale of the sparticle mass, \textit{i.e.}, at scales $Q$ typically between $\sim100$~GeV and a few TeV.

To simplify the discussion further, we begin by reintroducing our basic model of decoupling from Sec.~\ref{sec:broken}, in which there are only two thresholds. We assume in this simple model that the sfermions and the electroweak gauginos are all at a mass scale $\sim600$~GeV while the heavy Higgs bosons and gluinos have a mass $\sim2$~TeV. In other words, our theory is supersymmetric for $Q>2$~TeV, contains only sfermions, charginos and neutralinos together with SM particles for $600~\gev<Q<2$~TeV, and is the SM for $Q<600$~GeV. We use this model to examine the gauge and gaugino couplings before turning to quark Yukawa and Yukawa-like higgsino couplings where we will eventually consider more realistic scenarios.

\section{Gauge and Gaugino Couplings}

The RGEs of the gauge couplings, as in \eqref{eq:g1bet}-\eqref{eq:g3bet}, are flavour diagonal under renormalisation group evolution, even at higher loops. Although the Yukawa couplings enter at two-loop level, they only appear as `flavour-blind' traces. As a result, notwithstanding that the gauge couplings are affected by the location of the thresholds via the factors $N_{\mathcal{P}}$ and $\theta_{\mathcal{P}}$, these effects cannot contribute to additional flavour-violation.

On the other hand, we saw in Sec.~\ref{sec:broken} that the form of the RGEs for the (Yukawa-like) squark-quark-gaugino couplings  means that they develop a non-trivial matrix structure once the heaviest SUSY particle has been decoupled. Although these couplings \mbox{---} and similarly the higgsino couplings \mbox{---} cease to have any meaning below the lower threshold, there is still the opportunity in the intermediate regime for these couplings to develop significantly different values from their SM counterparts. Contributions to the $\tilde{\mathbf{g}}^{\Phi}$ running from terms involving Higgs boson and higgsino couplings, which cancel above all thresholds, result in non-zero off-diagonal elements when $Q<m_{H}$.

\begin{figure} \centering
\includegraphics[viewport=20 50 725 525, clip, scale=0.45]{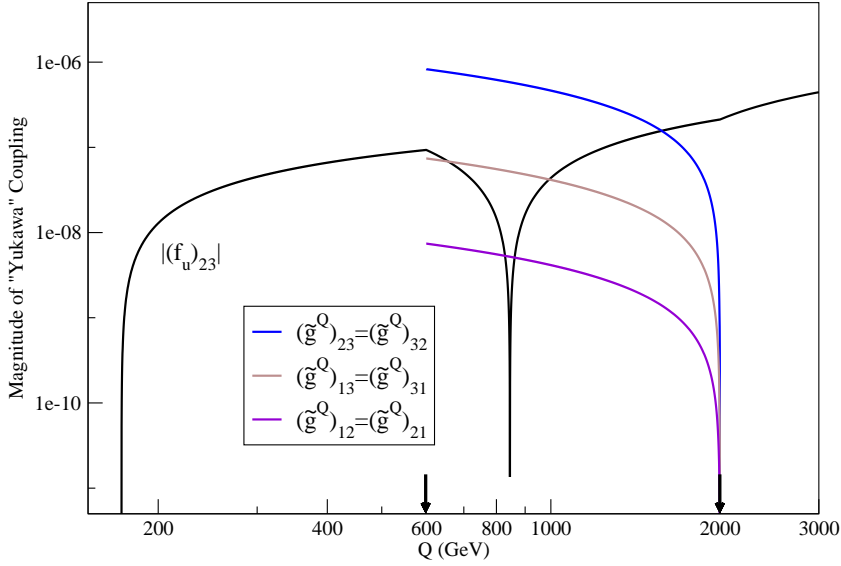} 
\caption[The evolution of the magnitudes of the off-diagonal elements of $\bgt^{Q}$ for our simplified threshold scenario.]{\small The evolution of the magnitudes of the off-diagonal elements of $\bgt^{Q}$ for our simplified threshold scenario where the thresholds (denoted by arrows) are clustered at $2\ $TeV and $600\ $GeV. Also shown for comparison is the running of the $(2,3)$ element of the up Yukawa coupling. The legend is in the same order (from top to bottom) as the curves.} \label{fig:gtoffdiag}
\end{figure}

As a supplement to Fig.~\ref{fig:gtpoffdiag}, where we showed the off-diagonal elements of the $U(1)$ gaugino couplings $\tilde{\mathbf{g}}^{\prime Q}$ and $\tilde{\mathbf{g}}^{\prime u_{R}}$, we display in Fig.~\ref{fig:gtoffdiag} the off-diagonal elements of $\tilde{\mathbf{g}}^{Q}$ (of course, the coupling $\tilde{\mathbf{g}}^{u_{R}}$ does not exist).
To construct these curves, we have solved the RGE in \eqref{app:bgtq} along with the rest of the gauge and Yukawa couplings, which form a closed set. As in Sec.~\ref{sec:broken}, the boundary condition on $\tilde{\mathbf{g}}^{Q}$ is
\begin{equation*}
\bgt^{\prime \Phi}_{ij}(Q=2\ \mathrm{TeV})=g'\delta_{ij}\;,
\end{equation*}
and the curves terminate at $600$~GeV when the squarks and gauginos decouple from the theory. Once again, we show the $(2,3)$ off-diagonal element of $\bdf_{u}$ for comparison.

We note the following features:
\begin{itemise}
\item The variation of the vertical scale is over a much smaller range than Fig.~\ref{fig:gtpoffdiag}. We see that the approximate scale of the elements of $\tilde{\mathbf{g}}^{Q}$ is the same as for $\tilde{\mathbf{g}}^{\prime Q}$ and therefore, as with the $U(1)$ coupling, the elements are of similar size to the largest of the off-diagonal Yukawa couplings. In models where the Yukawa couplings are the only source of flavour-violation, we would conclude that a study of flavour should necessarily include the matrix structure of the gaugino couplings.
\item The overall size of the left-squark gaugino couplings are several orders of magnitude larger than those of $\tilde{\mathbf{g}}^{\prime u_{R}}$. We see from \eqref{app:bgtpq}, \eqref{app:bgtpur} and \eqref{app:bgtq}, that the left-handed gaugino coupling RGEs depend on both the up-type and down-type Yukawa couplings, whereas the right-handed RGE depends only on the up-type Yukawa couplings. Since the scale is set by the Yukawa matrices, which (unlike the gaugino coupling matrices) have non-zero flavour structure at $Q=m_{H}$, we find that in our ``standard'' current basis (which was defined in Sec.~\ref{sec:basischoice}) the down-type Yukawa terms have larger contributions to the running.
\item Finally, as pointed out in Sec.~\ref{sec:broken}, the magnitudes of the off-diagonal elements of the gaugino coupling matrices are (almost) symmetric under interchange of the two indices. If we ignore the difference between the Higgs boson and higgsino coupling matrices, to be discussed next, we find that the right-hand side of the RGEs preserves the approximate hermiticity of the gaugino coupling matrices, which are proportional to the unit matrix at $Q=m_{H}$.
\end{itemise}

Before moving on to the Higgs boson and higgsino couplings, we briefly mention the diagonal entries of the gaugino couplings. Of course, these are equal to the usual gauge couplings above $Q=m_{H}$, but, as with the off-diagonal elements, can evolve differently below the scale of the highest threshold. Indeed, we find that their evolution below $Q=m_{H}$ in our simplified scenario is in the \textit{opposite direction} to the gauge couplings \cite{RGE1} because, for $Q<m_{\tilde{g}}$, the evolution of the diagonal gaugino couplings now depends on the much larger gluon coupling, even at the one loop level. When the gluino decouples from the theory, we can see from, for example, Eq.~\eqref{app:bgtpq}, that a cancellation which was taking place between $\tilde{\mathbf{g}}^{Q}_{s}$ and $g_{s}$ terms no longer occurs, causing a large change in the $\beta$-function as $\sgl\rightarrow0$. We note that, as can be seen in Fig.~4. of Ref.~\cite{RGE1}, although the splitting between the thresholds in this scenario is not large, the gaugino and gauge couplings develop a difference of $\sim4\%$. The existence of a difference in these couplings has been discussed in Refs.~\cite{cheng,*nojiri,feng,*nojiri2} where the possibility for this splitting to give an insight into the sparticle mass spectrum is explored.
 
\section{Quark Yukawa and Higgsino Couplings}\label{sec:Yukrun}

In this section we consider both the usual quark Yukawa couplings $\bdf_{u,d,e}$ and the higgsino Yukawa-like couplings, $\tilde{\bdf}^{\Phi}_{u,d,e}$. As an introduction to the more realistic scenarios to follow, we begin by exploring further our simplified threshold model.

\begin{figure}[t]\begin{centering}
\includegraphics[viewport=25 50 710 525, clip, scale=0.45]{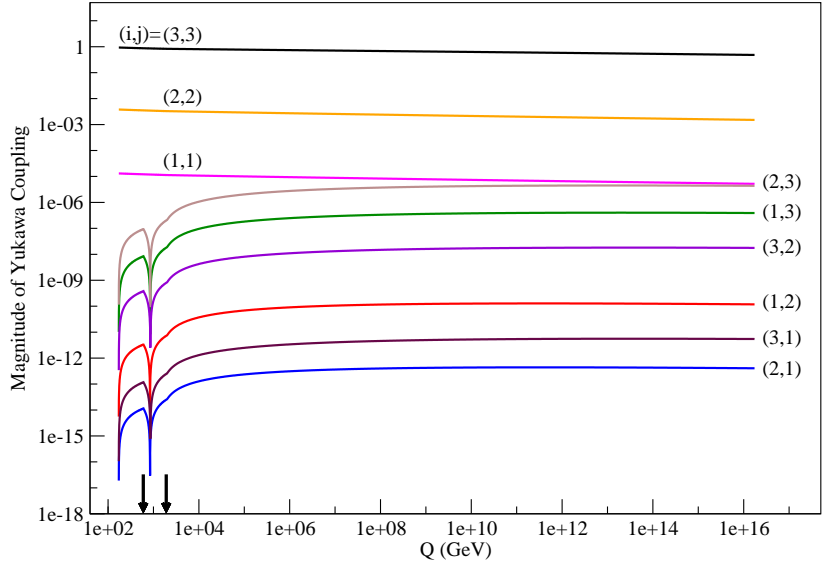} 
\caption[Evolution of the magnitudes of the elements of the up-quark Yukawa coupling matrix, $\bdf_{u}$, for our simplified threshold scenario.]{\small Evolution of the magnitudes of the elements of the up-quark Yukawa coupling matrix, $\bdf_{u}$, for the MSSM with thresholds, shown by arrows, clustered at $600$~GeV and $2$~TeV as discussed in the text. Above $m_{H}$ $(=2\ \mathrm{TeV})$ we plot $\left|(\bdf_{u})_{ij}\right|$ whereas below $m_{H}$, where the effective theory includes just one scalar Higgs doublet, we plot $\left|(\bm{\lambda}_{u})_{ij}\right|/\sin{\beta}$ which is equal to $\left|(\bdf_{u})_{ij}\right|$ at $Q=m_{H}=2$~TeV.}
\label{fig:absf}
\end{centering}
\end{figure}

The running of the elements of $\bdf_{u}$ is shown in Fig.~\ref{fig:absf}.
In order that all entries of the Yukawa matrix be displayed for the whole range of the running, we plot $\left|\left(\bdf_{u}\right)_{ij}\right|$ for $Q>m_{H}$ and $\left|\left(\bdl_{u}\right)_{ij}\right|/\sin{\beta}$ for $Q<m_{H}$. These two lines join continuously at $Q=m_{H}$ and, since we are plotting this figure in the basis where the up quark Yukawa couplings are diagonal at $m_{t}$, the off-diagonal entries are seen to go to zero at this scale.

The most striking feature of Fig.~\ref{fig:absf}, which we have seen before in Figs.~\ref{fig:gtpoffdiag} and~\ref{fig:gtoffdiag} but not specifically addressed, is the presence, in the off-diagonal entries, of a sharp dip in the range $Q\simeq850$~GeV, which similarly appears in the higgsino couplings. This feature is particularly important for our purposes since (in a more realistic scenario) it may appear close to the sale at which we decouple the squarks and directly enter the amplitudes for flavour-violating squark decays that will be calculated at this scale.

A detailed discussion of this feature can be found in Ref.~\cite{RGE1}. We simply note here that the off-diagonal up-type Yukawa couplings in our ``standard'' current basis of Sec.~\ref{sec:basischoice} obtain their overall scale from the `seed-terms' in the RGEs that depend on $\bdf_{d}$ or $\tilde{\bdf}^{\Phi}_{d}$. In contrast to the up-type Yukawa terms, for which all off-diagonal entries are small, these terms contain the KM matrix, whose presence allows for non-zero off-diagonal contributions. Given that the contribution to the RGE from these $\bdf_{d}$ terms is dominant, we find that the $\beta$-function undergoes a sign change in morphing from pure SM to pure MSSM running. This sign change ensures that the coupling in question alters direction and heads back towards zero, which was its starting value. Due to the fact that \mbox{---} as is the case in Fig.~\ref{fig:absf} \mbox{---} the real and imaginary parts of each specific off-diagonal coupling both head towards zero simultaneously, the magnitude of the Yukawa matrix entry will vanish at some specified point, producing a pronounced dip since we are plotting the logarithm of this magnitude. We have shown that in a more simplified threshold scenario, where all SUSY thresholds are located at a single point, the dips for all the off-diagonal entries occur at a single scale \cite{RGE1}, but the blurred out location of the zeros is a result of our using a more realistic spectrum. 

\begin{figure}[t]\begin{centering}
\includegraphics[viewport=20 50 710 525, clip, scale=0.45]{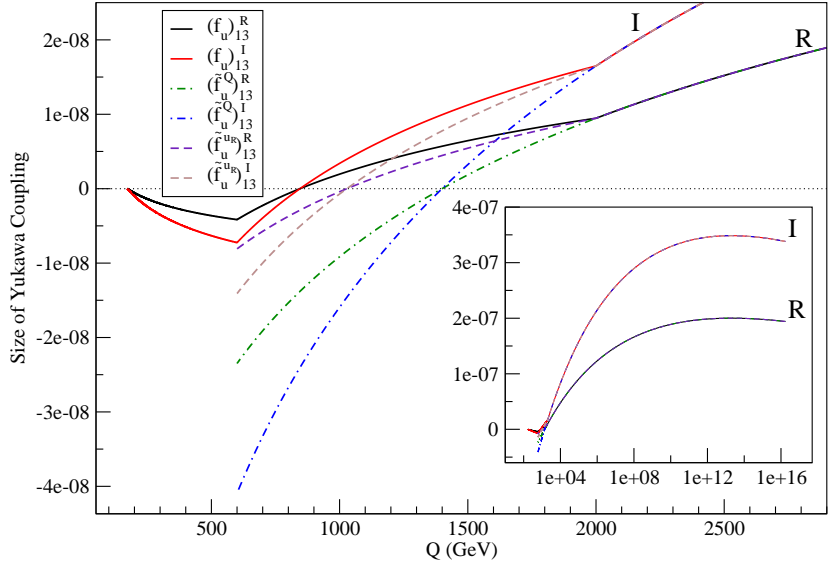} 
\caption[The real and imaginary part of the $(1,3)$ element of the up-quark Yukawa coupling matrix, $\bdf_{u}$, along with the corresponding elements of the matrices $\bftuqnb$ and $\bfturnb$ for the MSSM with the spectrum as in Fig.~\ref{fig:gtpoffdiag}.]{\small The real (R) and imaginary (I) part of the $(1,3)$ element of the up-quark Yukawa coupling matrix, $\bdf_{u}$, along with the corresponding elements of the matrices $\bftuqnb$ and $\bfturnb$ for the MSSM with the spectrum as in Fig.~\ref{fig:gtpoffdiag}. For the solid black and red lines, we plot the elements of the matrix $\bm{\lambda}_{u}/\sin{\beta}$ below $Q=m_{H}$, as in Fig.~\ref{fig:absf}. The main figure zooms in on the low end of the range of $Q$ where the Higgs boson and higgsino couplings differ from one another, while the inset shows the evolution all the way to $M_{\rm GUT}$.}
\label{fig:higgsino}
\end{centering}
\end{figure}

Having considered the running of the usual Yukawa couplings, we now turn to consider the off-diagonal elements of the $\tilde{\mathbf{f}}^{\Phi}_{u}$ matrices. We know that these will deviate from the elements of the corresponding Yukawa coupling matrix only below the thresholds at $Q=m_{H}$, but have magnitudes similar to the corresponding Yukawa coupling matrix elements from Fig.~\ref{fig:absf}. It is the \textit{difference} between the couplings of Higgs bosons and higgsinos that will be the main focus of our attention. In Fig.~\ref{fig:higgsino}, we show the evolution of the real and imaginary parts of the (1,3) element of (\textit{i.})~the Yukawa coupling matrix, $\bdf_u$ for $Q>m_H$ and $\bm{\lambda}_u/\sin\beta$ (which connects continuously to $\bdf_u$) for $Q<m_H$, (\textit{ii.})~the higgsino coupling matrices $\bftuqnb$, and $\bfturnb$ whose evolution is given in (\ref{app:bftuq}) and (\ref{app:bftur}), respectively, of Appendix~\ref{app:dlessRGEs}.
There is no particular reason for our choice of the $(1,3)$ element (which happens to have comparable real and imaginary pieces) for the illustration in the figure. Here we have focussed on the lower end of $Q$ where the $\mathbf{f}_{u}$ and $\tilde{\textbf{f}}^{\Phi}_{u}$ couplings are different, while the inset shows the evolution all the way to $M_{\rm GUT}$. Several points may be worthy of notice.
\begin{itemize}
\item For $Q>m_H$ where the effective theory is supersymmetric, we see that the real and imaginary parts of the Higgs boson and higgsino couplings separately come together as expected.
\item For $Q<m_H$, the higgsino couplings are split from the corresponding Higgs boson couplings as well as from one another by a  factor of several. For instance at the scale of squark masses, the real parts of the (1,3) element of both $\bftuqnb$ and $\bfturnb$ are quite different from the real parts of $({\bdf_u})_{13}$, and likewise for the imaginary parts. It seems to us that the use of the evolved Higgs boson coupling in place of the corresponding higgsino coupling could be a poor approximation.
\item Notice that while the real and imaginary parts of $\bftuq_{13}$ and $\bftur_{13}$ come to zero at the same point, the position of the zero differs for the two couplings. If the magnitude of the higgsino couplings were plotted on a logarithmic scale, as for the usual Yukawa couplings in Fig.~\ref{fig:absf}, we would see dips in their values at $\sim1400$~GeV and $\sim1000$~GeV respectively.
\end{itemize}

The variation of the diagonal entries of the usual Yukawa matrices and their higgsino counterparts can be found in Fig.~5. of Ref.~\cite{RGE1}. We find a large difference between the calculation of the complete RGEs in our simple threshold scenario compared to the two-loop calculation without thresholds. In addition, we see that the higgsino couplings can evolve to significantly different values from the usual Yukawa couplings over the range $600~\gev<Q<2$~TeV. As with the diagonal entries of the gauge and gaugino couplings, this difference is due to the appearance in the RGE of a dependence on the strong couplings when certain cancellations between SUSY and SM terms are no longer complete.

\subsection{mSUGRA}

Having considered the qualitative aspects of introducing thresholds into the dimensionless RGEs for the MSSM, we move on to continue some more specific models, where GUT scale parameters dictate the overall spectrum. Since the RGEs for the gauge and Yukawa couplings depend on the dimensionful parameters through the thresholds, we must use more realistic inputs constructed from the relations in Sec.~\ref{sec:gutins}.

\begin{figure}\begin{centering}
\includegraphics[viewport=20 45 710 560, clip, scale=0.45]{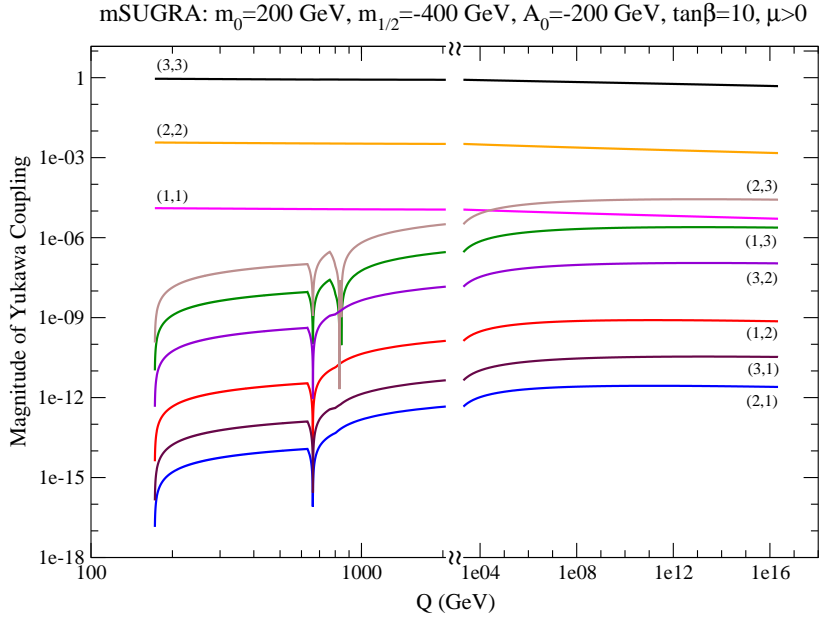} 
\caption[Evolution of the magnitudes of the complex elements of the up-quark Yukawa coupling matrix in the mSUGRA model]{\small Evolution of the magnitudes of the complex elements of the up-quark Yukawa coupling matrix in the mSUGRA model with $m_0=200$~GeV, $m_{1/2}=-400$~GeV, $A_0=-200$~GeV, $\tan\beta=10$ and $\mu >0$ in the basis where this matrix is diagonal at $Q=m_t$.  For $Q>m_{H}$ we plot $\left|(\bdf_{u})_{ij}\right|$ whereas for $Q<m_{H}$, we plot $\left|(\bm{\lambda}_{u})_{ij}\right|/\sin{\beta}$ which is equal to $\left|(\bdf_{u})_{ij}\right|$ at $Q=m_{H}$. In all the figures we take $m_{t}=172$~GeV.}
\label{fig:SUGRAfu}
\end{centering}
\end{figure}

To this end we choose an illustrative mSUGRA point with $m_0=200$~GeV, $m_{1/2}=-400$~GeV, $A_0=-200$~GeV, $\tan\beta=10$ and $\mu >0$ for which the particle spectrum is as in Table.~\ref{tab:SUGRAthresh}
\begin{table}
\centering
\begin{tabular*}{.7\textwidth}{@{\extracolsep{\fill}}ll}
$M_{\rm SUSY}$ & 703 GeV\\
Higginos ($\mu$)&538 GeV\\
Gluinos ($m_{\tilde{g}}$)&941 GeV\\
$\mathcal{H}$, $H^{\pm}$ ($m_{H}$)&631 GeV\\
Bino ($|M_{1}|$)&166 GeV\\
Winos ($|M_{2}|$)&315 GeV\\
$(\tilde{u}_{L},\tilde{d}_{L}),\;(\tilde{c}_{L},\tilde{s}_{L}),\;(\tilde{t}_{L},\tilde{b}_{L})$&837
GeV, 837 GeV, 763 GeV\\
$\tilde{u}_{R},\;\tilde{c}_{R},\;\tilde{t}_{R}$&809 GeV, 809 GeV, 645
GeV\\
$\tilde{d}_{R},\;\tilde{s}_{R},\;\tilde{b}_{R}$&806 GeV, 806 GeV,
801 GeV\\
$(\tilde{\nu}_{eL},\tilde{e}_{L}),\;(\tilde{\nu}_{\mu
L},\tilde{\mu}_{L}),\;(\tilde{\nu}_{\tau L},\tilde{\tau}_{L})$&331 GeV,
331 GeV, 329 GeV\\
$\tilde{e}_{R},\;\tilde{\mu}_{R},\;\tilde{\tau}_{R}$&249 GeV, 249 GeV,
245 GeV\\
\end{tabular*}
\caption[The approximate location of the thresholds for the mSUGRA case in Fig.~\ref{fig:SUGRAfu} and most subsequent mSUGRA examples.]{\small The approximate location of the thresholds for the mSUGRA case in Fig.~\ref{fig:SUGRAfu} and most subsequent mSUGRA examples.}
\label{tab:SUGRAthresh}
\end{table}
and begin by again showing the magnitude of the up Yukawa coupling matrix\footnote{We have broken the scale at $Q\sim2$~TeV to more clearly show features of the running at small $Q$. Note that this gives the impression of a change in slope of all the curves between the high $Q$ and low $Q$ regime, but this is just an artifact of the dramatic change in scale.}, in Fig.~\ref{fig:SUGRAfu}.
In a similar manner to Yukawa matrix in the simplified spectrum (Fig.~\ref{fig:absf}), there are (approximately) aligned dips at $Q\sim650$~GeV, common to all the off-diagonal elements, that occur because of the change in the sign of the coefficient of the $\bdf_{d}\bdf^{\dagger}_{d}$-type terms that drive the growth of these off-diagonal elements from zero at $Q=m_{t}$. The presence of the second zero at the higher value of $Q$, just in the (2,3) and (1,3) elements, is accidental. It occurs because of conspiracies between terms in the corresponding $\beta$-function as the left-type squarks are decoupled. Notice that the lowest four curves, though they do not have this additional dip, show kinks at these same values of $Q$, corresponding to the decoupling of these squarks. For several other mSUGRA cases, we have checked that while squark decoupling causes kinks in the curves, the coupling does not drop to even close to zero for a second time, in contrast to the behaviour in our illustrative example in the figure.

\begin{figure}
\begin{centering}
\includegraphics[viewport=15 45 710 560, clip, scale=0.45]{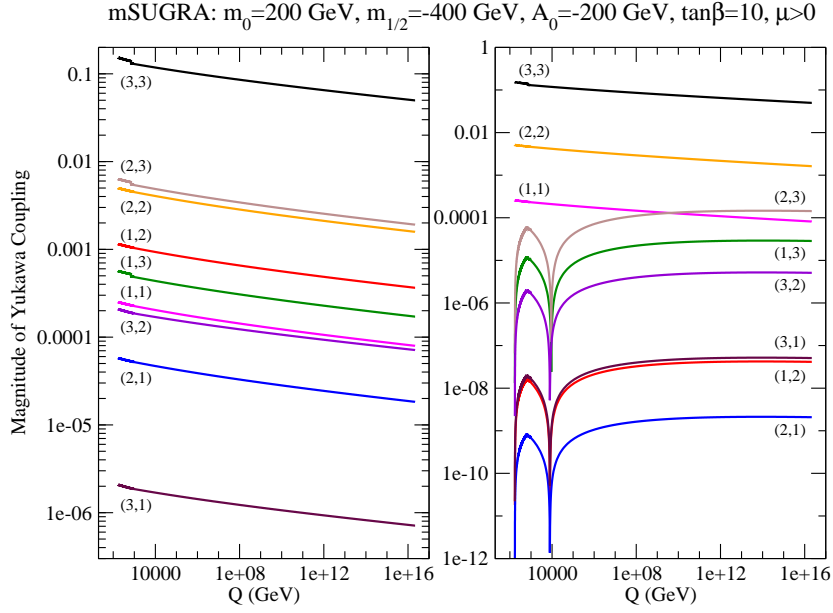} 
\caption[Evolution of the magnitudes of the complex elements of the down-quark Yukawa coupling matrix in the same mSUGRA model as Fig.~\ref{fig:SUGRAfu}.]{\small Evolution of the magnitudes of the complex elements of the down-quark Yukawa coupling matrix in the same mSUGRA model as Fig.~\ref{fig:SUGRAfu}.  In the left-frame, we show the elements of the matrix in the ``standard'' basis, whereas in the right frame, we show these elements in the basis where the $\bm{\lambda}_d$ is diagonal at $Q=m_t$. As with Fig.~\ref{fig:SUGRAfu}, for $Q>m_{H}$ we plot $\left|(\bdf_{d})_{ij}\right|$ whereas for $Q<m_{H}$ we plot $\left|(\bm{\lambda}_{d})_{ij}\right|/\cos{\beta}$.}
\label{fig:SUGRAfd} \end{centering} \end{figure}

The evolution of the down-quark Yukawa coupling matrix for the same  mSUGRA point is shown in Fig.~\ref{fig:SUGRAfd}.  
We have shown these couplings both in our ``standard'' current basis from Sec.~\ref{sec:basischoice} (where the up-type Yukawa couplings are diagonal at $Q=m_t$) in the left frame, and in the basis where the down-type Yukawa coupling matrix is diagonal at $Q=m_t$ in the right frame.  The matrices in the two bases are connected by the KM matrix, as given by \eqref{eq:yukbasis}.

The curves in the left frame are all smooth (except for the small kink in the curves for $|\left(\bdf_{d}\right)_{i3}|$ that occurs because of the SUSY correction to the bottom quark Yukawa coupling \cite{pierce}), and do not show the dip to zero that appeared in the previous figure. This is not surprising because underlying the explanation of this dip was the fact that the off-diagonal elements evolved from zero at $Q=m_t$ \cite{RGE1}.  Notice also that in this frame the off-diagonal elements are not necessarily smaller than the diagonal elements even for $Q$ below the region of $1$~TeV.

The magnitudes of the off-diagonal matrix elements in the frame on the right, which do start at zero at $Q=m_t$, show the anticipated aligned dips, except that the location of the dip is shifted considerably to the right relative to Fig.~\ref{fig:SUGRAfu}. This shift is not difficult to understand. The large top quark Yukawa coupling in the SM governs the evolution of the off-diagonal elements of the down-type Yukawa couplings, causing them to evolve much more rapidly from zero in Fig.~\ref{fig:SUGRAfd}. Therefore, in order to evolve back to zero after the sign flip in the $\beta$-function due to the additional Higgs and SUSY particles, a longer evolution distance is needed. Furthermore, beyond the Higgs boson threshold the off-diagonal elements of $\bdf_{u}$ in Fig.~\ref{fig:SUGRAfu} are accelerated to zero on account of the fact that the down-type Yukawa couplings $\bdf_{d}$, that enter in the evolution of these elements, are enhanced by a factor $\sim1/\cos{\beta}$, pushing the dip in this figure to a low value of $Q$.

\subsection{Additional GUT Scale Flavour-violating Terms}\label{sec:quarkyukadd}

We now turn to briefly discuss what happens when we allow non-universality of GUT scale squark mass parameters, so as to split the squarks more than in mSUGRA. Recall that if the squarks all decouple together, all that happens is a change in the slope of the $\beta$-function. If, however, the squark masses are not all the same, our decoupling procedure entails an additional rotation to the squark mass basis. If this basis differs significantly from our ``standard'' basis in which the quark Yukawa coupling matrix is diagonal (at $Q=m_t$), one may expect considerable deviation in the evolution of the off-diagonal elements from the mSUGRA case for $Q$ values in-between the highest and lowest squark thresholds. The introduction of non-universal squark mass parameters via non-vanishing values of the constants $R_{U,D,E}$ and $S_{U,D,E}$ in (\ref{eq:GUTboundsferm}) never leads to significant effects because the rotation from the ``standard'' basis to the squark mass basis is small by construction. 

\begin{figure}\begin{centering}
\includegraphics[viewport=20 45 710 580, clip, scale=0.45]{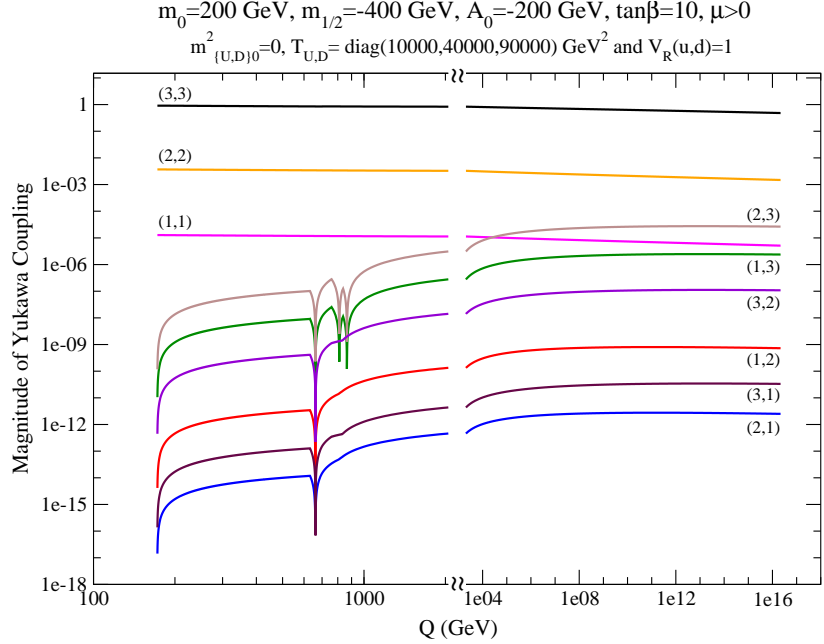} 
\caption[The same as Fig.~\ref{fig:SUGRAfu} except that the GUT scale right-handed squark mass matrices are now given by (\ref{eq:Tbc}) in the ``standard'' current basis.]{\small The same as Fig.~\ref{fig:SUGRAfu} except that the GUT scale right-handed squark mass matrices are now given by (\ref{eq:Tbc}) in the ``standard'' current basis.} \label{fig:T1fu} \end{centering} \end{figure}

The question then is whether we can have large deviations from Figs.~\ref{fig:SUGRAfu} and~\ref{fig:SUGRAfd} via the $\mathbf{T}_{U,D,E}$ (or $\mathbf{Z}_{u,d,e}$) matrices. To examine this, we set all GUT scale inputs to be the same as the mSUGRA case in Fig.~\ref{fig:SUGRAfu} except
\begin{subequations}\label{eq:Tbc}\begin{gather} m^{2}_{\{U,D\}0}=0\\
\mathbf{T}_{U,D}=\mathit{diag}\left\{10000,40000,90000\right\} {\rm GeV}^2\;.
\end{gather}\end{subequations}
We need, of course, to specify the basis in which the squark mass matrix is diagonal. If this is the ``standard'' current basis, with $\mathbf{V}_{L}(u)=\mathbf{V}_{R}(u,d)=\dblone$ at the weak scale, we see in Fig.~\ref{fig:T1fu} that although the right squark masses are now significantly split relative to the mSUGRA case, the Yukawa matrices are mostly unaltered.
That is to say, except for detailed changes in the evolution for $Q$ values in-between the squark thresholds (\textit{e.g.}, the additional structure in the dip at the higher value of $Q$ in the $(2,3)$ and $(1,3)$ elements), the evolution is not different in any important way at large values of $Q$. By choosing to enter $\mathbf{T}_{U}$ with diagonal (but unequal) entries, we are introducing only a mild amount of new flavour structure because the rotation between the ``standard'' basis and the squark mass basis is again small.

If we take instead that the right-handed squark matrices in (\ref{eq:Tbc}) are diagonal in a completely different basis specified by unitary matrices $\mathbf{V}_L(u)$, $\mathbf{V}_R(u)$ and $\mathbf{V}_R(d)$, entered in \inrge~using \eqref{eq:rotin} with randomly chosen values of $\alpha$, $\beta$, $\gamma$ and $\delta$, we may expect the evolution of Yukawa couplings to depart from the corresponding evolution in mSUGRA. We emphasize that in this model, in which $\bdm_{U,D}^2$ are diagonal in the basis where the corresponding Yukawa coupling matrices are very off-diagonal, includes potentially large flavour-violation in the singlet squark SSB mass matrices. 

\begin{figure}\begin{centering}
\includegraphics[viewport=20 45 710 580, clip, scale=0.45]{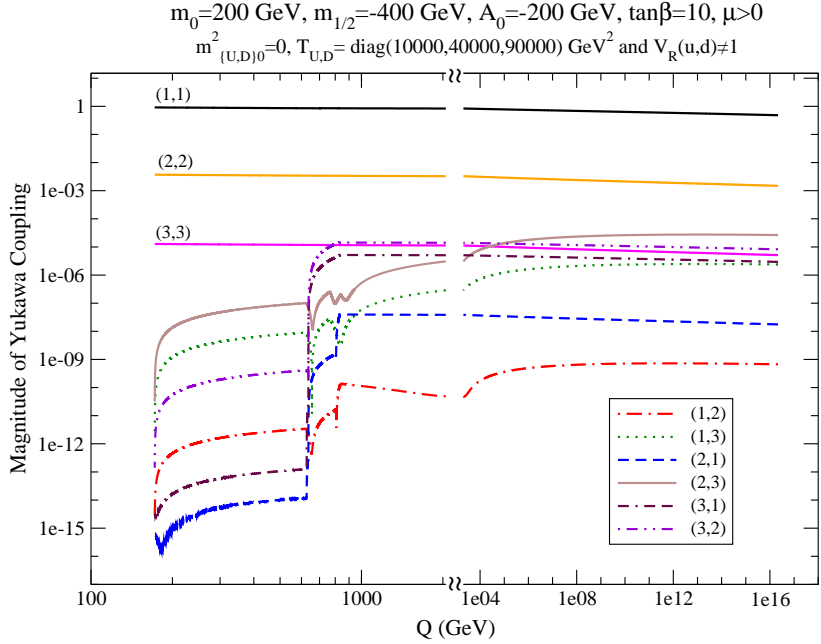} 
\caption[The same as Fig.~\ref{fig:SUGRAfu} except that the GUT scale right-handed squark mass matrices are now given by (\ref{eq:Tbc}) in the basis specified by (\ref{eq:genrot}) of the text.]{\small The same as Fig.~\ref{fig:SUGRAfu} except that the GUT scale right-handed squark mass matrices are now given by (\ref{eq:Tbc}) in the basis specified by (\ref{eq:genrot}) of the text. Note, however, that these matrix elements are shown in the ``standard'' current basis where the matrix is diagonal at $Q=m_t$.}
\label{fig:T2fu}
\end{centering}
\end{figure}

In Fig.~\ref{fig:T2fu} we illustrate the evolution of the magnitudes of the elements of the up-quark Yukawa coupling matrix with randomly chosen matrices,
\begin{subequations}\label{eq:genrot}\begin{align}
\mathbf{V}_{L}(u):&\;\alpha=2.053,\;\beta=0.254,\;\gamma=2.030,\;\delta_{\beta}=0.4829\\
\mathbf{V}_{R}(u):&\;\alpha=1.188,\;\beta=2.218,\;\gamma=0.763,\;\delta_{\beta}=0.87\\
\mathbf{V}_{R}(d):&\;\alpha=1.904,\;\beta=2.947,\;\gamma=1.847,\;\delta_{\beta}=1.14\;.
\end{align}\end{subequations}
%
Note that, just as in Fig.~\ref{fig:SUGRAfu}, we have plotted these elements in our ``standard'' basis where the Yukawa coupling matrix is diagonal at $Q=m_t$, and that in this basis the GUT scale matrices $\bdm_{U}^2$ and $\bdm_D^2$ will be random Hermitian matrices. We see that the evolution of the diagonal elements is not significantly altered from mSUGRA. This is because, although there is a large mismatch between the squark mass basis and our ``standard'' basis, this mismatch is operative over the small range of $Q$ between the highest and lowest thresholds and so has little impact on the largest elements. The two largest  off-diagonal entries (\textit{i.e.}, $(\bdf_{u})_{23}$ and $(\bdf_{u})_{13}$) similarly do not change significantly from Fig.~\ref{fig:SUGRAfu}, but all other entries are greatly altered. While it may seem that the values of these matrix elements at a large scale is quite irrelevant phenomenologically, the altered form of the Yukawa coupling matrix at the high scale could be of relevance to model-builders.

We have seen that the magnitudes of the off-diagonal elements in Fig.~\ref{fig:T2fu} remain small because the splitting in the squark spectrum is limited to  $\mathcal{O}(100-1000)$~GeV. The natural question then is whether we can get these to be larger by choosing extreme splitting between the squark eigenvalues. Even aside from potential flavour-changing effects, this is not easy. In general, such a GUT scale splitting also has a large value for $\mathcal{S}=m_{H_u}^2-m_{H_d}^2 +Tr\left[\bdm^2_Q-\bdm^2_L-2\bdm^2_U+\bdm^2_D+\bdm^2_E\right]$, which pulls the other squarks also to large masses, so the squark mass splitting is reduced by RGE effects. It may be possible to obtain split squarks by very carefully adjusting $\mathcal{S}$ to be zero, since $\mathcal{S}$ is then invariant under renormalisation group evolution, but we have not investigated this here.

\section{Gaugino Mass Parameters}

We have seen in Sec.~\ref{sec:GHP} that the evolution of the electroweak gaugino mass parameters acquires a dependence on the
$\mu$-parameter (and vice-versa) when threshold effects from splitting in the Higgs sector is taken into account. In models where $m_H \gg |\mu|, M_{1,2}$, the effect of the term explicitly dependent on $\mu$ in (\ref{app:rgem1})-(\ref{app:rgem2p}) (and the corresponding terms dependent on the gaugino mass parameters in the RGE for $\mu$) may be significant, so that the relation from Eq.~\eqref{eq:gauginounif}, $M_2/M_1=\alpha_2/\alpha_1$, which is expected in many models, is modified.  We should keep in mind that two loop terms will, in general, also alter this relation. Our point is that we should expect threshold corrections from the $\mu$ term in the RGE, as well as from the decoupling of sfermions, to be comparable to the two loop modifications, and so need to be included in a quantitative analysis.  Within the mSUGRA context we have small values of $|\mu|$, and hence $m_H \gg |\mu|$, in the hyperbolic branch/focus point (HB/FP) region \cite{focusI,*focusII} which occurs for large values of $m_0$, and is one of the regions favoured by the association of a light neutralino with the dark matter relic-density in connection with the comments in Sec.~\ref{sec:DM}. We mention here that {\em the location of this HB/FP region is significantly altered by the inclusion of the threshold corrections.}

\subsection{mSUGRA}

\begin{figure} \centering
\includegraphics[viewport=20 45 725 580, clip, scale=0.45]{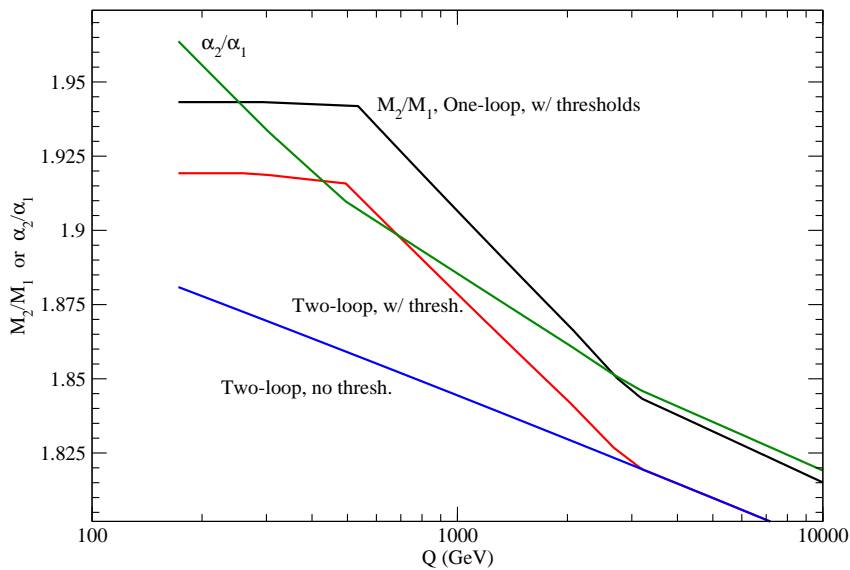} 
\caption[The evolution of the gaugino mass ratio, $M_2/M_1$, for an alternative mSUGRA point.]{\small The evolution of the gaugino mass ratio, $M_2/M_1$, for the mSUGRA point: $m_0=3075$~GeV, $m_{1/2}=-600$~GeV, $A_0=0$, $\tan\beta=10$ and $\mu>0$. We plot the ratio in three cases: one-loop with thresholds; two-loop with thresholds; and two-loop without thresholds. The gauge coupling ratio, $\alpha_{2}/\alpha_{1}$, is also shown for comparison.} \label{fig:m12compmS}
\end{figure}

The above considerations led us to examine the evolution of the gauge couplings and the electroweak gaugino parameters for a
relic-density-consistent mSUGRA model point in the HB/FP region with $m_0=3075$~GeV, $m_{1/2}=-600$~GeV, $A_0=0$, $\tan\beta=10$ and $\mu>0$, for which ($\mu, M_1, M_2) \simeq (304, -258, -495)$~GeV at the weak scale.\footnote{Without threshold corrections, ISAJET gives a similar spectrum for $m_0\simeq 3660$~GeV.} As can be seen in Fig.~\ref{fig:m12compmS}, at the two loop level with all threshold effects included, $M_2/M_1=1.919$ to be compared to $M_2/M_1=1.881$ obtained without including threshold effects.
Thus, although threshold effects actually bring us closer to ${\alpha_2/\alpha_1}\ (=1.964)$, their inclusion is clearly necessary for a quantitative analysis of mass parameters that may be extracted at an $e^+e^-$ linear collider, where a precision of better than 1\% will be possible if charginos are kinematically accessible.

\subsection{Split Supersymmetry Models}

The threshold corrections to gaugino mass parameters can be much larger in the so-called split SUSY model \cite{split1,*split2,*split3} that has received considerable attention in the recent literature. In these models, the naturalness of the scalar Higgs sector (which we view as one of the primary motivations for weak scale SUSY) is abandoned, while gauge coupling unification and the neutralino dark matter candidate of $R$-parity violating models are preserved. Gaugino mass parameters and $|\mu|$ are assumed to be at the weak scale, while scalar masses are at an intermediate scale. This means that sfermion masses as well as $m_H$ are very large (with the SM Higgs doublet fine-tuned to be light), so that charginos and neutralinos are the only new particles (other than a SM Higgs boson) at the weak scale.

\begin{figure}[t]\begin{centering}
\includegraphics[viewport=20 50 725 550, clip, scale=0.45]{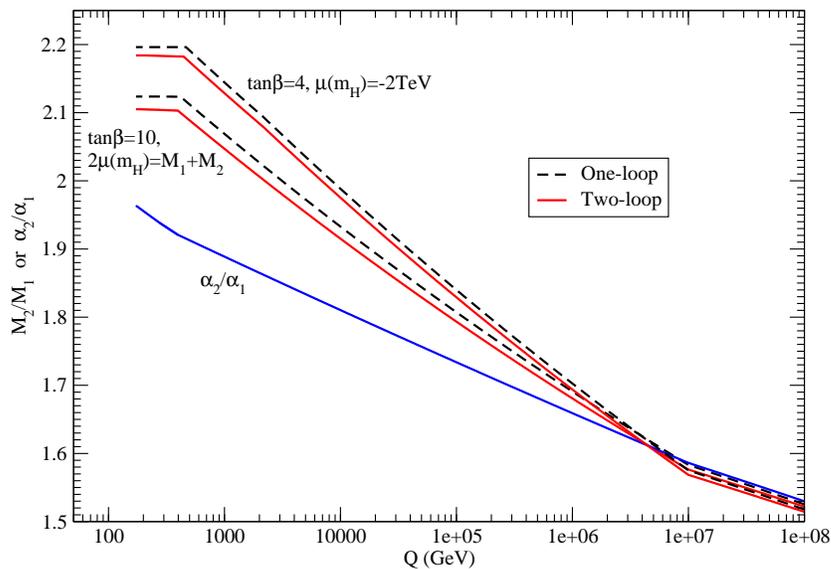} 
\caption[The evolution of the gaugino mass ratio for a split SUSY scenario.]{\small The evolution of the gaugino mass ratio, $M_2/M_1$, at the one (dashed) and two (solid) loop levels, along with the two-loop evolution of $\alpha_{2}/\alpha_{1}$, for a split SUSY model where scalars masses are around $10^7$~GeV, with gauginos and higgsinos at the weak scale. Gaugino mass unification is assumed and the gaugino mass parameters are negative. The other parameters are as mentioned in the text.}
\label{fig:split}
\end{centering}
\end{figure}

As an illustration, in Fig.~\ref{fig:split} we plot the variation of the ratio $M_{2}/M_{1}$ at the one-loop level (dashed), as well as at the two-loop level (solid), with the renormalisation scale $Q$, along with the two-loop value of $\alpha_{2}/\alpha_{1}$.
We show results, first where the value of $|\mu|$ is set exactly in-between $|M_1|$ and $|M_2|$ so that the lightest neutralino acquires a significant higgsino component, to qualitatively mimic mixed higgsino dark matter.\footnote{In the absence of a real theory of split SUSY, we should view this figure only as a qualitative illustration of potentially large threshold effects. Here, we take the sfermion mass parameters to be $10^7$~GeV at $Q=M_{\rm GUT}$, $m_{1/2}=-350$~GeV and $A_0=0$. Since it is not possible to satisfy the EWSB conditions except when $\tan\beta$ is hierarchically large -- this would cause down-type Yukawa couplings to become non-perturbative -- we treat $\mu$ and $\tan\beta$ as phenomenological parameters, and fix $m_H$ to be $10^7$~GeV in this figure. The parameters $m_{H_u}^2$ and $m_{H_d}^2$ (indeed all scalar mass parameters) are never needed since the RGEs for gaugino masses, $\mu$, the $\ba$-parameters, and the dimensionless couplings form a closed set even at the two-loop level. Sfermion masses only enter via the location of thresholds.} We have checked that the $M_2/M_1$ values do not change in a significant way for yet larger values of $\tan\beta$.  It is clear that the relation ${M_2}/{M_1}={\alpha_2}/{\alpha_1}$ is violated at the several percent level by the threshold corrections, without which the $M_2/M_1$ lines would have continued to low values with the same slope that they have above $10^7$~GeV.  For this point, the $2sc\times \mu$ term that explicitly appears in the RGE is very small so that the result is independent of the sign of $\mu$: most of the difference is an effect of the sfermion loop contributions being switched off below $10^7$~GeV.

To gain some idea of how large the effect of this $\mu$ term might be, we have also shown $M_2/M_1$ for $\mu=-2$~TeV, with the multiple $(\mu M_2)>0$ and $\tan\beta=4$.  Since such a large value of $\mu$ would be totally incompatible with the measured relic density, and small values of $\tan\beta$ are unnatural in these models without some modification to the EWSB sector, the reader should view these curves only as a guide to how much the gaugino mass ratio may deviate from its ``unification value''. The difference between the two cases is almost entirely due to the different choice of $\mu$. We see that the gaugino mass unification condition will, in this case, be violated by $\sim 10$\%. If instead we choose the opposite sign for $(\mu M_2)$, but keep $|\mu|=2$~TeV, this ``large $|\mu|$ line'' would be \textit{lower} than the line with $\mu = (M_1+M_2)/2$ by about the same amount that it is higher than this line  in the figure.  Clearly, increasing the splitting between the scalar and the gaugino/higgsino sector of the theory will cause even further violation of the unification condition, and $\sim 20$\% effects appear to be plausible if the scalars are instead at the $10^{11}$~GeV scale.

\section{SSB $\mathbf{a}$-parameters: mSUGRA and Non-universal Inputs}

\begin{figure}[t]\begin{centering}
\includegraphics[viewport=20 50 725 560, clip, scale=0.45]{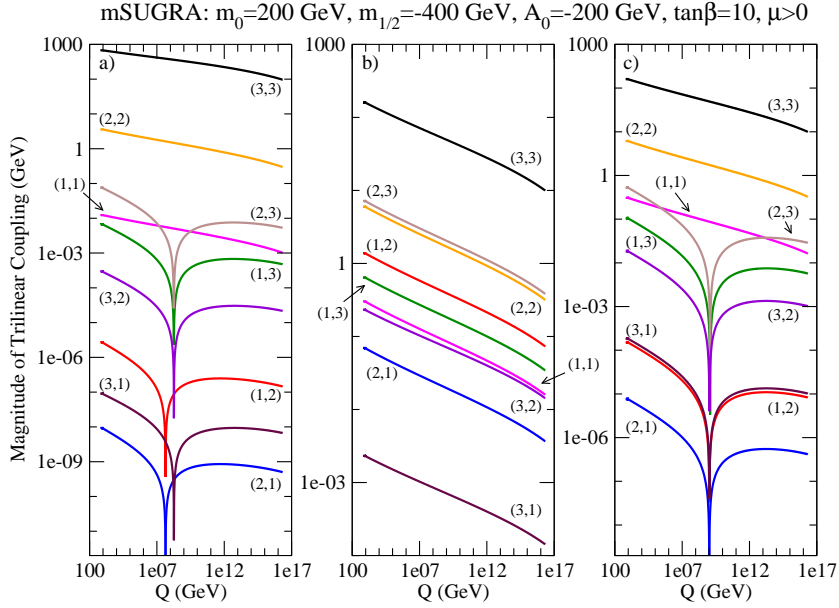} 
\caption[The magnitude of the elements of the trilinear coupling matrix $\ba_{u,d}$ for the mSUGRA model in Fig.~\ref{fig:SUGRAfu}.]{\small The magnitude of the elements of the trilinear coupling matrix $\ba_{u,d}$ for the mSUGRA model in Fig.~\ref{fig:SUGRAfu}. We show (\textit{a})~$|\left(\ba_u\right)_{ij}|$, (\textit{b})~$|\left(\ba_d\right)_{ij}|$, in the ``standard'' current basis, and (\textit{c})~$|\left(\ba_d\right)_{ij}|$  in the basis where down-quark Yukawa couplings are diagonal at $Q=m_t$. The curves extend between $Q=m_H$ and $Q=M_{\rm GUT}$.}
\label{fig:SUGRAtri}
\end{centering}
\end{figure}

Returning to the mSUGRA case considered in Figs.~\ref{fig:SUGRAfu} and~\ref{fig:SUGRAfd}, we show the magnitudes of the individual entries of the trilinear coupling matrices $\ba_u$ and $\ba_d$ in Fig.~\ref{fig:SUGRAtri}. There are three separate plots: (\textit{a})~$|\left(\ba_u\right)_{ij}|$ and (\textit{b})~$|\left(\ba_d\right)_{ij}|$ in our ``standard'' current basis where the up-type Yukawa coupling matrix is diagonal at $Q=m_t$, along with $|\left(\ba_d\right)_{ij}|$ in the basis where the down-type Yukawa coupling matrix is diagonal at $Q=m_t$ in frame (\textit{c}).
We terminate the curves at $Q=m_H$ since below this scale we have a single Higgs scalar doublet model, and the trilinear couplings ${\bf a}_{u,d}$ evolve only as part of a linear combination with $\tilde{\mu}^{*}\bdf^{h_{u,d}}_{u,d}$ as discussed in Sec.~\ref{sec:trideriv}.

We see that the curves in frame (\textit{a}) show a simple dip structure indicating that the real and imaginary parts of ${\bf a}_{u,d}$ are really monotonic functions of $Q$ that pass through zero together, in a manner similar to the elements of the Yukawa coupling matrix in Fig.~\ref{fig:higgsino}. The actual location of the zero is somewhat harder to analyse than in the case of the Yukawa couplings because even though ${\ba_{u}}$ obtains off-diagonal components only because the down-quark Yukawa matrix is not diagonal at $Q=m_t$, the matrix $\ba_u$ is off-diagonal even at the GUT scale.

The off-diagonal elements of $\ba_d$ in frame (\textit{b}) start off with a much larger magnitude in the ``standard'' basis at $Q=M_{\rm GUT}$ because the corresponding Yukawa coupling matrix has large off-diagonal pieces. In this case, the evolution of these off-diagonal elements receives significant contributions from \textit{all} entries in the RGE (unlike the evolution of the off-diagonal elements in frame (\textit{a}) or of the off-diagonal Yukawa couplings discussed earlier, where contributions from the off-diagonal down-type Yukawa matrices govern the evolution), and never go through zero; the situation is similar to that
in the first frame of Fig.~\ref{fig:SUGRAfd}.

We see from frame (\textit{c}) that the magnitudes of $\left({\bf a}_d\right)_{ij}$ in the basis that the down-type Yukawa coupling matrix is diagonal at $Q=m_t$ again show the characteristic dip structure indicating that the off-diagonal elements increase in magnitude from their value at $Q=M_{\rm GUT}$ to some maximum magnitude at an intermediate scale, but then smoothly reverse direction and thereafter evolve monotonically through zero to the low scale. The elements in frames (\textit{b}) and (\textit{c}) are, of course related by, (\ref{eq:aubasis}). We note that the off-diagonal elements in frame (\textit{c}), because these are driven by the larger ``up-type'' Yukawa couplings, are bigger than those in frame (\textit{a}) where it is the down-type Yukawa coupling matrix that largely determines the entries.

\begin{figure}[t]\begin{centering}
\includegraphics[viewport=20 50 710 585, clip, scale=0.45]{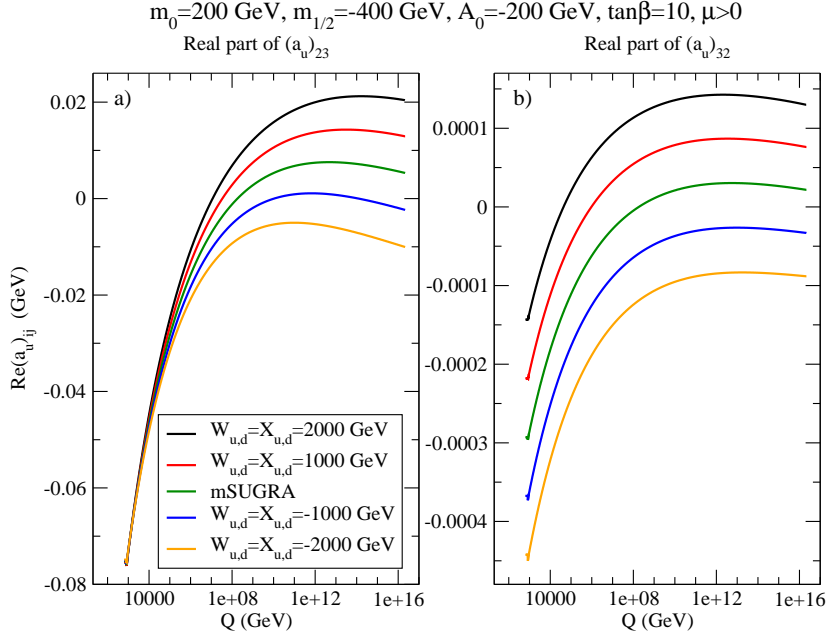} 
\caption[The evolution of the real parts of $\left({\bf a}_u\right)_{23}$ and $\left({\bf a}_u\right)_{32}$ for a model with non-universal inputs.]{\small The evolution of (\textit{a})~Re$\left(\ba_u\right)_{23}$ and (\textit{b})~Re$\left(\ba_u\right)_{32}$ for a model with the GUT scale values of $\ba$-parameters set as in \eqref{eq:GUTboundtri} for several values of $W$ and $X$ shown in the same order as the legend (with $A_{\{u,d\}0}=A_{0}$). The GUT scale SSB scalar and gaugino mass parameters are assumed to be universal, with the same value as in Fig.~\ref{fig:SUGRAtri}. We terminate the curves at $Q=m_H$.
}
\label{fig:au23wx}
\end{centering}
\end{figure}

Finally, in Fig.~\ref{fig:au23wx} we consider a model with non-universal values of $\ba$-parameters, but where these are not a new source of flavour-violation.
Specifically, we consider a model with the same values for mSUGRA parameters as in Fig.~\ref{fig:SUGRAtri}, but with non-zero values for $W_{u,d}$ and $X_{u,d}$ in (\ref{eq:GUTboundtri}) (with $A_{\{u,d\}0}=A_{0}$), and illustrate the evolution of (\textit{a})~Re$\left({\bf a}_u\right)_{23}$ and (\textit{b})~Re$\left({\bf a}_u\right)_{32}$. We have checked that the imaginary parts of these matrix elements are about four orders of magnitude smaller.    

The striking feature of frame~(\textit{a}) is that the various curves, which start with very different values of Re$\left({\bf a}_u\right)_{23}$ at $Q=M_{\rm GUT}$, appear to focus to a common value at the low scale. We have checked, however, that although they all cross at $Q\simeq 1.5$~TeV, they do not all converge at precisely the same value of $Q$. This apparent convergence, which persists for other values of mSUGRA parameters, is sensitively dependent on the special GUT scale boundary conditions for $\ba_u$ that we have used. We have checked that if instead we use a general matrix $\mathbf{Z}_u$ in (\ref{eq:GUTboundtri}), the corresponding evolution is completely different.  We do not have a good explanation for the seeming convergence in frame (\textit{a}), and only note that it is not generic to \textit{all} elements of ${\bf a}_u$ as evidenced, for example, by the corresponding evolution of Re$\left({\bf a}_u\right)_{32}$ in frame~(\textit{b}) of the figure.

\section{Soft Masses}

Our discussion of the evolution of the scalar mass SSB parameters will have a similar progression to the previous sections, beginning with an examination of our canonical mSUGRA point. We will then proceed to introduce non-universal GUT scale boundary conditions, first where there is no additional source of flavour-violation other than the superpotential Yukawa couplings, and then when we set $\mathbf{T}_{U,D}\neq\mathbf{0}$ in the same model as Fig.~\ref{fig:T2fu}.

\subsection{mSUGRA}

\begin{figure}[t]\begin{centering}
\includegraphics[viewport=5 50 710 580, clip, scale=0.45]{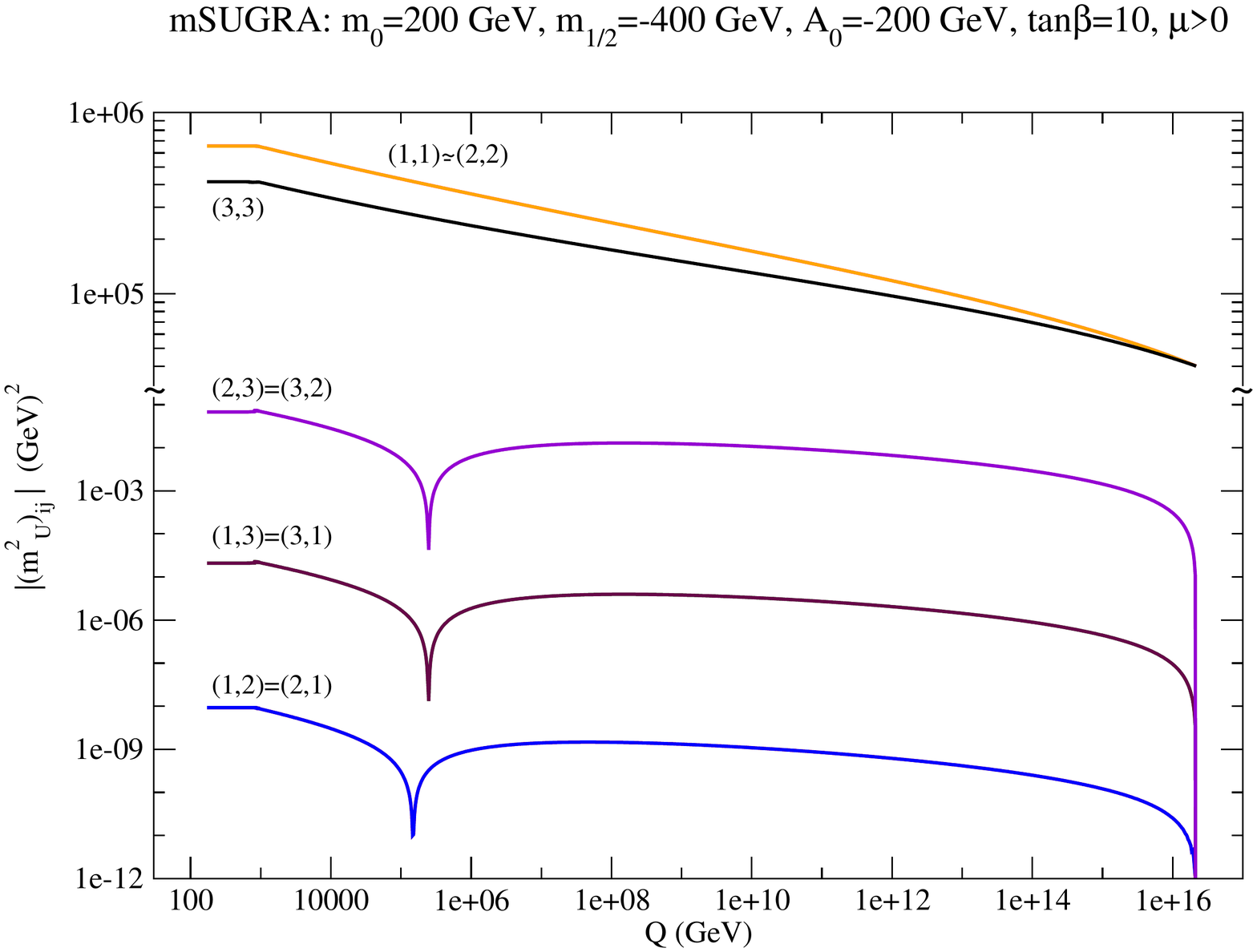} 
\caption[The scale dependence of the magnitudes of the matrix $\bdm_U^2$ for the mSUGRA model with parameters as in Fig.~\ref{fig:SUGRAfu}.]{\small The scale dependence of the magnitudes of the entries of $\bdm_U^2$ for the mSUGRA model with parameters as in Fig.~\ref{fig:SUGRAfu} in our ``standard'' basis. Note the we have broken the vertical scale at 0.1~GeV$^2$ and also used a different scale above this to better show the splitting of the (3,3) entry from the other diagonal entries.}
\label{fig:SUGRAmup}
\end{centering}
\end{figure}

We begin by showing in Fig.~\ref{fig:SUGRAmup} the evolution of the magnitudes of $\bdm_U^2$ in our ``standard'' current basis for the mSUGRA model with the same parameters as in Fig.~\ref{fig:SUGRAfu}. 
The diagonal matrix elements start from a common value $m_0$ and increase as we go to the weak scale because of gauge (and gaugino) interactions. Although the splitting between the (1,1) and (2,2) elements that occurs because of the Yukawa couplings is too small to be visible in the figure, the (3,3) element is reduced significantly on account of the large (3,3) entry in $\bdf_u$. Notice that the curves become flat once the squarks are all decoupled.

The magnitudes of the three independent off-diagonal elements of the Hermitian matrix $\bdm_U^2$ start from zero at $Q=M_{\rm GUT}$, and rapidly rise because $\bdf_u$ has off-diagonal entries at $Q=M_{\rm GUT}$. Note the break in the vertical scale in the figure. The ordering of the magnitudes of the off-diagonal elements of $\bdm_U^2$ can be simply gauged from the up-type Yukawa coupling matrix, since all SSB mass matrices at the GUT scale are proportional to the unit matrix. These off-diagonal elements start from zero at $Q=M_{\rm GUT}$, evolve to a maximum magnitude, smoothly reverse direction at an intermediate scale and then continue to evolve monotonically all the way to the weak scale. The dips in the figure occur where the real, and simultaneously the imaginary, part of $\left(\bdm_U^2\right)_{ij}$ changes sign during the course of its evolution to $Q=m_t$.

\begin{figure}[t]\begin{centering}
\includegraphics[viewport=5 50 710 580, clip, scale=0.45]{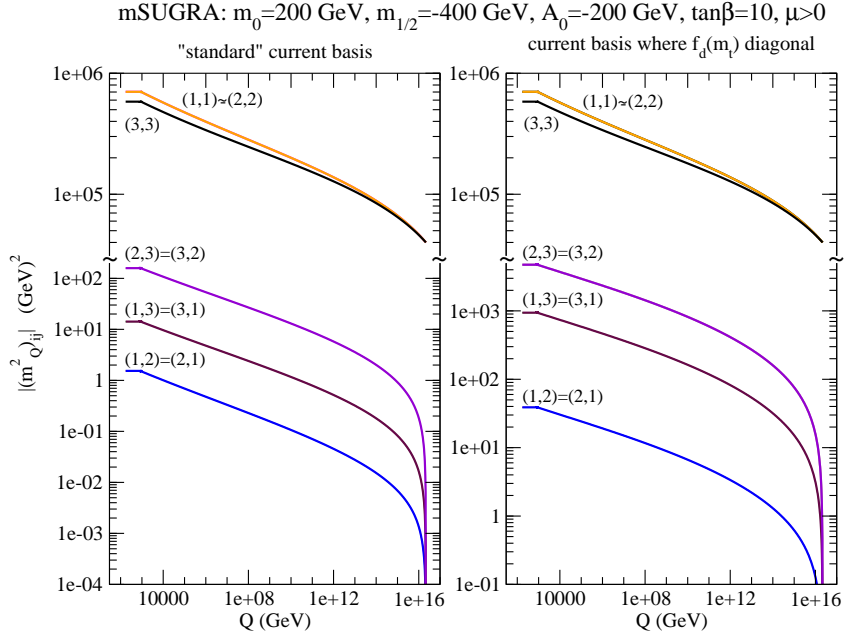} 
\caption[The scale dependence of the magnitudes of the matrix $\bdm_Q^2$ for the mSUGRA model with parameters as in Fig.~\ref{fig:SUGRAfu}, shown in two different bases.]{\small The scale dependence of the magnitudes of the entries of $\bdm_U^2$ for the mSUGRA model with parameters as in Fig.~\ref{fig:SUGRAfu} in the basis where the up-type (left frame), or the down-type (right frame), quark Yukawa coupling matrix is diagonal at $Q=m_t$. Note the we have broken the vertical scales to better the display the matrix elements.}
\label{fig:SUGRAmq}
\end{centering}
\end{figure}

Fig.~\ref{fig:SUGRAmq} shows the running of $\bdm^{2}_{Q}$ for the same mSUGRA point.
In the left frame, we show the magnitudes of the elements in our ``standard'' basis, where the up quark Yukawa coupling matrix is diagonal at $Q=m_{t}$. The frame on the right shows the magnitudes of the elements of this same matrix, but in the basis where the down-type quark Yukawa coupling matrix is diagonal at $m_{t}$. We have checked that the large difference in the size of the off-diagonal elements in the two frames is indeed accounted for by the fact that the corresponding matrices are related by \eqref{eq:mqbasis}. Unlike in Fig.~\ref{fig:SUGRAmup}, there is no dip in the magnitudes of the off-diagonal elements because they evolve monotonically from zero at the GUT scale, until the squarks are all decoupled. 

\subsection{Additional GUT Scale Flavour Structure}

\begin{figure}[t]\begin{centering}
\includegraphics[viewport=20 50 710 580, clip, scale=0.45]{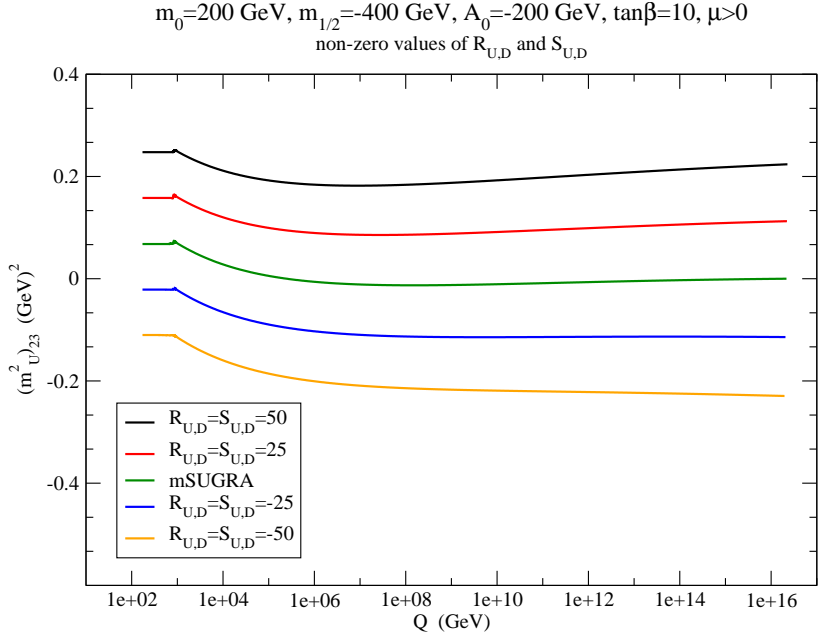} 
\caption[The scale dependence of $(\bdm^{2}_{U})_{23}$ for several sets of values of $R_{U,D}$ and $S_{U,D}$ that appear in \eqref{eq:GUTboundsferm} for the model with non-universal GUT scale SSB squark mass parameters.]{\small The scale dependence of $(\bdm^{2}_{U})_{23}$ for several sets of values of $R_{U,D}$ and $S_{U,D}$ that appear in (\ref{eq:GUTboundsferm}) for the model with non-universal GUT scale SSB squark mass parameters, but $c_{U,D}=1$. The curves are in the same order as the legend, and all the other parameters are set as in Fig.~\ref{fig:SUGRAmup}.}
\label{fig:rsmu}
\end{centering}
\end{figure}

To understand how the non-universal boundary conditions in \eqref{eq:GUTboundsferm} impact the evolution of the squark mass
matrices we examine a model with non-zero values of $R_{U,D}$ and $S_{U,D}$, but with universal gaugino masses and $\ba$-parameters. In Fig.~\ref{fig:rsmu} we plot the value of $(\bdm^{2}_{U})_{23}$ for the set of values of $R_{U,D}$ and $S_{U,D}$ shown in the figure, with all other parameters set as in Fig.~\ref{fig:SUGRAmup} (including $c_{U,D}=1$), so that $R_{U,D}=S_{U,D}=0$ corresponds to the mSUGRA model in this figure.

We see that with non-zero values of $R_{U,D}$ and $S_{U,D}$, $\left(\bdm_U^2\right)_{23}$ already starts off with a substantial value (positive or negative) at $Q=M_{\rm GUT}$, and evolves slowly with $Q$. This situation is qualitatively similar to that of the mSUGRA case shown in Fig.~\ref{fig:SUGRAmup} once $Q$ has evolved away somewhat from $M_{\rm GUT}$ so that the matrix element has had a chance to grow from zero. However, because the curves start of with rather large values at the GUT scale (except for the middle mSUGRA curve) the evolution does not take them through zero for any value of $Q>m_t$. As a result, the dip which was the most prominent feature of Fig.~\ref{fig:SUGRAmup} is absent, except in the middle curve which does cross zero for $Q\sim2.5\times 10^5$~GeV. 

\begin{figure}[t]\begin{centering}
\includegraphics[viewport=5 50 710 560, clip, scale=0.45]{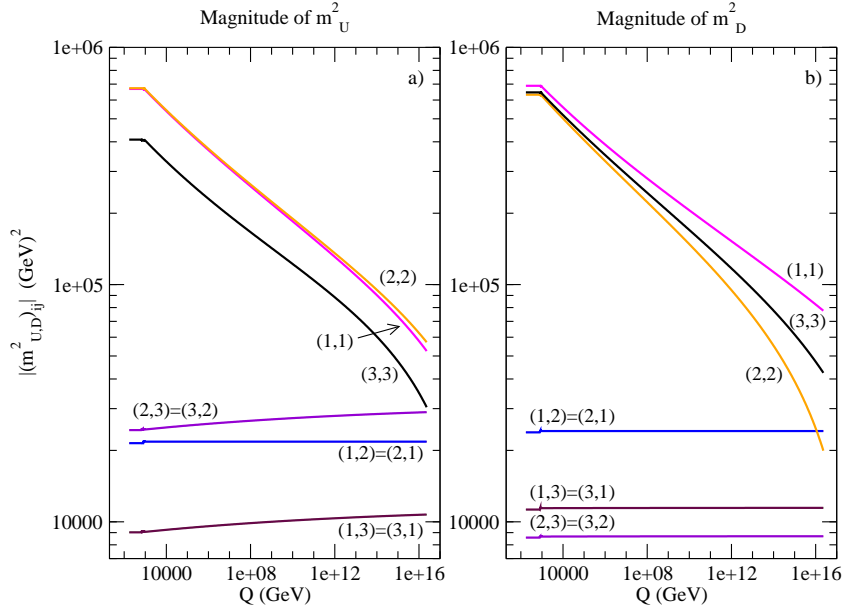} 
\caption[The magnitude of the elements of $\bdm^{2}_{U}$ and $\bdm^{2}_{D}$ for the non-universal model considered in Fig.~\ref{fig:T2fu}.]{\small (\textit{a})~The same as Fig.~\ref{fig:SUGRAmup} except for the non-universal model considered in Fig.~\ref{fig:T2fu}, and (\textit{b})~the magnitude of the corresponding elements of $\bdm_D^2$ for the same scenario. The elements are plotted in our ``standard'' current basis but are exactly the same in the basis where the down quark Yukawa couplings are diagonal.}
\label{fig:T2mu}
\end{centering}
\end{figure}

Finally, in Fig.~\ref{fig:T2mu} we return to the non-mSUGRA case that we considered in Fig.~\ref{fig:T2fu}.
We emphasize again that in this case ${\bf m}_U^2$ and ${\bf m}_D^2$ are diagonal only in the basis where the superpotential Yukawa coupling matrices have large off-diagonal elements so that we would expect this model to include new and potentially large sources of flavour-violation that may well already be excluded by data. As in Fig.~\ref{fig:SUGRAmup}, we show $|\left(\bdm^{2}_{U}\right)_{ij}|$ in the basis where the up-quark Yukawa coupling matrix is diagonal at $Q=m_t$.  We see from the figure that the matrix $\bdm_U^2$ has large off-diagonal entries at $Q=M_{GUT}$. As expected, the gauge (and gaugino) interactions cause the diagonal entries to increase, whereas the off-diagonal elements which do not ``feel'' these terms evolve much more slowly with $Q$. Note also that because $\left(\bdm^{2}_{U}\right)_{11}<\left(\bdm^{2}_{U}\right)_{22}$ at $Q=M_{\rm GUT}$, Yukawa coupling effects draw them closer as we go to low scales.  We also remark that the negative GUT scale value of $\mathcal{S}$ tends to reduce the diagonal elements of $\bdm^{2}_{U}$ as we go to low scales, but pulls up the corresponding elements of $\bdm^{2}_{D}$.

Although the renormalisation group evolution increases the gap between the off-diagonal and diagonal elements at the low scale, notice that the off-diagonal elements are separated by just one order of magnitude from the difference between the diagonal elements, so that we may expect large flavour-mixing between the $SU(2)$ singlet up squarks. This mixing, if anything, is even larger in the down squark sector as can be seen in frame (\textit{b}). A careful evaluation of inter-generation squark mixing is clearly necessary for any discussion of flavour-violation in squark decays, the subject of the next chapter.

%% file: stop.tex
\chapter{Decay of the Lighter Stop}\label{ch:stopdec}

As we alluded to in the introduction, the sfermions of the MSSM have interactions which permit the $\tilde{t}_{1}$ to decay according to $\tilde{t}_{1}\rightarrow t\tilde{Z}_{i}$ and $\tilde{t}_{1}\rightarrow b\tilde{W}_{i}$. However, the large top mass and Yukawa coupling leads to several factors which may preclude the lighter stop from decaying through either of these two options. As we have seen, in mSUGRA the large top Yukawa coupling depresses the $(3,3)$ entries of the SSB soft mass matrices $\bdm^{2}_{Q}$ and $\bdm^{2}_{U}$ during RGE evolution from the high scale, rendering the lighter stop potentially lighter than the chargino. In addition, there is the possibility for a large mass splitting between the stop states coming from the $\mathbf{a}$-parameters, further reducing the $\tilde{t}_{1}$ mass. In a scenario where the $\tilde{t}_{1}$ is sufficiently light, the two-body decays listed above will therefore be kinematically inaccessible.

In order to apply our RGE analysis to flavour-violating decays of squarks, we must find the relation between the current basis squark states and the mass eigenstates. Using the weak scale SSB and `Yukawa' coupling matrices, we will write out the mass terms in the Lagrangian and diagonalise this matrix to obtain the physical squark states. Rewriting the various squark interactions in terms of these physical states, we will then be able to find the partial width for flavour-changing stop decay \cite{RGE2}.

\section{Squark Mass Matrix}

As we saw in Eq.~\eqref{eq:umixmat} of Sec.~\ref{sec:sqmass}, the up-type squark mass terms can be collected into a $(6\times6)$ matrix so that the Lagrangian contains
\begin{equation}\label{eq:umixmatII}
\mathcal{L} \ni -\left( {\tilde{u}}^{\dag}_{Li}, {\tilde{u}}^{\dag}_{Ra} \right)\left(\begin{array}{cc}\left(\bm{\mathcal{M}}^{2}_{LL}\right)_{ij}&\left(\bm{\mathcal{M}}^{2}_{LR}\right)_{ib}\\\left(\bm{\mathcal{M}}^{2}_{LR}\right)^{\dagger}_{aj}&\left(\bm{\mathcal{M}}^{2}_{RR}\right)_{ab}\end{array}\right)\left( \begin{array}{c} \tilde{u}_{Lj} \\ \tilde{u}_{Rb} \end{array} \right)\;,
\end{equation}
where $i,j$ label left-handed squarks and $a,b$ label right-handed squarks. The physical states are found by diagonalising the $(6\times6)$ matrix, and the squark mass basis is defined to be $\left(\tilde{t}_{1},\tilde{t}_{2},\tilde{c}_{1},\tilde{c}_{2},\tilde{u}_{1},\tilde{u}_{2}\right)$ so that the $\tilde{t}_{1}$ is signified by $\tilde{u}^{\tilde{M}}_{1}$. 

The individual entries of \eqref{eq:umixmatII} were written in Sec.~\ref{sec:sqmass} and we repeat them here for convenience
\begin{subequations}\label{eq:msq}
\begin{align}
\label{eq:mllsq}\left(\bm{\mathcal{M}}^{2}_{LL}\right)_{ij} = & {(\bdm^2_{Q})}_{ij}+ v_u^{2} {\left({\bdf^{*}_u}\bdf^{T}_u\right)}_{ij}+\left(\frac{\glp^{
2}}{12}-\frac{\gtwl^{2}}{4}\right)\left(v_{u}^{2}-v_{d}^{2}\right)\bm{\delta}_{ij}\\
{(\bm{\mathcal{M}}^{2}_{RR})}_{ab} = & {(\bdm^2_{U})}_{ab}+ v_u^{2}{\left(\bdf^{T}_u{\bdf^{*}_u}\right)}_{ab}-\frac{\glp^{2}}{3}\left(v_{u}^{2}-v_{d}^{2}\right)\bm{\delta}_{ab}\\
\left(\bm{\mathcal{M}}^{2}_{LR}\right)_{ib}=&-v_{u}\left(\ba_{u}\right)^{*}_{ib}+v_{d}\mtfuhus_{ib}\;.
\end{align}
\end{subequations}
where we now keep track of the possibility for separation between $\mu$ and $\tilde{\mu}$, and $\bdf_{u}$ and $\tilde{\bdf}^{h_{u}}_{u}$. These relations should be used with caution. Recall from the discussion in Sec.~\ref{sec:trideriv} that we have a restricted group of operators when $Q<m_{H}$. If we are evaluating $\left(\bm{\mathcal{M}}^{2}_{LR}\right)_{ib}$ at a scale below $m_{H}$ we must use the combination $\left[-\sn\left(\ba_{u}\right)^{*}_{ib}+\cs\mtfuhus_{ib}\right]$ for which the RGE is \eqref{app:triu} in Appendix~\ref{app:dfulRGEs}.

\section{Calculation of the Width}

The two-body flavour-changing decay of the stop occurs due to non-zero couplings of squarks to quarks and neutralinos. These couplings appear in the Lagrangian from both gaugino interactions and superpotential terms and are
\begin{equation}\begin{split}
\mathcal{L}\ni&-\frac{1}{\sqrt{2}}\left\{\tilde{u}^{\dagger}_{Lj}\left(\bgtq_{ji}\bar{\lambda}_{3}+\frac{1}{3}\bgtpq_{ji}\bar{\lambda}_{0}\right)P_{L}u_{i}+\bar{u}_{b}\bgtpur_{ba}\left(-\frac{4}{3}\right)\tilde{u}_{Ra}P_{L}\lambda_{0}+\mathrm{h.c.}\right\}\\
&-\left\{\bar{\Psi}_{h^{0}_{u}}\tilde{u}^{\dagger}_{Rb}\bftur^{T}_{bi}P_{L}u_{i}+\bar{u}_{b}\bftuq^{T}_{bi}\tilde{u}_{Li}P_{L}\Psi_{h^{0}_{u}}+\mathrm{h.c.}\right\}\;.
\end{split}\end{equation}
Following the prescription in Ref.~\cite{wss}, we replace the gauginos and higgsinos with the lightest neutralino using the neutralino eigenvector, $v^{(i)}_{k}$, and inserting a $(i\gamma_{5})^{{\theta}_{i}}$ to allow for negative squared masses
\begin{subequations}\begin{align}
\lambda_{0}=&\sum_{i}v^{(i)}_{4}(i\gamma_{5})^{\theta_{i}}\tilde{Z}_{i}\\
\lambda_{3}=&\sum_{i}v^{(i)}_{3}(i\gamma_{5})^{\theta_{i}}\tilde{Z}_{i}\\
\Psi_{h^{0}_{u}}=&\sum_{i}v^{(i)}_{1}(i\gamma_{5})^{\theta_{i}}\tilde{Z}_{i}\;,
\end{align}\end{subequations}
in order to obtain
\begin{subequations}\begin{align}
\mathcal{L}\left(\tilde{u}_{Lj}c\tilde{Z}_{1}\right) =& {\tilde{u}_{Lj}}^{\dag}\overline{\tilde{Z}_{1}}\left(i\mathbf{A}_{j}P_L-{(i)}^{\theta_1}\bftuq^{*}_{j2}v^{(1)}_1P_R\right)c + \mathrm{h.c.} \\
\mathcal{L}\left(\tilde{u}_{Rb}c\tilde{Z}_{1}\right) =&{\tilde{u}_{Rb}}^{\dag}\overline{\tilde{Z}_{1}}\left(i\mathbf{B}_{b}P_R-{\left(-i\right)}^{\theta_1}\bftur^{T}_{b2}v^{(1)}_1P_L\right)c + \mathrm{h.c.} \;,
\end{align}\end{subequations}
where
\begin{subequations}\label{eq:AjBb}\begin{align}
\mathbf{A}_{j}\equiv&\frac{{\left(-i\right)}^{\theta_1-1}}{\sqrt{2}}\left[\bgtq_{j2}v^{(1)}_3+\frac{\bgtpq_{j2}}{3}v^{(1)}_4\right]\\
\mathbf{B}_{b}\equiv&\frac{4}{3\sqrt{2}}\bgtpur^{\dagger}_{b2}{\left(i\right)}^{\theta_1-1}v^{(1)}_4\;.
\end{align}\end{subequations}
Using (\ref{eq:squarkrot}) to write the squarks in terms of mass eigenstates and picking out the $\tilde{t}_{1}$, we obtain the Lagrangian of interest:
\begin{equation}
\mathcal{L}\left(\tilde{t}_1 c\tilde{Z}_{1}\right) \ni {\tilde{t}_1}^{\dag}\overline{\tilde{Z}_{1}}\left[\alpha P_L+\beta P_R\right]c+\mathrm{h.c.}
\end{equation}
The coefficients of $P_L$ and $P_R$ are
\begin{subequations}\label{eq:albet}\begin{align}
\alpha\equiv&i(\bm{\mathcal{U}}_{L})^{\dagger}_{1j}\mathbf{A}_{j}-{\left(-i\right)}^{\theta_1}v^{(1)}_1(\bm{\mathcal{U}}_{R})^{\dagger}_{1b}\bftur^{T}_{b2}\\
\beta\equiv&i(\bm{\mathcal{U}}_{R})^{\dagger}_{1b}\mathbf{B}_{b}-{\left(i\right)}^{\theta_1}v^{(1)}_1(\bm{\mathcal{U}}_{L})^{\dagger}_{1j}\bftuq^{*}_{j2}\;.
\end{align}\end{subequations}
where we assign the first index of $\bm{\mathcal{U}}_{L,R}^{\dagger}$ to be $\alpha=1$ since this is the index which represents the $\tilde{t}_{1}$ eigenstate. With this Lagrangian we find that the partial width of the $\tilde{t}_1$ to $c$ plus $\tilde{Z}_{1}$ is:
\begin{equation} \begin{split} \label{eq:twobodgamma}
\Gamma(\tilde{t}_{1}\rightarrow\mathit{c\tilde{Z}_{1}})=&\frac{1}{16\pi m^3_{\tilde{t}_1}}\left\{\left({\left|\alpha\right|}^2+{\left|\beta\right|}^2\right)\left(m^2_{\tilde{t}_1}-m^2_c-m^2_{\tilde{Z}_{1}}\right)-2m_cm_{\tilde{Z}_{1}}\left(\alpha\beta^*+\beta\alpha^*\right)\right\} \\
&\qquad\qquad\qquad\qquad\qquad\qquad\qquad\qquad\qquad\qquad\quad\times\lambda^{1/2}\left(m^2_{\tilde{t}_1},m^2_c,m^2_{\tilde{Z}_{1}}\right)\;,
\end{split}\end{equation}
with 
\begin{equation}
\nonumber \lambda\left( x,y,z \right)=x^2+y^2+z^2-2xy-2xz-2yz\;.
\end{equation}

Since the splitting between the diagonal entries of $\tilde{\mathbf{g}}^{(\prime)\Phi}$ and their non-tilde counterparts is small, the $j,b=2$ contributions to $\mathbf{A}_{j}$ and $\mathbf{B}_{b}$ in \eqref{eq:AjBb} are qualitatively approximated by using the standard gauge couplings. Similarly, the contributions from $\tilde{\bdf}^{\Phi}_{u}$ in \eqref{eq:albet} are qualitatively approximated by the usual Yukawa couplings. We have checked that the width calculated ignoring the tilde-couplings differs from the full calculation in general by a few percent.

Furthermore, we find that in the mSUGRA class of models, the contributions to $\Gamma(\tilde{t}_{1}\rightarrow\mathit{c\tilde{Z}_{1}})$ from the threshold-induced off-diagonal entries of $\tilde{\mathbf{g}}^{(\prime)\Phi}$ are generally two orders of magnitude smaller than the contribution from squark mixing (the size of which we will estimate in Sec.~\ref{sec:single}). To see this, we first write the RGEs for the gaugino couplings in terms of the mass-basis Yukawa matrices, using the KM matrix to convert $\bdf_{d}$ in our ``standard'' current basis into the basis where the down quark Yukawa couplings are diagonal at $m_{t}$. This means that the majority of the flavour structure in the down Yukawa coupling terms will be contained in the KM matrix. On the understanding that $\bdf_{d}$ is now in this basis, and taking all three flavours of sfermions in each group to decouple at the same point, we can write the seed term for off-diagonal entries to the $SU(2)$ gaugino-squark-quark coupling, $\tilde{\mathbf{g}}^{Q}$, whose RGE is found in \eqref{app:bgtq}, as
\begin{equation}\label{eq:gtseed}
\frac{d\bgtq_{ij}}{dt}\sim\frac{1}{16\pi^{2}}\left[(\bdf_{u})^{*}_{ik}(\bdf_{u})^{T}_{kl}\bgtq_{lj}+\mathbf{K}_{ik}(\bdf_{d})^{*}_{kl}(\bdf_{d})^{T}_{lm}\mathbf{K}^{\dagger}_{mn}\bgtq_{nj}\right]\;,
\end{equation}
where $i\ne j$. For the off-diagonal entries this relation holds at the threshold of the heaviest SUSY particle, where $\bgtq_{ij}=g\times\dblone$. To estimate the size of the off-diagonal elements of $\tilde{\mathbf{g}}^{Q}$ we therefore take them to be diagonal on the right-hand side of \eqref{eq:gtseed} and choose $k,l,m,n$ to find the maximum contribution. Referring to the general size of the off-diagonal elements of the Yukawa matrices in both Fig.~\ref{fig:SUGRAfu} and the right-hand frame of Fig.~\ref{fig:SUGRAfd}, in the region of the highest SUSY threshold, we find that the largest off-diagonal entry is approximately
\begin{equation}\label{eq:gtseednum}
\frac{d\bgtq_{23}}{dt}\sim\frac{1}{16\pi^{2}}\left[\left(1\times10^{-7}\right)(1)(0.1)+\left(4\times10^{-2}\right)(0.2)(0.2)(1)(0.1)\right]\;,
\end{equation}
so that this entry can at most become $\sim\mathit{few}\times10^{-6}$, coming from the down-type Yukawa coupling contribution. The $U(1)$ gaugino couplings will have much smaller entries since their RGEs do not contain KM dependent Yukawa coupling terms. If we now allow for additional flavour-violating terms arising in the SSB parameters, in a non-mSUGRA scenario, off-diagonal gaugino couplings are still unimportant for the width calculation. This is because, even if the sfermion splitting in quite large, the effect of their additional flavour-violation will only enter the RGEs for the gaugino couplings through the thresholds. Although with a completely general rotation between the sfermion mass basis and the original current basis there will be additional large contributions to the RGE, the effect will be limited due to the small size of the logarithm and will therefore tend to remain sub-dominant. For example, although in the case of large mixing in $\bdm^{2}_{U}$ the off-diagonal entries of the $U(1)$ gaugino couplings can become proportional to $(\bdf_{u})_{33}$, there will be corresponding increases in the mixing terms $\bm{\mathcal{U}}_{L,R}$ which will simultaneously increase the terms depending on diagonal entries of $\tilde{\mathbf{g}}^{\prime\Phi}$.

Turning to the Yukawa coupling contributions in \eqref{eq:albet}, we see that since $\left(\bm{\mathcal{U}}_{L}\right)^{\dagger}_{1j}$ and $\left(\bm{\mathcal{U}}_{R}\right)^{\dagger}_{1b}$ are largest when $j,b=3$ (\textit{i.e.}, when the squark current basis state is $\tilde{t}_{L}$ and $\tilde{t}_{R}$ respectively), the Yukawa coupling contributions are largest for $(\bdf_{u})_{32}$, which is $\sim10^{-7}$. For the $j,b=2$ case, the mixing is reduced, and $\bdf_{u}\propto m_{c}/M_{W}$, so that the Yukawa coupling terms are unimportant in mSUGRA. Note that in non-mSUGRA scenarios, since the squark mixing can become large, these terms are no longer sub-dominant, and may need to be taken into account.

In mSUGRA, we are left with the two gauge coupling terms, \eqref{eq:AjBb}, as the dominant contributions to \eqref{eq:albet}, so that the width is dependent upon the standard gauge coupling terms and the size of the squark mixing given by $\left(\bm{\mathcal{U}}_{L,R}\right)_{12}$ from \eqref{eq:squarkrot}. To a good approximation we can write the mixing as
\begin{subequations}
\begin{align}\label{eq:mixest}
\left(\bm{\mathcal{U}}_{L}\right)^{\dagger}_{12}&=\frac{\left(\bm{\mathcal{M}}_{\tilde{u}}\right)_{23}\left(\bm{\mathcal{U}}_L\right)^\dagger_{13}+\left(\bm{\mathcal{M}}_{\tilde{u}}\right)_{26}\left(\bm{\mathcal{U}}_R\right)^\dagger_{13}}{m^2_{\tilde{t}_1}-\left(\bm{\mathcal{M}}_{\tilde{u}}\right)_{22}}\;,\\[5pt]
\left(\bm{\mathcal{U}}_{R}\right)^{\dagger}_{12}&=\frac{\left(\bm{\mathcal{M}}_{\tilde{u}}\right)_{53}\left(\bm{\mathcal{U}}_L\right)^\dagger_{13}+\left(\bm{\mathcal{M}}_{\tilde{u}}\right)_{56}\left(\bm{\mathcal{U}}_R\right)^\dagger_{13}}{m^2_{\tilde{t}_1}-\left(\bm{\mathcal{M}}_{\tilde{u}}\right)_{55}}\;,
\end{align}
\end{subequations}
where the $\left(\bm{\mathcal{U}}_{L,R}\right)^\dagger_{13}$ can be approximately calculated using the $(2\times2)$ sub-matrix for the mixing between $\tilde{t}_{L}$ and $\tilde{t}_{R}$.

\section{Single-step Estimation of the Width}\label{sec:single}

As an alternative to the full calculation of the squark mixing outlined above, Hikasa and Kobayashi \cite{hkstop} developed a scheme for estimating the width without solving the RGEs precisely. Working within the mSUGRA framework with squarks of around $30$~GeV, and using the approximation that $\tilde{Z}_1\simeq \tilde{\gamma}$, they showed that if tree-level two-body decays of $\tilde{t}_1$ were kinematically forbidden, $\tilde{t}_1\to c\tilde{Z}_1$ would be the dominant decay mode of $\tilde{t}_1$. While their approximations and analysis are certainly valid for $m_{\tilde{t}_1}$ values that they considered fifteen years ago, this is not the case for top squark masses in the range of interest today. Since their approximate formula is widely used in the literature when calculating $\tilde{t}_{1}$ decay patterns \cite{wohr,*djouadi,boehm}, we estimate the decay width using their approach (including the straightforward extension to general mixing in the neutralino sector \cite{BDGGTstop,*han}) for comparisons with our full RGE calculation.

For models with no squark mixing at the GUT scale, in the limit that $m_{c}$ and $m_{u}\rightarrow 0$, $\tilde{c}_R$, $\tilde{u}_{L}$ and $\tilde{u}_{R}$ do not mix with the stops so that (for the $\tilde{t}_{1}$), $\left(\bm{\mathcal{U}}_{R}\right)^{\dagger}_{12}$ vanishes and we only need to find $\left(\bm{\mathcal{U}}_{L}\right)^{\dagger}_{12}$. We write the mass matrix for the scalars exactly in terms of the quantities $\Delta_L$ and $\Delta_R$ which are the $\tilde{t}_L - \tilde{c}_L$ and $\tilde{t}_R - \tilde{c}_L$ entries:
\begin{equation}\label{eq:defdeltas}
\left( {\tilde{t}_L}^{\dag}, {\tilde{t}_R}^{\dag}, {\tilde{c}_L}^{\dag} \right) \left( \begin{tabular}{ccc} \multicolumn{2}{c}{\multirow{2}{*}{${{\bm{\mathcal{M}}}}^2_{\tilde{t}}$}} & $\Delta_L$ \\ && $\Delta_R$ \\ ${\Delta^*_L}$ & ${\Delta^*_R}$ & $m^2_{\tilde{c}_L}$ \end{tabular} \right) \left( \begin{array}{c} \tilde{t}_L \\ \tilde{t}_R \\ \tilde{c}_L \end{array} \right)\;.
\end{equation}
We can find an approximate value for $\Delta_{L}$ and $\Delta_{R}$ at the scale $Q$ by approximating the RGE running to be a straight line and evaluating the right-hand side of the RGEs at a single scale. If all off-diagonal entries of the Yukawa and SSB matrices are ignored, $\Delta_L$ and $\Delta_R$ are given by
\begin{align}
\begin{split}
\Delta_L=&-\frac{1}{16\pi^{2}}\ln{\left[\frac{\mgut}{Q}\right]}\mathbf{K}^*_{cb}\mathbf{K}_{tb}\\
&\qquad\times\left\{\left[\left(\bdm^2_Q\right)_{22}+\left(\bdm^2_Q\right)_{33}+2\left(\bdm^2_D\right)_{33}+2m^2_{H_d}\right]\left(\bdf_{d}\right)^{*}_{33}\left(\bdf_{d}\right)^{T}_{33}+2\left(\ba_{d}\right)^{*}_{33}\left(\ba_{d}\right)^{T}_{33}\right\} 
\label{eq:deltal} \end{split}\\
\Delta_R=&\frac{2}{16\pi^{2}}\ln{\left[\frac{\mgut}{Q}\right]}\mathbf{K}^{*}_{cb}\mathbf{K}_{tb}v_{u}\left(\bdf_{u}\right)_{33}\left(\ba_{d}\right)_{33}\left(\bdf_{d}\right)^{\dagger}_{33}\;.
\label{eq:deltar} \end{align}
Since these equations do not take into account threshold effects, we do not use different relations in the $Q<m_{H}$ regime. If we are calculating the stop decay at a scale below $m_{H}$, we evaluate $\Delta_{L}$ and $\Delta_{R}$ at $Q=m_{H}$.

To derive these relations, we see from \eqref{eq:defdeltas} and \eqref{eq:mllsq} that $\Delta_{L}\simeq(\bdm^{2}_{Q})_{32}$ if we ignore the contribution of the Yukawa and gauge coupling terms, so we must find the weak scale value of $(\bdm^{2}_{Q})_{32}$. To this end, the running of ${\bf{m}}^2_Q$ is written down using a single step:
\begin{equation}
{\bf{m}}^2_Q(\mathrm{Q})\simeq{\bf{m}}^2_Q(\mathrm{GUT})-\ln{\left[\frac{M_{GUT}}{Q}\right]}\frac{d{\bf{m}}^2_Q}{dt}\;,
\end{equation}
where we take $Q$ to be somewhere near the weak scale. We choose to use our ``standard'' current basis so that the up Yukawa matrices are approximately diagonal at the scale $Q$. We then write the down Yukawa couplings in terms of the KM matrix using \eqref{eq:yukbasis} so that we have $\bdf_{d}(d)$, which is also approximately diagonal. In this basis we ignore all terms proportional to off-diagonal entries of the Yukawa and SSB matrices which we assume to be higher order corrections. To further simplify our equation for $\Delta_L$, the masses of the light quarks are set to zero, which makes both $\bdf_{u}(u)$ and $\bdf_{d}(d)$ identically zero except for $(\bdf_{u,d})_{33}$. Under these approximations, the $(3,2)$ entry of ${\bdm}^2_Q(\mathrm{Q})$ reduces to the form \eqref{eq:deltal}. The derivation of $\Delta_R$ is very similar except that the important quantity to run is the left-right coupling, ${(\ba_u)}_{23}$.

Next, we approximate the mixing to be a small perturbation to the eigenstates, so that the physical $\tilde{t}_{1}$, $\tilde{t}_{2}$ and $\tilde{c}_{L}$ states are only changed a small amount by the mixing. The mass matrix is approximately diagonalised to obtain
\begin{equation} \label{eq:hkmix}
{\left( \begin{array}{c} \tilde{t}_1 \\ \tilde{t}_2 \\ \tilde{c}_L \end{array} \right)}_{\mathit{phys}}
=
\left( \begin{array}{ccc}
1 & 0 & \epsilon \\
0 & 1 & {\epsilon}' \\
-\epsilon & -{\epsilon}' & 1 \\
\end{array} \right)
\left( \begin{array}{c} \tilde{t}_1 \\ \tilde{t}_2 \\ \tilde{c}_L \end{array} \right)\;,
\end{equation}
where the small perturbations are found to be
\begin{eqnarray}
\label{eq:hkepsilon}\epsilon & = & \frac{\Delta^{*}_L\cos{\theta_t}-\Delta^{*}_R\sin{\theta_t}}{m^2_{\tilde{t}_1}-m^2_{\tilde{c}_L}} \\
{\epsilon}' & = & \frac{\Delta^{*}_R\cos{\theta_t}+\Delta^{*}_L\sin{\theta_t}}{m^2_{\tilde{t}_2}-m^2_{\tilde{c}_L}}\;.
\end{eqnarray}
The most important relation to take from (\ref{eq:hkmix}) is that of the physical $\tilde{t}_{1}$ state
\begin{equation}
\label{eq:physstop}
\tilde{t}_1(\mathit{phys})=\tilde{t}_1+\epsilon\tilde{c}_L\;.
\end{equation}
The mixing factor, $\epsilon$, is therefore the extent of mixing of the $\tilde{t}_{1}$ with $\tilde{c}_{L}$ \mbox{---} which we denoted in \eqref{eq:squarkrot} by $\left(\bm{\mathcal{U}}_{L}\right)^{\dagger}_{12}$, and estimated in \eqref{eq:mixest}, the form of which is directly analogous to \eqref{eq:hkepsilon}. In mSUGRA scenarios, $\Delta_{L}$ has a size of approximately $$|\Delta_{L}|\sim\frac{8}{16\pi^{2}}\ln{\left[\frac{\mgut}{\msusy}\right]}\mathbf{K}^{*}_{cb}\mathbf{K}_{tb}m^{2}_{0}f^{2}_{b}\;,$$ and the $\tilde{t}_{R}-\tilde{c}_{L}$ mixing angle has a similar magnitude, so that $|\epsilon|\sim|\Delta_{L}|/(\mathit{few}\ m^{2}_{0})\sim\mathit{few}\times10^{-4}$.

Due to the approximations made in this method, the rate calculation in \eqref{eq:twobodgamma} simplifies dramatically to become
\begin{equation}\label{eq:hkgamma}
\Gamma=\frac{1}{16\pi m^{3}_{\tilde{t}_{1}}}\left|\epsilon\right|^{2}\left|A\right|^{2}\left(m^{2}_{\tilde{t}_{1}}-m^2_{\tilde{Z}_{1}}\right)^{2}\;,
\end{equation}
where
\begin{equation}
\left|A\right|^{2}=\frac{1}{2}\left|gv^{(1)}_{3}+\frac{1}{3}g'v^{(1)}_{4}\right|^{2}\;.
\end{equation}

\input{stdec}

\section{Sample Numerical Results}\label{sec:stopres}

\subsection{Single-step RGE Integration and the Stop Decay Rate}

As a first example, displaying the need for a complete solution of the full RGEs, we show results in Table~\ref{tab:msugcomp} for the mSUGRA point $m_0=250$~GeV, $m_{1/2}=-250$~GeV, $A_0=-930$~GeV, $\tan\beta=20$ and $\mu<0$.
\begin{table}
\centering
\begin{tabular}{llr}
&Method&Width (GeV)\\ \hline
$\Gamma(\mathit{\tilde{t}_{1}\rightarrow bW\tilde{Z}_{1}})$&&$8.6\times10^{-8}$\\
$\Gamma(\mathit{\tilde{t}_{1}\rightarrow c\tilde{Z}_{1}})$&``single-step'' estimate&$\sim41\times10^{-8}$\\
&full calculation&$3.3\times10^{-8}$\\
\end{tabular}
\caption[Partial widths for the two- and three-body decays of the $\tilde{t}_{1}$ for a sample mSUGRA point.]{\small Partial widths for the two- and three-body decays of the $\tilde{t}_{1}$ for the mSUGRA parameters: $m_0=250$~GeV, $m_{1/2}=-250$~GeV, $A_0=-930$~GeV, $\tan\beta=20$ and $\mu<0$. The two-body decay width is calculated using two methods, a `single-step' approximation, and our full RGE solution.}\label{tab:msugcomp}
\end{table}
For this point, $m_{\tilde{t}_{1}}\simeq181$~GeV, $m_{\tilde{Z}_{1}}=102$~GeV and $\tilde{W}_{1}=197$~GeV so that tree level two-body decays of the $\tilde{t}_{1}$ are kinematically forbidden but both the two-body decay that we are considering, and the three-body decay to $\mathit{bW\tilde{Z}_{1}}$ \cite{wohr} are open. We see that the `single-step' approximation of the decay width \textit{overestimates} the actual value by a factor $13.6$, the difference being entirely due to the error in calculating the value of $\epsilon$ $(=\left(\bm{\mathcal{U}}_{L}\right)^{\dagger}_{12})$. In the mSUGRA model with a neutral LSP, when the usual flavour-conserving two-body decays are kinematically forbidden, we find that the single-step approximation overestimates by a factor of between $10-25$.

If we used the single-step RGE approximation to find the branching ratios for this point we would conclude $\mathrm{BR}(\mathit{c\tilde{Z}_{1}})\simeq0.83$ and $\mathrm{BR}(\mathit{bW\tilde{Z}_{1}})\simeq0.17$ so that the three-body decay is subdominant. In contrast, when we use the full RGE calculation the three-body decay becomes dominant with $\mathrm{BR}(\mathit{bW\tilde{Z}_{1}})\simeq0.72$, completely changing the qualitative picture of top squark decays. Admittedly, the change in the qualitative picture in this case is due to the three-body decay being close to the kinematic boundary. However, if the three-body decay is closed, the two-body loop decay will compete with various four-body decays, in which case a similar rearranging of the dominant decay modes may result from our RGE calculation of the two-body width.

Note that the width obtained for the two-body decay depends on the choice of scale at which we calculate it. Since the $\tilde{t}_{1}$ is mainly $\tilde{t}_{R}$, we evaluate the decay rate using the parameters at a scale equal to the lightest right-handed squark threshold. We find that if we instead use the scale where the heaviest right-handed squark decouples as opposed to the lightest, there is just a $6\%$ change in the partial width.

\subsection{Model Dependence of $\protect\Gamma(\tilde{t}_{1}\rightarrow\mathit{c\tilde{Z}_{1}})$}

To illustrate the effect of non-universal GUT scale inputs we first return to our sample point from the previous chapter, namely $m_0=200$~GeV, $m_{1/2}=-400$~GeV, $A_0=-200$~GeV, $\tan\beta=10$ and $\mu>0$, and include non-zero $\mathbf{T}_{Q}$ or $\mathbf{T}_{U,D}$ (while simultaneously setting $m^{2}_{Q0}=0$ or $m^{2}_{\{U,D\}0}=0$ respectively) in a variety of scenarios. Our results are shown in Table~\ref{tab:widthTcomp}, beginning with Scenario $(1)$ which is our standard mSUGRA case.
\begin{table}[p]
\centering
\begin{tabular*}{0.95\textwidth}{@{\extracolsep{\fill}}cllc}
\multicolumn{3}{l}{Scenario}&Width\\ \hline
$(1)$&\multicolumn{2}{l}{mSUGRA \mbox{---} no dependence on specific $\mathbf{V}_{L,R}(u,d)$}&$2.2\times10^{-9}$~GeV\\[5pt]
$(2a)$&$\mathbf{V}_{R}(u)=\mathbf{V}_{R}(d)=\mathbf{V}_{L}(u)=\dblone$&$\mathbf{T}_{U,D}\neq\mathbf{0}$&$3.9\times10^{-9}$~GeV\\
$(2b)$&&$\mathbf{T}_{Q}\neq\mathbf{0}$&$1.6\times10^{-9}$~GeV\\[5pt]
$(3a)$&$\mathbf{V}_{L}(u)\neq\dblone,\; \mathbf{V}_{R}(u)=\mathbf{V}_{R}(d)=\dblone$&$\mathbf{T}_{U,D}\neq\mathbf{0}$&$3.9\times10^{-9}$~GeV\\
$(3b)$&&$\mathbf{T}_{Q}\neq\mathbf{0}$&$2.7\times10^{-5}$~GeV\\[5pt]
$(4a)$&$\mathbf{V}_{R}(d)\neq\dblone,\; \mathbf{V}_{R}(u)=\mathbf{V}_{L}(u)=\dblone$&$\mathbf{T}_{U,D}\neq\mathbf{0}$&$3.6\times10^{-9}$~GeV\\
$(4b)$&&$\mathbf{T}_{Q}\neq\mathbf{0}$&$1.6\times10^{-9}$~GeV\\[5pt]
$(5a)$&$\mathbf{V}_{R}(u)\neq\dblone,\; \mathbf{V}_{R}(d)=\mathbf{V}_{L}(u)=\dblone$&$\mathbf{T}_{U,D}\neq\mathbf{0}$&$5.8\times10^{-3}$~GeV\\
$(5b)$&&$\mathbf{T}_{Q}\neq\mathbf{0}$&$1.6\times10^{-9}$~GeV\\[5pt]
$(6a)$&$\mathbf{V}_{R}(u)\neq\dblone,\; \mathbf{V}_{R}(d)\neq\dblone,\; \mathbf{V}_{L}(u)\neq\dblone$&$\mathbf{T}_{U,D}\neq\mathbf{0}$&$5.8\times10^{-3}$~GeV\\
$(6b)$&&$\mathbf{T}_{Q}\neq\mathbf{0}$&$2.7\times10^{-5}$~GeV\\[5pt]
\end{tabular*}
\caption[A comparison of the two-body loop decay widths for six scenarios: mSUGRA, and five non-mSUGRA cases.]{\small A comparison of the two-body loop decay widths for six scenarios. For each scenario we list the basis used for our GUT scale inputs, with rotation matrices as specified by \eqref{eq:genrot} in the text. In case $(a)$ of each scenario, we take $\mathbf{T}_{Q}=\mathbf{0}$ and $\mathbf{T}_{U,D}=\mathit{diag}\{10000,40000,90000\}~\mathrm{GeV}^{2}$. In case $(b)$, $\mathbf{T}_{U,D}=\mathbf{0}$  and $\mathbf{T}_{Q}=\mathit{diag}\{10000,40000,90000\}~\mathrm{GeV}^{2}$. When $\mathbf{T}_{Q,U,D}\neq\mathbf{0}$, the corresponding $m^{2}_{\{Q,U,D\}0}=0$ and the mSUGRA case is the same as in Fig.~\ref{fig:SUGRAfu}. The value of $m_{\tilde{t}_{1}}$ is constant in all scenarios to within about 5\%, so its contribution to the variation of the partial width is small.}\label{tab:widthTcomp}
\end{table}
In mSUGRA, all dependence on the $\mathbf{V}_{L,R}(u,d)$ matrices is restricted to the KM combination, and therefore the result is independent of our range of choices in the other scenarios.  We mention that the scenarios we examine are unrealistic both from the perspective of observing $\tilde{t}_{1}$ decays \mbox{---} indeed, for this sample mSUGRA scenario, tree-level decays of $\tilde{t}_{1}$ are accessible \mbox{---} as well as for that of flavour-changing neutral currents. Our purpose here is to understand how $\Gamma(\tilde{t}_{1}\rightarrow c\tilde{Z}_{1})$ is altered (and also which quantities it is sensitive to) as we change the scenarios to systematically allow increased non-universality and/or flavour-violation in the SSB sector.

As expected for Scenario $(2)$, although there is a significant splitting in the squark masses, the stop width is of the same order of magnitude as the mSUGRA result. We saw in Sec.~\ref{sec:quarkyukadd} that for non-zero $\mathbf{T}_{U,D}$ entered in the ``standard'' current basis, there is little difference to the running of the usual Yukawa couplings and their tilde counterparts, and the same can be said for non-zero $\mathbf{T}_{Q}$ in this basis. The width sees only a small effect because the RGEs for the gauge and Yukawa couplings only depend on the squark masses through the thresholds, and also because the rotation to the squark mass basis is small.\footnote{We emphasise that the situation would be quite different for flavour-violating decays of down-type squarks, since the down-squark matrix is now not aligned with the corresponding Yukawa matrix.} Any deviation from mSUGRA shows up in our diagonalisation of the squark mass matrix, and the difference between the first two scenarios is due to a $\sim15-25\%$ change in the entry $\left(\bm{\mathcal{U}}_{L}\right)^{\dagger}_{12}$. Note that although the eigenvalues of the squark mass matrices at the GUT scale have large splitting, the matrices themselves remain close to diagonal all the way to the GUT scale.

 We can understand the effect of varying $\mathbf{V}_{L,R}(u,d)$, by considering rotating back from this general basis to our ``standard'' current basis. In effect, varying our choice of rotation matrices changes the GUT scale boundary conditions when viewed in our ``standard'' basis. We would therefore only expect to see a change in the width for GUT scale inputs that correspond to a change when viewed in this basis. For example, in Scenario $(3a)$ we have chosen a basis such that $\bdm^{2}_{U,D}$ are unchanged when rotated back into the ``standard'' basis. Therefore entering $\mathbf{T}_{U,D}\neq\mathbf{0}$ in the basis specified in Scenario $(3)$ is the same as entering the same $\mathbf{T}_{U,D}$ in the ``standard'' current basis, \textit{i.e.}, Scenario $(2)$, and we would expect this to be reflected in the resulting width. On the other hand, the boundary condition for $\bdm^{2}_{Q}$ in the current basis does depend on $\mathbf{V}_{L}(u)$, and we would therefore expect the width to be very different between Scenario $(2b)$ and Scenario $(3b)$.

Looking, therefore, at Scenario $(3)$ we see no change as a result of the $\mathbf{T}_{U,D}\neq\mathbf{0}$ input, but a four order of magnitude change as a result of setting $\mathbf{T}_{Q}\neq\mathbf{0}$. As already mentioned, since the GUT scale boundary conditions for $\bdm^{2}_{U,D}$ have no dependence on $\mathbf{V}_{L}(u)$ or $\mathbf{V}_{L}(d)$, but $\bdm^{2}_{Q}$ does, this difference in behaviour between $(3a)$ and $(3b)$ was expected. The large jump in the width for $(3b)$ can be understood as a result of a dramatic increase in the level of $\tilde{t}_{1}-\tilde{c}_{L}$ mixing. We find that the $\mathbf{T}_{Q}\neq\mathbf{0}$ input results in a change in both $\bdm^{2}_{Q}$ and $\bdm^{2}_{U}$. This is because the seed terms at the GUT scale for flavour off-diagonal entries of $\bdm^{2}_{U}$ depend on the structure\footnote{The RGEs in our ``standard'' current basis contain KM factors next to down-type Yukawa couplings only. Therefore, if $\bdm^{2}_{Q}$ is diagonal in the ``standard'' basis, we must use off-diagonal Yukawa coupling terms to obtain off-diagonal seed terms in the $\bdm^{2}_{U}$ RGE. On the other hand, if $\bdm^{2}_{Q}$ has all entries $\mathcal{O}(1)$, the off-diagonal seed terms depend on diagonal Yukawa matrix entries. The same goes for $\bdm^{2}_{U}$ in the $\bdm^{2}_{Q}$ RGE. In contrast, $\bdm^{2}_{D}$ always appears with KM terms in the $\bdm^{2}_{Q}$ RGE in the ``standard'' basis, and therefore we can obtain off-diagonal terms from diagonal Yukawa matrix entries regardless of the form of $\bdm^{2}_{D}$.} of the matrix $\bdm^{2}_{Q}$. However, although the change in the off-diagonal elements of $\bdm^{2}_{U}$ is in this case many orders of magnitude, the effect is still smaller than the increase in $\tilde{t}_{1}-\tilde{c}_{L}$ mixing coming from $\bdm^{2}_{Q}$.

Moving on to Scenario $(4)$, the rotation matrices used in this case mean that the boundary condition for Scenario $(4a)$, the $\mathbf{T}_{U,D}\neq\mathbf{0}$ case, is different only in $\bdm^{2}_{D}$, affecting the running of the $\bdm^{2}_{Q,U}$ matrices to a small extent through the RGEs. As a result, we see that the width is different from the other scenarios by a small factor ($<10\%$). On the other hand, the boundary condition on $\bdm^{2}_{Q}$ is not dependent on $\mathbf{V}_{R}(d)$, and the width in $(4b)$ is therefore unchanged from Scenario $(2b)$.

In Scenario $(5a)$, where we choose to enter the GUT scale matrices, $\mathbf{T}_{U,D}$, in a basis which depends on $\mathbf{V}_{R}(u)$ (thereby altering the GUT scale boundary condition on $\bdm^{2}_{U}$), the largest contribution to the width comes from the fact that the mixing between the right-handed squarks has been greatly increased. In particular, the $\tilde{c}_{R}$ forms a substantial fraction of the $\tilde{t}_{1}$ and couples strongly to the bino-like neutralino. This is due to the fact that the off-diagonal entries of $\bdm^{2}_{U}$ (expressed in our ``standard'' current basis) are non-zero at the GUT scale and remain large during the running to the weak scale as can be seen in Fig.~\ref{fig:T2mu}. Although one might expect the contribution to \eqref{eq:albet} from off-diagonal entries in the higgsino and gaugino `Yukawa' matrices to become significant, the enhancement that they receive is not sufficient to bring them to the same level as the contribution from the $\left(\bm{\mathcal{U}}_{R}\right)^{\dagger}_{12}\mathbf{B}_{2}$ term in $\beta$ which is now two orders of magnitude larger than any other entry in \eqref{eq:albet}. In contrast, Scenario $(5b)$ shows no change from $(2b)$, due to the fact, once again, that the boundary condition is the same in this basis.

Finally, Scenario $(6a)$ is equal to Scenario $(5a)$, with a small difference not shown in the table due to the effect of $\bdm^{2}_{D}$ entering the RGE for $\bdm^{2}_{Q}$. Scenarios $(6b)$ and $(3b)$ coincide because when $\mathbf{T}_{Q}\neq\mathbf{0}$ the GUT scale boundary conditions only differ for different values of $\mathbf{V}_{L}(u,d)$.

In summary, we find that when we take a large amount of splitting between diagonal entries of the SSB scalar mass at the GUT scale, our results are very sensitive to the choice of basis. Although it is possible to choose the rotation matrices carefully enough that the effect can be small, as in Scenarios $(2a)$, $(2b)$, $(3a)$, $(4a)$, $(4b)$ and $(5b)$, when the rotation is completely general as in the two final scenarios in Table~\ref{tab:widthTcomp}, there is the potential for large deviations from the mSUGRA result. These deviations are overwhelmingly due to an increase in the relevant mixings, rather than a result of the difference between the tilde-couplings and their counterparts.

We move on to take a small splitting between the diagonal elements \mbox{---} say, $\left(m_{0}\pm\varepsilon~\mathrm{GeV}\right)^{2}$ \mbox{---} so that we may consider flavour-violation effects in a more realistic scenario. For this case we take the compressed SUSY scenario \cite{martcomp,uscomp}, proposed by Martin, where efficient neutralino annihilation to top pairs via the exchange of a light squark leads to the observed cold dark matter relic density. We use mSUGRA-like GUT scale inputs, where $m_0=500$~GeV, $A_0=M_{1}$, $\tan\beta=10$ and $\mu>0$, but the gaugino masses are split so that $1.5M_{1}=M_{2}=3M_{3}$.

\begin{figure}[t]\begin{centering}
\includegraphics[viewport=15 45 710 555, clip, scale=0.45]{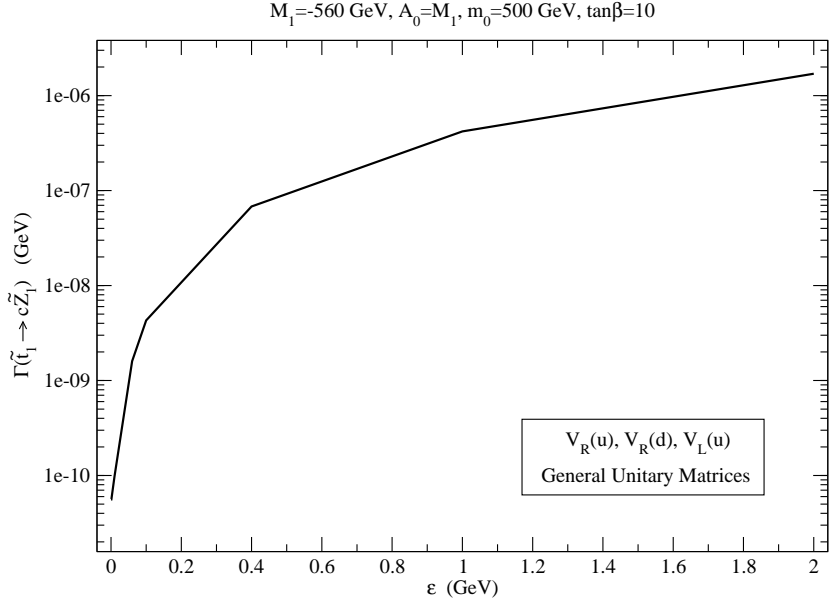}
\caption[Effect on the two-body decay width of giving $\mathbf{T}_{U,D}$ non-equal diagonal entries in the general basis given by \eqref{eq:genrot} for a variety of splittings in a ``compressed SUSY'' scenario.]{\small Effect on the two-body decay width of setting $\bdm^{2}_{U,D}=0$ and $\mathbf{T}_{U}=\mathbf{T}_{D}=\mathit{diag}\{(500-\varepsilon)^{2},500^{2},(500+\varepsilon)^{2}\}$ in the general basis given by \eqref{eq:genrot} in the text. We use the ``compressed SUSY'' GUT scale inputs where $m_0=500$~GeV, $M_1=-560$~GeV, $A_0=M_{1}$, $\tan\beta=10$ and $\mu>0$, and $1.5M_{1}=M_{2}=3M_{3}$. The width is plotted as a function of $\varepsilon$.}
\label{fig:splitinc}
\end{centering}
\end{figure}

Fig.~\ref{fig:splitinc} examines the effect of increasing the size, $\varepsilon$, of the splitting between the diagonal entries of $\mathbf{T}_{U,D}$, where $\mathbf{T}_{U}=\mathbf{T}_{D}=\mathit{diag}\{(500-\varepsilon)^{2},500^{2},(500+\varepsilon)^{2}\}$  (again setting $m^{2}_{\{U,D\}0}=0$) for a fixed value of $M_1=-560$~GeV.
We see that the effect of the general rotation, \eqref{eq:genrot}, on the size of the width rapidly increases once $\varepsilon$ is given a non-zero value, and that even a small splitting of $\varepsilon=1$~GeV increases the width by more than four orders of magnitude. This is \textit{not} the case if we enter such a small splitting in the ``standard'' current basis, where there is no additional flavour structure introduced. Indeed, the significant enhancement of the width is entirely due to our choice of rotation matrices $\mathbf{V}_{L,R}$.

\begin{figure}[t]\begin{centering}
\includegraphics[viewport=15 45 710 555, clip, scale=0.45]{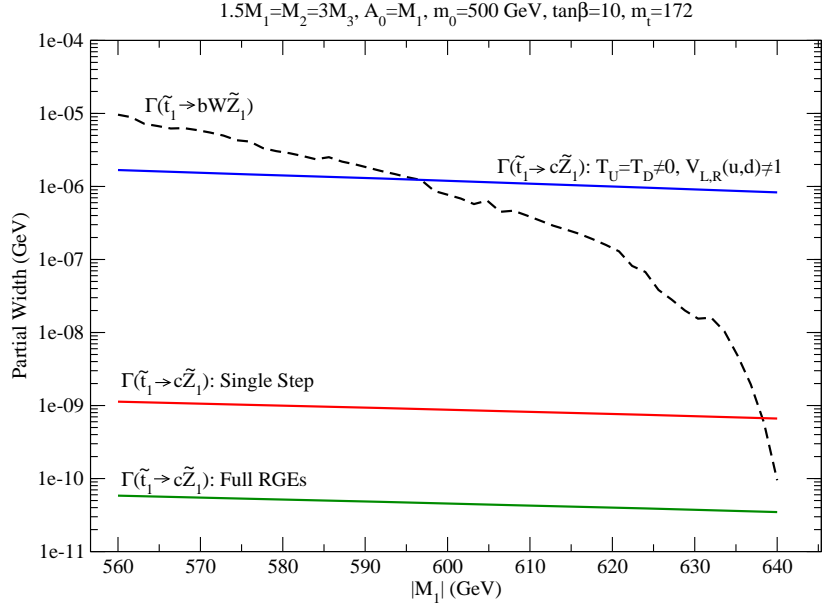}
\caption[Comparison of the partial width of the three-body decay of the stop with the two-body flavour-changing decay in a ``compressed SUSY'' scenario.]{\small Comparison of the partial width of the three-body decay of the stop with the two-body flavour-changing decay in a ``compressed SUSY'' scenario. The GUT scale inputs are the same as Fig.~\ref{fig:splitinc} except that we now vary $M_{1}$ and compare the partial width obtained with $\mathbf{T}_{U,D}=\mathbf{0}$ to that obtained with $\bdm^{2}_{U,D}=0$ and $\mathbf{T}_{U,D}=\mathit{diag}\{(498)^{2},500^{2},(502)^{2}\}$ in the general basis given by \eqref{eq:genrot} in the text. The dashed black line shows the three-body decay $\tilde{t}_{1}\rightarrow\mathit{bW\tilde{Z}_{1}}$ while the other lines represent the various two-body calculations.}
\label{fig:compscan}
\end{centering}
\end{figure}

To see how the model dependence of $\tilde{t}_{1}\rightarrow\mathit{bW\tilde{Z}_{1}}$ can have significant phenomenological implications, we again consider the compressed SUSY scenario of Fig.~\ref{fig:splitinc} and vary $|M_{1}|$ up to $640$~GeV. The result is shown in Fig.~\ref{fig:compscan}.
Over the whole of this region the two-body flavour-conserving decays of the stop, as well as its decays to sneutrinos, are kinematically forbidden, but the three-body decay $\tilde{t}_{1}\rightarrow\mathit{bW\tilde{Z}_{1}}$ is allowed until it reaches its kinematic limit at approx.~$640$~GeV. The black dashed line in the figure is the partial width for the three-body decay which clearly develops a phase space suppression towards the high $|M_{1}|$ end of the graph. The partial two-body width calculated using the single-step method is in the range $10^{-9}$~GeV and competes with the three-body decay only on the extreme edge of phase space. It is interesting to note, however, that the result using full RGEs (which is the lowest line at approx. $10^{-10}$) is even less important in the branching fraction. Although this difference between the single-step result and the full result is unimportant for the three-body decay, it may be important in the case that a four-body decay begins to compete at large $|M_{1}|$.

The highest horizontal two-body line in Fig.~\ref{fig:compscan} represents the result of our complete calculation taking $\mathbf{T}_{U,D}=\mathit{diag}\{498^{2},500^{2},502^{2}\}$ and $m^{2}_{\{U,D\}0}=0$ in the general basis of \eqref{eq:genrot}. We see the expected enhancement from taking a small splitting between diagonal entries in a general current basis, as predicted in Fig.~\ref{fig:splitinc}. In this case, the two-body decay is competitive with the decay $\tilde{t}_{1}\rightarrow\mathit{bW\tilde{Z}_{1}}$ over a much larger range than before, representing the phenomenological implications of introducing dependence in the RGEs on the specifics of the $\mathbf{V}_{L,R}(u,d)$ matrices. We should also mention that if we take this same splitting, but use $\mathbf{V}_{R}(u,d)=\mathbf{V}_{L}(u)=\dblone$, we recover the result with no splitting to within around $4\%$.

%% file: stdec.tex
\section{The Decay Subroutine, \progstd}\label{sec:std}

We pause here for a brief description of the numerical calculation of the two-body decay width, for which we have developed a subroutine called \progstd. As described in Chapter~\ref{ch:rgeflav}, the programme \progrge~runs the full set of RGEs for the dimensionless and dimensionful couplings from the high scale, which in mSUGRA is the GUT scale, to $m_{H}$, the scale at which the heavy Higgs bosons decouple. It does this using an iterative procedure in an attempt to fix the boundary conditions at three scales, $M_{Z}$, $\msusy$ and $\mgut$. At the end of the iteration the couplings are rotated into the basis where the up-type quark Yukawa couplings are diagonal at $m_{t}$\footnote{If the user has chosen for the output to be in the basis where the down-type quark Yukawa couplings are diagonal at $m_{t}$, \progstd~exits without calculating the partial width.} and then passed to \progstd.

\progstd~evolves the couplings to the scale at which the decay is to be calculated, which we choose to be the lightest of the eigenvalues of the left- and right-handed soft mass matrices for the up-type squarks, using the subroutine \texttt{DECRUN}. When the programme arrives at the required scale, the mass matrices are reconstructed using the saved eigenvalues and eigenvectors from the running procedure. In other words, we recreate the mass matrices in the basis where the squarks are diagonal using the eigenvalues of each squark at its decoupling scale. We then rotate from the squark mass basis to the ``standard'' current basis using the known eigenvectors in this basis.

Once we have all the couplings at the correct scale in the quark mass basis, we construct the $(6\times6)$ mass matrix using \eqref{eq:umixmatII} with \eqref{eq:msq} in the subroutine \texttt{CUSQM} \mbox{---} remembering to use the restricted set of couplings if we are calculating the decay at a scale $Q<m_{H}$. The mass matrix is then diagonalised in \texttt{USMMA}, by the \texttt{EISPACK} subroutine \texttt{CG}, to find the eigenvalue and eigenvector of the $\tilde{t}_{1}$.

Finally, the decay rate calculation from (\ref{eq:twobodgamma}) is carried out, unless the kinematics forbid this decay from occurring. To find the rate, we must know $\bgtpqnb$, $\bgtqnb$, $\bgtpurnb$, $\bftuqnb$, and $\bfturnb$, along with the masses and mixings of the squarks and neutralinos. The result is compared to the estimate obtained using the single-step estimate in \eqref{eq:hkgamma} and the partial width for the three-body decay \cite{wohr}, obtained by \progisa.

%% file: summary.tex
\chapter{Summary}\label{ch:summ}

One remarkable property of supersymmetric theories with widely disparate mass scales is that the hierarchy between the scales is stable to radiative corrections. Thus, we can construct models based at the high scale and legitimately calculate their implications at the weak scale. The RGEs provide us with the means to use our high scale theories to evaluate weak scale observables, and as a result of renormalisation group evolution, simple models can lead to a rich structure which will soon be probed at the LHC.

For realistic models in which SUSY is softly broken by an as yet unknown mechanism somewhere below the Planck scale, the fact that sparticle masses are different to their SM counterparts comes with some intriguing consequences. In a purely supersymmetric theory, the dimensionless couplings of gauginos and higgsinos to the fermion-sfermion system are, respectively, related by supersymmetry to the three gauge couplings and to the three Yukawa coupling matrices that specify the interactions of matter fermions with Higgs bosons. These relations, however, are no longer valid once SUSY is broken.

We have reexamined the complete set of RGEs for the MSSM (excluding the quartic scalar couplings), keeping track of threshold corrections and flavour-mixing effects from quarks and squarks. To implement the decoupling of heavy particles we consider a series of effective theories, where the particle content is dictated by the scale, so that the influence of individual particles on renormalisation group evolution is removed at the scale of their mass. It is found that once the heaviest SUSY particle decouples the evolution of gaugino and higgsino couplings is no longer the same as that of their SM counterparts. Indeed, the gaugino couplings to fermions and sfermions develop a matrix flavour structure, something forbidden in unbroken supersymmetry. We have also described how to treat the new complications to squark decoupling due to flavour mixing among the squarks, and to the electroweak symmetry breaking conditions (traditionally used to fix the magnitude of the superpotential $\mu$ from the observed value of $M_{Z}$) from the fact that below the scale of SUSY breaking it is not $\mu^{2}$ but a related parameter $\tilde{\mu}^{2}$ that enters the scalar potential in the Higgs sector.


Our complete set of MSSM RGEs can be found in Appendices~\ref{app:dlessRGEs}~and~\ref{app:dfulRGEs}. We do not list the RGEs for  quartic couplings as these are less important from a phenomenological perspective, though they do enter the RGEs of the SSB mass matrices and trilinear couplings at one-loop. In this case, we set them equal to their supersymmetric value. As mentioned above, the RGEs contain a number of `Yukawa' couplings in addition to those usually written for the MSSM. For these couplings we use the boundary condition that they are equal to the corresponding SM couplings at all scales above the highest SUSY threshold. We write the dimensionful couplings in two regimes \mbox{---} above and below the scale of the heavy Higgs bosons, $m_{H}$. This is because we have a restricted set of couplings and mass terms below $m_{H}$, where the heavy Higgs bosons decouple, and some linear combinations of couplings do not appear in the effective theory.

The full set of threshold RGEs have been assembled into a computer code, \progrge, which also includes two-loop MSSM and SM RGEs. Since we expect the effect of particle decoupling to be of the same order as two-loop sized effects (assuming $\mathcal{O}(1)$ splitting in the sparticle sector), we have placed these two contributions to the running on an equal footing. Our inputs for SSB parameters at the high scale consist of two pieces: the most general form for SSB parameters that does not introduce new flavour-violating structure (\textit{i.e.}, all flavour structure arises from the usual Yukawa coupling terms in the superpotential), and additional terms that allow the use of arbitrary values for the SSB parameters in a general current basis. The RGEs are then solved using an iterative procedure.

Sample results have been presented for a range of SUSY scenarios. We have shown the key features of the introduction of thresholds into the Yukawa and Yukawa-like couplings of Higgs bosons, higgsinos and gauginos to fermions and sfermions for a simple model where the SUSY thresholds are clustered at two distinct scales. We were able to verify our assertion that threshold effects are of the same order as two-loop corrections and also see that flavour off-diagonal entries in the gaugino couplings can be of comparable order or even larger than the corresponding Yukawa coupling terms, making the inclusion of these effects essential to obtain two-loop accuracy of results in the study of models where the Yukawa matrices are the sole source of flavour-violation.

These conclusions apply equally well to more realistic scenarios such as the sample mSUGRA point we have considered, where we have shown the dependence on scale of the SSB trilinear and scalar mass matrices. By introducing non-universal GUT scale boundary conditions, we found that it is easy to obtain weak scale SSB scalar mass matrices whose off-diagonal terms are less than an order of magnitude below the diagonal entries. However, since the Yukawa couplings only ``feel'' the effects of the SSB parameters through the thresholds, they are not greatly affected in the region of the scale of the SUSY masses.

We have also considered the renormalisation of the gaugino mass parameters, and possible deviation (in addition to two-loop effects) from the usual $M_{2}/M_{1}=\alpha_{2}/\alpha_{1}$ relation expected in models with gaugino mass unification. In mSUGRA, we find that the thresholds have an effect on this relation which is comparable to the two-loop contribution. This would be particularly important for any quantitative analysis at a possible $e^{+}e^{-}$ linear collider. Moving on to consider a more extreme model of SUSY, we find that when the particle spectrum is significantly separated out, as in the split SUSY scenario, there is the potential for violation from the $M_{2}/M_{1}$ condition at the $\sim20\%$ level.

The analysis presented here is intended only as a sample of the range of effects that, to our knowledge, have not been previously considered. The introduction of flavour structure in the gaugino couplings could be of particular importance in future studies of flavour-violation in the sparticle sector. In addition, the ability to introduce non-universal high scale inputs, while maintaining strict control over new sources of flavour-violation could be useful for model building.

As a sample of the phenomenological implications of our work, we have considered the two-body flavour-changing decay of the top squark, $\tilde{t}_{1}\rightarrow c\tilde{Z}_{1}$. We found that previous estimates for the partial width in mSUGRA have overestimated by between a factor of $10-25$ so that in regions of parameter space where the two-body decay may compete with three- or four-body decays, our analysis obtains a much smaller branching ratio for the flavour-violating mode. However, we have seen that an introduction of even small amounts of additional flavour structure at the GUT scale can enhance the two-body partial width by many orders of magnitude, resulting in an even greater competition with the kinematically suppressed modes.

As anticipated by the results presented in Ch.~\ref{ch:results}, the decay rates in Sec.~\ref{sec:stopres} are sensitive to the individual matrices, $\mathbf{V}_{L,R}(u,d)$, that enter via the diagonalization of the Yukawa coupling matrices, and not just to the KM matrix. Indeed, this dependence of physics on the separate matrices is the generic situation, while the dependence of physics on just the KM combination that we have become used to from studies within the SM or the mSUGRA frameworks, is true only in very special situations: see Table~\ref{tab:widthTcomp}.\footnote{Although this is well-known to many authors, we stress this  here because there has been occasional confusion about this issue. For a different example, see p. 215 of Ref.~\cite{wss}.}

In conclusion, we find that a prerequisite to a full two-loop level treatment of the RGEs is the introduction of threshold effects such as those presented here. The difference between the usual gauge/Yukawa couplings and their SUSY counterparts, which results from the decoupling of SUSY particles from the effective theory, is of importance in a wide range of phenomenological scenarios. In addition, threshold effects in models with a large misalignment between the SSB matrices and Yukawa coupling matrices can result in large changes to the off-diagonal entries of the Yukawa coupling matrices at the high scale. We have presented the full set of RGEs for the MSSM (excluding quartic couplings) and observed that a consistent treatment of flavour effects in the RGEs can make a significant difference to the widths of flavour-violating decays of supersymmetric particles.

%% file: fullrges.tex
\chapter{RGEs for Dimensionless Couplings}\label{app:dlessRGEs}
This appendix contains the RGEs with full thresholds for the dimensionless couplings of the MSSM with $R$-parity conservation. Note that these RGEs are valid only in the basis in which the SSB sfermion mass matrices are diagonal. In any other basis they must be modified to account for the rotation from this basis to the sfermion mass basis so that the sfermions can be properly decoupled. For more information, see Sec~\ref{sec:sqdec}.

\begin{equation}\begin{split}
{\left(4\pi\right)}^2\frac{d(\sn\bdf_u)_{ij}}{dt}=&\frac{\sn}{2}\left\{3\left[\sn^2\h+\cs^2\Hh\right](\bdf_u\bdf_u^\dagger)_{ik}+\left[\cs^2\h+\sn^2\Hh\right](\bdf_d\bdf_d^\dagger)_{ik}\right.\\
&\left.\quad\ +4\cs^2\left[-\h+\Hh\right](\bdf_d\bdf_d^\dagger)_{ik}\right\}(\bdf_u)_{kj}\\
&+\sn(\bdf_u)_{ik}\left[\sh\sql\bftuq^\dagger_{kl}\bftuq_{lj}+\frac{4}{9}\sbi\sul\bgtpur^*_{kl}\bgtpur^T_{lj}\right.\\
&\left.\qquad\qquad\ \ +\frac{4}{3}\sgl\sul\bgtsur^*_{kl}\bgtsur^T_{lj}\right]\\
&+\frac{\sn}{4}\left[2\sh\suk\bftur_{ik}\bftur^\dagger_{kl}+2\sh\sdk\bftdr_{ik}\bftdr^\dagger_{kl}+3\swi\sqk\bgtq^T_{ik}\bgtq^*_{kl}\right.\\
&\left.\qquad+\frac{1}{9}\sbi\sqk\bgtpq^T_{ik}\bgtpq^*_{kl}+\frac{16}{3}\sgl\sqk\bgtsq^T_{ik}\bgtsq^*_{kl}\right](\bdf_u)_{lj}\\
&+\sn\sh\sqk\left[-3\swi\gthus\bgtq^T_{ik}+\frac{1}{3}\sbi\gtphus\bgtpq^T_{ik}\right]\bftuq_{kj}\\
&-\frac{4}{3}\sn\sbi\sh\suk\gtphus\bftur_{ik}\bgtpur^T_{kj}\\
&+\sn(\bdf_u)_{ij}\left[\left(\sn^2\h+\cs^2\Hh\right)\mathrm{Tr}{\left\{3\bdf_u^\dagger\bdf_u\right\}}+\cs^2\left(\h-\Hh\right)\mathrm{Tr}{\left\{3\bdf_d^\dagger\bdf_d+\bdf_e^\dagger\bdf_e\right\}}\right]\\
&+\frac{\sn}{2}\sh(\bdf_u)_{ij}\left\{3\swi\left[\mgthusq\left(\sn^2\h+\cs^2\Hh\right)+\mgthdsq\left(\cs^2\h-\cs^2\Hh\right)\right]\right.\\
&\left.\qquad\qquad\qquad+\sbi\left[\mgtphusq\left(\sn^2\h+\cs^2\Hh\right)+\mgtphdsq\left(\cs^2\h-\cs^2\Hh\right)\right]\right\}\\
&-\sn(\bdf_u)_{ij}\left[\frac{17}{12}g'^2+\frac{9}{4}g^2_2+8g^2_s\right]\;,
\end{split}\end{equation}
\begin{equation}\begin{split}
{\left(4\pi\right)}^2\frac{d(\cs\bdf_d)_{ij}}{dt}=&\frac{\cs}{2}\left\{3\left[\cs^2\h+\sn^2\Hh\right](\bdf_d\bdf_d^\dagger)_{ik}+\left[\sn^2\h+\cs^2\Hh\right](\bdf_u\bdf_u^\dagger)_{ik}\right.\\
&\left.\quad\ +4\sn^2\left[-\h+\Hh\right](\bdf_u\bdf_u^\dagger)_{ik}\right\}(\bdf_d)_{kj}\\
&+\cs(\bdf_d)_{ik}\left[\sh\sql\bftdq^\dagger_{kl}\bftdq_{lj}+\frac{1}{9}\sbi\sdl\bgtpdr^*_{kl}\bgtpdr^T_{lj}\right.\\
&\left.\qquad\qquad\ \ +\frac{4}{3}\sgl\sdl\bgtsdr^*_{kl}\bgtsdr^T_{lj}\right]\\
&+\frac{\cs}{4}\left[2\sh\suk\bftur_{ik}\bftur^\dagger_{kl}+2\sh\sdk\bftdr_{ik}\bftdr^\dagger_{kl}+3\swi\sqk\bgtq^T_{ik}\bgtq^*_{kl}\right.\\
&\left.\qquad+\frac{1}{9}\sbi\sqk\bgtpq^T_{ik}\bgtpq^*_{kl}+\frac{16}{3}\sgl\sqk\bgtsq^T_{ik}\bgtsq^*_{kl}\right](\bdf_d)_{lj}\\
&+\cs\sh\sqk\left[-3\swi\gthds\bgtq^T_{ik}-\frac{1}{3}\sbi\gtphds\bgtpq^T_{ik}\right]\bftdq_{kj}\\
&-\frac{2}{3}\cs\sbi\sh\sdk\gtphds\bftdr_{ik}\bgtpdr^T_{kj}\\
&+\cs(\bdf_d)_{ij}\left[\sn^2\left(\h-\Hh\right)\mathrm{Tr}{\left\{3\bdf_u^\dagger\bdf_u\right\}}+\left(\cs^2\h+\sn^2\Hh\right)\mathrm{Tr}{\left\{3\bdf_d^\dagger\bdf_d+\bdf_e^\dagger\bdf_e\right\}}\right]\\
&+\frac{\cs}{2}\sh(\bdf_d)_{ij}\left\{3\swi\left[\mgthusq\left(\sn^2\h-\sn^2\Hh\right)+\mgthdsq\left(\cs^2\h+\sn^2\Hh\right)\right]\right.\\
&\left.\qquad\qquad\qquad+\sbi\left[\mgtphusq\left(\sn^2\h-\sn^2\Hh\right)+\mgtphdsq\left(\cs^2\h+\sn^2\Hh\right)\right]\right\}\\
&-\cs(\bdf_d)\left[\frac{5}{12}g'^2+\frac{9}{4}g^2_2+8g^2_s\right]\;,
\end{split}\end{equation}
\begin{equation}\begin{split}
{\left(4\pi\right)}^2\frac{d(\cs\bdf_e)_{ij}}{dt}=&\frac{3}{2}\cs\left[\cs^2\h+\sn^2\Hh\right](\bdf_e\bdf_e^\dagger\bdf_e)_{ij}\\
&+\cs(\bdf_e)_{ik}\left[\sh\sll\bftel^\dagger_{kl}\bftel_{lj}+\sbi\sel\bgtper^*_{kl}\bgtper^T_{lj}\right]\\
&+\frac{\cs}{4}\left[2\sh\sek\bfter_{ik}\bfter^\dagger_{kl}+3\swi\slk\bgtl^T_{ik}\bgtl^*_{kl}\right.\\
&\left.\qquad\ +\sbi\slk\bgtpl^T_{ik}\bgtpl^*_{kl}\right](\bdf_e)_{lj}\\
&+\cs\sh\slk\left[-3\swi\gthds\bgtl^T_{ik}+\sbi\gtphds\bgtpl^T_{ik}\right]\bftel_{kj}\\
&-2\cs\sbi\sh\sek\gtphds\bfter_{ik}\bgtper^T_{kj}\\
&+\cs(\bdf_e)_{ij}\left[\sn^2\left(\h-\Hh\right)\mathrm{Tr}{\left\{3\bdf_u^\dagger\bdf_u\right\}}+\left(\cs^2\h+\sn^2\Hh\right)\mathrm{Tr}{\left\{3\bdf_d^\dagger\bdf_d+\bdf_e^\dagger\bdf_e\right\}}\right]\\
&+\frac{\cs}{2}\sh(\bdf_e)_{ij}\left\{3\swi\left[\mgthusq\left(\sn^2\h-\sn^2\Hh\right)+\mgthdsq\left(\cs^2\h+\sn^2\Hh\right)\right]\right.\\
&\left.\qquad\qquad\qquad+\sbi\left[\mgtphusq\left(\sn^2\h-\sn^2\Hh\right)+\mgtphdsq\left(\cs^2\h+\sn^2\Hh\right)\right]\right\}\\
&-\cs(\bdf_e)_{ij}\left[\frac{15}{4}g'^2+\frac{9}{4}g^2_2\right]\;,
\end{split}\end{equation}
\begin{equation}\begin{split}\label{app:bftuq}
{\left(4\pi\right)}^2\frac{d\bftuq_{ij}}{dt}=&\left[\sn^2\h+\cs^2\Hh\right]\bftuq_{ik}(\bdf_u^\dagger\bdf_u)_{kj}\\
&+\bftuq_{ik}\left[\frac{4}{9}\sbi\sul\bgtpur^*_{kl}\bgtpur^T_{lj}+\frac{4}{3}\sgl\sul\bgtsur^*_{kl}\bgtsur^T_{lj}\right.\\
&\left.\qquad\qquad\ +\sh\sql\bftuq^\dagger_{kl}\bftuq_{lj}\right]\\
&+\frac{3}{2}\sh\bftuq_{ij}\left[\suk\bftur^\dagger_{kl}\bftur_{lk}+\sql\bftuq^\dagger_{kl}\bftuq_{lk}\right]\\
&+\frac{1}{4}\sh\left[\sn^2\h+\cs^2\Hh\right]\bftuq_{ij}\left\{3\swi\mgthusq+\sbi\mgtphusq\right\}\\
&+\left[\sn^2\h+\cs^2\Hh\right]\left\{-3\swi\gthu\bgtq^*_{ik}+\frac{1}{3}\sbi\gtphu\bgtpq^*_{ik}\right\}(\bdf_u)_{kj}\\
&-\frac{4}{9}\sbi\sul\bgtpq^*_{ik}\bftur_{kl}\bgtpur^T_{lj}-\frac{16}{3}\sgl\sul\bgtsq^*_{ik}\bftur_{kl}\bgtsur^T_{lj}\\
&+\sql\left[\frac{3}{2}\swi\bgtq^*_{ik}\bgtq^T_{kl}+\frac{1}{18}\sbi\bgtpq^*_{ik}\bgtpq^T_{kl}+\frac{8}{3}\sgl\bgtsq^*_{ik}\bgtsq^T_{kl}\right]\bftuq_{lj}\\
&+\sh\sql\left[\bftuq_{ik}\bftuq^\dagger_{kl}+\bftdq_{ik}\bftdq^\dagger_{kl}\right]\bftuq_{lj}\\
&-\bftuq_{ij}\left[\frac{25}{12}g'^2+\frac{9}{4}g^2_2+4g^2_s\right]\;,
\end{split}\end{equation}
\begin{equation}\begin{split}
{\left(4\pi\right)}^2\frac{d\bftdq_{ij}}{dt}=&\left[\cs^2\h+\sn^2\Hh\right]\bftdq_{ik}(\bdf_d^\dagger\bdf_d)_{kj}\\
&+\bftdq_{ik}\left[\frac{1}{9}\sbi\sdl\bgtpdr^*_{kl}\bgtpdr^T_{lj}+\frac{4}{3}\sgl\sdl\bgtsdr^*_{kl}\bgtsdr^T_{lj}\right.\\
&\left.\qquad\qquad\ +\sh\sql\bftdq^\dagger_{kl}\bftdq_{lj}\right]\\
&+\frac{1}{2}\sh\bftdq_{ij}\left[3\sdk\bftdr^\dagger_{kl}\bftdr_{lk}+\sek\bfter^\dagger_{kl}\bfter_{lk}\right.\\
&\left.\qquad\qquad\qquad+3\sql\bftdq^\dagger_{kl}\bftdq_{lk}+\sll\bftel^\dagger_{kl}\bftel_{lk}\right]\\
&+\frac{1}{4}\sh\left[\cs^2\h+\sn^2\Hh\right]\bftdq_{ij}\left\{3\swi\mgthdsq+\sbi\mgtphdsq\right\}\\
&+\left[\cs^2\h+\sn^2\Hh\right]\left\{-3\swi\gthd\bgtq^*_{ik}-\frac{1}{3}\sbi\gtphd\bgtpq^*_{ik}\right\}(\bdf_d)_{kj}\\
&+\frac{2}{9}\sbi\sdl\bgtpq^*_{ik}\bftdr_{kl}\bgtpdr^T_{lj}-\frac{16}{3}\sgl\sdl\bgtsq^*_{ik}\bftdr_{kl}\bgtsdr^T_{lj}\\
&+\sql\left[\frac{3}{2}\swi\bgtq^*_{ik}\bgtq^T_{kl}+\frac{1}{18}\sbi\bgtpq^*_{ik}\bgtpq^T_{kl}+\frac{8}{3}\sgl\bgtsq^*_{ik}\bgtsq^T_{kl}\right]\bftdq_{lj}\\
&+\sh\sql\left[\bftuq_{ik}\bftuq^\dagger_{kl}+\bftdq_{ik}\bftdq^\dagger_{kl}\right]\bftdq_{lj}\\
&-\bftdq_{ij}\left[\frac{13}{12}g'^2+\frac{9}{4}g^2_2+4g^2_s\right]\;,
\end{split}\end{equation}
\begin{equation}\begin{split}
{\left(4\pi\right)}^2\frac{d\bftel_{ij}}{dt}=&\left[\cs^2\h+\sn^2\Hh\right]\bftel_{ik}(\bdf_e^\dagger\bdf_e)_{kj}\\
&+\bftel_{ik}\left[\sbi\sel\bgtper^*_{kl}\bgtper^T_{lj}+\sh\sll\bftel^\dagger_{kl}\bftel_{lj}\right]\\
&+\frac{1}{2}\sh\bftel_{ij}\left[3\sdk\bftdr^\dagger_{kl}\bftdr_{lk}+\sek\bfter^\dagger_{kl}\bfter_{lk}\right.\\
&\left.\qquad\qquad\qquad+3\sql\bftdq^\dagger_{kl}\bftdq_{lk}+\sll\bftel^\dagger_{kl}\bftel_{lk}\right]\\
&+\frac{1}{4}\sh\left[\cs^2\h+\sn^2\Hh\right]\bftel_{ij}\left\{3\swi\mgthdsq+\sbi\mgtphdsq\right\}\\
&+\left[\cs^2\h+\sn^2\Hh\right]\left\{-3\swi\gthd\bgtl^*_{ik}+\sbi\gtphd\bgtpl^*_{ik}\right\}(\bdf_e)_{kj}\\
&-2\sbi\sel\bgtpl^*_{ik}\bfter_{kl}\bgtper^T_{lj}\\
&+\frac{1}{2}\sll\left[3\swi\bgtl^*_{ik}\bgtl^T_{kl}+\sbi\bgtpl^*_{ik}\bgtpl^T_{kl}\right]\bftel_{lj}\\
&+\sh\sll\bftel_{ik}\bftel^\dagger_{kl}\bftel_{lj}-\bftel_{ij}\left[\frac{15}{4}g'^2+\frac{9}{4}g^2_2\right]\;,
\end{split}\end{equation}
\begin{equation}\begin{split}\label{app:bftur}
{\left(4\pi\right)}^2\frac{d\bftur_{ij}}{dt}=&\frac{3}{2}\sh\left[\suk\bftur_{lk}\bftur^\dagger_{kl}+\sql\bftuq_{lk}\bftuq^\dagger_{kl}\right]\bftur_{ij}\\
&+\frac{1}{4}\sh\left[\sn^2\h+\cs^2\Hh\right]\bftur_{ij}\left\{3\swi\mgthusq+\sbi\mgtphusq\right\}\\
&+\frac{1}{2}\left\{\left[\sn^2\h+\cs^2\Hh\right](\bdf_u\bdf_u^\dagger)_{ik}+\left[\cs^2\h+\sn^2\Hh\right](\bdf_d\bdf_d^\dagger)_{ik}\right\}\bftur_{kj}\\
&+\frac{1}{2}\left[\sh\suk\bftur_{ik}\bftur^\dagger_{kl}+\sh\sdk\bftdr_{ik}\bftdr^\dagger_{kl}+\frac{3}{2}\swi\sqk\bgtq^T_{ik}\bgtq^*_{kl}\right.\\
&\left.\qquad\ +\frac{1}{18}\sbi\sqk\bgtpq^T_{ik}\bgtpq^*_{kl}+\frac{8}{3}\sgl\sqk\bgtsq^T_{ik}\bgtsq^*_{kl}\right]\bftur_{lj}\\
&-\frac{4}{9}\sbi\sqk\bgtpq^T_{ik}\bftuq_{kl}\bgtpur^*_{lj}-\frac{16}{3}\sgl\sqk\bgtsq^T_{ik}\bftuq_{kl}\bgtsur^*_{lj}\\
&-\frac{4}{3}\sbi\left[\sn^2\h+\cs^2\Hh\right]\gtphu(\bdf_u)_{ik}\bgtpur^*_{kj}+2\sh\suk\bftur_{ik}\bftur^\dagger_{kl}\bftur_{lj}\\
&+\suk\bftur_{ik}\left[\frac{8}{9}\sbi\bgtpur^T_{kl}\bgtpur^*_{lj}+\frac{8}{3}\sgl\bgtsur^T_{kl}\bgtsur^*_{lj}\right]\\
&-\bftur_{ij}\left[\frac{5}{6}g'^2+\frac{9}{2}g^2_2+4g^2_s\right]\;,
\end{split}\end{equation}
\begin{equation}\begin{split}
{\left(4\pi\right)}^2\frac{d\bftdr_{ij}}{dt}=&\frac{1}{2}\sh\left[3\sdk\bftdr_{lk}\bftdr^\dagger_{kl}+3\sql\bftdq_{lk}\bftdq^\dagger_{kl}+\sek\bfter_{lk}\bfter^\dagger_{kl}\right.\\
&\left.\qquad+\sll\bftel_{lk}\bftel^\dagger_{kl}\right]\bftdr_{ij}\\
&+\frac{1}{4}\sh\left[\cs^2\h+\sn^2\Hh\right]\bftdr_{ij}\left\{3\swi\mgthdsq+\sbi\mgtphdsq\right\}\\
&+\frac{1}{2}\left\{\left[\sn^2\h+\cs^2\Hh\right](\bdf_u\bdf_u^\dagger)_{ik}+\left[\cs^2\h+\sn^2\Hh\right](\bdf_d\bdf_d^\dagger)_{ik}\right\}\bftdr_{kj}\\
&+\frac{1}{2}\left[\sh\suk\bftur_{ik}\bftur^\dagger_{kl}+\sh\sdk\bftdr_{ik}\bftdr^\dagger_{kl}+\frac{3}{2}\swi\sqk\bgtq^T_{ik}\bgtq^*_{kl}\right.\\
&\left.\qquad\ +\frac{1}{18}\sbi\sqk\bgtpq^T_{ik}\bgtpq^*_{kl}+\frac{8}{3}\sgl\sqk\bgtsq^T_{ik}\bgtsq^*_{kl}\right]\bftdr_{lj}\\
&+\frac{2}{9}\sbi\sqk\bgtpq^T_{ik}\bftdq_{kl}\bgtpdr^*_{lj}-\frac{16}{3}\sgl\sqk\bgtsq^T_{ik}\bftdq_{kl}\bgtsdr^*_{lj}\\
&-\frac{2}{3}\sbi\left[\cs^2\h+\sn^2\Hh\right]\gtphd(\bdf_d)_{ik}\bgtpdr^*_{kj}+2\sh\sdk\bftdr_{ik}\bftdr^\dagger_{kl}\bftdr_{lj}\\
&+\sdk\bftdr_{ik}\left[\frac{2}{9}\sbi\bgtpdr^T_{kl}\bgtpdr^*_{lj}+\frac{8}{3}\sgl\bgtsdr^T_{kl}\bgtsdr^*_{lj}\right]\\
&-\bftdr_{ij}\left[\frac{5}{6}g'^2+\frac{9}{2}g^2_2+4g^2_s\right]\;,
\end{split}\end{equation}
\begin{equation}\begin{split}
{\left(4\pi\right)}^2\frac{d\bfter_{ij}}{dt}=&\frac{1}{2}\sh\left[3\sdk\bftdr_{lk}\bftdr^\dagger_{kl}+3\sql\bftdq_{lk}\bftdq^\dagger_{kl}+\sek\bfter_{lk}\bfter^\dagger_{kl}\right.\\
&\left.\qquad+\sll\bftel_{lk}\bftel^\dagger_{kl}\right]\bfter_{ij}\\
&+\frac{1}{4}\sh\left[\cs^2\h+\sn^2\Hh\right]\bfter_{ij}\left\{3\swi\mgthdsq+\sbi\mgtphdsq\right\}\\
&+\frac{1}{2}\left[\cs^2\h+\sn^2\Hh\right](\bdf_e\bdf_e^\dagger)_{ik}\bfter_{kj}\\
&+\frac{1}{2}\left[\sh\sek\bfter_{ik}\bfter^\dagger_{kl}+\frac{3}{2}\swi\slk\bgtl^T_{ik}\bgtl^*_{kl}\right.\\
&\left.\qquad\ +\frac{1}{2}\sbi\slk\bgtpl^T_{ik}\bgtpl^*_{kl}\right]\bfter_{lj}\\
&-2\sbi\slk\bgtpl^T_{ik}\bftel_{kl}\bgtper^*_{lj}-2\sbi\left[\cs^2\h+\sn^2\Hh\right]\gtphd(\bdf_e)_{ik}\bgtper^*_{kj}\\
&+2\sh\sek\bfter_{ik}\bfter^\dagger_{kl}\bfter_{lj}+2\sbi\sek\bfter_{ik}\bgtper^T_{kl}\bgtper^*_{lj}\\
&-\bfter_{ij}\left[\frac{3}{2}g'^2+\frac{9}{2}g^2_2\right]\;,
\end{split}\end{equation}
\begin{equation}\begin{split}\label{app:bgtpq}
{\left(4\pi\right)}^2\frac{d\bgtpq_{ij}}{dt}=&\frac{1}{2}\sbi\left[\frac{1}{3}\sql\bgtpq_{lk}\bgtpq^\dagger_{kl}+\sll\bgtpl_{lk}\bgtpl^\dagger_{kl}+\frac{8}{3}\suk\bgtpur_{lk}\bgtpur^\dagger_{kl}\right.\\
&\left.\qquad\ +\frac{2}{3}\sdk\bgtpdr_{lk}\bgtpdr^\dagger_{kl}+2\sek\bgtper_{lk}\bgtper^\dagger_{kl}\right]\bgtpq_{ij}\\
&+\frac{1}{2}\sh\sbi\bgtpq_{ij}\left\{\left[\sn^2\h+\cs^2\Hh\right]\mgtphusq+\left[\cs^2\h+\sn^2\Hh\right]\mgtphdsq\right\}\\
&+\frac{1}{2}\bgtpq_{ik}\left\{\left[\sn^2\h+\cs^2\Hh\right](\bdf_u^*\bdf_u^T)_{kj}+\left[\cs^2\h+\sn^2\Hh\right](\bdf_d^*\bdf_d^T)_{kj}\right\}\\
&+\frac{1}{2}\bgtpq_{ik}\left[\sh\sul\bftur^*_{kl}\bftur^T_{lj}+\sh\sdl\bftdr^*_{kl}\bftdr^T_{lj}\right.\\
&\qquad\qquad\quad\ +\frac{3}{2}\swi\sql\bgtq^\dagger_{kl}\bgtq_{lj}+\frac{1}{18}\sbi\sql\bgtpq^\dagger_{kl}\bgtpq_{lj}\\
&\left.\qquad\qquad\quad\ +\frac{8}{3}\sgl\sql\bgtsq^\dagger_{kl}\bgtsq_{lj}\right]\\
&+4\sh\left[-2\sul\bftuq^*_{ik}\bgtpur_{kl}\bftur^T_{lj}+\sdl\bftdq^*_{ik}\bgtpdr_{kl}\bftdr^T_{lj}\right]\\
&+6\sh\left\{\left[\sn^2\h+\cs^2\Hh\right]\gtphu\bftuq^*_{ik}(\bdf_u)^T_{kj}-\left[\cs^2\h+\sn^2\Hh\right]\gtphd\bftdq^*_{ik}(\bdf_d)^T_{kj}\right\}\\
&+\frac{1}{2}\sql\left[3\swi\bgtq_{ik}\bgtq^\dagger_{kl}+\frac{1}{9}\sbi\bgtpq_{ik}\bgtpq^\dagger_{kl}\right.\\
&\left.\qquad\qquad+\frac{16}{3}\sgl\bgtsq_{ik}\bgtsq^\dagger_{kl}\right]\bgtpq_{lj}\\
&+\sh\sql\left[\bftuq^*_{ik}\bftuq^T_{kl}+\bftdq^*_{ik}\bftdq^T_{kl}\right]\bgtpq_{lj}\\
&-\bgtpq_{ij}\left[\frac{1}{12}g'^2+\frac{9}{4}g^2_2+4g^2_s\right]\;,
\end{split}\end{equation}
\begin{equation}\begin{split}
{\left(4\pi\right)}^2\frac{d\bgtpl_{ij}}{dt}=&\frac{1}{2}\sbi\left[\frac{1}{3}\sql\bgtpq_{lk}\bgtpq^\dagger_{kl}+\sll\bgtpl_{lk}\bgtpl^\dagger_{kl}+\frac{8}{3}\suk\bgtpur_{lk}\bgtpur^\dagger_{kl}\right.\\
&\left.\qquad\ +\frac{2}{3}\sdk\bgtpdr_{lk}\bgtpdr^\dagger_{kl}+2\sek\bgtper_{lk}\bgtper^\dagger_{kl}\right]\bgtpl_{ij}\\
&+\frac{1}{2}\sh\sbi\bgtpl_{ij}\left\{\left[\sn^2\h+\cs^2\Hh\right]\mgtphusq+\left[\cs^2\h+\sn^2\Hh\right]\mgtphdsq\right\}\\
&+\frac{1}{2}\bgtpl_{ik}\left[\cs^2\h+\sn^2\Hh\right](\bdf_e^*\bdf_e^T)_{kj}\\
&+\frac{1}{2}\bgtpl_{ik}\left[\sh\sel\bfter^*_{kl}\bfter^T_{lj}+\frac{3}{2}\swi\sll\bgtl^\dagger_{kl}\bgtl_{lj}\right.\\
&\left.\qquad\qquad\quad\ +\frac{1}{2}\sbi\sll\bgtpl^\dagger_{kl}\bgtpl_{lj}\right]\\
&-4\sh\sel\bftel^*_{ik}\bgtper_{kl}\bfter^T_{lj}+2\sh\left[\cs^2\h+\sn^2\Hh\right]\gtphd\bftel^*_{ik}(\bdf_e)^T_{kj}\\
&+\frac{1}{2}\sll\left[3\swi\bgtl_{ik}\bgtl^\dagger_{kl}+\sbi\bgtpl_{ik}\bgtpl^\dagger_{kl}\right]\bgtpl_{lj}\\
&+\sh\sll\bftel^*_{ik}\bftel^T_{kl}\bgtpl_{lj}-\bgtpl_{ij}\left[\frac{3}{4}g'^2+\frac{9}{4}g^2_2\right]\;,
\end{split}\end{equation}
\begin{equation}\begin{split}\label{app:bgtpur}
{\left(4\pi\right)}^2\frac{d\bgtpur_{ij}}{dt}=&\left[\sn^2\h+\cs^2\Hh\right](\bdf_u^T\bdf_u^*)_{ik}\bgtpur_{kj}\\
&+\left[\frac{4}{9}\sbi\suk\bgtpur_{ik}\bgtpur^\dagger_{kl}+\frac{4}{3}\sgl\suk\bgtsur_{ik}\bgtsur^\dagger_{kl}\right.\\
&\left.\quad\ +\sh\sqk\bftuq^T_{ik}\bftuq^*_{kl}\right]\bgtpur_{lj}\\
&+\frac{1}{2}\sbi\bgtpur_{ij}\left[\frac{1}{3}\sql\bgtpq^\dagger_{kl}\bgtpq_{lk}+\sll\bgtpl^\dagger_{kl}\bgtpl_{lk}\right.\\
&\qquad\qquad\qquad\quad+\frac{8}{3}\suk\bgtpur^\dagger_{kl}\bgtpur_{lk}+\frac{2}{3}\sdk\bgtpdr^\dagger_{kl}\bgtpdr_{lk}\\
&\left.\qquad\qquad\qquad\quad+2\sek\bgtper^\dagger_{kl}\bgtper_{lk}\right]\\
&+\frac{1}{2}\sbi\sh\bgtpur_{ij}\left\{\left[\sn^2\h+\cs^2\Hh\right]\mgtphusq+\left[\cs^2\h+\sn^2\Hh\right]\mgtphdsq\right\}\\
&-3\sbi\sh\left[\sn^2\h+\cs^2\Hh\right]\gtphu(\bdf_u)^T_{ik}\bftur^*_{kj}\\
&-\sh\sqk\bftuq^T_{ik}\bgtpq_{kl}\bftur^*_{lj}+2\sh\suk\bgtpur_{ik}\bftur^T_{kl}\bftur^*_{lj}\\
&+\suk\bgtpur_{ik}\left[\frac{8}{9}\sbi\bgtpur^\dagger_{kl}\bgtpur_{lj}+\frac{8}{3}\sgl\bgtsur^\dagger_{kl}\bgtsur_{lj}\right]\\
&-\bgtpur_{ij}\left[\frac{4}{3}g'^2+4g^2_s\right]\;,
\end{split}\end{equation}
\begin{equation}\begin{split}
{\left(4\pi\right)}^2\frac{d\bgtpdr_{ij}}{dt}=&\left[\cs^2\h+\sn^2\Hh\right](\bdf_d^T\bdf_d^*)_{ik}\bgtpdr_{kj}\\
&+\left[\frac{1}{9}\sbi\sdk\bgtpdr_{ik}\bgtpdr^\dagger_{kl}+\frac{4}{3}\sgl\sdk\bgtsdr_{ik}\bgtsdr^\dagger_{kl}\right.\\
&\left.\quad\ +\sh\sqk\bftdq^T_{ik}\bftdq^*_{kl}\right]\bgtpdr_{lj}\\
&+\frac{1}{2}\sbi\bgtpdr_{ij}\left[\frac{1}{3}\sql\bgtpq^\dagger_{kl}\bgtpq_{lk}+\sll\bgtpl^\dagger_{kl}\bgtpl_{lk}\right.\\
&\qquad\qquad\qquad\quad+\frac{8}{3}\suk\bgtpur^\dagger_{kl}\bgtpur_{lk}+\frac{2}{3}\sdk\bgtpdr^\dagger_{kl}\bgtpdr_{lk}\\
&\left.\qquad\qquad\qquad\quad+2\sek\bgtper^\dagger_{kl}\bgtper_{lk}\right]\\
&+\frac{1}{2}\sbi\sh\bgtpdr_{ij}\left\{\left[\sn^2\h+\cs^2\Hh\right]\mgtphusq+\left[\cs^2\h+\sn^2\Hh\right]\mgtphdsq\right\}\\
&-6\sbi\sh\left[\cs^2\h+\sn^2\Hh\right]\gtphd(\bdf_d)^T_{ik}\bftdr^*_{kj}\\
&+2\sh\sqk\bftdq^T_{ik}\bgtpq_{kl}\bftdr^*_{lj}+2\sh\sdk\bgtpdr_{ik}\bftdr^T_{kl}\bftdr^*_{lj}\\
&+\sdk\bgtpdr_{ik}\left[\frac{2}{9}\sbi\bgtpdr^\dagger_{kl}\bgtpdr_{lj}+\frac{8}{3}\sgl\bgtsdr^\dagger_{kl}\bgtsdr_{lj}\right]\\
&-\bgtpdr_{ij}\left[\frac{1}{3}g'^2+4g^2_s\right]\;,
\end{split}\end{equation}
\begin{equation}\begin{split}
{\left(4\pi\right)}^2\frac{d\bgtper_{ij}}{dt}=&\left[\cs^2\h+\sn^2\Hh\right](\bdf_e^T\bdf_e^*)_{ik}\bgtper_{kj}\\
&+\left[\sbi\sek\bgtper_{ik}\bgtper^\dagger_{kl}+\sh\slk\bftel^T_{ik}\bftel^*_{kl}\right]\bgtper_{lj}\\
&+\frac{1}{2}\sbi\bgtper_{ij}\left[\frac{1}{3}\sql\bgtpq^\dagger_{kl}\bgtpq_{lk}+\sll\bgtpl^\dagger_{kl}\bgtpl_{lk}\right.\\
&\qquad\qquad\qquad\quad+\frac{8}{3}\suk\bgtpur^\dagger_{kl}\bgtpur_{lk}+\frac{2}{3}\sdk\bgtpdr^\dagger_{kl}\bgtpdr_{lk}\\
&\left.\qquad\qquad\qquad\quad+2\sek\bgtper^\dagger_{kl}\bgtper_{lk}\right]\\
&+\frac{1}{2}\sbi\sh\bgtper_{ij}\left\{\left[\sn^2\h+\cs^2\Hh\right]\mgtphusq+\left[\cs^2\h+\sn^2\Hh\right]\mgtphdsq\right\}\\
&-2\sbi\sh\left[\cs^2\h+\sn^2\Hh\right]\gtphd(\bdf_e)^T_{ik}\bfter^*_{kj}\\
&-2\sh\slk\bftel^T_{ik}\bgtpl_{kl}\bfter^*_{lj}+2\sh\sek\bgtper_{ik}\bfter^T_{kl}\bfter^*_{lj}\\
&+2\sbi\sek\bgtper_{ik}\bgtper^\dagger_{kl}\bgtper_{lj}-3\bgtper_{ij}g'^2\;,
\end{split}\end{equation}
\begin{equation}\begin{split}
{\left(4\pi\right)}^2\frac{d\left(\sn\gtphu\right)}{dt}=&\frac{2}{4}\sbi\left[\frac{1}{3}\sql\bgtpq_{lk}\bgtpq^\dagger_{kl}+\sll\bgtpl_{lk}\bgtpl^\dagger_{kl}+\frac{8}{3}\suk\bgtpur_{lk}\bgtpur^\dagger_{kl}\right.\\
&\left.\qquad\ +\frac{2}{3}\sdk\bgtpdr_{lk}\bgtpdr^\dagger_{kl}+2\sek\bgtper_{lk}\bgtper^\dagger_{kl}\right]\sn\gtphu\\
&+\frac{1}{2}\sh\left\{\left[\sn^2\h+\cs^2\Hh\right]\mgtphusq+\left[\cs^2\h+\sn^2\Hh\right]\mgtphdsq\right\}\sn\gtphu\\
&+\frac{1}{2}\sh\left[3\suk\sn\gtphu\bftur^*_{lk}\bftur^{T}_{kl}+3\sql\sn\gtphu\bftuq^*_{lk}\bftuq^T_{kl}\right]\\
&+\frac{1}{4}\sh\left[\sn^2\h+\cs^2\Hh\right]\sn\gtphu\left\{3\swi\mgthusq+\sbi\mgtphusq\right\}\\
&+2\sql\sn\bgtpq_{lk}(\bdf_u)^*_{km}\bftuq^T_{ml}-8\suk\sn\bftur^{T}_{km}(\bdf_u)^*_{ml}\bgtpur_{lk}\\
&+\sn\gtphu\left[\left(\sn^2\h+\cs^2\Hh\right)\mathrm{Tr}\{3\bdf_u^\dagger\bdf_u\}+\cs^2\left(\h-\Hh\right)\mathrm{Tr}\left\{3\bdf_d^\dagger\bdf_d+\bdf_e^\dagger\bdf_e\right\}\right]\\
&+\frac{1}{2}\sh\sn\gtphu\left\{3\swi\left[\left(\sn^2\h+\cs^2\Hh\right)\mgthusq+\cs^2\left(\h-\Hh\right)\mgthdsq\right]\right.\\
&\left.\qquad\qquad\qquad+\sbi\left[\left(\sn^2\h+\cs^2\Hh\right)\mgtphusq+\cs^2\left(\h-\Hh\right)\mgtphdsq\right]\right\}\\
&-\sn\gtphu\left[\frac{3}{4}g'^2+\frac{9}{4}g^2_2\right]\;,
\end{split}\end{equation}
\begin{equation}\begin{split}
{\left(4\pi\right)}^2\frac{d\left(\cs\gtphd\right)}{dt}=&\frac{2}{4}\sbi\left[\frac{1}{3}\sql\bgtpq_{lk}\bgtpq^\dagger_{kl}+\sll\bgtpl_{lk}\bgtpl^\dagger_{kl}+\frac{8}{3}\suk\bgtpur_{lk}\bgtpur^\dagger_{kl}\right.\\
&\left.\qquad\ +\frac{2}{3}\sdk\bgtpdr_{lk}\bgtpdr^\dagger_{kl}+2\sek\bgtper_{lk}\bgtper^\dagger_{kl}\right]\cs\gtphd\\
&+\frac{1}{2}\sh\left\{\left[\sn^2\h+\cs^2\Hh\right]\mgtphusq+\left[\cs^2\h+\sn^2\Hh\right]\mgtphdsq\right\}\cs\gtphd\\
&+\frac{1}{2}\sh\left[3\sdk\cs\gtphd\bftdr^*_{lk}\bftdr^{T}_{kl}+\sek\cs\gtphd\bfter^*_{lk}\bfter^{T}_{kl}\right.\\
&\left.\qquad\qquad+3\sql\cs\gtphd\bftdq^*_{lk}\bftdq^T_{kl}+\sll\cs\gtphd\bftel^*_{lk}\bftel^T_{kl}\right]\\
&+\frac{1}{4}\sh\left[\cs^2\h+\sn^2\Hh\right]\cs\gtphd\left\{3\swi\mgthdsq+\sbi\mgtphdsq\right\}\\
&+2\left[-\sql\cs\bgtpq_{lk}(\bdf_d)^*_{km}\bftdq^T_{ml}+\sll\cs\bgtpl_{lk}(\bdf_e)^*_{km}\bftel^T_{ml}\right]\\
&-4\left[\sdk\cs\bftdr^{T}_{km}(\bdf_d)^*_{ml}\bgtpdr_{lk}+\sek\cs\bfter_{km}(\bdf_e)^\dagger_{ml}\bgtper_{lk}\right]\\
&+\cs\gtphd\left[\sn^2\left(\h-\Hh\right)\mathrm{Tr}\{3\bdf_u^\dagger\bdf_u\}+\left(\cs^2\h+\sn^2\Hh\right)\mathrm{Tr}\left\{3\bdf_d^\dagger\bdf_d+\bdf_e^{\dagger}\bdf_e\right\}\right]\\
&+\frac{1}{2}\sh\cs\gtphd\left\{3\swi\left[\sn^2\left(\h-\Hh\right)\mgthusq+\left(\cs^2\h+\sn^2\Hh\right)\mgthdsq\right]\right.\\
&\left.\qquad\qquad\qquad+\sbi\left[\sn^2\left(\h-\Hh\right)\mgtphusq+\left(\cs^2\h+\sn^2\Hh\right)\mgtphdsq\right]\right\}\\
&-\cs\gtphd\left[\frac{3}{4}g'^2+\frac{9}{4}g^2_2\right]\;,
\end{split}\end{equation}
\begin{equation}\begin{split}\label{app:bgtq}
{\left(4\pi\right)}^2\frac{d\bgtq_{ij}}{dt}=&\frac{1}{2}\swi\left[3\sql\bgtq_{lk}\bgtq^\dagger_{kl}+\sll\bgtl_{lk}\bgtl^\dagger_{kl}\right]\bgtq_{ij}\\
&+\frac{1}{2}\sh\swi\bgtq_{ij}\left\{\left[\sn^2\h+\cs^2\Hh\right]\mgthusq+\left[\cs^2\h+\sn^2\Hh\right]\mgthdsq\right\}\\
&+\frac{1}{2}\bgtq_{ik}\left\{\left[\sn^2\h+\cs^2\Hh\right](\bdf_u^*\bdf_u^T)_{kj}+\left[\cs^2\h+\sn^2\Hh\right](\bdf_d^*\bdf_d^T)_{kj}\right\}\\
&+\frac{1}{2}\bgtq_{ik}\left[\sh\sul\bftur^*_{kl}\bftur^T_{lj}+\sh\sdl\bftdr^*_{kl}\bftdr^T_{lj}\right.\\
&\qquad\qquad\quad\ +\frac{3}{2}\swi\sql\bgtq^\dagger_{kl}\bgtq_{lj}+\frac{1}{18}\sbi\sql\bgtpq^\dagger_{kl}\bgtpq_{lj}\\
&\left.\qquad\qquad\quad\ +\frac{8}{3}\sgl\sql\bgtsq^\dagger_{kl}\bgtsq_{lj}\right]\\
&-2\sh\left\{\left[\sn^2\h+\cs^2\Hh\right]\gthu\bftuq^*_{ik}(\bdf_u)^T_{kj}+\left[\cs^2\h+\sn^2\Hh\right]\gthd\bftdq^*_{ik}(\bdf_d)^T_{kj}\right\}\\
&+\frac{1}{2}\sql\left[3\swi\bgtq_{ik}\bgtq^\dagger_{kl}+\frac{1}{9}\sbi\bgtpq_{ik}\bgtpq^\dagger_{kl}+\frac{16}{3}\sgl\bgtsq_{ik}\bgtsq^\dagger_{kl}\right]\bgtq_{lj}\\
&+\sh\sql\left[\bftuq^*_{ik}\bftuq^T_{kl}+\bftdq^*_{ik}\bftdq^T_{kl}\right]\bgtq_{lj}\\
&-\bgtq_{ij}\left[\frac{1}{12}g'^2+\frac{33}{4}g^2_2+4g^2_s\right]\;,
\end{split}\end{equation}
\begin{equation}\begin{split}
{\left(4\pi\right)}^2\frac{d\bgtl_{ij}}{dt}=&\frac{1}{2}\swi\left[3\sql\bgtq_{lk}\bgtq^\dagger_{kl}+\sll\bgtl_{lk}\bgtl^\dagger_{kl}\right]\bgtl_{ij}\\
&+\frac{1}{2}\sh\swi\bgtl_{ij}\left\{\left[\sn^2\h+\cs^2\Hh\right]\mgthusq+\left[\cs^2\h+\sn^2\Hh\right]\mgthdsq\right\}\\
&+\frac{1}{2}\left[\cs^2\h+\sn^2\Hh\right]\bgtl_{ik}(\bdf_e^*\bdf_e^T)_{kj}\\
&+\frac{1}{2}\bgtl_{ik}\left[\sh\sel\bfter^*_{kl}\bfter^T_{lj}+\frac{3}{2}\swi\sll\bgtl^\dagger_{kl}\bgtl_{lj}\right.\\
&\left.\qquad\qquad\quad\ +\frac{1}{2}\sbi\sll\bgtpl^\dagger_{kl}\bgtpl_{lj}\right]\\
&-2\sh\left[\cs^2\h+\sn^2\Hh\right]\gthd\bftel^*_{ik}(\bdf_e)^T_{kj}\\
&+\frac{1}{2}\sll\left[3\swi\bgtl_{ik}\bgtl^\dagger_{kl}+\sbi\bgtpl_{ik}\bgtpl^\dagger_{kl}\right]\bgtl_{lj}\\
&+\sh\sll\bftel^*_{ik}\bftel^T_{kl}\bgtl_{lj}-\bgtq_{ij}\left[\frac{3}{4}g'^2+\frac{33}{4}g^2_2\right]\;,
\end{split}\end{equation}
\begin{equation}\begin{split}
{\left(4\pi\right)}^2\frac{d\left(\sn\gthu\right)}{dt}=&\frac{1}{2}\swi\left[3\sql\bgtq_{lk}\bgtq^\dagger_{kl}+\sll\bgtl_{lk}\bgtl^\dagger_{kl}\right]\sn\gthu\\
&+\frac{1}{2}\sh\left\{\left[\sn^2\h+\cs^2\Hh\right]\mgthusq+\left[\cs^2\h+\sn^2\Hh\right]\mgthdsq\right\}\sn\gthu\\
&+\frac{1}{2}\sh\left[3\suk\sn\gthu\bftur^*_{lk}\bftur_{kl}+3\sql\sn\gthu\bftuq^*_{lk}\bftuq^T_{kl}\right]\\
&+\frac{1}{4}\sh\left[\sn^2\h+\cs^2\Hh\right]\sn\gthu\left\{3\swi\mgthusq+\sbi\mgtphusq\right\}\\
&-6\sql\sn\bgtq_{lk}(\bdf_u)^*_{km}\bftuq^T_{ml}\\
&+\sn\gthu\left[\left(\sn^2\h+\cs^2\Hh\right)\mathrm{Tr}\{3\bdf_u^\dagger\bdf_u\}+\cs^2\left(\h-\Hh\right)\mathrm{Tr}\left\{3\bdf_d^\dagger\bdf_d+\bdf_e^\dagger\bdf_e\right\}\right]\\
&+\frac{1}{2}\sh\sn\gthu\left\{3\swi\left[\left(\sn^2\h+\cs^2\Hh\right)\mgthusq+\cs^2\left(\h-\Hh\right)\mgthdsq\right]\right.\\
&\left.\qquad\qquad\qquad+\sbi\left[\left(\sn^2\h+\cs^2\Hh\right)\mgtphusq+\cs^2\left(\h-\Hh\right)\mgtphdsq\right]\right\}\\
&-\sn\gthu\left[\frac{3}{4}g'^2+\frac{33}{4}g^2_2\right]\;,
\end{split}\end{equation}
\begin{equation}\begin{split}
{\left(4\pi\right)}^2\frac{d\left(\cs\gthd\right)}{dt}=&\frac{1}{2}\swi\left[3\sql\bgtq_{lk}\bgtq^\dagger_{kl}+\sll\bgtl_{lk}\bgtl^\dagger_{kl}\right]\cs\gthd\\
&+\frac{1}{2}\sh\left\{\left[\sn^2\h+\cs^2\Hh\right]\mgthusq+\left[\cs^2\h+\sn^2\Hh\right]\mgthdsq\right\}\cs\gthd\\
&+\frac{1}{2}\sh\left[3\sdk\cs\gthd\bftdr^*_{lk}\bftdr^{T}_{kl}+\sek\cs\gthd\bfter^*_{lk}\bfter^{T}_{kl}\right.\\
&\left.\qquad\qquad+3\sql\cs\gthd\bftdq^*_{lk}\bftdq^T_{kl}+\sll\cs\gthd\bftel^*_{lk}\bftel^T_{kl}\right]\\
&+\frac{1}{4}\sh\left[\cs^2\h+\sn^2\Hh\right]\cs\gthd\left\{3\swi\mgthdsq+\sbi\mgtphdsq\right\}\\
&-2\left[3\sql\cs\bgtq_{lk}(\bdf_d)^*_{km}\bftdq^T_{ml}+\sll\cs\bgtl_{lk}(\bdf_e)^*_{km}\bftel^T_{ml}\right]\\
&+\cs\gthd\left[\sn^2\left(\h-\Hh\right)\mathrm{Tr}\{3\bdf_u^\dagger\bdf_u\}+\left(\cs^2\h+\sn^2\Hh\right)\mathrm{Tr}\left\{3\bdf_d^\dagger\bdf_d+\bdf_e^\dagger\bdf_e\right\}\right]\\
&+\frac{1}{2}\sh\cs\gthd\left\{3\swi\left[\sn^2\left(\h-\Hh\right)\mgthusq+\left(\cs^2\h+\sn^2\Hh\right)\mgthdsq\right]\right.\\
&\left.\qquad\qquad\qquad+\sbi\left[\sn^2\left(\h-\Hh\right)\mgtphusq+\left(\cs^2\h+\sn^2\Hh\right)\mgtphdsq\right]\right\}\\
&-\cs\gthd\left[\frac{3}{4}g'^2+\frac{33}{4}g^2_2\right]\;,
\end{split}\end{equation}
\begin{equation}\begin{split}
{\left(4\pi\right)}^2\frac{d\bgtsq_{ij}}{dt}=&\frac{1}{2}\sgl\left[2\sql\bgtsq_{lk}\bgtsq^\dagger_{kl}+\suk\bgtsur_{lk}\bgtsur^\dagger_{kl}+\sdk\bgtsdr_{lk}\bgtsdr^\dagger_{kl}\right]\bgtsq_{ij}\\
&+\frac{1}{2}\bgtsq_{ik}\left\{\left[\sn^2\h+\cs^2\Hh\right](\bdf_u)^*_{kl}(\bdf_u)^T_{lj}+\left[\cs^2\h+\sn^2\Hh\right](\bdf_d)^*_{kl}(\bdf_d)^T_{lj}\right\}\\
&+\frac{1}{2}\bgtsq_{ik}\left[\sh\sul\bftur^*_{kl}\bftur^T_{lj}+\sh\sdl\bftdr^*_{kl}\bftdr^T_{lj}\right.\\
&\qquad\qquad\quad\ +\frac{3}{2}\swi\sql\bgtq^\dagger_{kl}\bgtq_{lj}+\frac{1}{18}\sbi\sql\bgtpq^\dagger_{kl}\bgtpq_{lj}\\
&\left.\qquad\qquad\quad\ +\frac{8}{3}\sgl\sql\bgtsq^\dagger_{kl}\bgtsq_{lj}\right]\\
&-2\sh\left[\sul\bftuq^*_{ik}\bgtsur_{kl}\bftur^T_{lj}+\sdl\bftdq^*_{ik}\bgtsdr_{kl}\bftdr^T_{lj}\right]\\
&+\frac{1}{2}\sql\left[3\swi\bgtq_{ik}\bgtq^\dagger_{kl}+\frac{1}{9}\sbi\bgtpq_{ik}\bgtpq^\dagger_{kl}+\frac{16}{3}\sgl\bgtsq_{ik}\bgtsq^\dagger_{kl}\right]\bgtsq_{lj}\\
&+\sh\sql\left[\bftuq^*_{ik}\bftuq^T_{kl}+\bftdq^*_{ik}\bftdq^T_{kl}\right]\bgtsq_{lj}\\
&-\bgtsq_{ij}\left[\frac{1}{12}g'^2+\frac{9}{4}g^2_2+13g^2_s\right]\;,
\end{split}\end{equation}
\begin{equation}\begin{split}
{\left(4\pi\right)}^2\frac{d\bgtsur_{ij}}{dt}=&\left[\sn^2\h+\cs^2\Hh\right](\bdf_u^T\bdf_u^*)_{ik}\bgtsur_{kj}\\
&+\left[\frac{4}{9}\sbi\suk\bgtpur_{ik}\bgtpur^\dagger_{kl}+\frac{4}{3}\sgl\suk\bgtsur_{ik}\bgtsur^\dagger_{kl}\right.\\
&\left.\quad\ +\sh\sqk\bftuq^T_{ik}\bftuq^*_{kl}\right]\bgtsur_{lj}\\
&+\sgl\sql\bgtsur_{ij}\bgtsq^\dagger_{kl}\bgtsq_{lk}\\
&+\frac{1}{2}\sgl\bgtsur_{ij}\left[\suk\bgtsur^\dagger_{kl}\bgtsur_{lk}+\sdk\bgtsdr^\dagger_{kl}\bgtsdr_{lk}\right]\\
&-4\sh\sqk\bftuq^T_{ik}\bgtsq_{kl}\bftur^*_{lj}+2\sh\suk\bgtsur_{ik}\bftur^T_{kl}\bftur^*_{lj}\\
&+\suk\bgtsur_{ik}\left[\frac{8}{9}\sbi\bgtpur^\dagger_{kl}\bgtpur_{lj}+\frac{8}{3}\sgl\bgtsur^\dagger_{kl}\bgtsur_{lj}\right]\\
&-\bgtsur_{ij}\left[\frac{4}{3}g'^2+13g^2_s\right]\;,
\end{split}\end{equation}
\begin{equation}\begin{split}
{\left(4\pi\right)}^2\frac{d\bgtsdr_{ij}}{dt}=&\left[\cs^2\h+\sn^2\Hh\right](\bdf_d^T\bdf_d^*)_{ik}\bgtsdr_{kj}\\
&+\left[\frac{1}{9}\sbi\sdk\bgtpdr_{ik}\bgtpdr^\dagger_{kl}+\frac{4}{3}\sgl\sdk\bgtsdr_{ik}\bgtsdr^\dagger_{kl}\right.\\
&\left.\quad\ +\sh\sqk\bftdq^T_{ik}\bftdq^*_{kl}\right]\bgtsdr_{lj}\\
&+\sgl\sql\bgtsdr_{ij}\bgtsq^\dagger_{kl}\bgtsq_{lk}\\
&+\frac{1}{2}\sgl\bgtsdr_{ij}\left[\suk\bgtsur^\dagger_{kl}\bgtsur_{lk}+\sdk\bgtsdr^\dagger_{kl}\bgtsdr_{lk}\right]\\
&-4\sh\sqk\bftdq^T_{ik}\bgtsq_{kl}\bftdr^*_{lj}+2\sh\sdk\bgtsdr_{ik}\bftdr^T_{kl}\bftdr^*_{lj}\\
&+\sdk\bgtsdr_{ik}\left[\frac{2}{9}\sbi\bgtpdr^\dagger_{kl}\bgtpdr_{lj}+\frac{8}{3}\sgl\bgtsdr^\dagger_{kl}\bgtsdr_{lj}\right]\\
&-\bgtsdr_{ij}\left[\frac{1}{3}g'^2+13g^2_s\right]\;.
\end{split}\end{equation}

\chapter{RGEs for Dimensionful Couplings}\label{app:dfulRGEs}
This appendix contains the RGEs with full thresholds for the dimensionful couplings of the MSSM with $R$-parity conservation. 
Note that we write these RGEs in the current basis in which the SSB sfermion mass matrices, but not the quark Yukawa matrices, are diagonal. In any other basis they must be modified to account for the rotation from this basis to the sfermion mass basis so that the sfermions can be properly decoupled. In this case the squark $\theta_{\tilde{q}_k}$'s become matrices $\bm{\Theta}_{q}$ as discussed in Sec~\ref{sec:sqdec}, where further details may be found. The RGEs for the superpotential parameter $\mu$ and the gaugino SSB mass parameters are,
\begin{equation}\label{app:rgemu}\begin{split}
{\left(4\pi\right)}^2\frac{d\mu}{dt}=&\frac{1}{2}\mu\sh\left[3\suk\bftur^\dagger_{kl}\bftur_{lk}+3\sdk\bftdr^\dagger_{kl}\bftdr_{lk}+\sek\bfter^\dagger_{kl}\bfter_{lk}\right.\\
&\left.\qquad\quad+3\sqk\bftuq_{kl}\bftuq^\dagger_{lk}+3\sqk\bftdq_{kl}\bftdq^\dagger_{lk}+\slk\bftel_{kl}\bftel^\dagger_{lk}\right]\\
&+\frac{1}{4}\mu\sh\left[\left(3\swi\mgthusq+\sbi\mgtphusq\right)\left(\sn^2\h+\cs^2\Hh\right)\right.\\
&\left.\qquad\qquad+\left(3\swi\mgthdsq+\sbi\mgtphdsq\right)\left(\cs^2\h+\sn^2\Hh\right)\right]\\
&+\sn\cs\left(-\h+\Hh\right)\left[3\swi\gthu\left(M_2+iM'_{2}\right)\gthd+\sbi\gtphu\left(M_1+iM'_{1}\right)\gtphd\right]\\
&-\mu\sh\left(\frac{3}{2}g'^2+\frac{9}{2}g^2_2\right)\;,
\end{split}\end{equation}
\begin{equation}\label{app:rgem1}\begin{split}
{\left(4\pi\right)}^2\frac{dM_1}{dt}=&M_1\sbi\left[\frac{1}{3}\sqk\bgtpq_{kl}\bgtpq^\dagger_{lk}+\slk\bgtpl_{kl}\bgtpl^\dagger_{lk}+\frac{8}{3}\suk\bgtpur^\dagger_{kl}\bgtpur_{lk}\right.\\
&\left.\qquad\quad+\frac{2}{3}\sdk\bgtpdr^\dagger_{kl}\bgtpdr_{lk}+2\sek\bgtper^\dagger_{kl}\bgtper_{lk}\right.\\
&\left.\qquad\quad+\sh\mgtphusq\left(\sn^2\h+\cs^2\Hh\right)+\sh\mgtphdsq\left(\cs^2\h+\sn^2\Hh\right)\right]\\[5pt]
&+2\sn\cs\left(-\h+\Hh\right)\sh\left[\gtphd\mu^*\gtphu+(\gtphd)^{*}\mu(\gtphu)^{*}\right]\;,
\end{split}\end{equation}
\begin{equation}\begin{split}
{\left(4\pi\right)}^2\frac{dM'_1}{dt}=&M'_1\sbi\left[\frac{1}{3}\sqk\bgtpq_{kl}\bgtpq^\dagger_{lk}+\slk\bgtpl_{kl}\bgtpl^\dagger_{lk}+\frac{8}{3}\suk\bgtpur^\dagger_{kl}\bgtpur_{lk}\right.\\
&\left.\qquad\quad+\frac{2}{3}\sdk\bgtpdr^\dagger_{kl}\bgtpdr_{lk}+2\sek\bgtper^\dagger_{kl}\bgtper_{lk}\right.\\
&\left.\qquad\quad+\sh\mgtphusq\left(\sn^2\h+\cs^2\Hh\right)+\sh\mgtphdsq\left(\cs^2\h+\sn^2\Hh\right)\right]\\[5pt]
&+2i\sn\cs\left(-\h+\Hh\right)\sh\left[\gtphd\mu^*\gtphu-(\gtphd)^{*}\mu(\gtphu)^{*}\right]\;,
\end{split}\end{equation}
\begin{equation}\label{app:rgem2}\begin{split}
{\left(4\pi\right)}^2\frac{dM_2}{dt}=&M_2\swi\left[3\sqk\bgtq_{kl}\bgtq^\dagger_{lk}+\slk\bgtl_{kl}\bgtl^\dagger_{lk}+\sh\mgthusq\left(\sn^2\h+\cs^2\Hh\right)\right.\\
&\left.\qquad\quad+\sh\mgthdsq\left(\cs^2\h+\sn^2\Hh\right)\right]\\[5pt]
&+2\sn\cs\left(-\h+\Hh\right)\sh\left[\gthd\mu^*\gthu+(\gthd)^{*}\mu(\gthu)^{*}\right]-12\swi M_2g^2_2\;,
\end{split}\end{equation}
\begin{equation}\label{app:rgem2p}\begin{split}
{\left(4\pi\right)}^2\frac{dM'_2}{dt}=&M'_2\swi\left[3\sqk\bgtq_{kl}\bgtq^\dagger_{lk}+\slk\bgtl_{kl}\bgtl^\dagger_{lk}+\sh\mgthusq\left(\sn^2\h+\cs^2\Hh\right)\right.\\
&\left.\qquad\quad+\sh\mgthdsq\left(\cs^2\h+\sn^2\Hh\right)\right]\\[5pt]
&+2i\sn\cs\left(-\h+\Hh\right)\sh\left[\gthd\mu^*\gthu-(\gthd)^{*}\mu(\gthu)^{*}\right]-12\swi M'_2g^2_2\;,
\end{split}\end{equation}
\begin{equation}\begin{split}
{\left(4\pi\right)}^2\frac{dM_3}{dt}=&M_3\sgl\left[2\sqk\bgtsq_{kl}\bgtsq^\dagger_{lk}+\suk\bgtsur^\dagger_{kl}\bgtsur_{lk}+\sdk\bgtsdr^\dagger_{kl}\bgtsdr_{lk}-18g^2_2\right]\;,
\end{split}\end{equation}
\begin{equation}\label{app:rgem3p}\begin{split}
{\left(4\pi\right)}^2\frac{dM'_3}{dt}=&M'_3\sgl\left[2\sqk\bgtsq_{kl}\bgtsq^\dagger_{lk}+\suk\bgtsur^\dagger_{kl}\bgtsur_{lk}+\sdk\bgtsdr^\dagger_{kl}\bgtsdr_{lk}-18g^2_2\right]\;.
\end{split}\end{equation}

\newpage
The following RGEs are only valid above $Q=m_{H}$, where $\theta_{h}=\theta_{H}=1$. We separate the two regimes of different Higgs boson content to simplify the resulting formulae, and to make explicit the parameters which remain in the theory below the heavy Higgs decoupling scale, $Q=m_{H}$.
\begin{equation}\begin{split}\label{eq:RGEau}
{\left(4\pi\right)}^2\frac{d\au_{ij}}{dt}=&\suk\au_{ik}\left[-\frac{2\glp^{2}}{3}\delta_{kj}+2\left[\fuul^{\dagger}\fuul\right]_{kj}\right]\\
&+\sul\sqk\left[-2\left(\frac{\glp^{2}}{9}+\frac{4\gthl^{2}}{3}\right)\delta_{ik}\delta_{lj}+6\fuhu_{ij}\fuhu^{\dagger}_{lk}\right]\au_{kl}\\
&+\sqk\left[\left(\frac{\glp^{2}}{6}-\frac{3\gtwl^{2}}{2}\right)\delta_{ik}+4\left[\fuur\fuur^{\dagger}\right]_{ik}\right]\au_{kj}\\
&+2\sdk\ad_{ik}\left[\fdul^{\dagger}\fuul\right]_{kj}\\
&+\frac{2}{3}\sbi\left(M_{1}-iM'_{1}\right)\left(\sh(\gtphu)^{*}\bgtpq^{*}_{ik}\bftur_{kj}-\frac{4}{3}\bgtpq^{*}_{ik}(\bdf_{u})_{kl}\bgtpur^{*}_{kj}\right.\\
&\left.\qquad\qquad\qquad\qquad\quad-4\sh\bftuq_{ik}\bgtpur^{*}_{kj}(\gtphu)^{*}\right)\\
&-6\swi\sh\left(M_{2}-iM'_{2}\right)(\gthu)^{*}\bgtq^{*}_{ik}\bftur_{kj}\\
&-\frac{32}{3}\sgl\left(M_{3}-iM'_{3}\right)\bgtsq^{*}_{ik}(\bdf_{u})_{kl}\bgtsur^{*}_{kj}\\
&+\suk\au_{ik}\left[\frac{8}{9}\sbi\bgtpur^{T}_{kl}\bgtpur^{*}_{lj}+\frac{8}{3}\sgl\bgtsur^{T}_{kl}\bgtsur^{*}_{lj}+2\sh\bftur^{\dagger}_{kl}\bftur_{lj}\right]\\
&+\left[3(\bdf^{\dagger}_{u})_{kl}(\bdf_{u})_{lk}+\frac{1}{2}\sh\sbi\mgtphusq+\frac{3}{2}\sh\swi\mgthusq\right]\au_{ij}\\
&+\sql\left[\sh\bftuq_{ik}\bftuq^{\dagger}_{kl}+\sh\bftdq_{ik}\bftdq^{\dagger}_{kl}+\frac{1}{18}\sbi\bgtpq^{*}_{ik}\bgtpq^{T}_{kl}\right.\\
&\left.\qquad\quad+\frac{3}{2}\swi\bgtq^{*}_{ik}\bgtq^{T}_{kl}+\frac{8}{3}\sgl\bgtsq^{*}_{ik}\bgtsq^{T}_{kl}\right]\au_{lj}\\
&-3\left\{\left(\frac{1}{36}\sqi+\frac{4}{9}\suj+\frac{1}{4}\right)g'^{2}+\frac{3}{4}\left(\sqi+1\right)g^{2}_{2}\right.\\
&\left.\qquad\qquad\qquad\qquad\qquad\qquad\qquad\qquad\qquad+\frac{4}{3}\left(\sqi+\suj\right)g^{2}_{3}\right\}\au_{ij}\;,
\end{split}\end{equation}
\begin{equation}\begin{split}\label{eq:RGEmtsfuhu}
{\left(4\pi\right)}^2\frac{d\mtsfuhu_{ij}}{dt}=&\frac{2\glp^{2}}{3}\suk\delta_{kj}\mtsfuhu_{ik}\\
&+\sul\sqk\left[-2\left(\frac{\glp^{2}}{9}+\frac{4\gthl^{2}}{3}\right)\delta_{ik}\delta_{lj}+6\fuhu_{ij}\fuhu^{\dagger}_{lk}\right]\mtsfuhu_{kl}\\
&-\sqk\left[\left(\frac{\glp^{2}}{6}-\frac{3\gtwl^{2}}{2}\right)\delta_{ik}+2\left[\fddr\fddr^{\dagger}\right]_{ik}\right]\mtsfuhu_{kj}\\
&-2\sdk\mtsfdhd_{ik}\left[\fddl^{\dagger}\fudl\right]_{kj}\\
&-\frac{2}{3}\sbi\sh\mu^{*}\gtphd\left(4\bftuq_{ik}\bgtpur^{*}_{kj}-\bgtpq^{*}_{ik}\bftur_{kj}\right)\\
&-6\sh\swi\mu^{*}\gthd\bgtq^{*}_{ik}\bftur_{kj}+4\sh\mu^{*}\bftdq_{ik}(\bdf^{\dagger}_{d})_{kl}\bftur_{lj}\\
&+\suk\mtsfuhu_{ik}\left[\frac{8}{9}\sbi\bgtpur^{T}_{kl}\bgtpur^{*}_{lj}+\frac{8}{3}\sgl\bgtsur^{T}_{kl}\bgtsur^{*}_{lj}\right.\\
&\left.\qquad\qquad\qquad\quad+2\sh\bftur^{\dagger}_{kl}\bftur_{lj}\right]\\
&+\left[3(\bdf^{\dagger}_{d})_{kl}(\bdf_{d})_{lk}+(\bdf^{\dagger}_{e})_{kl}(\bdf_{e})_{lk}+\frac{1}{2}\sbi\mgtphdsq+\frac{3}{2}\swi\mgthdsq\right]\mtsfuhu_{ij}\\
&+\sql\left[\sh\bftuq_{ik}\bftuq^{\dagger}_{kl}+\sh\bftdq_{ik}\bftdq^{\dagger}_{kl}+\frac{1}{18}\sbi\bgtpq^{*}_{ik}\bgtpq^{T}_{kl}\right.\\
&\left.\qquad\quad+\frac{3}{2}\swi\bgtq^{*}_{ik}\bgtq^{T}_{kl}+\frac{8}{3}\sgl\bgtsq^{*}_{ik}\bgtsq^{T}_{kl}\right]\mtsfuhu_{lj}\\
&-3\left\{\left(\frac{1}{36}\sqi+\frac{4}{9}\suj+\frac{1}{4}\right)g'^{2}+\frac{3}{4}\left(\sqi+1\right)g^{2}_{2}\right.\\
&\qquad\qquad\qquad\qquad\qquad\qquad\qquad\qquad\left.+\frac{4}{3}\left(\sqi+\suj\right)g^{2}_{3}\right\}\mtsfuhu_{ij}\;,
\end{split}\end{equation}
\begin{equation}\begin{split}
{\left(4\pi\right)}^2\frac{d\ad_{ij}}{dt}=&\sqk\left[-\left(\frac{\glp^{2}}{6}+\frac{3\gtwl^{2}}{2}\right)\delta_{ik}+4\left[\fddr\fddr^{\dagger}\right]_{ik}\right]\ad_{kj}\\
&+\sdl\sqk\left[2\left(\frac{\glp^{2}}{18}-\frac{4\gthl^{2}}{3}\right)\delta_{ik}\delta_{lj}+6\fdhd_{ij}\fdhd^{\dagger}_{lk}\right]\ad_{kl}\\
&+2\sel\slk\fdhd_{ij}\fehd^{\dagger}_{lk}\bae_{kl}\\
&+\sdk\ad_{ik}\left[-\frac{\glp^{2}}{3}\delta_{kj}+2\left[\fddl^{\dagger}\fddl\right]_{kj}\right]+2\suk\au_{ik}\left[\fudl^{\dagger}\fddl\right]_{kj}\\
&+\frac{2}{3}\sbi\left(M_{1}-iM'_{1}\right)\left(-\sh(\gtphd)^{*}\bgtpq^{*}_{ik}\bftdr_{kj}+\frac{2}{3}\bgtpq^{*}_{ik}(\bdf_{d})_{kl}\bgtpdr^{*}_{lj}\right.\\
&\left.\qquad\qquad\qquad\qquad\quad-2\sh\bftdq_{ik}\bgtpdr^{*}_{kj}(\gtphd)^{*}\right)\\
&-6\swi\sh\left(M_{2}-iM'_{2}\right)(\gthd)^{*}\bgtq^{*}_{ik}\bftdr_{kj}\\
&-\frac{32}{3}\sgl\left(M_{3}-iM'_{3}\right)\bgtsq^{*}_{ik}(\bdf_{d})_{kl}\bgtsdr^{*}_{lj}\\
&+\sql\left[\sh\bftuq_{ik}\bftuq^{\dagger}_{kl}+\sh\bftdq_{ik}\bftdq^{\dagger}_{kl}+\frac{1}{18}\sbi\bgtpq^{*}_{ik}\bgtpq^{T}_{kl}\right.\\
&\left.\qquad\quad+\frac{3}{2}\swi\bgtq^{*}_{ik}\bgtq^{T}_{kl}+\frac{8}{3}\sgl\bgtsq^{*}_{ik}\bgtsq^{T}_{kl}\right]\ad_{lj}\\
&+\left[3(\bdf_{d})_{kl}(\bdf^{\dagger}_{d})_{lk}+(\bdf_{e})_{kl}(\bdf^{\dagger}_{e})_{lk}+\frac{1}{2}\sbi\sh\mgtphdsq+\frac{3}{2}\swi\sh\mgthdsq\right]\ad_{ij}\\
&+\sdk\ad_{ik}\left[\frac{2}{9}\sbi\bgtpdr^{T}_{kl}\bgtpdr^{*}_{lj}+\frac{8}{3}\sgl\bgtsdr^{T}_{kl}\bgtsdr^{*}_{lj}+2\sh\bftdr^{\dagger}_{kl}\bftdr_{lj}\right]\\
&-3\left\{\left(\frac{1}{36}\sqi+\frac{1}{9}\sdj+\frac{1}{4}\right)g'^{2}+\frac{3}{4}\left(\sqi+1\right)g^{2}_{2}+\frac{4}{3}\left(\sqi+\sdj\right)g^{2}_{3}\right\}\ad_{ij}\;,
\end{split}\end{equation}
\begin{equation}\begin{split}
{\left(4\pi\right)}^2\frac{d\mtsfdhd_{ij}}{dt}=&\frac{\glp^{2}}{3}\sdk\delta_{kj}\mtsfdhd_{ik}\\
&+\sdl\sqk\left[2\left(\frac{\glp^{2}}{18}-\frac{4\gthl^{2}}{3}\right)\delta_{ik}\delta_{lj}+6\fdhd_{ij}\fdhd^{\dagger}_{lk}\right]\mtsfdhd_{kl}\\
&+2\sel\slk\mtsfdhd_{ij}\fehd^{\dagger}_{lk}\fehd_{kl}\\
&+\sqk\left[\left(\frac{\glp^{2}}{6}+\frac{3\gtwl^{2}}{2}\right)\delta_{ik}-2\left[\fuur\fuur^{\dagger}\right]_{ik}\right]\mtsfdhd_{kj}\\
&-2\suk\mtsfuhu_{ik}\left[\fuul^{\dagger}\fdul\right]_{kj}\\
&-\frac{2}{3}\sbi\sh\mu^{*}\gtphu\left(2\bftdq_{ik}\bgtpdr^{*}_{kj}+\bgtpq^{*}_{ik}\bftdr_{kj}\right)\\
&-6\sh\swi\mu^{*}\gthu\bgtq^{*}_{ik}\bftdr_{kj}+4\sh\mu^{*}\bftuq_{ik}(\bdf^{\dagger}_{u})_{kl}\bftdr_{lj}\\
&+\sdk\mtsfdhd_{ik}\left[\frac{2}{9}\sbi\bgtpdr^{T}_{kl}\bgtpdr^{*}_{lj}+\frac{8}{3}\sgl\bgtsdr^{T}_{kl}\bgtsdr^{*}_{lj}\right.\\
&\left.\qquad\qquad\qquad\quad+2\sh\bftdr^{\dagger}_{kl}\bftdr_{lj}\right]\\
&+\left[3(\bdf_{u})_{kl}(\bdf^{\dagger}_{u})_{lk}+\frac{1}{2}\sbi\mgtphusq+\frac{3}{2}\swi\mgthusq\right]\mtsfdhd_{ij}\\
&+\sql\left[\sh\bftuq_{ik}\bftuq^{\dagger}_{kl}+\sh\bftdq_{ik}\bftdq^{\dagger}_{kl}+\frac{1}{18}\sbi\bgtpq^{*}_{ik}\bgtpq^{T}_{kl}\right.\\
&\left.\qquad\quad+\frac{3}{2}\swi\bgtq^{*}_{ik}\bgtq^{T}_{kl}+\frac{8}{3}\sgl\bgtsq^{*}_{ik}\bgtsq^{T}_{kl}\right]\mtsfdhd_{lj}\\
&-3\left\{\left(\frac{1}{36}\sqi+\frac{1}{9}\sdj+\frac{1}{4}\right)g'^{2}+\frac{3}{4}\left(\sqi+1\right)g^{2}_{2}\right.\\
&\qquad\qquad\qquad\qquad\qquad\qquad\qquad\qquad\left.+\frac{4}{3}\left(\sqi+\sdj\right)g^{2}_{3}\right\}\mtsfdhd_{ij}\;,
\end{split}\end{equation}
\begin{equation}\begin{split}
{\left(4\pi\right)}^2\frac{d\bae_{ij}}{dt}=&\slk\left[\left(\frac{\glp^{2}}{2}-\frac{3\gtwl^{2}}{2}\right)\delta_{ik}+4\left[\feer\feer^{\dagger}\right]_{ik}\right]\bae_{kj}\\
&+\sel\slk\left[-\glp^{2}\delta_{ik}\delta_{lj}+2\fehd_{ij}\fehd^{\dagger}_{lk}\right]\bae_{kl}+6\sdl\sqk\fehd_{ij}\fdhd^{\dagger}_{lk}\ad_{kl}\\
&+\sek\bae_{ik}\left[-\glp^{2}\delta_{kj}+2\left[\feel^{\dagger}\feel\right]_{kj}\right]\\
&+2\sbi\left(M_{1}-iM'_{1}\right)\left(\sh(\gtphd)^{*}\bgtpl^{*}_{ik}\bfter_{kj}-2\bgtpl^{*}_{ik}(\bdf_{e})_{kl}\bgtper^{*}_{lj}\right.\\
&\left.\qquad\qquad\qquad\qquad\quad-2\sh\bftel_{ik}\bgtper^{*}_{kj}(\gtphd)^{*}\right)\\
&-6\swi\sh\left(M_{2}-iM'_{2}\right)(\gthd)^{*}\bgtl^{*}_{ik}\bfter_{kj}\\
&+\sll\left[\sh\bftel_{ik}\bftel^{\dagger}_{kl}+\frac{1}{2}\sbi\bgtpl^{*}_{ik}\bgtpl^{T}_{kl}+\frac{3}{2}\swi\bgtl^{*}_{ik}\bgtl^{T}_{kl}\right]\bae_{lj}\\
&+\left[3(\bdf_{d})_{kl}(\bdf^{\dagger}_{d})_{lk}+(\bdf_{e})_{kl}(\bdf^{\dagger}_{e})_{lk}+\frac{1}{2}\sbi\sh\mgtphdsq+\frac{3}{2}\swi\sh\mgthdsq\right]\bae_{ij}\\
&+\sek\bae_{ik}\left[2\sbi\bgtper^{T}_{kl}\bgtper^{*}_{lj}+2\sh\bfter^{\dagger}_{kl}\bfter_{lj}\right]\\
&-3\left\{\left(\frac{1}{4}\sLi+\sej+\frac{1}{4}\right)g'^{2}+\frac{3}{4}\left(\sLi+1\right)g^{2}_{2}\right\}\bae_{ij}\;,
\end{split}\end{equation}
\begin{equation}\begin{split}
{\left(4\pi\right)}^2\frac{d\mtsfehd_{ij}}{dt}=&\glp^{2}\sek\delta_{kj}\mtsfehd_{ik}\\
&+\sel\slk\left[-\glp^{2}\delta_{ik}\delta_{lj}+2\fehd_{ij}\fehd^{\dagger}_{lk}\right]\mtsfehd_{kl}\\
&+6\sdl\sqk\mtsfehd_{ij}\fdhd^{\dagger}_{lk}\fdhd_{kl}\\
&-\slk\left[\left(\frac{\glp^{2}}{2}-\frac{3\gtwl^{2}}{2}\right)\delta_{ik}\right]\mtsfehd_{kj}\\
&-2\sbi\sh\mu^{*}\gtphu\left(2\bftel_{ik}\bgtper^{*}_{kj}-\bgtpl^{*}_{ik}\bfter_{kj}\right)\\
&-6\sh\swi\mu^{*}\gthu\bgtl^{*}_{ik}\bfter_{kj}\\
&+\sek\mtsfehd_{ik}\left[2\sbi\bgtper^{T}_{kl}\bgtper^{*}_{lj}+2\sh\bfter^{\dagger}_{kl}\bfter_{lj}\right]\\
&+\left[3(\bdf_{u})_{kl}(\bdf^{\dagger}_{u})_{lk}+\frac{1}{2}\sbi\mgtphusq+\frac{3}{2}\swi\mgthusq\right]\mtsfehd_{ij}\\
&+\sel\left[\sh\bftel_{ik}\bftel^{\dagger}_{kl}+\frac{1}{2}\sbi\bgtpl^{*}_{ik}\bgtpl^{T}_{kl}\right.\\
&\left.\qquad\quad+\frac{3}{2}\swi\bgtl^{*}_{ik}\bgtl^{T}_{kl}\right]\mtsfehd_{lj}\\
&-3\left\{\left(\frac{1}{4}\sLi+\sej+\frac{1}{4}\right)g'^{2}+\frac{3}{4}\left(\sLi+1\right)g^{2}_{2}\right\}\mtsfehd_{ij}\;,
\end{split}\end{equation}
\begin{equation}\begin{split}
{\left(4\pi\right)}^{2}\frac{d\left(m^{2}_{H_{u}}+\left|\tilde{\mu}\right|^{2}\right)}{dt}=&\frac{3}{2}\left[\glp^{2}+\gtwl^{2}\right]\left(m^{2}_{H_{u}}+\left|\tilde{\mu}\right|^{2}\right)-\glp^{2}\left(m^{2}_{H_{d}}+\left|\tilde{\mu}\right|^{2}\right)\\
&+\suk\sul\left[-2\glp^{2}\delta_{lk}+6\left[\fuult\fuuls\right]_{lk}\right]\musq_{kl}\\
&+\sqk\sql\left[\glp^{2}\delta_{lk}+6\left[\fuurs\fuurt\right]_{lk}\right]\mqsq_{kl}\\
&+\sdk\sdl\glp^{2}\delta_{lk}\mdsq_{kl}-\slk\sll\glp^{2}\delta_{lk}\mlsq_{kl}\\
&+\sek\sel\glp^{2}\delta_{lk}\mesq_{kl}+6\suk\sql\aus_{lk}\aut_{kl}\\
&+6\sqk\sdl\mtsq\tfdhds_{lk}\tfdhdt_{kl}+2\slk\sel\mtsq\tfehds_{lk}\tfehdt_{kl}\\
&-2\sh\left|\mu\right|^{2}\left\{\sbi\mgtphusq+3\swi\mgthusq\right\}\\
&-2\sh\left\{\sbi\left(M^{2}_{1}+M'^{2}_{1}\right)\mgtphusq+3\swi\left(M^{2}_{2}+M'^{2}_{2}\right)\mgthusq\right\}\\
&-\left(\frac{3g'^{2}}{2}+\frac{9g^{2}_{2}}{2}\right)\left(m^{2}_{H_{u}}+\left|\tilde{\mu}\right|^{2}\right)\\
&+\left\{\left[6\bdf^{*}_{u}\bdf^{T}_{u}\right]_{kk}+\sbi\sh\mgtphusq+3\swi\sh\mgthusq\right\}\left(m^{2}_{H_{u}}+\left|\tilde{\mu}\right|^{2}\right)\;,
\end{split}\end{equation}
\begin{equation}\begin{split}
{\left(4\pi\right)}^{2}\frac{d\left(m^{2}_{H_{d}}+\left|\tilde{\mu}\right|^{2}\right)}{dt}=&-\glp^{2}\left(m^{2}_{H_{u}}+\left|\tilde{\mu}\right|^{2}\right)+\frac{3}{2}\left[\glp^{2}+\gtwl^{2}\right]\left(m^{2}_{H_{d}}+\left|\tilde{\mu}\right|^{2}\right)\\
&+2\suk\sul\glp^{2}\delta_{lk}\musq_{kl}\\
&+\sqk\sql\left[-\glp^{2}\delta_{lk}+6\left[\fddrs\fddrt\right]_{lk}\right]\mqsq_{kl}\\
&+\sdk\sdl\left[-\glp^{2}\delta_{lk}+6\left[\fddlt\fddls\right]_{lk}\right]\mdsq_{kl}\\
&+\slk\sll\left[\glp^{2}\delta_{lk}+2\left[\feers\feert\right]_{lk}\right]\mlsq_{kl}\\
&+\sek\sel\left[-\glp^{2}\delta_{lk}+2\left[\feelt\feels\right]_{lk}\right]\mesq_{kl}\\
&+6\suk\sql\mtsq\tfuhus_{lk}\tfuhut_{kl}+6\sqk\sdl\ads_{lk}\adt_{kl}\\
&+2\slk\sel\baes_{lk}\baet_{kl}\\
&-2\sh\left|\mu\right|^{2}\left\{\sbi\mgtphdsq+3\swi\mgthdsq\right\}\\
&-2\sh\left\{\sbi \left(M^{2}_{1}+M'^{2}_{1}\right)\mgtphdsq+3\swi\left(M^{2}_{2}+M'^{2}_{2}\right)\mgthdsq\right\}\\
&-\left(\frac{3g'^{2}}{2}+\frac{9g^{2}_{2}}{2}\right)\left(m^{2}_{H_{d}}+\left|\tilde{\mu}\right|^{2}\right)\\
&+\left\{\left[6\bdf^{*}_{d}\bdf^{T}_{d}+2\bdf^{*}_{e}\bdf^{T}_{e}\right]_{kk}+\sbi\sh\mgtphdsq+3\swi\sh\mgthdsq\right\}\\
&\qquad\qquad\qquad\qquad\qquad\qquad\qquad\qquad\qquad\qquad\times\left(m^{2}_{H_{d}}+\left|\tilde{\mu}\right|^{2}\right)\;,
\end{split}\end{equation}
\begin{equation}\begin{split}\label{app:mqamh}
{\left(4\pi\right)}^2\frac{d\mqsq_{ij}}{dt}=&\left\{\frac{1}{3}\glp^{2}\delta_{ij}+2\left[\fuurs\fuurt\right]_{ij}\right\}\left(m^{2}_{H_{u}}+\left|\tilde{\mu}\right|^{2}\right)\\
&+\left\{-\frac{1}{3}\glp^{2}\delta_{ij}+2\left[\fddrs\fddrt\right]_{ij}\right\}\left(m^{2}_{H_{d}}+\left|\tilde{\mu}\right|^{2}\right)\\
&-\frac{2}{3}\suk\glp^{2}\delta_{ij}\musq_{kk}+\frac{1}{3}\sqk\glp^{2}\delta_{ij}\mqsq_{kk}\\
&+\sqk\sql\left(\frac{\glp^{2}}{18}+\frac{3\gtwl^{2}}{2}+\frac{8\gthl^{2}}{3}\right)\delta_{ik}\delta_{lj}\mqsq_{kl}\\
&+\frac{1}{3}\sdk\glp^{2}\delta_{ij}\mdsq_{kk}-\frac{1}{3}\slk\glp^{2}\delta_{ij}\mlsq_{kk}+\frac{1}{3}\sek\glp^{2}\delta_{ij}\mesq_{kk}\\
&+2\suk\sul\fuhus_{ik}\musq_{kl}\fuhut_{lj}+2\sdk\sdl\fdhds_{ik}\mdsq_{kl}\fdhdt_{lj}\\
&+2\suk\aus_{ik}\aut_{kj}+2\suk\mtsq\tfuhus_{ik}\tfuhut_{kj}\\
&+2\sdk\ads_{ik}\adt_{kj}+2\sdk\mtsq\tfdhds_{ik}\tfdhdt_{kj}\\
&-\frac{2}{9}\sbi\left(M^{2}_{1}+M'^{2}_{1}\right)\bgtpq_{ik}\bgtpq^{\dagger}_{kj}-6\swi\left(M^{2}_{2}+M'^{2}_{2}\right)\bgtq_{ik}\bgtq^{\dagger}_{kj}\\
&-\frac{32}{3}\sgl\left(M^{2}_{3}+M'^{2}_{3}\right)\bgtsq_{ik}\bgtsq^{\dagger}_{kj}\\
&-4\sh\left|\mu\right|^{2}\left[\bftuq^{*}_{ik}\bftuq^{T}_{kj}+\bftdq^{*}_{ik}\bftdq^{T}_{kj}\right]\\
&-3\left(\sqi+\sqj\right)\left(\frac{1}{36}g'^{2}+\frac{3}{4}g^{2}_{2}+\frac{4}{3}g^{2}_{3}\right)\mqsq_{ij}\\
&+\sql\left[\frac{1}{18}\sbi\bgtpq_{ik}\bgtpq^{\dagger}_{kl}+\frac{3}{2}\swi\bgtq_{ik}\bgtq^{\dagger}_{kl}+\frac{8}{3}\sgl\bgtsq_{ik}\bgtsq^{\dagger}_{kl}\right.\\
&\left.\qquad\quad+\sh\bftuq^{*}_{ik}\bftuq^{T}_{kl}+\sh\bftdq^{*}_{ik}\bftdq^{T}_{kl}\right]\mqsq_{lj}\\
&+\sqk\mqsq_{ik}\left[\frac{1}{18}\sbi\bgtpq_{kl}\bgtpq^{\dagger}_{lj}+\frac{3}{2}\swi\bgtq_{kl}\bgtq^{\dagger}_{lj}\right.\\
&\left.\qquad\qquad\qquad\ +\frac{8}{3}\sgl\bgtsq_{kl}\bgtsq^{\dagger}_{lj}+\sh\bftuq^{*}_{kl}\bftuq^{T}_{lj}+\sh\bftdq^{*}_{kl}\bftdq^{T}_{lj}\right]\;,
\end{split}\end{equation}
\begin{equation}\begin{split}\label{app:muamh}
{\left(4\pi\right)}^2\frac{d\musq_{ij}}{dt}=&\left\{-\frac{4}{3}\glp^{2}\delta_{ij}+4\left[\fuult\fuuls\right]_{ij}\right\}\left(m^{2}_{H_{u}}+\left|\tilde{\mu}\right|^{2}\right)+\frac{4}{3}\glp^{2}\delta_{ij}\left(m^{2}_{H_{d}}+\left|\tilde{\mu}\right|^{2}\right)\\
&+\frac{8}{3}\suk\glp^{2}\delta_{ij}\musq_{kk}+\frac{8}{3}\suk\sul\left[\frac{1}{3}\glp^{2}+\gthl^{2}\right]\delta_{ik}\delta_{lj}\musq_{kl}\\
&-\frac{4}{3}\sqk\glp^{2}\delta_{ij}\mqsq_{kk}-\frac{4}{3}\sdk\glp^{2}\delta_{ij}\mdsq_{kk}+\frac{4}{3}\slk\glp^{2}\delta_{ij}\mlsq_{kk}\\
&-\frac{4}{3}\sek\glp^{2}\delta_{ij}\mesq_{kk}+4\sqk\sql\fuhut_{ik}\fuhus_{lj}\mqsq_{kl}\\
&+4\sqk\aut_{ik}\aus_{kj}+4\sqk\mtsq\tfuhut_{ik}\tfuhus_{kj}\\
&-\frac{32}{9}\sbi\left(M^{2}_{1}+M'^{2}_{1}\right)\bgtpur^{\dagger}_{ik}\bgtpur_{kj}-\frac{32}{3}\sgl\left(M^{2}_{3}+M'^{2}_{3}\right)\bgtsur^{\dagger}_{ik}\bgtsur_{kj}\\
&-8\sh\left|\mu\right|^{2}\bftur^{T}_{ik}\bftur^{*}_{kj}-3\left(\sui+\suj\right)\left(\frac{4}{9}g'^{2}+\frac{4}{3}g^{2}_{3}\right)\musq_{ij}\\
&+\sul\left[\frac{8}{9}\sbi\bgtpur^{\dagger}_{ik}\bgtpur_{kl}+\frac{8}{3}\sgl\bgtsur^{\dagger}_{ik}\bgtsur_{kl}\right.\\
&\left.\qquad\quad+2\sh\bftur^{T}_{ik}\bftur^{*}_{kl}\right]\musq_{lj}\\
&+\suk\musq_{ik}\left[\frac{8}{9}\sbi\bgtpur^{\dagger}_{kl}\bgtpur_{lj}+\frac{8}{3}\sgl\bgtsur^{\dagger}_{kl}\bgtsur_{lj}\right.\\
&\left.\qquad\qquad\qquad\ \ +2\sh\bftur^{T}_{kl}\bftur^{*}_{lj}\right]\;,
\end{split}\end{equation}
\begin{equation}\begin{split}
{\left(4\pi\right)}^2\frac{d\mdsq_{ij}}{dt}=&\frac{2}{3}\glp^{2}\delta_{ij}\left(m^{2}_{H_{u}}+\left|\tilde{\mu}\right|^{2}\right)+\left\{-\frac{2}{3}\glp^{2}\delta_{ij}+4\left[\fddlt\fddls\right]_{ij}\right\}\left(m^{2}_{H_{d}}+\left|\tilde{\mu}\right|^{2}\right)\\
&-\frac{4}{3}\suk\glp^{2}\delta_{ij}\musq_{kk}+\frac{2}{3}\sqk\glp^{2}\delta_{ij}\mqsq_{kk}+\frac{2}{3}\sdk\glp^{2}\delta_{ij}\mdsq_{kk}\\
&+\frac{2}{3}\sdk\sdl\left[\frac{1}{3}\glp^{2}+4\gthl^{2}\right]\delta_{ik}\delta_{lj}\mdsq_{kl}-\frac{2}{3}\slk\glp^{2}\delta_{ij}\mlsq_{kk}\\
&+\frac{2}{3}\sek\glp^{2}\delta_{ij}\mesq_{kk}+4\sqk\sql\fdhdt_{ik}\fdhds_{lj}\mqsq_{kl}\\
&+4\sqk\adt_{ik}\ads_{kj}+4\sqk\mtsq\tfdhdt_{ik}\tfdhds_{kj}\\
&-\frac{8}{9}\sbi\left(M^{2}_{1}+M'^{2}_{1}\right)\bgtpdr^{\dagger}_{ik}\bgtpdr_{kj}-\frac{32}{3}\sgl\left(M^{2}_{3}+M'^{2}_{3}\right)\bgtsdr^{\dagger}_{ik}\bgtsdr_{kj}\\
&-8\sh\left|\mu\right|^{2}\bftdr^{T}_{ik}\bftdr^{*}_{kj}-3\left(\sdi+\sdj\right)\left(\frac{1}{9}g'^{2}+\frac{4}{3}g^{2}_{3}\right)\mdsq_{ij}\\
&+\sdl\left[\frac{2}{9}\sbi\bgtpdr^{\dagger}_{ik}\bgtpdr_{kl}+\frac{8}{3}\sgl\bgtsdr^{\dagger}_{ik}\bgtsdr_{kl}\right.\\
&\left.\qquad\quad+2\sh\bftdr^{T}_{ik}\bftdr^{*}_{kl}\right]\mdsq_{lj}\\
&+\sdk\mdsq_{ik}\left[\frac{2}{9}\sbi\bgtpdr^{\dagger}_{kl}\bgtpdr_{lj}+\frac{8}{3}\sgl\bgtsdr^{\dagger}_{kl}\bgtsdr_{lj}\right.\\
&\left.\qquad\qquad\qquad\ \ +2\sh\bftdr^{T}_{kl}\bftdr^{*}_{lj}\right]\;,
\end{split}\end{equation}
\begin{equation}\begin{split}
{\left(4\pi\right)}^2\frac{d\mlsq_{ij}}{dt}=&-\glp^{2}\delta_{ij}\left(m^{2}_{H_{u}}+\left|\tilde{\mu}\right|^{2}\right)+\left\{\glp^{2}\delta_{ij}+2\left[\feers\feert\right]_{ij}\right\}\left(m^{2}_{H_{d}}+\left|\tilde{\mu}\right|^{2}\right)\\
&+2\suk\glp^{2}\delta_{ij}\musq_{kk}-\sqk\glp^{2}\delta_{ij}\mqsq_{kk}-\sdk\glp^{2}\delta_{ij}\mdsq_{kk}\\
&+\slk\glp^{2}\delta_{ij}\mlsq_{kk}+\slk\sll\left(\frac{\glp^{2}}{2}+\frac{3\gtwl^{2}}{2}\right)\delta_{ik}\delta_{lj}\mlsq_{kl}\\
&-\sek\glp^{2}\delta_{ij}\mesq_{kk}+2\sek\sel\fehds_{ik}\mesq_{kl}\fehdt_{lj}\\
&+2\sek\baes_{ik}\baet_{kj}+2\sek\mtsq\tfehds_{ik}\tfehdt_{kj}\\
&-2\sbi\left(M^{2}_{1}+M'^{2}_{1}\right)\bgtpl_{ik}\bgtpl^{\dagger}_{kj}-6\swi\left(M^{2}_{2}+M'^{2}_{2}\right)\bgtl_{ik}\bgtl^{\dagger}_{kj}\\
&-4\sh\left|\mu\right|^{2}\bftel^{*}_{ik}\bftel^{T}_{kj}-3\left(\sLi+\sLj\right)\left(\frac{1}{4}g'^{2}+\frac{3}{4}g^{2}_{2}\right)\mlsq_{ij}\\
&+\sll\left[\frac{1}{2}\bgtpl_{ik}\bgtpl^{\dagger}_{kl}\sbi+\frac{3}{2}\bgtl_{ik}\bgtl^{\dagger}_{kl}\swi+\bftel^{*}_{ik}\bftel^{T}_{kl}\sh\right]\mlsq_{lj}\\
&+\slk\mlsq_{ik}\left[\frac{1}{2}\bgtpl_{kl}\bgtpl^{\dagger}_{lj}\sbi+\frac{3}{2}\bgtl_{kl}\bgtl^{\dagger}_{lj}\swi+\bftel^{*}_{kl}\bftel^{T}_{lj}\sh\right]\;,
\end{split}\end{equation}
\begin{equation}\begin{split}\label{app:meamh}
{\left(4\pi\right)}^2\frac{d\mesq_{ij}}{dt}=&2\glp^{2}\delta_{ij}\left(m^{2}_{H_{u}}+\left|\tilde{\mu}\right|^{2}\right)+\left\{-2\glp^{2}\delta_{ij}+4\left[\feelt\feels\right]_{ij}\right\}\left(m^{2}_{H_{d}}+\left|\tilde{\mu}\right|^{2}\right)\\
&-4\suk\glp^{2}\delta_{ij}\musq_{kk}+2\sqk\glp^{2}\delta_{ij}\mqsq_{kk}+2\sdk\glp^{2}\delta_{ij}\mdsq_{kk}\\
&-2\slk\glp^{2}\delta_{ij}\mlsq_{kk}+2\sek\glp^{2}\delta_{ij}\mesq_{kk}\\
&+2\sek\sel\glp^{2}\delta_{lj}\delta_{ik}\mesq_{kl}+4\slk\sll\fehdt_{ik}\mlsq_{kl}\fehds_{lj}\\
&+4\slk\baet_{ik}\baes_{kj}+4\slk\mtsq\tfehdt_{ik}\tfehds_{kj}\\
&-8\sbi\left(M^{2}_{1}+M'^{2}_{1}\right)\bgtper^{\dagger}_{ik}\bgtper_{kj}-8\sh\left|\mu\right|^{2}\bfter^{T}_{ik}\bfter^{*}_{kj}\\
&-3\left(\sei+\sej\right)g'^{2}\mesq_{ij}\\
&+\sel\left[2\bgtper^{\dagger}_{ik}\bgtper_{kl}\sbi+2\bfter^{T}_{ik}\bfter^{*}_{kl}\sh\right]\mesq_{lj}\\
&+\sek\mesq_{ik}\left[2\bgtper^{\dagger}_{kl}\bgtper_{lj}\sbi+2\bfter^{T}_{kl}\bfter^{*}_{lj}\sh\right]\;.
\end{split}\end{equation}
\newpage
Below the scale $Q=m_{H}$, as discussed in Secs.~\ref{sec:trideriv} and \ref{sec:higgmassderiv}, the trilinear couplings to the doublet $\mathsf{h}$, and the mass parameter $m^{2}_{\mathsf{h}}$ remain in the theory, with RGEs given by,
\vspace{0.5cm}

\noindent${\left(4\pi\right)}^2\frac{d\left[\sn\au_{ij}-\cs\mtsfuhu_{ij}\right]}{dt}$
\begin{equation}\label{app:triu}\begin{split}
=&\h\suk\left[\sn\au_{ik}-\cs\mtsfuhu_{ik}\right]\left[\frac{2\glp^{2}}{3}\left(\cs^{2}-\sn^{2}\right)\delta_{kj}+2\sn^{2}\left[\fuul^{\dagger}\fuul\right]_{kj}\right]\\
&+\sul\sqk\left[-2\left(\frac{\glp^{2}}{9}+\frac{4\gthl^{2}}{3}\right)\delta_{ik}\delta_{lj}+6\fuhu_{ij}\fuhu^{\dagger}_{lk}\right]\left[\sn\au_{kl}-\cs\mtsfuhu_{kl}\right]\\
&+2\h\sqk\left[\left(\frac{\glp^{2}}{12}-\frac{3\gtwl^{2}}{4}\right)\left(\sn^{2}-\cs^{2}\right)\delta_{ik}+2\sn^{2}\left[\fuur\fuur^{\dagger}\right]_{ik}-\cs^{2}\left[\fddr\fddr^{\dagger}\right]_{ik}\right]\\
&\qquad\qquad\qquad\qquad\qquad\qquad\qquad\qquad\qquad\qquad\qquad\qquad\qquad\times\left[\sn\au_{kj}-\cs\mtsfuhu_{kj}\right]\\
&+\frac{2}{3}\sbi\sn\left(M_{1}-iM'_{1}\right)\\
&\qquad\quad\times\left(\sh(\gtphu)^{*}\bgtpq^{*}_{ik}\bftur_{kj}-\frac{4}{3}\bgtpq^{*}_{ik}(\bdf_{u})_{kl}\bgtpur^{*}_{kj}-4\sh\bftuq_{ik}\bgtpur^{*}_{kj}(\gtphu)^{*}\right)\\
&-\frac{32}{3}\sgl\sn\left(M_{3}-iM'_{3}\right)\bgtsq^{*}_{ik}(\bdf_{u})_{kl}\bgtsur^{*}_{lj}-6\swi\sh\sn\left(M_{2}-iM'_{2}\right)(\gthu)^{*}\bgtq^{*}_{ik}\bftur_{kj}\\
&+\frac{2}{3}\sbi\sh\cs\mu^{*}\gtphd\left(4\bftuq_{ik}\bgtpur^{*}_{kj}-\bgtpq^{*}_{ik}\bftur_{kj}\right)+6\sh\swi\cs\mu^{*}\gthd\bgtq^{*}_{ik}\bftur_{kj}\\
&-4\sh\cs\mu^{*}\bftdq_{ik}(\bdf^{\dagger}_{d})_{kl}\bftur_{lj}\\
&+\suk\left[\sn\au_{ik}-\cs\mtsfuhu_{ik}\right]\\
&\qquad\qquad\qquad\times\left[\frac{8}{9}\sbi\bgtpur^{T}_{kl}\bgtpur^{*}_{lj}+\frac{8}{3}\sgl\bgtsur^{T}_{kl}\bgtsur^{*}_{lj}+2\sh\bftur^{\dagger}_{kl}\bftur_{lj}\right]\\
&+\h\left[3\sn^{2}(\bdf^{\dagger}_{u})_{kl}(\bdf_{u})_{lk}+\cs^{2}\left\{3(\bdf^{\dagger}_{d})_{kl}(\bdf_{d})_{lk}+(\bdf^{\dagger}_{e})_{kl}(\bdf_{e})_{lk}\right\}\right]\left[\sn\au_{ij}-\cs\mtsfuhu_{ij}\right]\\
&+\frac{1}{2}\h\sh\left[\cs^{2}\left\{\sbi\mgtphdsq+3\swi\mgthdsq\right\}+\sn^{2}\left\{\sbi\mgtphusq+3\swi\mgthusq\right\}\right]\\
&\qquad\qquad\qquad\qquad\qquad\qquad\qquad\qquad\qquad\qquad\qquad\times\left[\sn\au_{ij}-\cs\mtsfuhu_{ij}\right]\\
&+\sql\left[\sh\bftuq_{ik}\bftuq^{\dagger}_{kl}+\sh\bftdq_{ik}\bftdq^{\dagger}_{kl}+\frac{1}{18}\sbi\bgtpq^{*}_{ik}\bgtpq^{T}_{kl}+\frac{3}{2}\swi\bgtq^{*}_{ik}\bgtq^{T}_{kl}\right.\\
&\left.\qquad\quad+\frac{8}{3}\sgl\bgtsq^{*}_{ik}\bgtsq^{T}_{kl}\right]\left[\sn\au_{lj}-\cs\mtsfuhu_{lj}\right]\\
&-3\left\{\left(\frac{1}{36}\sqi+\frac{4}{9}\suj+\frac{1}{4}\h\right)g'^{2}+\frac{3}{4}\left(\sqi+\h\right)g^{2}_{2}+\frac{4}{3}\left(\sqi+\suj\right)g^{2}_{3}\right\}\\
&\qquad\qquad\qquad\qquad\qquad\qquad\qquad\qquad\qquad\qquad\qquad\qquad\qquad\times\left[\sn\au-\cs\mtsfuhu\right]_{ij}\;,
\end{split}\end{equation}
\noindent${\left(4\pi\right)}^2\frac{d\left[\cs\ad_{ij}-\sn\mtsfdhd_{ij}\right]}{dt}$
\begin{equation}\begin{split}
=&2\h\sqk\left[-\left(\frac{\glp^{2}}{12}+\frac{3\gtwl^{2}}{4}\right)\left(\cs^{2}-\sn^{2}\right)\delta_{ik}-\sn^{2}\left[\fuur\fuur^{\dagger}\right]_{ik}+2\cs^{2}\left[\fddr\fddr^{\dagger}\right]_{ik}\right]\\
&\qquad\qquad\qquad\qquad\qquad\qquad\qquad\qquad\qquad\qquad\qquad\qquad\qquad\times\left[\cs\ad_{kj}-\sn\mtsfdhd_{kj}\right]\\
&+\sdl\sqk\left[2\left(\frac{\glp^{2}}{18}-\frac{4\gthl^{2}}{3}\right)\delta_{ik}\delta_{lj}+6\fdhd^{\dagger}_{lk}\fdhd_{ij}\right]\left[\cs\ad_{kl}-\sn\mtsfdhd_{kl}\right]\\
&+2\sel\slk\fdhd_{ij}\fehd^{\dagger}_{lk}\left[\cs\bae_{kl}-\sn\mtsfehd_{kl}\right]\\
&+\h\sdk\left[\cs\ad_{ik}-\sn\mtsfdhd_{ik}\right]\left[-\frac{\glp^{2}}{3}\left(\cs^{2}-\sn^{2}\right)\delta_{kj}+2\cs^{2}\left[\fddl^{\dagger}\fddl\right]_{kj}\right]\\
&-4\sh\sn\mu^{*}\bftuq_{ik}(\bdf^{\dagger}_{u})_{kl}\bftdr_{lj}+\frac{2}{3}\sbi\sh\sn\mu^{*}\gtphu\left(2\bftdq_{ik}\bgtpdr^{*}_{kj}+\bgtpq^{*}_{ik}\bftdr_{kj}\right)\\
&+6\sh\swi\sn\mu^{*}\gthu\bgtq^{*}_{ik}\bftdr_{kj}\\
&+\frac{2}{3}\sbi\cs\left(M_{1}-iM'_{1}\right)\\
&\qquad\times\left(-\sh(\gtphd)^{*}\bgtpq^{*}_{ik}\bftdr_{kj}+\frac{2}{3}\bgtpq^{*}_{ik}(\bdf_{d})_{kl}\bgtpdr^{*}_{lj}-2\sh\bftdq_{ik}\bgtpdr^{*}_{kj}(\gtphd)^{*}\right)\\
&-\frac{32}{3}\sgl\cs\left(M_{3}-iM'_{3}\right)\bgtsq^{*}_{ik}(\bdf_{d})_{kl}\bgtsdr^{*}_{lj}-6\swi\sh\cs\left(M_{2}-iM'_{2}\right)(\gthd)^{*}\bgtq^{*}_{ik}\bftdr_{kj}\\
&+\sql\left[\sh\bftuq_{ik}\bftuq^{\dagger}_{kl}+\sh\bftdq_{ik}\bftdq^{\dagger}_{kl}+\frac{1}{18}\sbi\bgtpq^{*}_{ik}\bgtpq^{T}_{kl}+\frac{3}{2}\swi\bgtq^{*}_{ik}\bgtq^{T}_{kl}\right.\\
&\left.\qquad\quad +\frac{8}{3}\sgl\bgtsq^{*}_{ik}\bgtsq^{T}_{kl}\right]\left[\cs\ad_{lj}-\sn\mtsfdhd_{lj}\right]\\
&+\h\left[3\sn^{2}(\bdf_{u})_{kl}(\bdf^{\dagger}_{u})_{lk}+\cs^{2}\left\{3(\bdf_{d})_{kl}(\bdf^{\dagger}_{d})_{lk}+(\bdf_{e})_{kl}(\bdf^{\dagger}_{e})_{lk}\right\}\right]\left[\cs\ad_{ij}-\sn\mtsfdhd_{ij}\right]\\
&+\frac{1}{2}\h\sh\left[\cs^{2}\left\{\sbi\mgtphdsq+3\swi\mgthdsq\right\}+\sn^{2}\left\{\sbi\mgtphusq+3\swi\mgthusq\right\}\right]\\
&\qquad\qquad\qquad\qquad\qquad\qquad\qquad\qquad\qquad\qquad\qquad\times\left[\cs\ad_{ij}-\sn\mtsfdhd_{ij}\right]\\
&+\sdk\left[\cs\ad_{ik}-\sn\mtsfdhd_{ik}\right]\\
&\qquad\qquad\qquad\qquad\times\left[\frac{2}{9}\sbi\bgtpdr^{T}_{kl}\bgtpdr^{*}_{lj}+\frac{8}{3}\sgl\bgtsdr^{T}_{kl}\bgtsdr^{*}_{lj}+2\sh\bftdr^{\dagger}_{kl}\bftdr_{lj}\right]\\
&-3\left\{\left(\frac{1}{36}\sqi+\frac{1}{9}\sdj+\frac{1}{4}\h\right)g'^{2}+\frac{3}{4}\left(\sqi+\h\right)g^{2}_{2}+\frac{4}{3}\left(\sqi+\sdj\right)g^{2}_{3}\right\}\\
&\qquad\qquad\qquad\qquad\qquad\qquad\qquad\qquad\qquad\qquad\qquad\qquad\qquad\times\left[\cs\ad-\sn\mtsfdhd\right]_{ij}\;,
\end{split}\end{equation}
\noindent${\left(4\pi\right)}^2\frac{d\left[\cs\bae_{ij}-\sn\mtsfehd_{ij}\right]}{dt}$
\begin{equation}\begin{split}
=&2\h\slk\left[\left(\frac{\glp^{2}}{4}-\frac{3\gtwl^{2}}{4}\right)\left(\cs^{2}-\sn^{2}\right)\delta_{ik}+2\cs^{2}\left[\feer\feer^{\dagger}\right]_{ik}\right]\left[\cs\bae_{kj}-\sn\mtsfehd_{kj}\right]\\
&+\sel\slk\left[-\glp^{2}\delta_{ik}\delta_{lj}+2\fehd^{\dagger}_{lk}\fehd_{ij}\right]\left[\cs\bae_{kl}-\sn\mtsfehd_{kl}\right]\\
&+6\sdl\sqk\fehd_{ij}\fdhd^{\dagger}_{lk}\left[\cs\ad_{kl}-\sn\mtsfdhd_{kl}\right]\\
&+\h\sek\left[\cs\bae_{ik}-\sn\mtsfehd_{ik}\right]\left[-\glp^{2}\left(\cs^{2}-\sn^{2}\right)\delta_{kj}+2\cs^{2}\left[\feel^{\dagger}\feel\right]_{kj}\right]\\
&+2\sbi\sh\sn\mu^{*}\gtphu\left(2\bftel_{ik}\bgtper^{*}_{kj}-\bgtpl^{*}_{ik}\bfter_{kj}\right)+6\sh\swi\sn\mu^{*}\gthu\bgtl^{*}_{ik}\bfter_{kj}\\
&+2\sbi\cs\left(M_{1}-iM'_{1}\right)\\
&\qquad\qquad\times\left(\sh(\gtphd)^{*}\bgtpl^{*}_{ik}\bfter_{kj}-2\bgtpl^{*}_{ik}(\bdf_{e})_{kl}\bgtper^{*}_{lj}-2\sh\bftel_{ik}\bgtper^{*}_{kj}(\gtphd)^{*}\right)\\
&-6\swi\sh\cs\left(M_{2}-iM'_{2}\right)(\gthd)^{*}\bgtl^{*}_{ik}\bfter_{kj}\\
&+\sll\left[\sh\bftel_{ik}\bftel^{\dagger}_{kl}+\frac{1}{2}\sbi\bgtpl^{*}_{ik}\bgtpl^{T}_{kl}+\frac{3}{2}\swi\bgtl^{*}_{ik}\bgtl^{T}_{kl}\right]\left[\cs\bae_{lj}-\sn\mtsfehd_{lj}\right]\\
&+\h\left[3\sn^{2}(\bdf_{u})_{kl}(\bdf^{\dagger}_{u})_{lk}+\cs^{2}\left\{3(\bdf_{d})_{kl}(\bdf^{\dagger}_{d})_{lk}+(\bdf_{e})_{kl}(\bdf^{\dagger}_{e})_{lk}\right\}\right]\left[\cs\bae_{ij}-\sn\mtsfehd_{ij}\right]\\
&+\frac{1}{2}\h\sh\left[\cs^{2}\left\{\sbi\mgtphdsq+3\swi\mgthdsq\right\}+\sn^{2}\left\{\sbi\mgtphusq+3\swi\mgthusq\right\}\right]\\
&\qquad\qquad\qquad\qquad\qquad\qquad\qquad\qquad\qquad\qquad\qquad\times\left[\cs\bae_{ij}-\sn\mtsfehd_{ij}\right]\\
&+\sek\left[\cs\bae_{ik}-\sn\mtsfehd_{ik}\right]\left[2\sbi\bgtper^{T}_{kl}\bgtper^{*}_{lj}+2\sh\bfter^{\dagger}_{kl}\bfter_{lj}\right]\\
&-3\left\{\left(\frac{1}{4}\sLi+\sej+\frac{1}{4}\h\right)g'^{2}+\frac{3}{4}\left(\sLi+\h\right)g^{2}_{2}\right\}\left[\cs\bae-\sn\mtsfehd\right]_{ij}\;,
\end{split}\end{equation}
\noindent${\left(4\pi\right)}^{2}\frac{d\left[\sn^{2}\left(m^{2}_{H_{u}}+\left|\tilde{\mu}\right|^{2}\right)+\cs^{2}\left(m^{2}_{H_{d}}+\left|\tilde{\mu}\right|^{2}\right)-\sn\cs\left(b+b^{*}\right)\right]}{dt}$
\begin{equation}\begin{split}
=&\frac{3}{2}\h\left[\glp^{2}+\gtwl^{2}\right]\left(\cs^{2}-\sn^{2}\right)^{2}\left[\sn^{2}\left(m^{2}_{H_{u}}+\left|\tilde{\mu}\right|^{2}\right)+\cs^{2}\left(m^{2}_{H_{d}}+\left|\tilde{\mu}\right|^{2}\right)-\sn\cs\left(b+b^{*}\right)\right]\\
&+\suk\sul\left[-2\glp^{2}\left(\sn^{2}-\cs^{2}\right)\delta_{lk}+6\sn^{2}\left[\fuult\fuuls\right]_{lk}\right]\musq_{kl}\\
&+\sqk\sql\left[\glp^{2}\left(\sn^{2}-\cs^{2}\right)\delta_{lk}+6\sn^{2}\left[\fuurs\fuurt\right]_{lk}+6\cs^{2}\left[\fddrs\fddrt\right]_{lk}\right]\mqsq_{kl}\\
&+\sdk\sdl\left[\glp^{2}\left(\sn^{2}-\cs^{2}\right)\delta_{lk}+6\cs^{2}\left[\fddlt\fddls\right]_{lk}\right]\mdsq_{kl}\\
&+\slk\sll\left[-\glp^{2}\left(\sn^{2}-\cs^{2}\right)\delta_{lk}+2\cs^{2}\left[\feers\feert\right]_{lk}\right]\mlsq_{kl}\\
&+\sek\sel\left[\glp^{2}\left(\sn^{2}-\cs^{2}\right)\delta_{lk}+2\cs^{2}\left[\feelt\feels\right]_{lk}\right]\mesq_{kl}\\
&+6\suk\sql\left[\sn\au_{lk}-\cs\mtsfuhu_{lk}\right]\left[\sn\au^{\dagger}_{kl}-\cs\mtsfuhu^{\dagger}_{kl}\right]\\
&+6\sqk\sdl\left[\cs\ad_{lk}-\sn\mtsfdhd_{lk}\right]\left[\cs\ad^{\dagger}_{kl}-\sn\mtsfdhd^{\dagger}_{kl}\right]\\
&+2\slk\sel\left[\cs\bae_{lk}-\sn\mtsfehd_{lk}\right]\left[\cs\bae^{\dagger}_{kl}-\sn\mtsfehd^{\dagger}_{kl}\right]\\
&-2\sh\left|\mu\right|^{2}\left\{\sbi\left[\sn^{2}\mgtphusq+\cs^{2}\mgtphdsq\right]+3\swi\left[\sn^{2}\mgthusq+\cs^{2}\mgthdsq\right]\right\}\\
&-2\sh\left\{\sbi\left(M^{2}_{1}+M'^{2}_{1}\right)\left[\sn^{2}\mgtphusq+\cs^{2}\mgtphdsq\right]\right.\\
&\left.\qquad\quad\ +3\swi\left(M^{2}_{2}+M'^{2}_{2}\right)\left[\sn^{2}\mgthusq+\cs^{2}\mgthdsq\right]\right\}\\
&-\frac{1}{2}\left\{-4\sh\sbi\sn\cs\mu^{*}\gtphu\gtphd\left(M_{1}+iM'_{1}\right)-12\sh\swi\sn\cs\mu^{*}\gthu\gthd\left(M_{2}+iM'_{2}\right)\right\}\\
&-\frac{1}{2}\left\{-4\sh\sbi\sn\cs\mu(\gtphu)^{*}(\gtphd)^{*}\left(M_{1}-iM'_{1}\right)-12\sh\swi\sn\cs\mu(\gthu)^{*}(\gthd)^{*}\left(M_{2}-iM'_{2}\right)\right\}\\
&-\h\left(\frac{3g'^{2}}{2}+\frac{9g^{2}_{2}}{2}\right)\left[\sn^{2}\left(m^{2}_{H_{u}}+\left|\tilde{\mu}\right|^{2}\right)+\cs^{2}\left(m^{2}_{H_{d}}+\left|\tilde{\mu}\right|^{2}\right)-\sn\cs\left(b+b^{*}\right)\right]\\
&+\h\left[\sn^{2}\left\{\left[6\bdf^{*}_{u}\bdf^{T}_{u}\right]_{kk}+\sbi\sh\mgtphusq+3\swi\sh\mgthusq\right\}\right.\\
&\left.\qquad\qquad\qquad+\cs^{2}\left\{\left[6\bdf^{*}_{d}\bdf^{T}_{d}+2\bdf^{*}_{e}\bdf^{T}_{e}\right]_{kk}+\sbi\sh\mgtphdsq+3\swi\sh\mgthdsq\right\}\right]\\
&\qquad\qquad\qquad\qquad\qquad\qquad\qquad\times\left[\sn^{2}\left(m^{2}_{H_{u}}+\left|\tilde{\mu}\right|^{2}\right)+\cs^{2}\left(m^{2}_{H_{d}}+\left|\tilde{\mu}\right|^{2}\right)-\sn\cs\left(b+b^{*}\right)\right]\;,
\end{split}\end{equation}
\newpage
With $\theta_{H}=0$, RGEs for the remaining SSB scalar mass parameters take the form,
\begin{equation}\begin{split}\label{app:mqbmh}
\noindent{\left(4\pi\right)}^2\frac{d\mqsq_{ij}}{dt}=&\h\left\{-\frac{1}{3}\left(\cs^{2}-\sn^{2}\right)\glp^{2}\delta_{ij}+2\sn^{2}\left[\fuurs\fuurt\right]_{ij}+2\cs^{2}\left[\fddrs\fddrt\right]_{ij}\right\}\\
&\qquad\qquad\qquad\quad\times\left[\sn^{2}\left(m^{2}_{H_{u}}+\left|\tilde{\mu}\right|^{2}\right)+\cs^{2}\left(m^{2}_{H_{d}}+\left|\tilde{\mu}\right|^{2}\right)-\sn\cs\left(b+b^{*}\right)\right]\\
&-\frac{2}{3}\suk\glp^{2}\delta_{ij}\musq_{kk}+\frac{1}{3}\sqk\glp^{2}\delta_{ij}\mqsq_{kk}\\
&+\sqk\sql\left(\frac{\glp^{2}}{18}+\frac{3\gtwl^{2}}{2}+\frac{8\gthl^{2}}{3}\right)\delta_{ik}\delta_{lj}\mqsq_{kl}\\
&+\frac{1}{3}\sdk\glp^{2}\delta_{ij}\mdsq_{kk}-\frac{1}{3}\slk\glp^{2}\delta_{ij}\mlsq_{kk}+\frac{1}{3}\sek\glp^{2}\delta_{ij}\mesq_{kk}\\
&+2\suk\sul\fuhus_{ik}\musq_{kl}\fuhut_{lj}+2\sdk\sdl\fdhds_{ik}\mdsq_{kl}\fdhdt_{lj}\\
&+2\suk\h\left[\sn\aus_{ik}-\cs\mtfuhus_{ik}\right]\left[\sn\aut_{kj}-\cs\mtsfuhut_{kj}\right]\\
&+2\sdk\h\left[\cs\ads_{ik}-\sn\mtfdhds_{ik}\right]\left[\cs\adt_{kj}-\sn\mtsfdhdt_{kj}\right]\\
&-\frac{2}{9}\sbi\left(M^{2}_{1}+M'^{2}_{1}\right)\bgtpq_{ik}\bgtpq^{\dagger}_{kj}-6\swi\left(M^{2}_{2}+M'^{2}_{2}\right)\bgtq_{ik}\bgtq^{\dagger}_{kj}\\
&-\frac{32}{3}\sgl\left(M^{2}_{3}+M'^{2}_{3}\right)\bgtsq_{ik}\bgtsq^{\dagger}_{kj}\\
&-4\sh\left|\mu\right|^{2}\left[\bftuq^{*}_{ik}\bftuq^{T}_{kj}+\bftdq^{*}_{ik}\bftdq^{T}_{kj}\right]\\
&-3\left(\sqi+\sqj\right)\left(\frac{1}{36}g'^{2}+\frac{3}{4}g^{2}_{2}+\frac{4}{3}g^{2}_{3}\right)\mqsq_{ij}\\
&+\sql\left[\frac{1}{18}\sbi\bgtpq_{ik}\bgtpq^{\dagger}_{kl}+\frac{3}{2}\swi\bgtq_{ik}\bgtq^{\dagger}_{kl}+\frac{8}{3}\sgl\bgtsq_{ik}\bgtsq^{\dagger}_{kl}\right.\\
&\qquad\quad\left.+\sh\bftuq^{*}_{ik}\bftuq^{T}_{kl}+\sh\bftdq^{*}_{ik}\bftdq^{T}_{kl}\right]\mqsq_{lj}\\
&+\sqk\mqsq_{ik}\left[\frac{1}{18}\sbi\bgtpq_{kl}\bgtpq^{\dagger}_{lj}+\frac{3}{2}\swi\bgtq_{kl}\bgtq^{\dagger}_{lj}\right.\\
&\qquad\qquad\qquad\ \ \left.+\frac{8}{3}\sgl\bgtsq_{kl}\bgtsq^{\dagger}_{lj}x+\sh\bftuq^{*}_{kl}\bftuq^{T}_{lj}+\sh\bftdq^{*}_{kl}\bftdq^{T}_{lj}\right]\;,
\end{split}\end{equation}
\begin{equation}\begin{split}
\hspace{-1cm}{\left(4\pi\right)}^2\frac{d\musq_{ij}}{dt}=&\h\left\{\frac{4}{3}\left(\cs^{2}-\sn^{2}\right)\glp^{2}\delta_{ij}+4\sn^{2}\left[\fuult\fuuls\right]_{ij}\right\}\\
&\qquad\qquad\qquad\qquad\quad\times\left[\sn^{2}\left(m^{2}_{H_{u}}+\left|\tilde{\mu}\right|^{2}\right)+\cs^{2}\left(m^{2}_{H_{d}}+\left|\tilde{\mu}\right|^{2}\right)-\sn\cs\left(b+b^{*}\right)\right]\\
&+\frac{8}{3}\suk\glp^{2}\delta_{ij}\musq_{kk}+\frac{8}{3}\suk\sul\left[\frac{1}{3}\glp^{2}+\gthl^{2}\right]\delta_{ik}\delta_{lj}\musq_{kl}\\
&-\frac{4}{3}\sqk\glp^{2}\delta_{ij}\mqsq_{kk}-\frac{4}{3}\sdk\glp^{2}\delta_{ij}\mdsq_{kk}+\frac{4}{3}\slk\glp^{2}\delta_{ij}\mlsq_{kk}\\
&-\frac{4}{3}\sek\glp^{2}\delta_{ij}\mesq_{kk}+4\sqk\sql\fuhut_{ik}\fuhus_{lj}\mqsq_{kl}\\
&+4\sqk\h\left[\sn\aut_{ik}-\cs\mtsfuhut_{ik}\right]\left[\sn\aus_{kj}-\cs\mtfuhus_{kj}\right]\\
&-\frac{32}{9}\sbi\left(M^{2}_{1}+M'^{2}_{1}\right)\bgtpur^{\dagger}_{ik}\bgtpur_{kj}-\frac{32}{3}\sgl\left(M^{2}_{3}+M'^{2}_{3}\right)\bgtsur^{\dagger}_{ik}\bgtsur_{kj}\\
&-8\sh\left|\mu\right|^{2}\bftur^{T}_{ik}\bftur^{*}_{kj}-3\left(\sui+\suj\right)\left(\frac{4}{9}g'^{2}+\frac{4}{3}g^{2}_{3}\right)\musq_{ij}\\
&+\sul\left[\frac{8}{9}\sbi\bgtpur^{\dagger}_{ik}\bgtpur_{kl}+\frac{8}{3}\sgl\bgtsur^{\dagger}_{ik}\bgtsur_{kl}+2\sh\bftur^{T}_{ik}\bftur^{*}_{kl}\right]\musq_{lj}\\
&+\suk\musq_{ik}\left[\frac{8}{9}\sbi\bgtpur^{\dagger}_{kl}\bgtpur_{lj}+\frac{8}{3}\sgl\bgtsur^{\dagger}_{kl}\bgtsur_{lj}+2\sh\bftur^{T}_{kl}\bftur^{*}_{lj}\right]\;,
\end{split}\end{equation}
\begin{equation}\begin{split}
\hspace{-1cm}{\left(4\pi\right)}^2\frac{d\mdsq_{ij}}{dt}=&\h\left\{-\frac{2}{3}\left(\cs^{2}-\sn^{2}\right)\glp^{2}\delta_{ij}+4\cs^{2}\left[\fddlt\fddls\right]_{ij}\right\}\\
&\qquad\qquad\qquad\qquad\quad\times\left[\sn^{2}\left(m^{2}_{H_{u}}+\left|\tilde{\mu}\right|^{2}\right)+\cs^{2}\left(m^{2}_{H_{d}}+\left|\tilde{\mu}\right|^{2}\right)-\sn\cs\left(b+b^{*}\right)\right]\\
&-\frac{4}{3}\suk\glp^{2}\delta_{ij}\musq_{kk}+\frac{2}{3}\sqk\glp^{2}\delta_{ij}\mqsq_{kk}+\frac{2}{3}\sdk\glp^{2}\delta_{ij}\mdsq_{kk}\\
&+\frac{2}{3}\sdk\sdl\left[\frac{1}{3}\glp^{2}+4\gthl^{2}\right]\delta_{ik}\delta_{lj}\mdsq_{kl}-\frac{2}{3}\slk\glp^{2}\delta_{ij}\mlsq_{kk}\\
&+\frac{2}{3}\sek\glp^{2}\delta_{ij}\mesq_{kk}+4\sqk\sql\fdhdt_{ik}\fdhds_{lj}\mqsq_{kl}\\
&+4\sqk\h\left[\cs\adt_{ik}-\sn\mtsfdhdt_{ik}\right]\left[\cs\ads_{kj}-\sn\mtfdhds_{kj}\right]\\
&-\frac{8}{9}\sbi\left(M^{2}_{1}+M'^{2}_{1}\right)\bgtpdr^{\dagger}_{ik}\bgtpdr_{kj}-\frac{32}{3}\sgl\left(M^{2}_{3}+M'^{2}_{3}\right)\bgtsdr^{\dagger}_{ik}\bgtsdr_{kj}\\
&-8\sh\left|\mu\right|^{2}\bftdr^{T}_{ik}\bftdr^{*}_{kj}-3\left(\sdi+\sdj\right)\left(\frac{1}{9}g'^{2}+\frac{4}{3}g^{2}_{3}\right)\mdsq_{ij}\\
&+\sdl\left[\frac{2}{9}\sbi\bgtpdr^{\dagger}_{ik}\bgtpdr_{kl}+\frac{8}{3}\sgl\bgtsdr^{\dagger}_{ik}\bgtsdr_{kl}+2\sh\bftdr^{T}_{ik}\bftdr^{*}_{kl}\right]\mdsq_{lj}\\
&+\sdk\mdsq_{ik}\left[\frac{2}{9}\sbi\bgtpdr^{\dagger}_{kl}\bgtpdr_{lj}+\frac{8}{3}\sgl\bgtsdr^{\dagger}_{kl}\bgtsdr_{lj}+2\sh\bftdr^{T}_{kl}\bftdr^{*}_{lj}\right]\;,
\end{split}\end{equation}
\begin{equation}\begin{split}
\hspace{-1cm}{\left(4\pi\right)}^2\frac{d\mlsq_{ij}}{dt}=&\h\left\{\left(\cs^{2}-\sn^{2}\right)\glp^{2}\delta_{ij}+2\cs^{2}\left[\feers\feert\right]_{ij}\right\}\\
&\qquad\qquad\qquad\qquad\quad\times\left[\sn^{2}\left(m^{2}_{H_{u}}+\left|\tilde{\mu}\right|^{2}\right)+\cs^{2}\left(m^{2}_{H_{d}}+\left|\tilde{\mu}\right|^{2}\right)-\sn\cs\left(b+b^{*}\right)\right]\\
&+2\suk\glp^{2}\delta_{ij}\musq_{kk}-\sqk\glp^{2}\delta_{ij}\mqsq_{kk}-\sdk\glp^{2}\delta_{ij}\mdsq_{kk}\\
&+\slk\glp^{2}\delta_{ij}\mlsq_{kk}+\slk\sll\left(\frac{\glp^{2}}{2}+\frac{3\gtwl^{2}}{2}\right)\delta_{ik}\delta_{lj}\mlsq_{kl}\\
&-\sek\glp^{2}\delta_{ij}\mesq_{kk}+2\sek\sel\fehds_{ik}\mesq_{kl}\fehdt_{lj}\\
&+2\sek\h\left[\cs\baes_{ik}-\sn\mtfehds_{ik}\right]\left[\cs\baet_{kj}-\sn\mtsfehdt_{kj}\right]\\
&-2\sbi\left(M^{2}_{1}+M'^{2}_{1}\right)\bgtpl_{ik}\bgtpl^{\dagger}_{kj}-6\swi\left(M^{2}_{2}+M'^{2}_{2}\right)\bgtl_{ik}\bgtl^{\dagger}_{kj}\\
&-4\sh\left|\mu\right|^{2}\bftel^{*}_{ik}\bftel^{T}_{kj}-3\left(\sLi+\sLj\right)\left(\frac{1}{4}g'^{2}+\frac{3}{4}g^{2}_{2}\right)\mlsq_{ij}\\
&+\sll\left[\frac{1}{2}\bgtpl_{ik}\bgtpl^{\dagger}_{kl}\sbi+\frac{3}{2}\bgtl_{ik}\bgtl^{\dagger}_{kl}\swi+\bftel^{*}_{ik}\bftel^{T}_{kl}\sh\right]\mlsq_{lj}\\
&+\slk\mlsq_{ik}\left[\frac{1}{2}\bgtpl_{kl}\bgtpl^{\dagger}_{lj}\sbi+\frac{3}{2}\bgtl_{kl}\bgtl^{\dagger}_{lj}\swi+\bftel^{*}_{kl}\bftel^{T}_{lj}\sh\right]\;,
\end{split}\end{equation}
\begin{equation}\begin{split}\label{app:mebmh}
\hspace{-1cm}{\left(4\pi\right)}^2\frac{d\mesq_{ij}}{dt}=&\h\left\{-2\left(\cs^{2}-\sn^{2}\right)\glp^{2}\delta_{ij}+4\cs^{2}\left[\feelt\feels\right]_{ij}\right\}\\
&\qquad\qquad\qquad\qquad\quad\times\left[\sn^{2}\left(m^{2}_{H_{u}}+\left|\tilde{\mu}\right|^{2}\right)+\cs^{2}\left(m^{2}_{H_{d}}+\left|\tilde{\mu}\right|^{2}\right)-\sn\cs\left(b+b^{*}\right)\right]\\
&-4\suk\glp^{2}\delta_{ij}\musq_{kk}+2\sqk\glp^{2}\delta_{ij}\mqsq_{kk}+2\sdk\glp^{2}\delta_{ij}\mdsq_{kk}\\
&-2\slk\glp^{2}\delta_{ij}\mlsq_{kk}+2\sek\glp^{2}\delta_{ij}\mesq_{kk}\\
&+2\sek\sel\glp^{2}\delta_{lj}\delta_{ik}\mesq_{kl}+4\slk\sll\fehdt_{ik}\mlsq_{kl}\fehds_{lj}\\
&+4\slk\h\left[\cs\baet_{ik}-\sn\mtsfehdt_{ik}\right]\left[\cs\baes_{kj}-\sn\mtfehds_{kj}\right]\\
&-8\sbi\left(M^{2}_{1}+M'^{2}_{1}\right)\bgtper^{\dagger}_{ik}\bgtper_{kj}-8\sh\left|\mu\right|^{2}\bfter^{T}_{ik}\bfter^{*}_{kj}\\
&-3\left(\sei+\sej\right)g'^{2}\mesq_{ij}\\
&+\sel\left[2\bgtper^{\dagger}_{ik}\bgtper_{kl}\sbi+2\bfter^{T}_{ik}\bfter^{*}_{kl}\sh\right]\mesq_{lj}\\
&+\sek\mesq_{ik}\left[2\bgtper^{\dagger}_{kl}\bgtper_{lj}\sbi+2\bfter^{T}_{kl}\bfter^{*}_{lj}\sh\right]\;.
\end{split}\end{equation}

%% file: drsmrge.tex
\chapter{$\protect\overline{\bm{\mathrm{DR}}}$ RGEs in the SM}\label{app:drbar}

The following are the RGEs for the gauge couplings, Yukawa matrices, quartic Higgs coupling and Higgs vacuum expectation value of the Standard Model in the \drb~renormalisation scheme. The conventions are as in Ref.~\cite{MVrgeI,*MVrgeII,*MVrgeIII} such that to convert to the conventions used in the body of the text we must use $\bdl_{u,d,e}=\mathbf{Y}^{T}_{u,d,e}$.

The gauge couplings run as follows:
\begin{equation}\label{app:gsmrge}
\frac{dg_i}{dt}=-b_i \frac{g^3_i}{16\pi^2}-\sum_j b_{ij}\frac{g^3_ig^2_j}{{(16\pi^2)}^2}-\frac{g^3_i}{{(16\pi^2)}^2}\sum_{k=u,d,e}c_{ik}\mathrm{Tr}\left[{\boldy}^{\dag}_k\boldy_k\right]\;,
\end{equation}
with $t=\ln[Q]$ and $\{i,j\}=1,2,3$ corresponding to the individual gauge groups. Using $n_g=\tfrac{1}{2}n_f$ ($n_f$ is the number of active flavours), the constants are defined by
\begin{equation}
\left(\begin{array}{c}b_1\\[4pt]b_2\\[4pt]b_3\end{array}\right)
=
\left(\begin{array}{c}0\\[4pt]\tfrac{22}{3}\\[4pt]11\end{array}\right)-n_g
\left(\begin{array}{c}\frac{4}{3}\\[4pt]\frac{4}{3}\\[4pt]\frac{4}{3}\end{array}\right)-
\left(\begin{array}{c}\frac{1}{10}\\[4pt]\frac{1}{6}\\[4pt]0\end{array}\right)
\end{equation}
\begin{equation}
b_{ij}=
\left(\begin{array}{ccc}0&0&0\\[4pt]0&\frac{136}{3}&0\\[4pt]0&0&102\end{array}\right)-n_g
\left(\begin{array}{ccc}\frac{19}{15}&\frac{3}{5}&\frac{44}{15}\\[4pt]\frac{1}{5}&\frac{49}{3}&4\\[4pt]\frac{11}{30}&\frac{3}{2}&\frac{76}{3}\end{array}\right)-
\left(\begin{array}{ccc}\frac{9}{50}&\frac{9}{10}&0\\[4pt]\frac{3}{10}&\frac{13}{6}&0\\[4pt]0&0&0\end{array}\right)\;,
\end{equation}
and
\begin{equation}
c_{ik}=\left(\begin{array}{ccc}\frac{17}{10}&\frac{1}{2}&\frac{3}{2}\\[4pt]\frac{3}{2}&\frac{3}{2}&\frac{1}{2}\\[4pt]2&2&0\end{array}\right)\;.
\end{equation}

The running of the Yukawa couplings is
\begin{equation}\label{app:ysmrge}
\frac{d\boldy_k}{dt}=\frac{1}{16\pi^2}\bbeta^{(1)}_{\boldy_k}+\frac{1}{{(16\pi^2)}^2}\bbeta^{(2)}_{\boldy_k}
\end{equation}
The $\bbeta^{(1)}$ are
\begin{align}
\bbeta^{(1)}_{\boldy_u}&=\boldy_u\left\{\tfrac{3}{2}\left(\boldyd_u\boldy_u-\boldyd_d\boldy_d\right)+Y_2\bident-\left(\tfrac{17}{20}g^2_1+\tfrac{9}{4}g^2_2+8g^2_3\right)\bident\right\} \\[2pt]
\bbeta^{(1)}_{\boldy_d}&=\boldy_d\left\{\tfrac{3}{2}\left(\boldyd_d\boldy_d-\boldyd_u\boldy_u\right)+Y_2\bident-\left(\tfrac{1}{4}g^2_1+\tfrac{9}{4}g^2_2+8g^2_3\right)\bident\right\} \\[2pt]
\bbeta^{(1)}_{\boldy_e}&=\boldy_e\left\{\tfrac{3}{2}\boldyd_e\boldy_e+Y_2\bident-\tfrac{9}{4}\left(g^2_1+g^2_2\right)\bident\right\}\;,
\end{align}
where
\begin{equation}
Y_2=3\tr[\boldyd_u\boldy_u]+3\tr[\boldyd_d\boldy_d]+\tr[\boldyd_e\boldy_e]\;.
\end{equation}
and the $\bbeta^{(2)}$ are
\begin{align}
\begin{split}
\bbeta^{(2)}_{\boldy_u}=\boldy_u\{&\tfrac{3}{2}\boldyd_u\boldy_u\boldyd_u\boldy_u-\boldyd_u\boldy_u\boldyd_d\boldy_d-\tfrac{1}{4}\boldyd_d\boldy_d\boldyd_u\boldy_u+\tfrac{11}{4}\boldyd_d\boldy_d\boldyd_d\boldy_d \\[2pt]
& +Y_2\left(\tfrac{5}{4}\boldyd_d\boldy_d-\tfrac{9}{4}\boldyd_u\boldy_u\right)-\chi_4\bident+\tfrac{3}{2}\lambda^2\bident-2\lambda\left(3\boldyd_u\boldy_u+\boldyd_d\boldy_d\right) \\[2pt]
& +\left(\tfrac{221}{80}g^2_1+\tfrac{117}{16}g^2_2+20g^2_3\right)\boldyd_u\boldy_u-\left(\tfrac{17}{80}g^2_1-\tfrac{27}{16}g^2_2+20g^2_3\right)\boldyd_d\boldy_d \\[2pt]
& +Y_4\bident+\left[\left(\tfrac{7}{150}+\tfrac{2}{3}n_g\right)g^4_1-\tfrac{9}{20}g^2_1g^2_2+\tfrac{19}{15}g^2_1g^2_3-\left(\tfrac{101}{8}-2n_g\right)g^4_2 \right.\\[2pt]
& \left.+9g^2_2g^2_3-\left(\tfrac{292}{3}-\tfrac{16}{3}n_g\right)g^4_3\right]\bident\}
\end{split} \\[2pt]
\begin{split}
\bbeta^{(2)}_{\boldy_d}=\boldy_d\{&\tfrac{3}{2}\boldyd_d\boldy_d\boldyd_d\boldy_d-\boldyd_d\boldy_d\boldyd_u\boldy_u-\tfrac{1}{4}\boldyd_u\boldy_u\boldyd_d\boldy_d+\tfrac{11}{4}\boldyd_u\boldy_u\boldyd_u\boldy_u \\[2pt]
& +Y_2\left(\tfrac{5}{4}\boldyd_u\boldy_u-\tfrac{9}{4}\boldyd_d\boldy_d\right)-\chi_4\bident+\tfrac{3}{2}\lambda^2\bident-2\lambda\left(3\boldyd_d\boldy_d+\boldyd_u\boldy_u\right) \\[2pt]
& +\left(\tfrac{161}{80}g^2_1+\tfrac{117}{16}g^2_2+20g^2_3\right)\boldyd_d\boldy_d-\left(\tfrac{77}{80}g^2_1-\tfrac{27}{16}g^2_2+20g^2_3\right)\boldyd_u\boldy_u \\[2pt]
& +Y_4\bident+\left[-\left(\tfrac{37}{300}-\tfrac{4}{15}n_g\right)g^4_1-\tfrac{27}{20}g^2_1g^2_2+\tfrac{31}{15}g^2_1g^2_3-\left(\tfrac{101}{8}-2n_g\right)g^4_2 \right.\\[2pt]
& \left.+9g^2_2g^2_3-\left(\tfrac{292}{3}-\tfrac{16}{3}n_g\right)g^4_3\right]\bident\}
\end{split} \\[2pt]
\begin{split}
\bbeta^{(2)}_{\boldy_e}=\boldy_e\{&\tfrac{3}{2}\boldyd_e\boldy_e\boldyd_e\boldy_e-\tfrac{9}{4}Y_2\boldyd_e\boldy_e-\chi_4\bident+\tfrac{3}{2}\lambda^2\bident-6\lambda\boldyd_e\boldy_e+\left(\tfrac{441}{80}g^2_1 \right.\\[2pt]
& \left.+\tfrac{117}{16}g^2_2\right)\boldyd_e\boldy_e+Y_4\bident+\left[\left(\tfrac{21}{100}+\tfrac{8}{5}n_g\right)g^4_1+\tfrac{27}{20}g^2_1g^2_2 \right.\\[2pt]
& \left.-\left(\tfrac{101}{8}-2n_g\right)g^4_2\right]\bident\}\;,
\end{split}
\end{align}
with
\begin{align}
\begin{split}
\chi_4=&\tfrac{9}{4}\left\{3\tr\left[\boldyd_u\boldy_u\boldyd_u\boldy_u\right]+3\tr\left[\boldyd_d\boldy_d\boldyd_d\boldy_d\right]+\tr\left[\boldyd_e\boldy_e\boldyd_e\boldy_e\right]\right.\\[2pt]
&\quad\left.-\tfrac{2}{3}\left[\boldyd_u\boldy_u\boldyd_d\boldy_d\right]\right\} \end{split}\\[2pt]
\begin{split}
Y_4=&\left(\tfrac{83}{40}g^2_1+\tfrac{27}{8}g^2_2+28g^2_3\right)\tr\left[\boldyd_u\boldy_u\right]+\left(-\tfrac{1}{40}g^2_1+\tfrac{27}{8}g^2_2+28g^2_3\right)\tr\left[\boldyd_d\boldy_d\right] \\[2pt]
& \left(\tfrac{93}{40}g^2_1+\tfrac{9}{8}g^2_2\right)\tr\left[\boldyd_e\boldy_e\right]\;.
\end{split}
\end{align}

The Lagrangian term for the Higgs quartic coupling is
\begin{equation}
\mathcal{L}_{SM}\ni-\frac{\lambda}{2}\left(\phi^{\dagger}\phi\right)^{2}\;,
\end{equation}
where $\phi$ is the SM Higgs field. It is not necessary to find the running of the \drb~coupling, since the difference between \drb~and \msb~$\lambda$ will be equivalent to a higher order effect in the Yukawa coupling running. However the gauge and Yukawa couplings are in the \drb~scheme for ease of computation. The RGE in the \msb~scheme is
\begin{equation}
\frac{d{\lambda}_{\overline{\mathrm{MS}}}}{dt}=\frac{1}{16\pi^2}\beta^{(1)}_{{\lambda}_{\overline{\mathrm{MS}}}}+\frac{1}{{(16\pi^2)}^2}\beta^{(2)}_{{\lambda}_{\overline{\mathrm{MS}}}}\;,
\end{equation}
with the one-loop $\beta$s given by
\begin{align}\begin{split}
\beta^{(1)}_{{\lambda}_{\overline{\mathrm{MS}}}}=&12\lambda^2_{\overline{\mathrm{MS}}}-\left(\tfrac{9}{5}g^2_1+9g^2_2\right)\lambda_{\overline{\mathrm{MS}}}+\tfrac{9}{4}\left(\tfrac{3}{25}g^4_1+\tfrac{2}{5}g^2_1g^2_2+g^4_2\right)+4Y_2\lambda_{\overline{\mathrm{MS}}}\\[2pt]
& -4H \end{split}\end{align}
and the two-loop $\beta$s given by:
\begin{align}\begin{split}
\beta^{(2)}_{{\lambda}_{\overline{\mathrm{MS}}}}=&-78\lambda^3_{\overline{\mathrm{MS}}}+18\left(\tfrac{3}{5}g^2_1+3g^2_2\right)\lambda^2_{\overline{\mathrm{MS}}}-\left[\left(\tfrac{265}{8}-10n_g\right)g^4_2-\tfrac{117}{20}g^2_1g^2_2 \right.\\[2pt]
& \left.-\tfrac{9}{25}\left(\tfrac{229}{24}+\tfrac{50}{9}n_g\right)g^4_1\right]\lambda_{\overline{\mathrm{MS}}}+\left(\tfrac{473}{8}-8n_g\right)g^6_2-\tfrac{3}{5}\left(\tfrac{121}{24}+\tfrac{8}{3}n_g\right)g^2_1g^4_2 \\[2pt]
& -\tfrac{9}{25}\left(\tfrac{239}{24}+\tfrac{40}{9}n_g\right)g^4_1g^2_2-\tfrac{27}{125}\left(\tfrac{59}{24}+\tfrac{40}{9}n_g\right)g^6_1 \\[2pt]
& +\left(-\tfrac{14}{5} g^2_1+18g^2_2-128g^2_3\right)\tr\left[\boldyd_u\boldy_u\boldyd_u\boldy_u\right] \\[2pt]
&+\left(\tfrac{34}{5}g^2_1+18g^2_2-128g^2_3\right)\tr\left[\boldyd_d\boldy_d\boldyd_d\boldy_d\right] \\[2pt]
& +\left(-\tfrac{42}{5}g^2_1+6g^2_2\right)\tr\left[\boldyd_e\boldy_e\boldyd_e\boldy_e\right]-\tfrac{3}{2}g^4_2Y_2 \\[2pt]
& +\lambda_{\overline{\mathrm{MS}}}\left\{\left(\tfrac{83}{10}g^2_1+\tfrac{27}{2}g^2_2+112g^2_3\right)\tr\left[\boldyd_u\boldy_u\right]+\left(-\tfrac{1}{10}g^2_1+\tfrac{27}{2}g^2_2\right.\right. \\[2pt]
&\left.\left.+112g^2_3\right)\tr\left[\boldyd_d\boldy_d\right]+\left(\tfrac{93}{10}g^2_1+\tfrac{9}{2}g^2_2\right)\tr\left[\boldyd_e\boldy_e\right]\right\} \\[2pt]
& +\tfrac{3}{5}g^2_1\left\{\left(-\tfrac{57}{10}g^2_1+21g^2_2\right)\tr\left[\boldyd_u\boldy_u\right]+\left(\tfrac{3}{2}g^2_1+9g^2_2\right)\tr\left[\boldyd_d\boldy_d\right]\right. \\[2pt]
& \left. +\left(-\tfrac{15}{2}g^2_1+11g^2_2\right)\tr\left[\boldyd_e\boldy_e\right]\right\}-24\lambda^2_{\overline{\mathrm{MS}}}Y_2-\lambda_{\overline{\mathrm{MS}}} H \\[2pt] 
&+6\lambda_{\overline{\mathrm{MS}}}\tr\left[\boldyd_u\boldy_u\boldyd_d\boldy_d\right]+20\tr\left[3\left(\boldyd_u\boldy_u\boldyd_u\boldy_u\boldyd_u\boldy_u\right)\right.\\[2pt]
&\left.+3\left(\boldyd_d\boldy_d\boldyd_d\boldy_d\boldyd_d\boldy_d\right)+\left(\boldyd_e\boldy_e\boldyd_e\boldy_e\boldyd_e\boldy_e\right)\right]\\[2pt]
&-12\tr\left[\boldyd_u\boldy_u\left(\boldyd_u\boldy_u+\boldyd_d\boldy_d \right)\boldyd_d\boldy_d\right]\;,
\end{split}
\end{align}
where
\begin{align}
H=& \,3\tr\left[\boldyd_u\boldy_u\boldyd_u\boldy_u\right]+3\tr\left[\boldyd_d\boldy_d\boldyd_d\boldy_d\right]+\tr\left[\boldyd_e\boldy_e\boldyd_e\boldy_e\right]\;.
\end{align}
To find the boundary condition on the Higgs quartic coupling, we must consider the matching of MSSM Lagrangian with the SM Lagrangian at the scale where the heavy Higgs particles decouple. Given that
\begin{equation}
\mathcal{L}_{\mathrm{MSSM}}\ni-\frac{1}{8}\left(g'^{2}+g^{2}\right)\left(\sn^{2}-\cs^{2}\right)^{2}\left(\left|\mathsf{h}\right|^{2}\right)^{2}\;,
\end{equation}
for the rotated SM-like MSSM Higgs field, $\mathsf{h}$, defined in \eqref{eq:hHbasis}, the value of $\lambda$ at the scale $m_{H}$ can be found by setting $\mathsf{h}=\phi$.

The Higgs vacuum expectation value ($v_{SM}$) is presented in the same way, with all terms in the \drb~scheme except $v_{SM}$ and $\lambda$ which are in the \msb~scheme:

\begin{align}
\frac{d\ln{v_{SM}}}{dt}=\frac{1}{16\pi^2}\gamma^{(1)}_{{v}_{\overline{\mathrm{MS}}}}+\frac{1}{{(16\pi^2)}^2}\gamma^{(2)}_{{v}_{\overline{\mathrm{MS}}}}\;,
\end{align}
with
\begin{align}
\gamma^{(1)}_{{v}_{\overline{\mathrm{MS}}}}=&\tfrac{9}{4}\left(\tfrac{1}{5}g^2_1+g^2_2\right)-Y_2 \\[2pt]
\begin{split}
\gamma^{(2)}_{{v}_{\overline{\mathrm{MS}}}}=&-\tfrac{3}{2}\lambda^2_{\overline{\mathrm{MS}}}-\left(\tfrac{83}{40}g^2_1+\tfrac{27}{8}g^2_2+28g^2_3\right)\tr\left[\boldyd_u\boldy_u\right] \\[2pt]
&-\left(-\tfrac{1}{40}g^2_1+\tfrac{27}{8}g^2_2+28g^2_3\right)\tr\left[\boldyd_d\boldy_d\right]-\left(\tfrac{93}{40}g^2_1+\tfrac{9}{8}g^2_2\right)\tr\left[\boldyd_e\boldy_e\right]+\chi_4 \\[2pt]
&-\tfrac{27}{80}g^2_1g^2_2-\left(\tfrac{93}{800}+\tfrac{1}{2}n_g\right)g^4_1+\left(\tfrac{463}{32}-\tfrac{5}{2}n_g\right)g^4_2\;.
\end{split}
\end{align}

%% file: smrgecheck.tex
\chapter{Non-Trivial Check on the Coefficient of $\protect g_3$ in the $\protect\overline{\bm{\mathrm{DR}}}$ SM RGEs}\label{app:smrgecheck}

This is a non-trivial check on one aspect of the SM RGEs in the \drb~scheme. We will compare our result for the \drb~SM RGE of the bottom Yukawa coupling to the \drb~running from Ref.~\cite{BFMTdr2ms}, but in principle our procedure can also be used to check the top and tau Yukawa couplings. In this appendix we only check the coefficient of $g_3$, however it is expected that this is the dominant contribution to the SM RGEs in this region.

If we set $g_1=g_2=\bdl_u=\bdl_d=\bdl_e=0$, the general form of the running of the strong coupling is
\begin{equation}
\frac{dg_3}{dt}=\gamma g^3_3 + \delta g^5_3\;.
\end{equation}
Similarly, the running of a general Yukawa coupling under the same conditions can be represented as
\begin{equation} \label{eq:yrge}
\frac{1}{\lambda_a}\frac{d\lambda_a}{dt}=a g^2_3 + b g^4_3\;.
\end{equation}
We need to rearrange this to obtain the coefficients $P$ and $X$ in
\begin{equation} \label{eq:javy}
\lambda_a=\lambda^0_a\left[g^2_3(t)\right]^P\left[1+Xg^2_3\right]\;,
\end{equation}
where $\lambda_a$ is the coupling at some scale related to $\lambda^0_a$, the coupling at another scale. This form is the same as that in Ref.~\cite{BFMTdr2ms}.

Rearranging (\ref{eq:yrge}) we get
\begin{align}
\nonumber \frac{d\lambda_a}{\lambda_a}=&\left[ag^2_3+bg^4_3\right]\frac{dt}{dg_3}dg_3 \\[2pt]
\nonumber =&\left[\frac{ag^2_3+bg^4_3}{\gamma g^3_3+\delta g^5_3}\right]dg_3 \\[2pt]
=&\left\{\frac{1}{g_3}\left[\frac{a}{\gamma +\delta g^2_3}+\frac{bg^2_3}{\gamma +\delta g^2_3}\right]\right\}dg_3\;.
\end{align}
The two terms in the square brackets can be expanded as power series in $g^2_3$ to obtain (after multiplying through the $1/g_3$)
\begin{equation} \label{eq:simyuk}
\frac{d\lambda_a}{\lambda_a}=\left\{\frac{a}{\gamma}\left(\frac{1}{g_3}-\frac{\delta}{\gamma}g_3+{\mathcal{O}}(g^3_3)\right)+\frac{b}{\gamma}\left(g_3-{\mathcal{O}}(g^3_3)\right)\right\}dg_3\;.
\end{equation}
The next term in the series ($cg^6_3$) would have contributed \[\frac{1}{g_3}\left[\frac{cg^4_3}{\gamma+\delta g^2_3}\right]={\mathcal{O}}\!\left(g^3_3\right)\;,\] so this expansion is only complete up to ${\mathcal{O}}\!\left(g_3\right)$ since we do not keep three-loop terms.

Going back to (\ref{eq:javy}), we can take the natural log of either side
\begin{equation}
\ln{\lambda_a}=\ln{\lambda^0_a} + P\ln{g^2_3(t)}+\ln{(1+Xg^2_3)}\;,
\end{equation}
which leads to an expression which has the same form as (\ref{eq:simyuk}):
\begin{equation}
\frac{d\lambda_a}{\lambda_a}=P\frac{dg^2_3}{g^2_3}+X\frac{dg^2_3}{1+Xg^2_3}=\left[\frac{2P}{g_3}+2g_3X(1-Xg^2_3)\right]dg_3\;.
\end{equation}
Comparing this to (\ref{eq:simyuk}) it can be seen that
\begin{equation} \label{eq:resp}
\boxed{2P=\frac{a}{\gamma}}\;,
\end{equation}
and
\begin{equation} \label{eq:resx}
\boxed{2X=-\frac{a\delta}{\gamma^2}+\frac{b}{\gamma}}\;.
\end{equation}
These are valid for all $(3,3)$ entries of the Yukawa couplings and are also independent of the renormalisation scheme.

Focusing on the bottom Yukawa coupling, and taking $n_g=\frac{5}{2}$ in the RGEs \eqref{app:gsmrge} and \eqref{app:ysmrge}, leads to
\begin{align}
\label{eq:resa} 16\pi^2a =&-8 \\[2pt]
\label{eq:resb} {\left(16\pi^2\right)}^2b=&-\left(\frac{292}{3}-\frac{16}{3}\cdot\frac{5}{2}\right)=-84 \\[2pt]
\label{eq:resgamma} 16\pi^2 \gamma=&-11+\frac{4}{3}\cdot\frac{5}{2}=-\frac{23}{3} \\[2pt] 
\label{eq:resdelta} {\left(16\pi^2\right)}^2\delta=&-102+\frac{76}{3}\cdot\frac{5}{2}=-\frac{116}{3}\;,
\end{align}
which, due to the choice of $n_g$, are correct at a scale between $m_t$ and $m_b$. Combining (\ref{eq:resp}) with (\ref{eq:resa}) and (\ref{eq:resgamma}) gives
\begin{equation} \label{eq:finp}
2P=8\cdot\frac{3}{23} \Rightarrow P=\frac{12}{23}\;.
\end{equation}
Similarly, (\ref{eq:resx}) combined with (\ref{eq:resa})\mbox{--}(\ref{eq:resdelta}) gives
\begin{equation} \label{eq:finx}
2X=\frac{1}{16\pi^2}\cdot8\cdot\frac{116}{3}\cdot\frac{9}{529}+\frac{1}{16\pi^2}\cdot84\cdot\frac{3}{23} \Rightarrow X=\frac{1}{4\pi^2}\frac{753}{1058}\;.
\end{equation}
The equation from Ref.~\cite{BFMTdr2ms} that is equivalent to (\ref{eq:javy}), and is valid in the region chosen above is
\begin{equation} \label{eq:javyb}
\lambda_a=\lambda^0_a\left[g^2_3(t)\right]^{12/23}\left[1+\frac{1}{4\pi^2}\frac{753}{1058}g^2_3\right]\;.
\end{equation}
Both $P$ and $X$ are seen to be the same as derived in (\ref{eq:finp}) and (\ref{eq:finx}).

It is possible to go back to the boxed equations, (\ref{eq:resp}) and (\ref{eq:resx}), and calculate these again in the \msb~scheme, the RGEs of which are well known \cite{ACMPRWrge}. The only difference to (\ref{eq:resa})\mbox{--}(\ref{eq:resdelta}) is that
\begin{equation}
{(16\pi^2)}^2b=-\frac{1012}{9}.
\end{equation}
This leads to the same $P$ and a new X which is
\begin{equation}
X=\frac{1}{4\pi^2}\frac{3731}{3174}.
\end{equation}
Correspondingly, the \msb~running of the bottom Yukawa coupling according to Ref.~\cite{BFMTdr2ms} is
\begin{equation}
\lambda_a=\lambda^0_a\left[g^2_3(t)\right]^{12/23}\left[1+\frac{1}{4\pi^2}\frac{3731}{3174}g^2_3\right],
\end{equation}
and this supports the method used to check the \drb~RGEs.

%% file: infile.tex
\chapter{A Sample Input File for \progrge} \label{app:inrgeeg}

The text that follows is a sample input file for \progrge. In this example, the code will be run with a non-zero phase for the KM matrix (Question 2.) so that most numerical inputs must be in complex notation, \textit{i.e.}, \verb+(a,b)+, where \verb+a+ and \verb+b+ are real numbers, as can be seen:\\

\begin{Verbatim}[numbers=left,fontsize=\footnotesize]
1. USE FULL TWO LOOP RUNNING? YES=1, NO=0
1

2. USE COMPLEX RUNNING WITH COMPLEX KM MATRIX? YES=1, NO=0
1

3. IF COMPLEX RUNNING, COMPLEX PHASE OF MU? YES=1, NO=0
0

 IF YES, THETA=: I.E. MU=|MU|*EXP(iTHETA)
3.142D0

4. USE ISAJET'S THREHSOLDS? YES=1, NO=0
0

5. USE ALL mSUGRA GUT CONDITIONS? YES=1, NO=0
1

 IF YES GO TO 17, IF NO:

6. UNIFIED GUT SCALE? YES=1, NO=0
1

 IF NO, VALUE OF HIGH SCALE FOR SUSY INPUT:
1.D19

7. USE mSUGRA GUT CONDITIONS FOR M1, M2, M3? YES=1, NO=0
1

 IF NO, M1, M2, M3 ARE: NB FOR COMPLEX RUNNING THESE ARE (M,M')
(0.D0,0.D0) (0.D0,0.D0) (0.D0,0.D0)

8. USE mSUGRA GUT CONDITIONS FOR M_{H_U}, M{H_D}? YES=1, NO=0
1

 IF NO, M_{H_U}, M_{H_D} ARE:
(0.D0,0.D0) (0.D0,0.D0)

9. USE mSUGRA GUT CONDITIONS FOR M_Q? YES=1, NO=0
1

 IF NO, USE M^2_Q = M^2_{Q0} \1 + T_Q.
 M_{Q0} IS REAL. IT IS:
0.D0

 T_Q MUST BE HERMITIAN. IT IS:
(0.D0,0.D0) (0.D0,0.D0) (0.D0,0.D0)
(0.D0,0.D0) (0.D0,0.D0) (0.D0,0.D0)
(0.D0,0.D0) (0.D0,0.D0) (0.D0,0.D0)

10. USE mSUGRA GUT CONDITIONS FOR M_U? YES=1, NO=0
0

 IF NO, USE 
  M^2_U = M^2_{U0} [ c_U \1 + R_U f^T_u f^*_u + S_U (f^T_u f^*_u)^2 ] + T_U.
 M_{U0} IS REAL. IT IS:
0.D0

 C_U IS EITHER 1 OR 0. IT IS:
0

 R_U IS REAL. IT IS:
0.D0

 S_U IS REAL. IT IS:
0.D0

 T_U MUST BE HERMITIAN. IT IS:
(0.D0,0.D0) (0.D0,0.D0) (0.D0,0.D0)
(0.D0,0.D0) (0.D0,0.D0) (0.D0,0.D0)
(0.D0,0.D0) (0.D0,0.D0) (0.D0,0.D0)

11. USE mSUGRA GUT CONDITIONS FOR M_D? YES=1, NO=0
0

 IF NO, USE 
  M^2_D = M^2_{D0} [ c_D \1 + R_D f^T_d f^*_d + S_D (f^T_d f^*_d)^2 ] + T_D.
 M_{D0} IS REAL. IT IS:
0.D0

 C_D IS EITHER 1 OR 0. IT IS:
0

 R_D IS REAL. IT IS:
0.D0

 S_D IS REAL. IT IS:
0.D0

 T_D MUST BE HERMITIAN. IT IS:
(0.D0,0.D0) (0.D0,0.D0) (0.D0,0.D0)
(0.D0,0.D0) (0.D0,0.D0) (0.D0,0.D0)
(0.D0,0.D0) (0.D0,0.D0) (0.D0,0.D0)

12. USE mSUGRA GUT CONDITIONS FOR M_L? YES=1, NO=0
1

 IF NO, USE M^2_L = M^2_{L0} \1 + T_L.
 M_{L0} IS REAL. IT IS:
0.D0

 T_L MUST BE HERMITIAN. IT IS:
(0.D0,0.D0) (0.D0,0.D0) (0.D0,0.D0)
(0.D0,0.D0) (0.D0,0.D0) (0.D0,0.D0)
(0.D0,0.D0) (0.D0,0.D0) (0.D0,0.D0)

13. USE mSUGRA GUT CONDITIONS FOR M_E? YES=1, NO=0
1

 IF NO, USE 
  M^2_E = M^2_{E0} [ c_E \1 + R_E f^T_e f^*_e + S_E (f^T_e f^*_e)^2 ] + T_E.
 M_{E0} IS REAL. IT IS:
0.D0

 C_E IS EITHER 1 OR 0. IT IS:
0

 R_E IS REAL. IT IS:
0.D0

 S_E IS REAL. IT IS:
0.D0

 T_E MUST BE HERMITIAN. IT IS:
(0.D0,0.D0) (0.D0,0.D0) (0.D0,0.D0)
(0.D0,0.D0) (0.D0,0.D0) (0.D0,0.D0)
(0.D0,0.D0) (0.D0,0.D0) (0.D0,0.D0)

14. USE mSUGRA GUT CONDITIONS FOR a_u? YES=1, NO=0
1

 IF NO, USE a_u = f_u [ A_{u0} \1 + W_u f^\dagger_u f_u
                         + X_u ( f^\dagger_u f_u )^2 ] + Z_u.
 A_{u0} IS:
(0.D0,0.D0)

 W_u IS:
(0.D0,0.D0)

 X_u IS:
(0.D0,0.D0)

 AND Z_u IS:
(0.D0,0.D0) (0.D0,0.D0) (0.D0,0.D0)
(0.D0,0.D0) (0.D0,0.D0) (0.D0,0.D0)
(0.D0,0.D0) (0.D0,0.D0) (0.D0,0.D0)

15. USE mSUGRA GUT CONDITIONS FOR a_d? YES=1, NO=0
1

 IF NO, USE a_d = f_d [ A_{D0} \1 + W_d f^\dagger_d f_d
                         + X_d ( f^\dagger_d f_d )^2 ] + Z_d.
 A_{D0} IS:
(0.D0,0.D0)

 W_d IS:
(0.D0,0.D0)

 X_d IS:
(0.D0,0.D0)

 IF NO, Z_d IS:
(0.D0,0.D0) (0.D0,0.D0) (0.D0,0.D0)
(0.D0,0.D0) (0.D0,0.D0) (0.D0,0.D0)
(0.D0,0.D0) (0.D0,0.D0) (0.D0,0.D0)

16. USE mSUGRA GUT CONDITIONS FOR a_e? YES=1, NO=0
1

 IF NO, USE a_e = f_e [ A_{e0} \1 + W_e f^\dagger_e f_e
                         + X_e ( f^\dagger_e f_e )^2 ] + Z_e.
 A_{e0} IS:
(0.D0,0.D0)

 W_e IS:
(0.D0,0.D0)

 X_e IS:
(0.D0,0.D0)

 Z_e IS:
(0.D0,0.D0) (0.D0,0.D0) (0.D0,0.D0)
(0.D0,0.D0) (0.D0,0.D0) (0.D0,0.D0)
(0.D0,0.D0) (0.D0,0.D0) (0.D0,0.D0)

17. OUTPUT IN BASIS WHERE UP QUARKS ARE DIAGONAL?
1

18. USER INPUTS ROTATION MATRICES ? YES=1, NO=0
1

19. IF YES, GO TO 20, IF NO, V^U_L = KM ? YES=1, NO=0
1

20. ARE V^U_R AND V^D_R IDENTITY? YES=1, NO=0
1

21. IF USER INPUTS V'S, THEN THE ANGLES AND PHASE IN THE UNITARY
    MATRICES ARE: (NB IF KM IS NOT COMPLEX THE PHASE IS IGNORED)

 V^U_L (ALPHA, BETA, GAMMA, DELTA)
2.053D0 0.254D0 2.03D0 0.4829D0

 V^U_R (ALPHA, BETA, GAMMA, DELTA)
1.188D0 2.218D0 .763D0 0.87D0

 V^D_R (ALPHA, BETA, GAMMA, DELTA)
1.904D0 2.947D0 1.847D0 1.14D0

\end{Verbatim}

%% file: orthfix.tex
\chapter{Numerical Instabilities Associated with Matrix Diagonalisation}\label{sec:orthfix}

\subsubsection{The problem}
We have emphasized that in order to properly implement particle decoupling into the RGEs, we have to be in the mass basis of the
particles being decoupled.  Our procedure for decoupling squarks, therefore, requires us to evaluate the unitary transformation from the given current basis to a new current basis that coincides with the squark mass basis (approximated, as discussed in the main text, to be the basis in which the SSB squark mass squared matrices are diagonal). Below the scale $Q$ where at least one squark has decoupled, we not only have the rotations $\mathbf{V}_{L,R}(u,d)$ (unitary matrices by construction) which connect the current basis with the basis in which the Yukawas are diagonal at $m_{t}$, but also the squark rotations $\mathbf{R}_{\bullet}$ which connect the current basis to the ``squark mass basis'' that are obtained by numerically diagonalizing the SSB matrices ${\bf m}_\bullet^2$. $\mathbf{R}_{\bullet}$ is of course the matrix of the orthogonal eigenvectors of ${\bf m}^2_\bullet$.  If there is a degeneracy of eigenvalues, the orthogonal eigenvectors are not uniquely defined. This leads to a practical problem when we {\it numerically} solve for the eigenvectors in the case that two eigenvalues of any SSB squark mass matrix with large off-diagonal components are degenerate to within $\sim 1$\%. In this case, the corresponding eigenvectors, because of (system-dependent) numerical errors are not exactly orthogonal, and the corresponding matrix $\mathbf{R}_{\bullet}$ is not precisely
unitary.\footnote{Using the \texttt{g77 FORTRAN} compiler with \textsf{Macintosh Intel Macbook}, together with the subroutine \texttt{CG} in the \texttt{EISPACK} collection of subroutines, we found that $\mathbf{R}_{\bullet}^\dagger \mathbf{R}_{\bullet}$ deviated from $\dblone$ to about one part in $10^{10}$ compared to a part in $10^{18}$ for $\mathbf{V}_{L,R}(u,d)^\dagger\mathbf{V}_{L,R}(u,d)$ for $\mathbf{V}_{L,R}(u,d)$ of the form (\ref{eq:rotin}). We obtain a similar size deviation from identity using the subroutine \texttt{ZGEEV} in the \texttt{LAPACK} collection of subroutines}

The deviation from unitarity is very small, a part in $10^{10}$ in our case, but is nonetheless orders of magnitude larger than what we can tolerate when calculating the smallest off-diagonal elements of ${\bf m}^2_\bullet$. To understand why our calculation is sensitive to this seemingly tiny level of noise, let us imagine what would happen if we attempted to evolve the off-diagonal elements of ${\bf m}_U^2$ from $Q=M_{\rm GUT}$ in a basis where the Yukawa coupling matrices all have large off-diagonal elements at the GUT scale. (This is not what we actually do, but we could imagine doing so since we know that we are well above all SUSY and Higgs field thresholds where the choice of basis should be irrelevant.) In the mSUGRA framework, the squark mass matrices are all given by ${\bf m}_\bullet^2=m_0^2\dblone$ at $Q=M_{GUT}$ in any basis. Then, from (\ref{app:muamh}) we see that these would develop off-diagonal components $\sim \textit{few}\times f^2 \times m_0^2  \simeq f^2 \times 4\times 10^4$ for the case shown in Fig.~\ref{fig:SUGRAmup}, where $f^2$ denotes the size of the off-diagonal element of ${\bf f}_u^T{\bf f}_u^*$. In this rough estimate we have assumed that the loop factor $1/(16\pi^2)$ is compensated for by the large logarithm.  In the general current basis where ${\bf f}_u$ has comparable off-diagonal and diagonal elements, $f^2\sim 1$, and the magnitude of the off-diagonal elements of ${\bf m}^2_U$ are ${\mathcal O}(10^4)$~GeV$^2$. Rotating to our standard current basis should yield the result in Fig.~\ref{fig:SUGRAmup}. In particular, we should obtain $\left|{\bf m}^2_U\right|_{12} \sim 10^{-9}$~GeV$^2$ because there would be large cancellations arising from the unitarity of $\mathbf{R}_U$ that would suppress this matrix element. If instead the unitarity of $\mathbf{R}$ holds only to a part in $10^{10}$ because of numerical errors in obtaining the eigenvectors, we will find that because the cancellations are not perfect all off-diagonal elements of ${\bf m}_U^2$ will have a magnitude that is at least $\textit{few} \times 100^{2}\times 10^{-10} \sim \textit{few} \times 10^{-6}$~GeV$^2$, much larger than the magnitude of the $(1,2)$ element in Fig.~\ref{fig:SUGRAmup}. We note here that the noise that leads to the non-unitarity of ${\bf V}_{L,R}(u,d)$ matrices at the $10^{-18}$ level is completely irrelevant. 

\subsubsection{The solution}
The non-unitarity of $\mathbf{R}_\bullet$ is only an issue when the off-diagonal entries of the squark mass matrix, in the basis where the Yukawas are diagonal at $m_{t}$, are small compared to the diagonal entries. Since we, therefore, only need to consider matrices that are already approximately diagonal, we can associate the eigenvectors, $(\mathbf{e}_{1},\mathbf{e}_{2},\mathbf{e}_{3})$, with the approximate eigenvalues $((\bdm^{2}_{\bullet})_{11},(\bdm^{2}_{\bullet})_{22}, (\bdm^{2}_{\bullet})_{33})$, respectively. As an illustration, let us take a case where the $(\bdm^{2}_{\bullet})_{23}$ entry is the off-diagonal entry with the largest magnitude, and $(\bdm^{2}_{\bullet})_{12}$ the one with the smallest. We know the ordering quite unambiguously because above all squark thresholds we do not need to rotate by the matrices $\mathbf{R}_\bullet$ that potentially are the origin of the noise. We need to ensure that the $(\bdm^{2}_{\bullet})_{12}$ entry does not suffer from any numerical noise due to the diagonalisation. This leads us to fix $\mathbf{e}_{1}\cdot \mathbf{e}_{2}=0$ and move any non-orthogonality of the eigenvectors into $\mathbf{e}_{2}\cdot \mathbf{e}_{3}$ so that the noise moves to $(\bdm^{2}_{\bullet})_{23}$, the off-diagonal element with the largest magnitude. To accomplish this, we slightly modify (by parts in $10^{10}$, the limit of accuracy of the diagonalization routines) {\it only} the eigenvector $\mathbf{e}_{2}$ from its value as given by the diagonalization routines, thereby leaving $\mathbf{e}_{1}\cdot\mathbf{e}_{3}$ unaffected. 

\begin{figure}[t]\begin{centering}
\includegraphics[viewport=5 50 710 580, clip, scale=0.41]{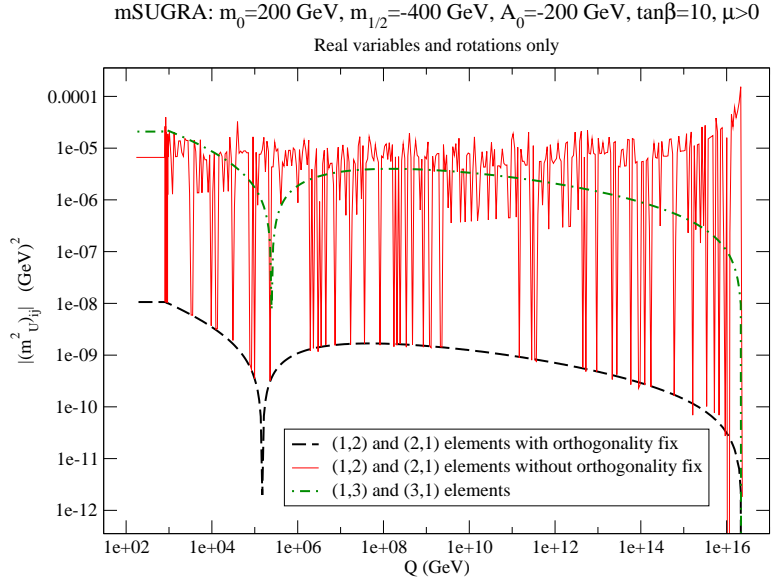} 
\caption[The scale-dependence of the $(1,2)$ and $(2,1)$ elements of the Hermitian $\bdm^{2}_{U}$ matrix for our sample mSUGRA point.]{\footnotesize The scale-dependence of the $(1,2)$ and $(2,1)$ elements of the Hermitian $\bdm^{2}_{U}$ matrix for our sample mSUGRA point (in the basis specified by (\ref{eq:genrot})) calculated using two different procedures discussed in the text. The dashed (black) line shows the magnitude of the smallest element, \textit{i.e.}, $|\left({\bf m}^2_U\right)_{12}| =  |\left({\bf m}^2_U\right)_{21}|$, when we have used our procedure to ensure that the corresponding eigenvectors are orthogonal. The solid (red) 
line shows the same element when we do not pay attention to the orthogonality of the eigenvectors of ${\bf m}_U^2$. The lighter dot-dashed (green) line shows the magnitude of the $(1,3)$ element, and provides a scale for the size of the numerical noise discussed in the text. The noise in this curve is too small to be visible. All elements are zero at the GUT scale.}
\label{fig:mu12bad}
\end{centering}
\end{figure}

To completely clarify what we have just described, although (as we have already stated) we do not need to rotate to the squark mass basis until we reach the highest squark threshold, we plot, in Fig.~\ref{fig:mu12bad}, the result for $|\bdm^{2}_{U}|_{12}$ obtained by the two different methods mentioned above, in the basis where the up-type Yukawas are diagonal at $m_{t}$, over the whole range $M_{Z}<Q<M_{\mathrm{GUT}}$.
The dashed (black) line shows the result where we have no rotation by $\mathbf{R}_U$ from $Q=M_{\rm GUT}$ until the highest squark threshold, beyond which we implement our method for ensuring that the error from the non-orthogonality of the eigenvectors of ${\bf m}^2_U$ only shows up in the (2,3) element. Indeed we see that this curve is smooth over its entire range of $Q$. The solid (red) line shows the result of carrying out the squark rotation without our fix of the eigenvectors over the entire range of $Q$. Note that there is significant noise all the way down to the low $Q$ region where only some of the squarks have decoupled. This noise is largest at $Q=M_{\rm GUT}$ where the eigenvalues of ${\bf m}_U^2$ are degenerate, and settles down to $10^{-5}$~GeV$^2$, not far from our estimate above. The important thing is that the frozen value of this element is significantly different in the two cases, as a result of this noise, just before squark decoupling. It is for this range of $Q$ (where the mass matrices that enter flavour-changing processes involving squarks will be evaluated) that we must reduce the numerical error as far as possible. The magnitude of the $(1,3)$ element of ${\bf m}_U^2$ is shown for comparison by the dot-dashed (green) curve. It has no visible noise because the corresponding eigenvalues are sufficiently split, and the corresponding eigenvectors are orthogonal to a very high accuracy. 

The reader will be struck by the fact that the random downward fluctuations in the solid curve are roughly bounded by the dashed (black) curve which shows the correct magnitude of the matrix element. The reason for this is that the {\it fluctuations} whose typical magnitude is $\sim 10^{-5}$~GeV$^2$ need to {\it randomly} fluctuate down by four orders of magnitude to even reach the dashed (black) line, and even more to go below, the chance for which is very small. Indeed it is because we have shown results for the case where the SSB squark mass matrices are real that we see these fluctuations go down to even the level of the dashed (black) line. For the more general case the chance for both the real and the imaginary part of any matrix element to {\it simultaneously} fluctuate downward by this large magnitude is very small, so that the calculated magnitude (not shown here) is always larger than $10^{-6}$~GeV$^2$.

\begin{figure}[t]\begin{centering}
\includegraphics[viewport=5 50 710 580, clip, scale=0.45]{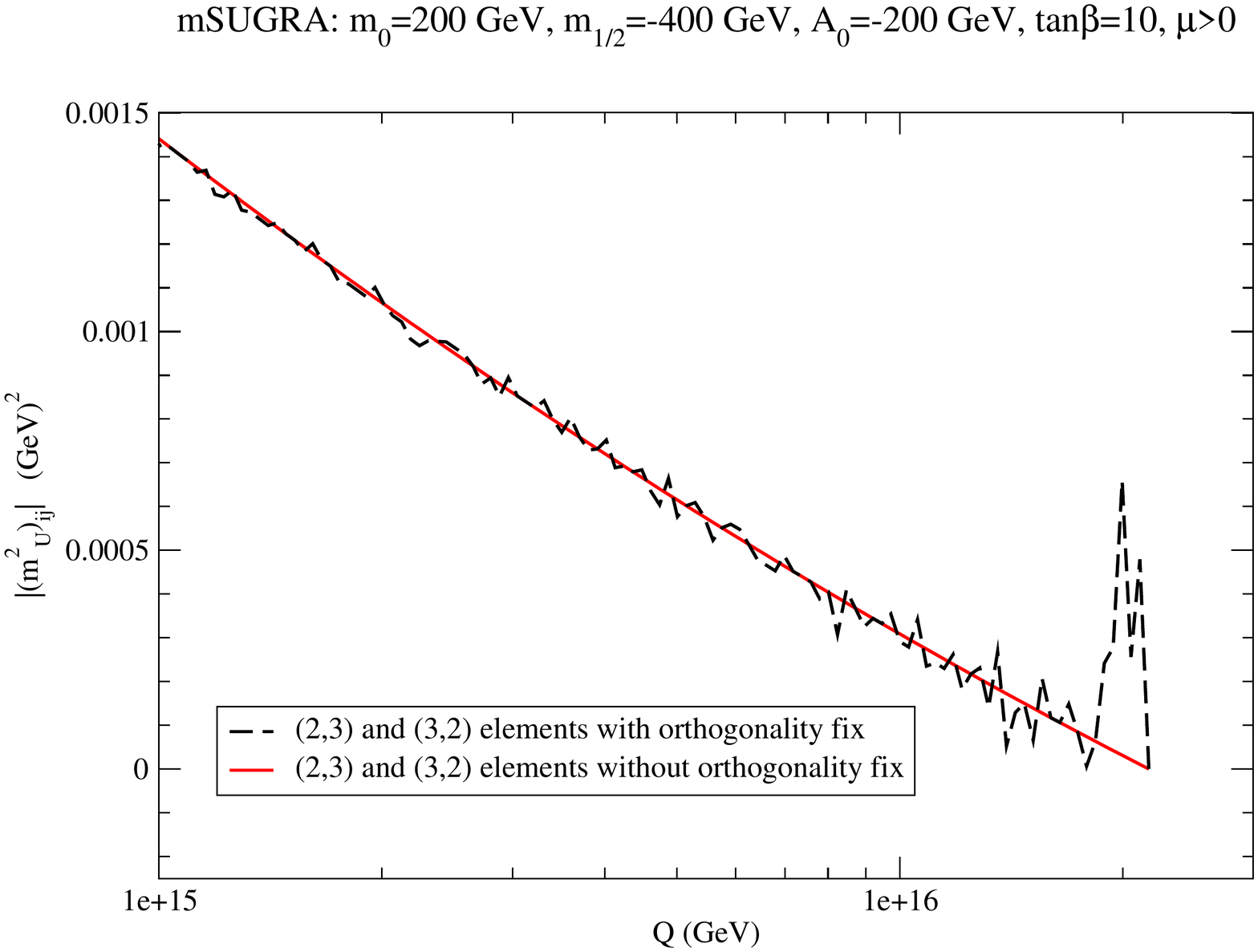} 
\caption[Scale dependence of the magnitude of the $(2,3)$ and $(3,2)$ entries of $\bdm^{2}_{U}$ for the same mSUGRA point as Fig.~\ref{fig:mu12bad}]{\footnotesize Scale dependence of the magnitude of the $(2,3)$ and $(3,2)$ entries of $\bdm^{2}_{U}$ for the same mSUGRA point as Fig.~\ref{fig:mu12bad}. We focus on the running at the extreme high scale, and compare the noise in the magnitude of just this element, both before and after fixing the orthogonality of the eigenvectors as described in the text. As in Fig.~\ref{fig:mu12bad}, the solid (red) line shows the result before the orthogonality fix while the dashed (black) line shows the result after this fix when this error has been moved to the $(2,3)$ element which now randomly fluctuates close to $M_{\rm GUT}$ where the eigenvalues of ${\bf m}_U^2$ are roughly degenerate.}
\label{fig:mu23bad}
\end{centering}
\end{figure}

After our fix of the eigenvectors, any error from the non-unitarity of ${\bf R}_U$ is shifted to the largest off-diagonal element, and the only residue of the resulting noise that remains is in the magnitude of this element for scales close to $M_{\rm GUT}$ \mbox{---} where the eigenvalues are closest \mbox{---} as seen in Fig.~\ref{fig:mu23bad}.
At lower scales, the eigenvalues split, and the noise level (whose magnitude remains the same as in Fig.~\ref{fig:mu12bad}) becomes insignificant.

Again, the solid (red) line shows the evolution of the magnitude of the $(2,3)$ element before fixing the eigenvectors, while the dashed black  line shows the same thing after the orthogonality fix. We see that the numerical noise has indeed moved to the $(2,3)$ element which now shows fluctuations, but only close to $M_{\rm GUT}$, where the eigenvalues of ${\bf m}_U^2$ are roughly degenerate.